\documentclass[reprint,superscriptaddress,amsmath,amssymb,aps,prx,
a4paper,twocolumn, floatfix
]{revtex4-2}
\usepackage{graphicx}
\usepackage{dcolumn}
\usepackage{bm}
\usepackage{braket}
\usepackage{booktabs, multirow}
\usepackage[colorlinks=true,urlcolor=blue,citecolor=blue,linkcolor=blue]{hyperref}
\usepackage{amsthm}
\usepackage{xcolor}
\usepackage{subfigure}
\usepackage{afterpage}
\usepackage{quantikz}
\usepackage{mdframed}
\usepackage{verbatim}
\usepackage{faktor}
\usepackage{mathrsfs}
\usepackage{cancel}
\usepackage[normalem]{ulem}

\newtheorem{theorem}{Theorem}
\newtheorem*{theorem*}{Theorem}
\newtheorem{proposition}{Proposition}
\newtheorem*{proposition*}{Proposition}
\newtheorem{corollary}{Corollary}
\newtheorem*{corollary*}{Corollary}
\newtheorem{lemma}{Lemma}
\newtheorem*{lemma*}{Lemma}

\newtheorem*{assumption*}{Assumption}
\newtheorem{definition}{Definition}
\newtheorem*{definition*}{Definition}

\newcommand{\nocontentsline}[3]{}
\let\origcontentsline\addcontentsline
\newcommand\stoptoc{\let\addcontentsline\nocontentsline} 
\newcommand\resumetoc{\let\addcontentsline\origcontentsline} 

\begin{document}
\renewcommand{\tocname}{Appendix} 
\stoptoc

\title{{Gradient Scalability and Taylor Surrogation of Quantum Cost Landscapes}}

\author{Sabri Meyer}
\email{sabri.meyer@unibas.ch}
\affiliation{Department of Mathematics and Computer Science, University of Basel, 4051 Basel, Switzerland}

\author{Francesco Scala}
\affiliation{Department of Mathematics and Computer Science, University of Basel, 4051 Basel, Switzerland}

\author{Francesco Tacchino}
\affiliation{IBM Quantum, IBM Research Europe -- Zurich, 8803 Rüschlikon, Switzerland}

\author{Aurelien Lucchi}
\affiliation{Department of Mathematics and Computer Science, University of Basel, 4051 Basel, Switzerland}

\date{\today}

\begin{abstract}
Variational Quantum Algorithms (VQAs) are promising candiadates for near-term quantum computing, yet they face scalability challenges due to barren plateaus, where gradients vanish exponentially in the system size. Recent conjectures suggest that avoiding barren plateaus might inherently lead to classical simulability, thus limiting the opportunities for quantum advantage. In this work, we advance the theoretical understanding of the relationship between {gradient scalability at initialization} and computational complexity of VQAs. {We first present the \emph{Taylor surrogate}, a classical simulation technique matching Pauli path runtime guarantees on near-Clifford regions, with runtime advantages in specific regimes.} {Leveraging the Taylor surrogate, we prove that beyond the previously established classically simulable regions the computational complexity is at least super-polynomial.} {Next, we introduce the \emph{Linear Clifford Encoder (LCE)}, a classically efficient ansatz modifier that ensures constant-scaling gradients within landscape regions close to Clifford circuits. Finally, numerical experiments on LCE-modified landscapes provide {preliminary empirical} evidence of a transition zone where the constant-scaling gradients may start to decay polynomially in super-polynomially complex regions rather than exponentially. These findings {suggest speculative} instances where non-vanishing gradients and super-polynomial complexity could potentially coexist, {vindicating the need for future formal proofs.}}
\end{abstract}

\maketitle

\section{Introduction}

{Variational Quantum Algorithms (VQAs)~\cite{Cerezo2021VQA} are promising tools for near-term quantum computers with a wide variety of possible applications in quantum chemistry, condensed matter physics and Quantum Machine Learning (QML)}~\cite{Peruzzo2014VQE, kuzmin2020var_state_prep, matos2021vqe,Gyawali2021Adaptive, Farrell2023Scalable,kottmann2024chemistry,2014QMLWittek,Biamonte2017quantum,cerezo2022challenges}. {However, the scalability of these algorithms is severely hindered by the phenomenon of barren plateaus~\cite{mcclean2018barren, Larocca2025review}, where gradients vanish exponentially with the number of qubits, thus requiring an exponential number of shots for gradient estimation during optimization. As a result, demonstrating indisputable advantages of VQAs over their classical counterparts remains an open question. A critical challenge arises when attempting to mitigate this issue: it has been conjectured that avoiding barren plateaus may inherently lead to classical simulability and thus limits quantum advantage~\cite{cerezo2023does}. Despite considerable efforts in this direction~\cite{zhang2024absence, deshpande2024dynamic, park2024hardware} it remains unknown whether barren plateaus can be avoided for arbitrary ansätze. Specifically, it is unclear if gradient scalability can be maintained without restricting the PQC architecture or limiting observables to local operators, both of which are often necessary to ensure the system remains classically hard to simulate~\cite{cerezo2023does}.}

{This work focuses on studying the interplay between gradient scalability at initialization and the computational complexity of quantum cost functions for arbitrary PQC ansätze. By Taylor-expanding the optimization landscape around Clifford circuits, we establish a rigorous mathematical framework to analyze both the cost of classical simulation and the behavior of gradients. Specifically, we propose a generalization of the quadratic Clifford expansion~\cite{mitarai2022quadratic} and leverage the Gottesman-Knill theorem~\cite{gottesman1998heisenberg, aaronson2004improved} to efficiently compute higher-order Taylor coefficients. This approach yields what we call the \emph{Taylor surrogate}~\cite{schreiber2023classical} for classically simulating expectations with near-Clifford PQCs. Our method is capable of surrogating the same regimes as the Pauli path technique~\cite{lerch2024efficient} while offering provable computational speedups in special cases. Furthermore, we analyze the cost of simulating circuits beyond known classically tractable limits~\cite{bravyi2016improved, bravyi2019simulation, beguvsic2023simulating}.} {By deriving super-polynomial lower bounds as a function of the model parameters, we offer new theoretical insights into the boundaries of classical hardness.}

{The efficiency of computing derivatives near Clifford points provides a direct path to managing gradient scalability. We introduce the \emph{Linear Clifford Encoder (LCE)}, a classically efficient ansatz modifier that systematically manages linear Taylor coefficients to ensure constant-scaling gradient norms at initialization. This result holds within local ``patches'' of the landscape near Clifford circuits, even for large-scale models, at the cost of incorporating a small number of Clifford gates into the ansatz.}

{Our Taylor framework further reveals a phase transition in computational complexity as the size of these surrogate patches increases. This threshold separates regimes of polynomial complexity, where LCE ensures constant-scaling gradients at initialization, from those exhibiting super-polynomial hardness. While we do not prove theoretical guarantees for the absence of barren plateaus beyond this threshold, numerical experiments indicate the existence of a ``transition patch'' where gradients may begin to decay polynomially rather than exponentially. We present this regime strictly as a numerical indication and an open hypothesis for future rigorous proof. Furthermore, for certain global observables known to suffer from barren plateaus~\cite{cerezo2021cost}, LCE-assisted VQA experiments provide encouraging evidence of significantly enhanced training dynamics with faster convergence. Although the value for practical problems remains a subject for future study, these experiments hint at the possibility of barren-plateau-free landscape regimes that nonetheless possess high computational complexity.}

The rest of this document is structured as follows. In Sec.~\ref{sec:preliminaries} we introduce the necessary concepts to lay the foundation for our main theoretical results, presented in Sec.~\ref{sec:theory}. There, we mathematically address the challenge of balancing {gradient scalability} and computational complexity. Sec.~\ref{sec:experiments} discusses numerical experiments, providing evidence of large landscape patches without barren plateaus. Finally, we summarize our findings in Sec.~\ref{sec:conclusion}. A compact overview comparing our theoretical results with related works is found in Appendix~\ref{sec:comparison}.

\section{Preliminaries \label{sec:preliminaries}}

In this section, we establish the four key ideas that we will leverage to yield our main results: the stabilizer formalism, the expressive power of PQCs, barren plateau mitigation, and a novel classical Taylor surrogate.

\subsection{Stabilizer Formalism}

The stabilizer formalism efficiently describes a large class of quantum states, named \emph{stabilizer states}, by tracking the Pauli operators that leave them invariant~\cite{nielsen2010quantum}. This underlies the Gottesman-Knill theorem~\cite{gottesman1998heisenberg, aaronson2004improved} which guarantees efficient classical simulation for circuits made solely of Clifford operators. Here, we briefly recall the definitions of the Pauli and Clifford groups; see also Refs.~\cite{mastel2023clifford, nielsen2010quantum}.

Let $N\geq1$ denote the number of qubits. Here, the $N$-qubit Pauli group is defined as the unitary subgroup
\begin{equation}
    \mathcal{P}_N:=\left\{\pm P, \pm i P: P\in\{I,X,Y,Z\}^{\otimes N}\right\},
\end{equation}
and its elements (modulo global phase) are referred to as (unsigned) Pauli strings. The $N$-qubit Clifford group is the subgroup of the unitary group $\operatorname{U}(2^N)$ that normalizes $\mathcal{P}_N$ (modulo global phases):
\begin{equation}
    \operatorname{Clifford}(N):=\{Q\in\operatorname{U}(2^N):Q\,\mathcal{P}_N\,Q^\dagger = \mathcal{P}_N\}.
\end{equation}
It is generated by the ensemble $\{H,S,\operatorname{CX}\}$. The space of stabilizer states is then
\begin{equation}
    \operatorname{Stab}(N):=\{Q|0\rangle:Q\in\operatorname{Clifford}(N)\},
\end{equation}
where $|0\rangle\equiv |0\rangle^{\otimes N}$. For further details, including a review of related simulation techniques, see Appendices~\ref{sec:gottesman-knill} and~\ref{sec:simulation}.

\subsection{Variational Quantum Algorithms}

An ideal $N$-qubit quantum system resides in a Hilbert space $\mathcal{H}_N$ spanned by $\{|b\rangle: b\in\{0,1\}^N\}$. We focus on a Parameterized Quantum Circuit (PQC)~\cite{Benedetti2019pqc} defined by {the reversed product of unitaries,}
\begin{equation}
\label{eq:PQC}
    U(\theta) = \prod_{k=D}^1 R_{V_k}(\theta_k)\, W_k {\,:= R_{V_D}(\theta_D)W_D\cdots R_{V_1}(\theta_1)W_1},
\end{equation}
with $D\geq1$ parameterized single-qubit rotation gates and Clifford operators $W_k$. In particular, $D$ corresponds to the circuit depth. Here, $R_{V_k}(\theta_k):=\exp\left(-i\frac{\theta_k}{2}V_k\right)$ with $V_k\in\{X,Y,Z\}$ acting on single qubits, using the notation $V_j:=I_{2^{j-1}}\otimes V\otimes I_{2^{N-j}}$. We then define $|\psi(\theta)\rangle:=U(\theta)|0\rangle$. Note that for $\theta=0$, $U(0)$ is Clifford and $|\psi(0)\rangle\in\operatorname{Stab}(N)$. Without loss of generality, we may only consider expanding around $\theta=0$.
 
We consider a non-identity observable
\begin{equation}
    H=\sum_{i\in\mathcal{I}}c_iP_i, \label{eq:observable}
\end{equation}
with Pauli strings $P_i\in\mathcal{P}_N\setminus\{I\}$ and coefficients $c_i\in\mathbb{R}$ satisfying $c_i\in\mathcal{O}(1)$ as $N\to\infty$ with indexing set $\mathcal{I}\subseteq\{I,X,Y,Z\}^{\otimes N}\setminus\{I_{2^N}\}$ uniquely identifying $P_i$. The expectation value (cost function) is
\begin{equation}
    C(\theta):=\langle\psi(\theta)|H|\psi(\theta)\rangle=\sum_{i\in\mathcal{I}}c_i\langle\psi(\theta)|P_i|\psi(\theta)\rangle, \label{eq:expectation}
\end{equation}
which is smooth, $2\pi$-periodic in each parameter, and bounded by $\|H\|$. A typical VQA~\cite{Cerezo2021VQA} utilizes a quantum processor to estimate the cost $C(\theta)$, which a classical optimizer {(such as gradient descent)} uses to update the parameters $\theta$, e.g., in Variational Quantum Eigensolver (VQE)~\cite{Peruzzo2014VQE} tasks that seek the minimizer,
\begin{equation}
    \theta_*=\arg\min_{\theta\in[-\pi,\pi]^D}C(\theta). \label{eq:minimizer}
\end{equation}

If the ansatz $U(\cdot)$ is sufficiently expressive, then $C(\theta_*)$ provides a good estimate of the lowest eigenvalue of $H$. The expressivity of a PQC refers to its ability to explore $\mathcal{H}_N$ in an unbiased manner ~\cite{sim2019expressibility}. Notably, expressivity is also closely linked to the so-called barren plateau phenomenon~\cite{mcclean2018barren, holmes2022connecting}.

\begin{figure*}[t]
    \centering
    \includegraphics[trim={0cm 5cm 0cm 4cm},clip,width=.99\textwidth]{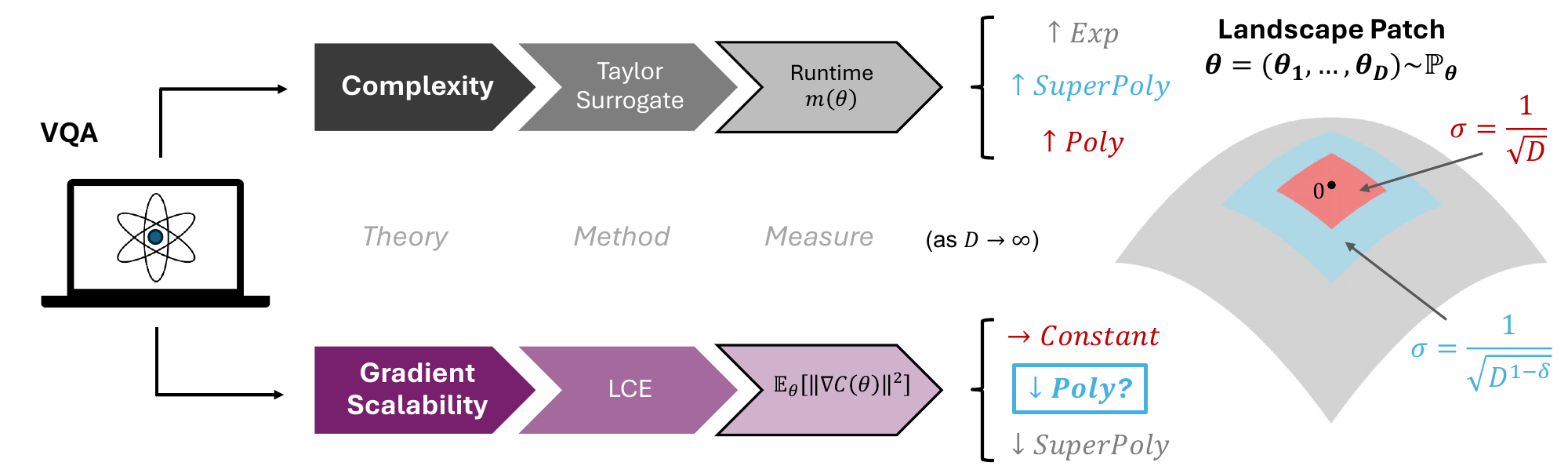}
    \caption{Illustration of our decomposition of the theoretical link between complexity and {gradient scalability} in VQAs. These two theories relate to each other with the landscape patch size $\sigma>0$ around $\theta=0$. The Linear Clifford Encoder (LCE) and the classical Taylor surrogate reveal how gradient scalability and computational complexity interact, exposing a super-polynomially complex transition patch (blue), where for sufficiently small $\delta>0$, gradients {are conjectured to} decay polynomially with the model size, {a hypothesis currently supported solely} by our numerical experiments. The scaling results hold true for observables satisfying $\|H\|=\mathcal{O}(1)$ and $c_{i_0}=\Omega(1)$ for some $i_0\in\mathcal{I}$.
    }
    \label{fig:intro}
\end{figure*}

\subsection{Barren Plateaus}
{Here and in the following, we make use of the standard Landau notations to describe asymptotic behavior of functions. Specifically, asymptotically as $D\to\infty$, $\mathcal{O}$ denotes an upper bound, $\Omega$ a lower bound, and $\Theta$ a tight bound. These notations are formally introduced at the beginning of the Appendix.}

We typically initialize $\theta\sim\mathbb{P}_\theta$, and the choice of the probability distribution $\mathbb{P}_\theta$ is crucial to avoid the barren plateau problem~\cite{mcclean2018barren, Larocca2025review, grant2019initialization, zhang2022escaping, wang2024trainability}. Barren plateaus occur whenever the expected squared gradient norm vanishes as
\begin{equation}
    \mathbb{E}_\theta[\|\nabla_\theta\,C(\theta)\|^2] = \mathcal{O}(2^{-\eta N}) \label{eq:barren-plateau}
\end{equation}
for some $\eta>0$, while the expected gradient satisfies $\mathbb{E}_\theta[\nabla_\theta\,C(\theta)]=0$, making optimization intractable. {Intuitively, high expressivity of PQCs is known to be one of the major causes of barren plateaus}~\cite{sim2019expressibility, holmes2022connecting}{, while it is also known to potentially limit classical simulability}~\cite{bravyi2019simulation, lerch2024efficient, goh2023lie}. The gradient is sufficient to characterize {an essentially  flat loss landscape (i.e. no curvature on average):} {indeed, by combining Chebychev's inequality with the higher-order parameter-shift rule, the probability of any $\mathcal{O}(1)$-order derivative being at least a certain constant is bounded by the variance of the gradient and thus exponentially suppressed for all constants if Eq.~\eqref{eq:barren-plateau} holds true~\cite{cerezo2021higher}}.

Various mitigation strategies~\cite{zhang2021toward, letcher2024tight, park2024hamiltonian, park2024hardware, shi2024avoiding, skolik2021layerwise, patti2021entanglement, volkoff2021large, pesah2021absence, mhiri2025unifying} exist that establish polynomial lower bounds of the form
\begin{equation}
    \mathbb{E}_\theta[\|\nabla_\theta\,C(\theta)\|^2]= \Omega(\operatorname{Poly}(N)^{-1}), \label{eq:no-barren-plateau}
\end{equation}
{but consistently appear to result into classical simulability~\cite{cerezo2023does}}. Our contribution introduces the \emph{Linear Clifford Encoder (LCE)} technique, which ensures constant scaling gradients on classically simulable landscape patches. In this work, we assume $D=\Theta(\operatorname{Poly}(N))$, meaning that the number of parameters $D$ scales polynomially with the number of qubits $N$. The polynomial is determined by the specifics of the circuit architecture, such as its layer depth.

\subsection{Classical Taylor Surrogate Patches}

For complexity benchmarking, we consider a \emph{classical surrogate} of a quantum learning model~\cite{schreiber2023classical} that classically reproduces its input-output relations, with high accuracy, efficiency and low probability of failure. Following ideas from Pauli path methods~\cite{ beguvsic2023simulating, lerch2024efficient, angrisani2024classically, rudolph2025pauli}, small-angle techniques~\cite{zhang2022escaping, wang2024trainability}, and quadratic Clifford expansion~\cite{mitarai2022quadratic}, we approximate $C(\theta)$ using a truncated Taylor expansion of arbitrary order $m-1$ around $\theta=0$:
\begin{equation}
    C(\theta) \approx \sum_{\|\alpha\|_1 < m}(D_\theta^\alpha C)(0)\,\frac{\theta^\alpha}{\alpha!} =: C_m(\theta), \label{eq:taylor}
\end{equation}
where for each multi-index $\alpha=(\alpha_1,\dots,\alpha_D){\,\in\mathbb{N}^D_0}$ we use the notations $D_\theta^\alpha:=\partial_{\theta_1}^{\alpha_1}\cdots\partial_{\theta_D}^{\alpha_D}$, $\theta^\alpha:=\theta_1^{\alpha_1}\cdots\theta_D^{\alpha_D}$, $\alpha!:=\alpha_1!\cdots\alpha_D!$, and $\|\alpha\|_1:=|\alpha_1|+\cdots+|\alpha_D|$ denoting the $\ell_1$-norm of $\alpha$.

As we will show below, the Taylor coefficients $(D_\theta^\alpha C)(0)$ can be classically simulated by combining higher-order parameter-shift rules~\cite{mitarai2018quantum, schuld2019evaluating, cerezo2021higher} and the Gottesman-Knill theorem~\cite{gottesman1998heisenberg, aaronson2004improved}, with a complexity of at most $\mathcal{O}(|\mathcal{I}|\,D^{\|\alpha\|_1})$. Thus, $C_m(\theta)$ serves as an efficient polynomial surrogate of $C(\theta)$ for near-Clifford patches, where parameters are initialized as
\begin{equation}
    \mathbb{P}_\theta\in\{\mathcal{N}(0,\sigma^2 I_D), \operatorname{Unif}([-\sigma,\sigma]^D)\},
\end{equation}
with $\sigma>0$ determining the patch size.

While small patches limit PQC expressivity (since a Clifford ensemble covers only a discrete subset of $\mathcal{H}_N$), increasing non-Clifford resources eventually render classical simulation intractable {for stabilizer-based techniques~\cite{lerch2024efficient, bravyi2016improved, bravyi2019simulation}}.

{We argue that other popular classical simulation methods are in general unsuitable for near-Clifford simulation, especially since we allow arbitrarily deep and complex PQCs, which is the reason why we focus on applying Gottesman-Knill in this work: since Clifford circuits allow an arbitrary amount of entanglement, this generally causes tensor network simulations to be hard due to an exponential growth of the bond dimension in the entanglement entropy~\cite{vidal2003efficient, eisert2010colloquium, eisert2013entanglement}. Likewise, an arbitrarily deep PQC ansatz generally leads to an exponentially large dynamical Lie algebra, thus causing Lie algebraic simulation techniques to be infeasible~\cite{goh2023lie}. 
To the best of our knowledge, it is unknown whether there exist more efficient simulation techniques to classically surrogate near-Clifford landscape patches than the Pauli path surrogate~\cite{lerch2024efficient} or our proposed Taylor surrogate. A brief comparison with the Pauli path method is provided in Appendix~\ref{sec:comparison}.}

Our framework aims to maximize the landscape patch size while maintaining both {gradient scalability} and classical intractability. An illustration of our approach is summarized in Fig.~\ref{fig:intro}.

\section{Theoretical Results \label{sec:theory}}

This section presents the main theoretical contributions of our work. The central idea is to employ the Gottesman-Knill theorem~\cite{gottesman1998heisenberg, aaronson2004improved} to classically simulate the Taylor coefficients of our Taylor surrogate. These can be used to systematically adapt the Clifford structure of the PQC, ensuring constant gradient norms within the surrogate patch. This technique, termed \emph{Linear Clifford Encoder (LCE)}, can be implemented efficiently using classical resources for large-scale PQCs.

After establishing the LCE transformation, we present complexity-theoretical insights showing that our Taylor surrogate simulation technique can also be leveraged to measure the circuit computational complexity as a function of the initialization strategy. We further identify a critical patch size where runtime complexity transitions from polynomial to super-polynomial, indicating a phase transition in computational complexity. Combined with the LCE theory, {we argue that these results could in principle be leveraged to prove the existence of super-polynomially complex landscape patches where gradient norms still decay only polynomially. While these results outline a clear theoretical pathway, we explicitly emphasize that the existence of a transition patch with polynomially decaying gradients remains a conjecture supported by our numerical evidence rather than a mathematically proven guarantee.

\subsection{Backpropagation of High-Order Clifford structures in the Heisenberg Picture}

We begin by outlining the core technical contribution and intuition behind our method. Since our proposed Taylor surrogate expands around Clifford circuits (recalling from Eq.~\eqref{eq:PQC} that $U(0)$ is Clifford), we introduce the notion of \emph{Clifford structures}—specific ensembles of Clifford operators that arise from the evaluation of the Taylor coefficients $(D_\theta^\alpha C)(0)$. These structures generate, via Heisenberg backpropagation~\cite{fuller2025improved}, a set of efficiently computable Pauli strings and can be used to classically benchmark~\cite{gottesman1998heisenberg, aaronson2004improved} optimization landscape patches prior to the deployment of the corresponding quantum algorithm on quantum processors.

As shown in Appendix~\ref{sec:parameter-shift}, the Taylor coefficients take the explicit parameter-shift~\cite{mitarai2018quantum, schuld2019evaluating, cerezo2021higher} form:
\begin{equation}
    (D_\theta^\alpha C)(0) = \frac{1}{2^{\|\alpha\|_1}}\sum_{j\leq \alpha} (-1)^{\|j\|_1}\binom{\alpha}{j}\, C(\xi_{j,\alpha}), \label{eq:taylor-coeff2}
\end{equation}
where {$\xi_{j,\alpha}:=\frac{\pi}{2}(\alpha-2j)$ {are equidistant parameter points} for fixed multi-indices $\alpha\in\mathbb{N}_0^D$, and $0\leq j\leq\alpha$, giving rise to Clifford configurations exclusively}. Indeed, for each shift $j\leq \alpha$, we establish in Appendix~\ref{sec:complexity} that evaluating the PQC in Eq.~\eqref{eq:PQC} at $\xi_{j,\alpha}$ results in a Clifford operator $U(\xi_{j,\alpha})$, enabling Gottesman-Knill-based simulation~\cite{gottesman1998heisenberg, aaronson2004improved}. This leads to an algorithm evaluating Eq.~\eqref{eq:taylor-coeff2} in polynomial time $\mathcal{O}(|\mathcal{I}|\,D^{\|\alpha\|_1})$ {if $\|\alpha\|_1=\mathcal{O}(1)$}, motivating the following:

\begin{definition}[High-Order Clifford Structures]\label{def:Clifford structure}
    For a variational ansatz $U(\theta)$ composed of Clifford gates and single-qubit Pauli rotations with $D\geq 1$ parameters, the Clifford structure up to order $m-1$ is the collection defined as
    \begin{equation*}
        \{U(\xi_{j,\alpha}):j\leq\alpha,\, \|\alpha\|_1< m\},
    \end{equation*}
    where $\xi_{j,\alpha}:=\frac{\pi}{2}(\alpha-2j)$, and each $U(\xi_{j,\alpha})$ is a Clifford unitary transformation.
\end{definition}

Since we expand around $\theta=0$, $U(0)$ always belongs to the Clifford structure. For instance, the \emph{linear Clifford structure} derived from the standard parameter-shift rule~\cite{mitarai2018quantum, schuld2019evaluating} is given by the Clifford ensemble $\left\{U(0), U\left(\pm\frac{\pi}{2}e_k\right):k\in[D]\right\}$ with unit vectors $e_k$. The linear Clifford structure is used to build the LCE in the next section.

Due to the transitive action of the Clifford group on $\mathcal{P}_N$, we obtain for each $j\leq \alpha$:
\begin{align}
    C(\xi_{j,\alpha}) &= \sum_{i\in\mathcal{I}}c_i\,\langle0|U(\xi_{j,\alpha})^\dagger P_i U(\xi_{j,\alpha})|0\rangle \nonumber \\
    &= \sum_{i\in\mathcal{I}}c_i\, \langle0|P_{i}^{j,\alpha}|0\rangle, \label{eq:shifted-cost2}
\end{align}
where $P_i^{j,\alpha} := U(\xi_{j,\alpha})^\dagger P_i U(\xi_{j,\alpha}) \in \mathcal{P}_N\setminus\{I\}$ defines a list of Heisenberg-evolved Pauli strings. These evolved strings are efficiently computable via the Aaronson-Gottesman check matrix algorithm~\cite{aaronson2004improved}, producing Pauli strings~\cite{nielsen2010quantum} indexed by $i\in\mathcal{I}$ (cf. Eq.~\eqref{eq:observable}).
{In Appendix~\ref{sec:parameter-shift}, we also provide three different visual representations of the Clifford structure as per Definition~\ref{def:Clifford structure} in order to help the intuition of the reader.}
\begin{figure*}[t]
    \centering
    \begin{subfigure}
        \centering
        \includegraphics[trim={9cm 5.9cm 8cm 4cm},clip,width=.490\textwidth]{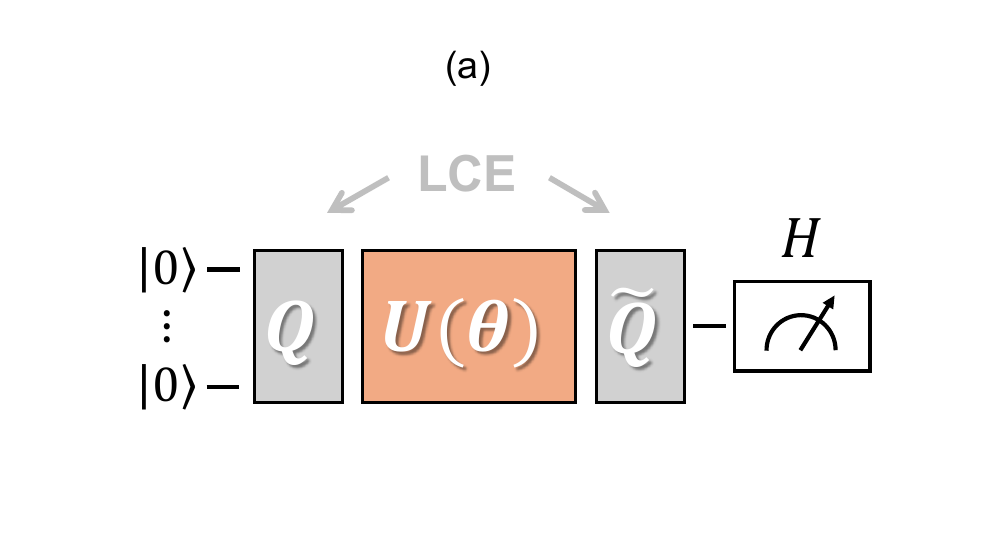}
    \end{subfigure}
    \hspace{0.0\textwidth}
    \begin{subfigure}
        \centering
        \includegraphics[trim={0cm 1cm 0cm 0cm},clip,width=.490\textwidth]{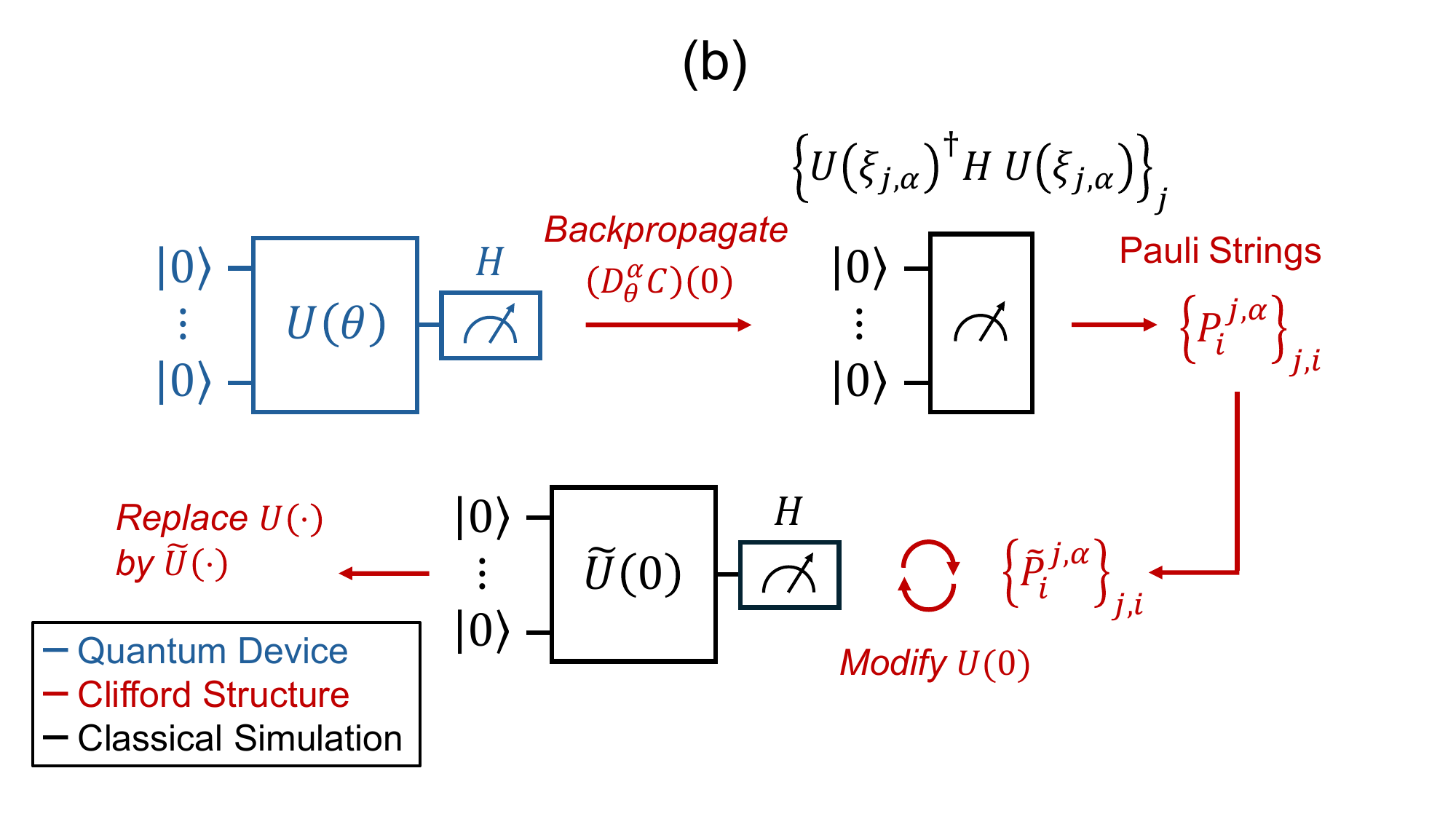}
    \end{subfigure}
    \caption{(a) {An illustration of the pair $(Q,\tilde{Q})$ of Clifford operators used in the LCE construction. The Clifford $Q$ translates the initial state, while $\tilde{Q}$ performs a change of basis on $H$.} (b) Classical framework for engineering Clifford structures.}
    \label{fig:Clifford structure}
\end{figure*}

To analyze the influence of $U(\cdot)$ on the optimization landscape, we combine Eqs.~\eqref{eq:taylor-coeff2} and~\eqref{eq:shifted-cost2}:
\begin{align}
    (D_\theta^\alpha C)(0) &= \frac{1}{2^{\|\alpha\|_1}}\sum_{j\leq \alpha} (-1)^{\|j\|_1}\binom{\alpha}{j}\, C(\xi_{j,\alpha}) \nonumber \\
    &= \frac{1}{2^{\|\alpha\|_1}}\sum_{j\leq \alpha} (-1)^{\|j\|_1}\binom{\alpha}{j}\, \sum_{i\in\mathcal{I}}c_i\,\langle0|P_{i}^{j,\alpha}|0\rangle. \label{eq:explicity-taylor-coeff2}
\end{align}
Thus, each Taylor coefficient depends on Pauli coefficients $\{c_i\}$, and Heisenberg-evolved Pauli strings $P_i^{j,\alpha}$ generated by the Clifford structure. Together with prefactors $(-1)^{\|j\|_1}\binom{\alpha}{j}$, manipulating the set of Pauli strings
\begin{equation}
    \{P_i^{j,\alpha}:j\leq\alpha,\,\|\alpha\|_1<m,\,i\in\mathcal{I}\},
\end{equation}
e.g., by adding or removing Clifford gates in $U(\cdot)$, allows one to influence summation cancellations and contributions in Eq.~\eqref{eq:explicity-taylor-coeff2}, for example to ensure a non-zero gradient component $(\partial_{\theta_k}C)(0)\neq0$. Since Eq.~\eqref{eq:taylor-coeff2} can be evaluated using Gottesman-Knill~\cite{gottesman1998heisenberg, aaronson2004improved}, this establishes an efficient classical framework for engineering the Clifford architecture of $U(\cdot)$ in order to achieve better gradients at $\theta=0$.

Fig.~\ref{fig:Clifford structure}b illustrates this framework: each Taylor coefficient $(D_\theta^\alpha C)(0)$ induces an $\alpha$-order Clifford structure that backpropagates through the observable~\cite{fuller2025improved}, producing a list of Pauli strings $\{P_i^{j,\alpha}\}_{i,j}$ which encode the $\alpha$-order contribution to the surrogate patch.

\subsection{The Linear Clifford Encoder (LCE) \label{subsec:LCE}}

Having introduced the mathematical framework, this prompts a central question: can we systematically engineer the Clifford architecture of $U(\cdot)$ to enhance landscape properties (e.g., {gradient scalability}) without sacrificing expressivity of the ansatz? In the following, we demonstrate that this is, in principle, achievable. We now propose a novel technique that leverages the use of first-order Clifford structures (cf. Definition~\ref{def:Clifford structure}) to ensure trainability on near-Clifford patches. Proofs are deferred to Appendix~\ref{sec:LCE}.

The previously discussed Taylor surrogation framework allows the explicit construction of a pair of Clifford operators $(Q,\tilde{Q})$ defining the transformation $\tilde{U}(\cdot):=\tilde{Q} U(\cdot) Q$ and yielding a \emph{LCE transformed cost} $\tilde{C}(\cdot)$, where LCE stands for \emph{Linear Clifford Encoder}, {as shown in Fig.~\ref{fig:Clifford structure}a}. {Even though this transformation inherently changes the Clifford architecture of the PQC ansatz, we prove in Appendix~\ref{sec:state-expressivity} that LCE maintains the minimizer in Eq.~\eqref{eq:minimizer} (possibly with another parameter configuration) given that the PQC is sufficiently expressive.}

{In brief, the LCE procedure consists in the following three key steps:}

\begin{enumerate}
    \item {Choose an arbitrary parameter coordinate $k\in[D]$ and a Pauli decomposition index $i_0\in\mathcal{I}$ with $c_{i_0}=\Omega(1)$.}
    \item {Construct the first LCE Clifford $\tilde{Q}$ that induces a relative phase of $\pm1$ between the two Pauli expectation in the parameter-shift in order to avoid cancellation. This can be viewed as a change of basis of the observable $H\mapsto \tilde{Q}^\dagger H\tilde{Q}$ in the Heisenberg picture.}
    \item {Construct the second LCE Clifford $Q$ that diagonalizes the previously $\tilde{Q}$-backpropagated Pauli term $\tilde{P}_{i_0}\mapsto Q^\dagger \tilde{P}_{i_0} Q \in\langle Z\rangle_N$. This can be viewed as a change of the initial state $|0\rangle\mapsto Q|0\rangle$ in the Schrödinger picture.}
\end{enumerate}

As a matter of fact, the two Clifford operators of LCE are used to explicitly control arbitrary linear Taylor coefficient $(\partial_{\theta_k}C)(0)$ (i.e., $\alpha=e_k$ for some unit vector $e_k$). Hence, given an arbitrary $k\in[D]$, our first main result shows how to use the first-order Clifford structure to transform any cost $C(\cdot)\mapsto\tilde{C}(\cdot)$ such that the gradient component satisfies $(\partial_{\theta_k}\tilde{C})(0)=\Omega(1)$, independently from $D$ or $N$.

The following theorem employs LCE to control the summation in Eq.~\eqref{eq:explicity-taylor-coeff2} and isolate a fixed coefficient $c_{i_0}$ of the observable $H$. Observe that the operators $(Q,\tilde{Q})$ depend on the choice of $k$ and $i_0$.
\begin{theorem}\label{thm:linear-clifford-encoder}
    Consider an arbitrary observable $H$ indexed by $\mathcal{I}$ in Eq.~\eqref{eq:observable}. Then, for every $k\in[D]$ and $i_0\in\mathcal{I}$, there exists a pair of Clifford operators $(Q,\tilde{Q})$ such that the gradient component $(\partial_{\theta_k}\tilde{C})(0)$ is equal to 
    \begin{align*}
        \beta_{i_0}(H) := \pm c_{i_0} + \frac{1}{2} \sum_{i \in \mathcal{I} \setminus \{i_0\}} &c_i \big\{ 
        \langle 0|\tilde{P}^{0,e_k}_i|0\rangle - \langle 0| \tilde{P}^{e_k,e_k}_i |0\rangle \big\},
    \end{align*}
    defining $\tilde{P}^{j,\alpha}_i:=Q^\dagger U(\xi_{j,\alpha})^\dagger \tilde{Q}^\dagger P_i \tilde{Q} U(\xi_{j,\alpha})Q$ and $\xi_{j,\alpha}:=\frac{\pi}{2}(\alpha - 2j)$ for $0\leq j\leq \alpha$. 
\end{theorem}

{Note that the choice of the coefficient $c_{i_0}$ is independent of $k$, thus allowing a total of $D\cdot\#\{i_0\in\mathcal{I}:c_{i_0}=\Omega(1)\}$ possible candidates for the LCE construction.} The proof of the theorem (cf.~Appendix~\ref{sec:LCE}) shows how to construct $(Q,\tilde{Q})$ explicitly using only few Clifford gates. {Explicit examples illustrating how LCE is used to modify the PQC are provided in Appendix~\ref{sec:ansatz}}. {The key idea is to choose the first Clifford $\tilde{Q}$ such that it ensures a relative phase between the difference of the Pauli expectations in the parameter-shift in order to avoid them from canceling to zero if they are both non-zero. Next, one constructs the second Clifford $Q$ such that both Pauli strings in the parameter-shift are diagonal giving rise to non-zero Pauli expectations $\pm1$.} 
The full gradient is then evaluated by classically computing the expectations of the transformed Pauli strings by using the stabilizer formalism as shown in Appendix~\ref{sec:gottesman-knill}. Furthermore, multiple gradient components may simultaneously benefit from the LCE, provably up to $\frac{D}{2}$ many components for structured PQCs, as discussed in Appendix~\ref{sec:LCE}.

Importantly, it is possible to explicitly construct the LCE such that $\beta_{i_0}(H)=1$ if $H=P$ is a single Pauli string (cf. Theorem~\ref{thm:linear-clifford-encoder-special} in Appendix~\ref{sec:LCE}). For general observables $H$, understanding the scaling behavior of $\beta_{i_0}(H)$ is more subtle due to potential cancellations among the Pauli terms. The lemma below shows that the probability of complete cancellation, that is $\beta_{i_0}(H) = 0$, decreases exponentially with the number of qubits $N$:

\begin{lemma} \label{lemma:observable-prob}
    Let $H=\sum_{i\in\mathcal{I}}c_iP_i$ be a random observable with $P_i\stackrel{i.i.d.}{\sim}\operatorname{Unif}(\mathcal{P}_N\setminus\{I\})$. Then, for each $i_0\in\mathcal{I}$, we have the conditional probability that
    \begin{equation*}
        \mathbb{P}\left(\beta_{i_0}(H)^2 = c_{i_0}^2 \mid P_{i_0}) \geq 1-\mathcal{O}(|\mathcal{I}|\,2^{-N}\right),
    \end{equation*}
    as $N\to\infty$.
\end{lemma}

Intuitively, LCE is applied to isolate the Pauli coefficient $c_{i_0}$, while the remaining terms in $\beta_{i_0}(H)$ {vanish with exponentially high probability}. We condition the probability on $P_{i_0}$ since this Pauli string is fixed for the LCE construction as shown in the proof of Theorem~\ref{thm:linear-clifford-encoder}. If $c_{i_0}=\Omega(1)$, then this almost surely guarantees a constant gradient lower bound, as long as $|\mathcal{I}|=\mathcal{O}(\operatorname{Poly}(N))$. Numerical experiments in Sec.~\ref{sec:experiments} (cf.~Fig.~\ref{fig:trainability}) and Appendix~\ref{sec:appendix-experiments} demonstrate that the probability of complete cancellation is indeed negligible, and the expected behavior closely matches the case $H = P$ where $\beta_{i_0}(H)=1$, even if we additionally assume i.i.d. coefficients $c_i\sim\operatorname{Unif}([-1,1])$.

\subsection{Escaping Barren Plateaus with the Linear Clifford Encoder (LCE) \label{sec:escaping-bp-lce}}

{We now combine the preceding results to show that the Clifford operators $(Q,\tilde{Q})$ of LCE guarantee constant-scaling gradients in a neighborhood of the origin. Specifically, when combined with the Taylor surrogate in Eq.~\eqref{eq:taylor}, Theorem~\ref{thm:linear-clifford-encoder} yields a constant lower bound in Eq.~\eqref{eq:no-barren-plateau} for arbitrary Pauli decompositions $H$ that contain at least one coefficient with constant scaling, as specified in Eq.~\eqref{eq:observable}.}

\begin{theorem}\label{thm:general-escaping-barren-plateau}
    Let $H$ be an arbitrary, uniformly randomly sampled observable as in Eq.~\eqref{eq:observable} with Pauli terms indexed by $\mathcal{I}$. Let $\delta>0$ and initialize $\theta \sim \mathbb{P}_\theta \in \{\mathcal{N}(0,\sigma^2 I_D), \operatorname{Unif}([-\sigma,\sigma]^D)\}$ with i.i.d. components and $\sigma = \mathcal{O}(\|H\|^{-1}D^{-\frac{1+\delta}{2}})$. Then, for each $k\in[D]$ and $i_0\in\mathcal{I}$ with $c_{i_0}=\Omega(1)$, there exists a pair of Clifford operators $(Q,\tilde{Q})$ such that, with probability at least $1-\mathcal{O}(|\mathcal{I}|\,2^{-N})$ (conditioned on $P_{i_0}$), as $D\to\infty$,
    \begin{equation*}
        \mathbb{E}_{\theta\sim\mathbb{P}_\theta}[(\partial_{\theta_k}\tilde{C}(\theta))^2] = \Omega(1),
    \end{equation*}
    where 
    $\tilde{C}(\cdot)$ denotes the LCE transformed cost.
\end{theorem}

Theorem~\ref{thm:general-escaping-barren-plateau} shows how to apply LCE (cf. Theorem~\ref{thm:linear-clifford-encoder}) in order to achieve a constant lower bound on the expected gradient square norm, independent of the circuit depth of $U(\theta)$. This is achieved by initializing the parameters on a sufficiently small patch, with a size $\sigma$ that depends on the number of parameters $D$ and the operator norm $\|H\|$ of the observable. 
In the case where $\|H\|=\mathcal{O}(1)$, the effective patch size $\sigma$ for escaping barren plateaus thus scales as $\mathcal{O}(D^{-1/2})$, which is consistent with the findings of previous works~\cite{wang2024trainability, lerch2024efficient, puig2025variational, mhiri2025unifying}. Otherwise, if $\|H\|$ depends on $D$ (or $N$), then the patch size $\sigma$ scales in proportion to $\|H\|^{-1}$ as shown in Theorem~\ref{thm:general-escaping-barren-plateau}. 
{That is, our analysis only assumes that
\begin{equation}
    \text{$c_{i}=\Omega(1)$ for some $i\in\mathcal{I}$,} \label{eq:condition-observable}
\end{equation}
and we additionally assume $\|H\|=\mathcal{O}(1)$ only to make our results (particularly the patch sizes) comparable with Refs.~\cite{wang2024trainability, lerch2024efficient, puig2025variational, mhiri2025unifying}, even though our results hold true for arbitrary $\|H\|$. Furthermore, note that rescaling the observable (such that $\|H\|=\mathcal{O}(1)$) does not affect its commutative structure and only rescales the energy levels, and the energy gaps accordingly}. In Appendix~\ref{appendix:Hamiltonian} we show that Eq.~\eqref{eq:condition-observable} contains a wide range of Hamiltonians, including cases for which no efficient (probabilistic) classical algorithm exists that can solve the underlying Hamiltonian ground energy decision problem~\cite{kempe2006complexity}, unless $\texttt{BPP}=\texttt{QMA}$. When combined with Theorem~\ref{thm:general-escaping-barren-plateau}, this particularly implies that any non-classical observable of the form in Eq.~\eqref{eq:condition-observable} remains compatible with LCE to explore all directions with $\Omega(1)$ initial descent magnitudes within $\mathcal{O}(D^{-\frac{1}{2}})$ patches. Crucially, this remains effective for arbitrary model sizes $D,N\to\infty$ of the parameterized circuit $U(\theta)$.

{Unlike previous studies limited to specific circuit geometries and local operators~\cite{zhang2024absence}, our proposed framework works for arbitrary ansätze. It remains valid for any observable with a constant-scaling Pauli term, providing a more universal theoretical foundation. Related work by Park et al.~\cite{park2024hardware} achieves constant scaling gradients as well, but on smaller patches, and their analysis relies on stronger assumptions on the observable.} A summary that compares additional barren plateau mitigation methods on small landscape regions with the LCE technique is provided in Table~\ref{tab:comparing-bp-techniques} in Appendix~\ref{sec:comparison}.

Since the solution $\theta_*$ is a priori unknown, a reasonable initialization strategy is to randomly sample $k\in[D]$ and $i\in\mathcal{I}$ for the LCE construction within $\mathbb{P}_\theta$-patches. Inside $\mathcal{O}(D^{-\frac{1}{2}})$ patches specifically, the total number of effective LCE samples is equal to $D\cdot\#\{i_0\in\mathcal{I}:c_{i_0}=\Omega(1)\}$; 
one for each direction $k\in[D]$, and one for each Pauli coefficient $c_{i_0}=\Omega(1)$ isolated from the remaining $i\in\mathcal{I}$. In general, the descent magnitude of such samples may depend on the specific choice of $k$ and $i_0$ before training.

Importantly, however, while a random walk lacks directional guidance, LCE uses structured gradient information to drive efficient optimization from the start. In fact, note that for any given $k\in[D]$ and $i_0\in\mathcal{I}$, {LCE has the effect of globally reshaping the loss landscape itself (via a translation and change of basis as elaborated in the next paragraph) so that the gradient at initialization captures a stronger, more informative descent signal. More intuitively, LCE ensures parameter initialization at a point with a steep slope along that direction.} This effect of LCE on the loss landscape is visually demonstrated in Appendix~\ref{sec:LCE}.

Crucially, we must ensure that the LCE transformation $C(\cdot)\mapsto\tilde{C}(\cdot)$ does not compromise expressivity of the PQC and does not significantly change the value of the minimum $\tilde{C}(\theta_*)$ (while the minimizer in Eq.~\eqref{eq:minimizer} may change). Heuristically, this is plausible: the operator $Q$ merely changes the initial stabilizer state $|0\rangle\mapsto|\psi_0\rangle=Q|0\rangle$, corresponding to a landscape translation, while $\tilde{Q}^\dagger H \tilde{Q}$ constitutes a basis change, {mapping $H$ to another Pauli string if $H\in\mathcal{P}_N$}. 
More formally, we prove in Appendix~\ref{sec:state-expressivity} that both expressivity and global optimality are preserved under LCE transformations via a geometric notion of \emph{$\varepsilon$-state-expressivity}. This implies that the LCE transformed PQC ansatz is able to explore the same regions as the original Hilbert space $\{|\psi(\theta)\rangle:\theta\in[-\pi,\pi]^D\}$, even though the optimal point $\theta_*$ might be different.

While LCE preserves the global minima, the risk of getting trapped in local minima remains. In such cases, Quantum Optimal Control (QOC)~\cite{magann2021pulses} and the theory of overparameterization~\cite{larocca2023theory} may help transform local minima into global ones~\cite{rabitz2004quantum}. We leave the study of this topic to future work.

Having introduced the LCE technique, which allows us to circumvent barren plateaus on $\mathcal{O}(D^{-1/2})$-sized surrogate patches, we now ask whether larger patch sizes are achievable. The next section explores how computational complexity scales with increasing parameter norms $\|\theta\|_1$, providing a measure of non-Cliffordness, relating patch sizes to classical simulability, and revealing a phase transition in computational complexity.

\subsection{Phase Transitions in Computational Complexity}

Let us now consider again the task of approximating a target PQC with the Taylor surrogate $C_m(\theta)$ from Eq.~\eqref{eq:taylor}, with truncation threshold $m = m(\theta)$. The truncation threshold directly determines its computational complexity, as $m$ directly determines the number of terms in the summation in Eq.~\eqref{eq:taylor} (cf. Lemma~\ref{lemma:stars-and-bars} in Appendix~\ref{sec:truncation}). We analyze the minimal threshold required to satisfy a surrogate error tolerance via two approaches:
\begin{enumerate}
    \item \emph{Worst-case error analysis:} $m = m(\theta)$ is deterministic for a fixed configuration $\theta\in[-\pi,\pi]^D$;
    \item \emph{Mean-squared error analysis:} $m = m(\mathbb{P}_\theta)$ is probabilistic, depending on an initialization strategy $\mathbb{P}_\theta$.
\end{enumerate}

This separation, previously explored by Lerch et al.~\cite{lerch2024efficient}, reveals distinct trade-offs in simulability and {gradient scalability}. We show in Appendix~\ref{sec:truncation} that the worst-case analysis characterizes computational complexity for arbitrary patch sizes, while mean-squared error is limited to $\mathcal{O}(D^{-1/2})$-sized patches. Notably, this threshold marks the boundary between polynomial and super-polynomial regimes (cf. Appendix~\ref{sec:complexity}) and identifies the largest known patch where barren plateaus can be provably mitigated~\cite{wang2024trainability, lerch2024efficient, puig2025variational, mhiri2025unifying} (cf. Theorem~\ref{thm:general-escaping-barren-plateau}).

\subsubsection{Worst-Case Error Analysis}

In Appendix~\ref{sec:truncation} we show the worst-case error obeys
\begin{equation}
    |C(\theta)-C_m(\theta)|\leq \frac{\|H\|}{m!}\|\theta\|_1^m. \label{eq:taylor-error2}
\end{equation}
Thus, for any fixed $\theta\in[-\pi,\pi]^D$, one can determine the minimal $m=m(\theta)$ needed to satisfy the error bound in Eq.~\eqref{eq:taylor-error2} as a function of $\|\theta\|_1$, which in this context serves as a way to measure the level of non-Cliffordness of the configuration. For the Taylor surrogate, this leads to the following $\ell_1$-characterization of a super-polynomial computational phase transition (cf. Appendix~\ref{sec:complexity}):

\begin{theorem}\label{thm:complexity}
    {If $\|\theta\|_1 = \mathcal{O}(1)$, the runtime of evaluating $C_m(\theta)$ is at most polynomial, namely $\mathcal{O}(|\mathcal{I}|D^m)$. If $\|\theta\|_1=o(D)$ strictly increases sub-linearly as $D\to\infty$, and $\#\{k\in[D]:\exists \xi\in\frac{\pi}{2}\mathbb{Z}:|\theta_k-\xi|> \Omega(D^{-1})\}=\Omega(D)$, then the runtime becomes at least super-polynomial,}
    \begin{equation*}
        {\operatorname{Time}C_m(\theta) = \Omega\left(\operatorname{Poly}(D)^{-1}\left(\frac{D}{\|\theta\|_1}\right)^{e\|\theta\|_1}\right),}
    \end{equation*}
   {and is at most $\mathcal{O}(|\mathcal{I}|D\,2^{2D})$ in the worst case.}
\end{theorem}

Appendix~\ref{sec:taylor-complexity} (cf. Fig.~\ref{fig:polynomial-complexity}) demonstrates how to compare the polynomial complexity across various PQC models, and how to identify computational overheads.

For $\theta\sim\mathbb{P}_\theta\in\{\mathcal{N}(0,\sigma^2I_D), \operatorname{Unif}([-\sigma,\sigma]^D)\}$ with i.i.d.\ components and $\sigma = D^{-r}$ for $r\in[0,1)$, the law of large numbers yields
\begin{equation}
    \mathbb{E}_\theta[m(\theta)] = \Theta(D^{1-r}) \label{eq:expected-truncation}
\end{equation}
as $D\to\infty$. Thus, the Taylor surrogate captures computational complexity across all patch sizes $\mathcal{O}(D^{-r})$ parametrized by $r\geq 0$. Combining the lower bound $m(\theta)=\Omega(\|\theta\|_1)$ as $D\to\infty$ (cf. Lemma~\ref{lemma:truncation-threshold} in Appendix~\ref{sec:truncation}) with Theorem~\ref{thm:complexity} thus reveals a \emph{phase transition in computational complexity} at $r=1$, separating polynomial from super-polynomial regimes. As $r\to 0$, the runtime saturates to exponential scaling.

The complexity of computing the Taylor surrogate $C_m(\theta)$ is essentially determined by the number of Taylor coefficients appearing in the truncated series in Eq.~\eqref{eq:taylor}. However, the \emph{effective} complexity of $C_m(\theta)$ is generally smaller since many of the Taylor coefficients could vanish. In practice, it is \emph{a priori unknown} which coefficients vanish, meaning that all of them have to be computed in any case, thus ruling out the possibility of exploiting this fact for obtaining a more powerful surrogate.

\begin{figure}[h]
    \centering
    \includegraphics[width=\linewidth]{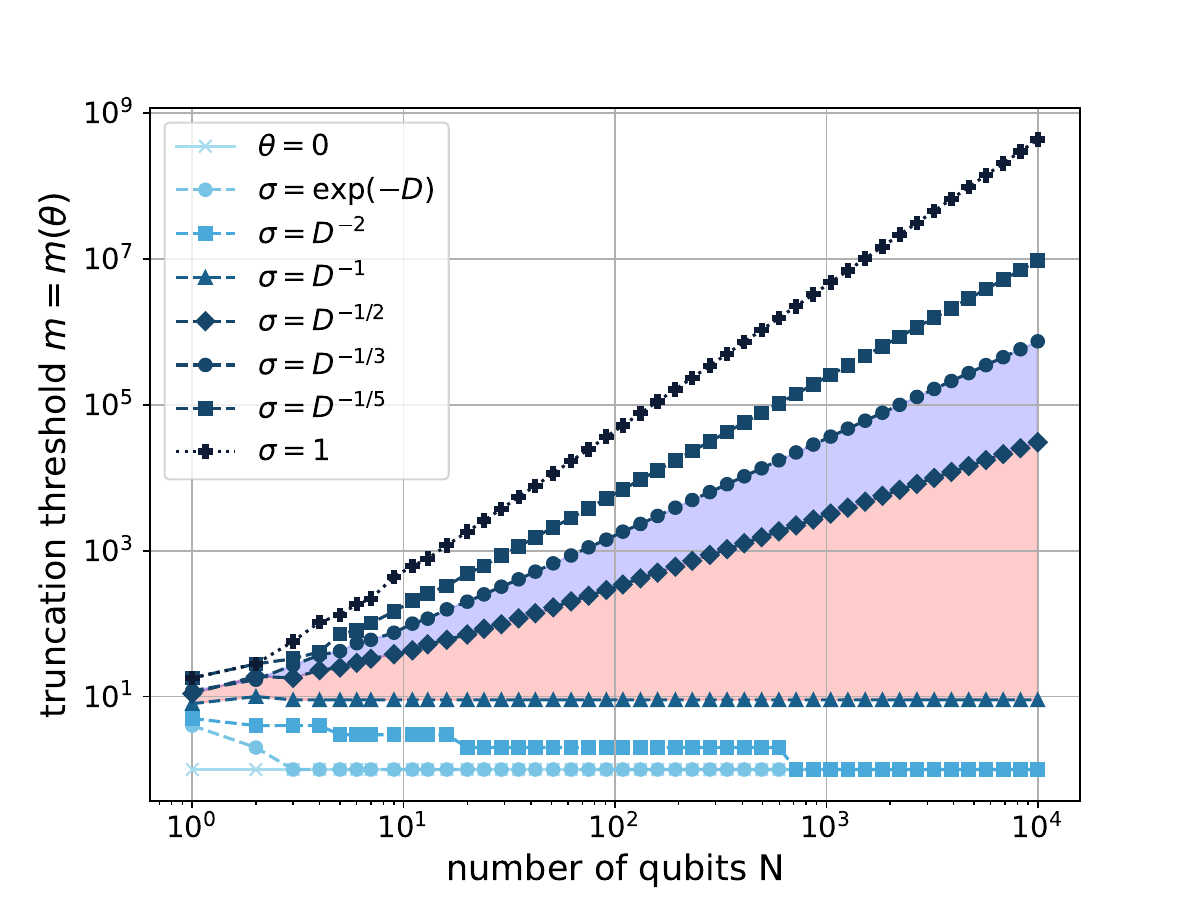}
    \caption{Worst-case analysis truncation threshold $m=m(\theta)$ of a model with $D=2N(N+1)$ parameters, as predicted by the expected truncation threshold in Eq.~\eqref{eq:expected-truncation}. Here, we set $\|H\|=1$ and $\varepsilon=10^{-6}$. The red region is easy to simulate classically (cf. Theorem~\ref{thm:probabilistic-complexity}), while the blue region is not (cf. Theorem~\ref{thm:complexity}).}
    \label{fig:truncation-statistics}
\end{figure}

\subsubsection{Mean-Squared Error Anaylsis}

For $\theta\sim\mathbb{P}_\theta\in\{\mathcal{N}(0,\sigma^2I_D), \operatorname{Unif}([-\sigma,\sigma]^D)\}$ with i.i.d.\ components, the mean-squared error satisfies
\begin{equation}
    \mathbb{E}_\theta[(C(\theta)-C_m(\theta))^2] \leq {c\,\|H\|^2\frac{(D\sigma^2)^m}{m^m}}, \label{eq:mean-squared-error2}
\end{equation}
for some constant $c > 0$ if $m=\mathcal{O}(1)$ (cf. Appendix~\ref{sec:truncation}). This implies $\sigma$ cannot exceed $\Omega(D^{-1/2})$ without incurring significant error. Thus, mean-squared analysis fails beyond $\mathcal{O}(D^{-1/2})$-patches, while worst-case analysis remains valid across the full parameter space.

Fig.~\ref{fig:truncation-statistics} illustrates the scaling of $m(\theta)$ when simulating a model with $D=2N(N+1)$ parameters, such as the linear-depth \texttt{mHEA} (minimalistic Hardware Efficient Ansatz~\cite{leone2024HEA, efficientsu2}) defined in Appendix~\ref{sec:ansatz}, with patch sizes $\mathcal{O}(\sigma)$ under various distributions $\mathbb{P}_\theta$. 
The red-shaded region corresponds to the regime where barren plateaus are provably avoided (cf. Theorem~\ref{thm:general-escaping-barren-plateau}). 
{The blue-shaded region marks the regime requiring at least super-polynomial computational resources for the Taylor surrogate}. As further shown in Theorem~\ref{thm:probabilistic-complexity}, the mean-squared error analysis collapses the red-shaded region to constant scaling $m(\mathbb{P}_\theta)=\mathcal{O}(1)$, implying classical simulability of $\mathcal{O}(D^{-1/2})$-patches—consistent with the findings of Lerch et al.~\cite{lerch2024efficient}.

To ensure a small mean-squared error, we distinguish between two initialization strategies:

\begin{itemize}
    \item[(i)] \emph{Asymptotic Local Initialization:} \\ For $\sigma=\mathcal{O}(\|H\|^{-1/m}D^{-(1+\delta)/2})$ with $\delta>0$, the error decays as $\mathcal{O}(D^{-\delta m})$.
    
    \item[(ii)] \emph{Non-Asymptotic Local Initialization:} \\ For {$\sigma = \mathcal{O}(q\|H\|^{-1/m}D^{-1/2})$} with $0<q<1$, the error decays as $\mathcal{O}(q^{2m})$.
\end{itemize}

{The first initialization strategy mitigates the problem of barren plateaus on small landscape patches (cf. Theorem~\ref{thm:general-escaping-barren-plateau}), while the second facilitates runtime comparison with state-of-the-art classical simulation methods~\cite{lerch2024efficient}. Both yield polynomial-time surrogates since the first initialization strategy results in a smaller patch than the second}:

\begin{theorem}\label{thm:probabilistic-complexity}
    {Let $\theta\sim\mathbb{P}_\theta\in\{\mathcal{N}(0,\sigma^2I_D), \operatorname{Unif}([-\sigma,\sigma]^D)\}$ with i.i.d.\ components, according to Eq.~\eqref{eq:mean-squared-error2} with constant $c>0$. Assume further that $\|H\|=\mathcal{O}(1)$. Then, with probability at least $1-\rho$, the computational cost of simulating $C(\theta)$ via $C_m(\theta)$ up to an error $\varepsilon>0$, is of polynomial order,
    \begin{equation*}
        \mathcal{O}\left(|\mathcal{I}|\,D^{\frac{1}{2\log \left(\frac{1}{q}\right)}\log\left(\frac{c}{\rho\,\varepsilon^2}\right)}\right)
    \end{equation*}
    for non-asymptotic local initialization, i.e., $\sigma = \mathcal{O}(qD^{-1/2})$ for some $0<q<1$.}
\end{theorem}

For general $\|H\|$, Theorem~\ref{thm:probabilistic-complexity} {establishes a classical surrogate for all patches $\sigma=\mathcal{O}(\|H\|^{-1}D^{-1/2})$}. In the special case where $\|H\|=\Theta(1)$ is normalized, this aligns with the main results of Lerch et al.~\cite{lerch2024efficient}, who study a classical surrogate based on the Pauli path formalism~\cite{beguvsic2023simulating}. Notably, as shown in Appendix~\ref{sec:complexity} (cf. Corollary~\ref{corr:faster-runtime}), our Taylor surrogate outperforms such a method on sufficiently small surrogate patches (i.e., $0.35<q<1$) under non-asymptotic local initialization. 
{A tabular comparison of the two surrogate complexities in terms of $\|\theta\|_1$ (which characterizes the patch size via $\sigma = \mathcal{O}(D^{-r})$ as per Eq.~\eqref{eq:expected-truncation}) is shown in Table~\ref{table:complexity-summary}}. {The Taylor surrogate outperforms the Pauli surrogate on sufficiently small patches $\sigma=qD^{-1/2}$ as per Corollary~\ref{corr:faster-runtime} whereas the Pauli path technique is preferable beyond these patches.} {While Ref.~\cite{lerch2024efficient} does not contain a complexity result as a continuous function of $\|\theta\|_1$ and instead characterizes} complexity only at $r\in\{0,\frac{1}{2},1\}$ for $\theta\sim\operatorname{Unif}([-\sigma,\sigma]^D)$ with $\sigma=\mathcal{O}(D^{-r})$, our analysis extends to the entire spectrum $r\geq 0$ via Eq.~\eqref{eq:expected-truncation}, {and additionally include super-polynomial complexity lower bounds beyond the classical patch (cf. Theorem~\ref{thm:complexity})}. Table~\ref{tab:comparing-surrogates} in Appendix~\ref{sec:comparison} provides a more detailed comparison with the Pauli path surrogate~\cite{lerch2024efficient} of small landscape patches.

\begin{table}[h]
    \centering
    \renewcommand{\arraystretch}{1.5} 
    \begin{tabular}{|c|c|}
        \hline
        \textbf{Norm $\|\theta\|_1$} & \textbf{Times} $t_{\text{Taylor}},t_{\text{Pauli}}$ \\ \hline \hline
        $\Theta(1)$ & $\mathcal{O}(\operatorname{Poly}(D)): \mathbf{t_{\textbf{Taylor}}<t_{\textbf{Pauli}}}$ \\ \hline
        $qD^{1/2}\text{ for some } q\in(0,1)$ & $\mathcal{O}(\operatorname{Poly}(D)): \mathbf{t_{\textbf{Taylor}}<t_{\textbf{Pauli}}}$ \\ \hline
        $cD^{1/2}\text{ for all }c>q$ & $\mathcal{O}(\operatorname{Poly}(D)): t_{\text{Taylor}}>t_{\text{Pauli}}$ \\ \hline
        $\Theta(D)$ & $t_{\text{Taylor}}=\mathcal{O}(2^{2D})$, $t_{\text{Pauli}}=\mathcal{O}(2^{D})$ \\ \hline
    \end{tabular}
    \caption{{Comparing the computational complexity of the Taylor surrogate and the Pauli path surrogate of Lerch et al.~\cite{lerch2024efficient} as a function of the measure $\|\theta\|_1$ of non-Cliffordness, with provable speedups on sufficiently small patches.}}
    \label{table:complexity-summary}
\end{table}

\section{Numerical Experiments \label{sec:experiments}}

We now present numerical experiments supporting our theoretical results. First, we benchmark the {gradient scalability} of LCE transformed PQCs using the Taylor surrogate to confirm that proper Clifford design yields constant-scaling gradients (cf. Theorem~\ref{thm:general-escaping-barren-plateau}). We then extend these insights to LCE-assisted Variational Quantum Eigensolver (VQE)~\cite{Peruzzo2014VQE, matos2021vqe} optimization, leveraging improved gradients on near-Clifford patches.

\subsection{Initial Gradient Scalability}

\begin{figure*}[t]
    \centering
    \includegraphics[width=1\linewidth]{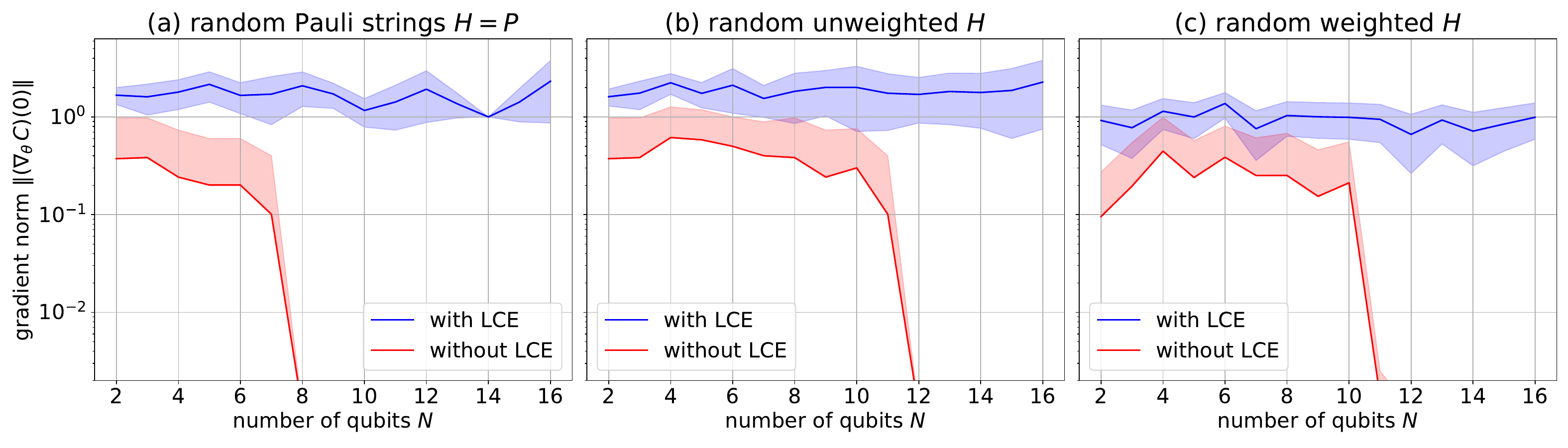}

    \caption{Scaling behavior of the initial gradient norm at $\theta=0$ for the \texttt{mHEA}-model with $L=N$ layers. The panels vary across types of random observables. The experiment highlights the effect of the LCE transformation. We average over 10 independent runs.}
    \label{fig:trainability}
\end{figure*}

We evaluate the scaling of the gradient norm $\|(\nabla_\theta\,C)(0)\|$ for increasing qubit numbers $N$ focusing on the \texttt{mHEA}~\cite{leone2024HEA} model—a minimalistic Hardware-Efficient Ansatz based on Qiskit’s \texttt{EfficientSU2}~\cite{efficientsu2}, comprising $Y$/$Z$-rotations and circular entanglement. 
Appendix~\ref{sec:ansatz} shows additional PQC models with their respective LCE modifications. Note that Theorem~\ref{thm:general-escaping-barren-plateau} applies in general even to deep PQCs.

\begin{figure}[h]
    \centering
    \includegraphics[width=\linewidth]{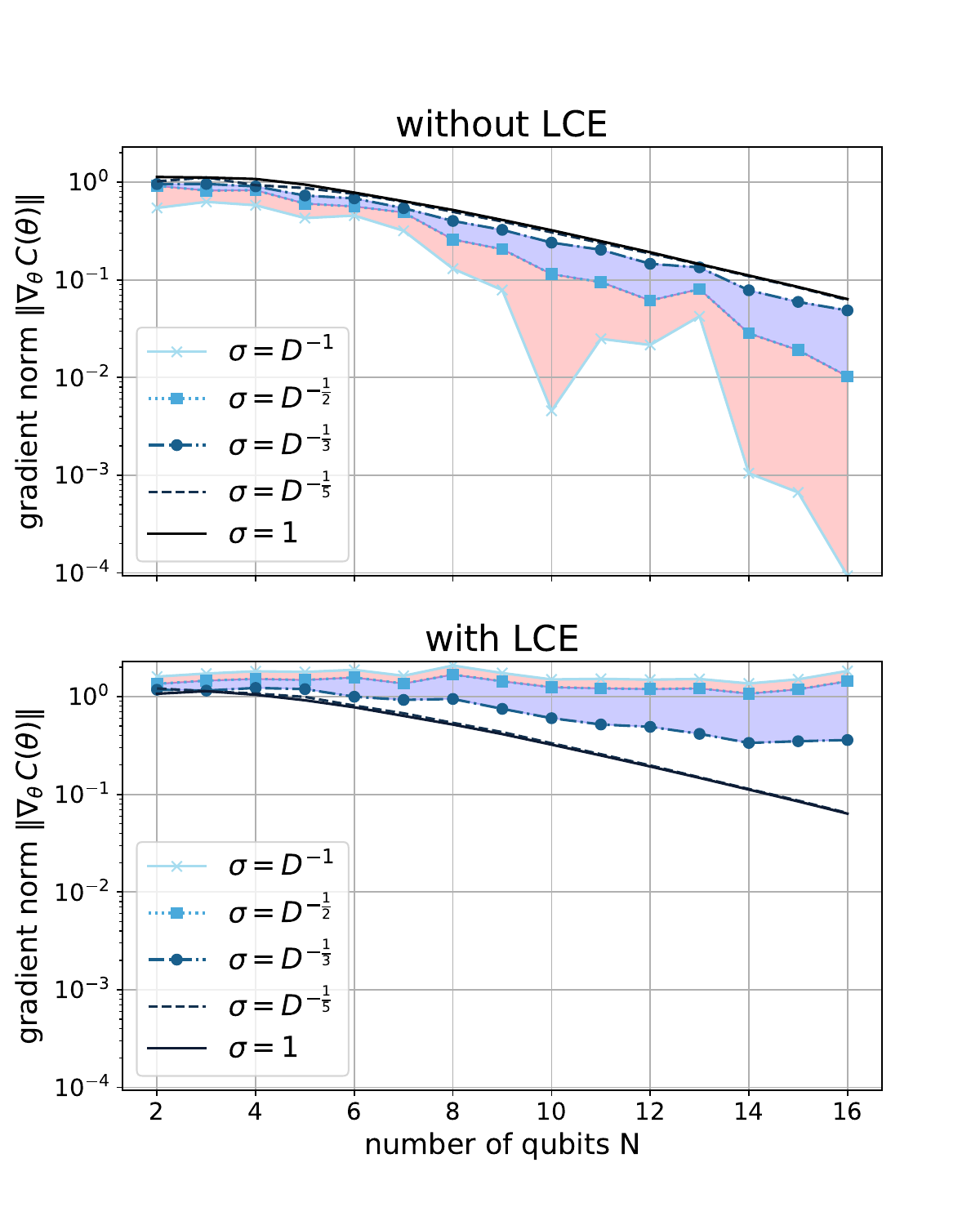}
    \caption{Scaling behavior of the gradient norm on various landscape patches for the \texttt{mHEA}-model with $L=N$ layers. We average over 50 independent runs of random $\theta\sim\mathbb{P}_\theta$ and uniformly sampled Pauli strings $H=P$.}
    \label{fig:escaping-BP}
\end{figure}

Fig.~\ref{fig:trainability} compares initial gradient norms at $\theta=0$ with and without LCE, employing \texttt{mHEA} with linear circuit depth $L=N$. Gradients are computed analytically via Eq.~\eqref{eq:explicity-taylor-coeff2}. Each run uses a randomly sampled observable $H$, and a random LCE construction. Panel (a) of Fig.~\ref{fig:trainability} focus on uniformly random single Pauli strings $H=P$. Without LCE, constructive contributions in Eq.~\eqref{eq:explicity-taylor-coeff2} vanish with increasing $N$, leading to zero expected gradient for sufficiently large $N$. With LCE, gradient norms remain constant in $N$. Panels (b) and (c) of Fig.~\ref{fig:trainability} extend results to random observables with $|\mathcal{I}|=\lfloor\sqrt{N}\rfloor$ Pauli terms. For observables with uniformly sampled Pauli strings $P_i$ and $c_i=1$ for all $i\in\mathcal{I}$ (called random unweighted $H$), the scaling compares to the single Pauli case. For observables with uniformly sampled Pauli strings $P_i$ and i.i.d. $c_i\sim\operatorname{Unif}([-1,1])$ for all $i\in\mathcal{I}$ (called random weighted $H$), gradient norms tend to behave more randomly due to random Pauli coefficients. This confirms that the LCE transformation remains effective for arbitrary observables. As previously mentioned, evaluation at $\theta=0$ is classically efficient by Gottesman-Knill~\cite{gottesman1998heisenberg, aaronson2004improved}, allowing simulation even for large $N$. Specifically, evaluating the gradient norm $\|(\nabla_\theta\,C)(0)\|$ takes $\mathcal{O}(|\mathcal{I}|D^2)$ operations: the vanilla parameter-shift rule~\cite{mitarai2018quantum, schuld2019evaluating} requires $\mathcal{O}(|\mathcal{I}|D)$ operations to evaluate $(\partial_{\theta_k}C)(0)$ (cf. Lemma~\ref{lemma:parameter-shift} in Appendix~\ref{sec:parameter-shift}), of which there are exactly $D$ distinct gradient components $k\in[D]$. Appendix~\ref{sec:appendix-experiments} contains gradient norm scaling experiments with up to $N=32$ qubits for \texttt{mHEA} with $L=1$ layer.

\begin{figure*}[]
    \centering
    \includegraphics[width=1\textwidth]{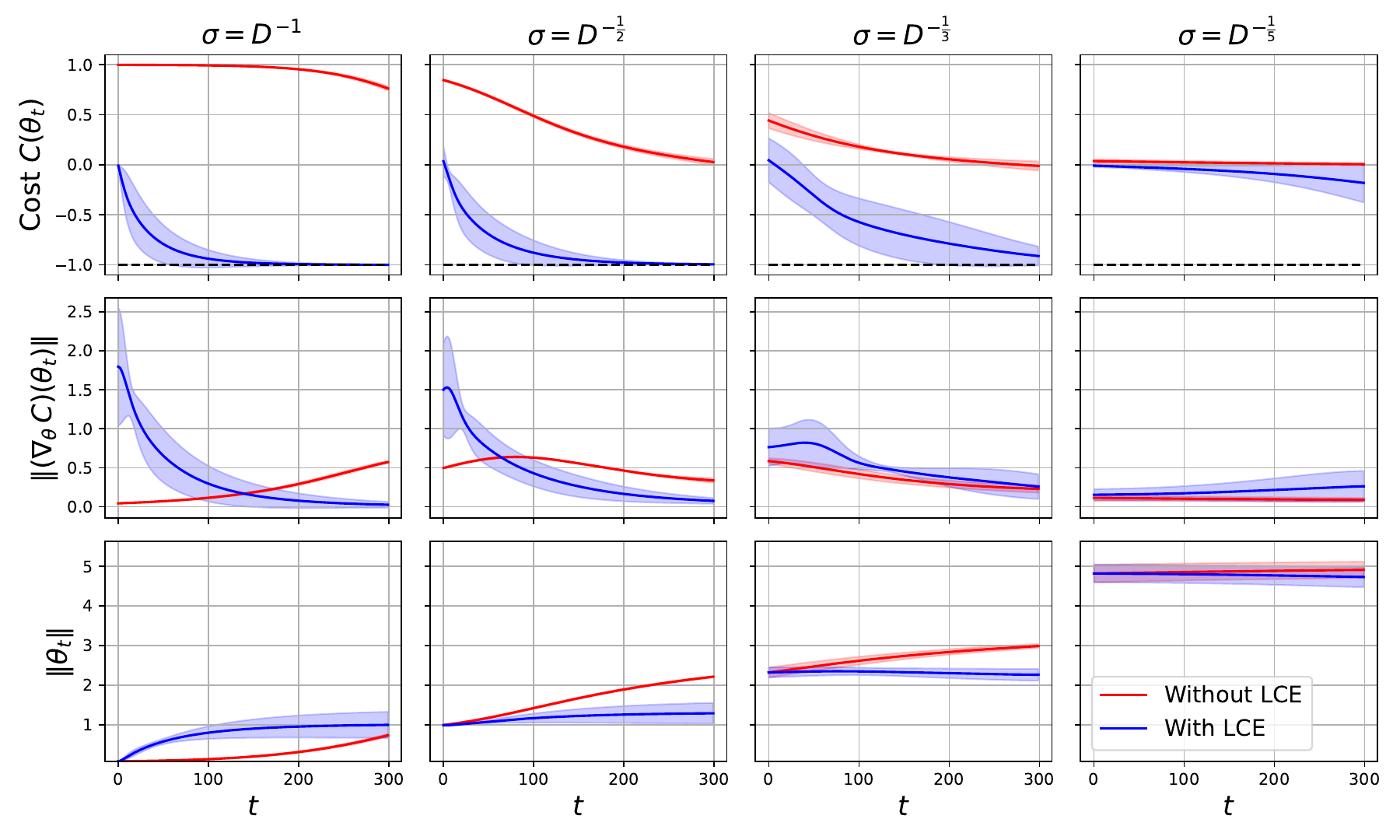}
    \caption{VQE Optimization using vanilla gradient descent with a learning rate of 0.01, averaging over 5 runs of random Gaussian initialization across varying patch sizes. We find the ground energy of the global observable $Z^{\otimes N}$ with $N=15$, employing the \texttt{mHEA} ansatz with $L=5$ layers.}
    \label{fig:VQE-experiment}
\end{figure*}

Fig.~\ref{fig:escaping-BP} repeats the experiment of the left panel in Fig.~\ref{fig:trainability} on near-Clifford patches, varying patch sizes via Gaussian initialization $\sigma=D^{-r}$. Each run uses a randomly sampled parameter, a random single Pauli string observable $H=P$, and a random LCE construction. The red-shaded region corresponds to classically efficient regimes (cf. Fig.~\ref{fig:truncation-statistics}); blue-shaded to super-polynomial regimes as per Theorem~\ref{thm:complexity}, where gradients {may decay polynomially in a transition zone just beyond the red regime}. {Within the constant-scaling regime (red-shaded), LCE ensures {gradient scalability independent of the model size}. For {too} large $\sigma$, gradients decay exponentially. This {empirically suggests} the existence of a transition patch (contained in the blue-shaded region) where gradients asymptotics smoothly transitions from constant to polynomial decay (cf. Eq.~\eqref{eq:no-barren-plateau})}

Interestingly, the improved gradient scaling behavior due to LCE closely resembles that shown in Fig.~\ref{fig:truncation-statistics}, suggesting a deeper link between {gradient scalability} and complexity. {Note that the non-LCE trends appearing noisier is likely an artefact of the log-scale on fluctuations of smaller orders of magnitude}. We refer to Appendix~\ref{sec:appendix-experiments} for comparisons across additional hardware-efficient PQC models.

\subsection{Variational Quantum Eigensolvers}

We now perform an ablation study on VQE optimization dynamics. 
{While the theory provided in the previous sections focuses solely on initialization, here we provide experiments investigating the effects of LCE during optimization.}
{When LCE is applied at initialization, we select a random parameter $k\in [D]$ and a random Pauli index $i_0\in\mathcal{I}$ with $c_{i_0}=\Omega(1)$ to construct a single LCE transformation (as elaborated in Sec.~\ref{subsec:LCE}) that will be kept fixed during the entire optimization procedure. This transformation can be viewed as a transformation of the optimization landscape $C(\cdot)\mapsto\tilde{C}(\cdot)$ where the minima are close in value to the one in the original landscape $C(\cdot)$, but with different parameter coordinates (see Appendix~\ref{sec:state-expressivity} and Fig.~\ref{fig:LCE-heightmap} in Appendix~\ref{sec:LCE}). In particular, we do not apply a new LCE transformation at each iteration during optimization.} Since the LCE technique applies to initialization only, we empirically test whether improved gradients translate to practical optimization benefits—especially under barren plateau-prone observables like global ones~\cite{cerezo2021cost} such as $H_{\operatorname{global}}:=Z^{\otimes N}$.

Fig.~\ref{fig:VQE-experiment} shows VQE optimization of $H_{\text{global}}$ employing the \texttt{mHEA} ansatz ($L=5$, $N=15$) via vanilla gradient descent (update rules defined as $\theta_{t+1}=\theta_t-\eta\nabla_\theta\,C(\theta_t)$ with constant learning rate $\eta>0$) over 5 runs with Gaussian initializations $\theta\sim\mathcal{N}(0,\sigma^2I_D)$ across varying patch sizes. The top row displays the cost minimization, highlighting faster convergence with LCE across all patch dimensions. 
The central row shows gradient norms, reflecting improved trainability both at initialization (cf. Theorem~\ref{thm:general-escaping-barren-plateau}) and throughout the minimization. 
The bottom row tracks the parameter norm $\|\theta_t\|_2$ {(which takes into account $2\pi$-periodicity of $C(\cdot)$ since the parameter trajectories remain within the hypercube $[-\pi,\pi]^D$ far from the periodic boundaries)}, indicative of classical hardness (cf. Theorem~\ref{thm:complexity}) showing that indeed this increases when enlarging the patch size. 
The numerical simulations continue to demonstrate successful convergence on the larger patch size $\sigma=\mathcal{O}(D^{-1/3})$ which is likely due to the improved initial gradients arising from LCE. As the patch size $\sigma$ increases, the gap between standard VQE and VQE with LCE gradually decreases, eventually approaching the situation where convergence rates appear to become more similar (cf. $\sigma=\mathcal{O}(D^{-1/5})$), {as for this patch size LCE seems to not have an impact on the gradients, see Fig.~\ref{fig:escaping-BP}}. This confirms that the benefits of LCE are most significant when employed on patches of moderate size. Additional experiments are compiled in Appendix~\ref{sec:appendix-experiments}, where we show that changing learning rates leaves the effect of LCE invariant, that LCE is compatible with finite sampling noise~\cite{kreplin2024reduction}.

{While the ground state of the $Z^{\otimes N}$ observable is not of direct practical interest, it offers a suitable testbed for barren plateaus mitigation~\cite{cerezo2021cost,Larocca2025review, letcher2024tight}. In Appendix~\ref{sec:appendix-experiments}, we provide VQE experiments on the Heisenberg model~\cite{bonechi1992heisenberg} which is a more structured observable composed of local terms. In such a case, the benefits of LCE tend to disappear, most likely due to the absence of barren plateaus implied by the locality of the observable when employing shallow circuits~\cite{cerezo2021cost}. Further investigation is required to find a class of practically relevant observables with a Pauli basis structure benefiting from LCE gradient improvement.}

\section{Conclusion \label{sec:conclusion}}

{In this work, we investigated the interplay between gradient scalability at initialization and computational complexity in Variational Quantum Algorithms (VQAs)~\cite{Cerezo2021VQA}. By leveraging our Taylor surrogate framework, we connected these two concepts and provided a rigorous analysis of the regimes where classical simulability and favorable gradient statistics coexist.}

{To address the challenge of barren plateaus, we introduced the \emph{Linear Clifford Encoder (LCE)}. This technique utilizes the stabilizer formalism~\cite{gottesman1998heisenberg, aaronson2004improved, nielsen2010quantum} to modify the first-order structure of PQCs, ensuring constant-scaling gradient norms within local, classically simulable near-Clifford landscape patches. The LCE is implemented as an efficient classical algorithm that modifies the PQC ansatz, making it a practical tool for improving the gradient statistics of large-scale models at initialization. Our numerical results provide encouraging evidence that the LCE can enhance training dynamics in Variational Quantum Eigensolver (VQE) tasks~\cite{Peruzzo2014VQE} for certain global observables that are typically prone to vanishing gradients~\cite{cerezo2021cost}.}

{On the complexity front, we introduced a \emph{Taylor surrogate} that generalized the quadratic Clifford expansion~\cite{mitarai2022quadratic} and leveraged the Gottesman-Knill theorem~\cite{gottesman1998heisenberg,rabitz2004quantum} to efficiently compute the high-order Taylor coefficients, thus establishing a classical surrogate~\cite{schreiber2023classical} for near-Clifford landscape patches. We showed that the Taylor surrogate matches Pauli path runtime guarantees~\cite{lerch2024efficient} with provable speedups on sufficiently small patches. Furthermore, we identified a \emph{computational phase transition} as PQCs deviate from the near-Clifford regime. This threshold, characterized by the $\ell_1$-norm of the parameters (or the random initialization patch size), identifies the boundary where our Taylor surrogate shifts from polynomial to super-polynomial complexity, as established by our lower bound analysis.}

{Below this threshold, where the landscape patch is classically simulable, LCE ensures constant-scaling gradients. Beyond that threshold, numerical experiments provide evidence that the gradients may begin to decay polynomially rather than exponentially. Notably, we introduce a framework that applies to arbitrary PQCs and all observables with at least one constant-scaling Pauli term. Therefore, this extends beyond the restricted circuit geometries and local operators discussed in recent literature~\cite{zhang2024absence}.}

{Despite these insights, several open questions remain. A theoretical understanding of the training dynamics with LCE, particularly the model behavior beyond initialization~\cite{liu2022representation,liu2023analytic, Abedi2023quantumlazytraining, larocca2023theory,you2023analyzing, scala2025towards}, is lacking. Importantly, we do not yet have convergence guarantees for finding the ground energy of Hamiltonians. In particular, it remains unclear whether LCE can be applied successfully for practically relevant observables. Additionally, the LCE technique is currently incompatible with quantum machine learning (QML) tasks that rely on quantum feature maps~\cite{abbas2021power}, which typically disrupt the Clifford structure. Extending our framework to such settings—possibly by encoding data into observables~\cite{tiblias2025efficient}—may extend some of our analytical guarantees to data-driven settings. Moreover, while LCE addresses first-order structure, the role of higher-order Clifford encoders remains unexplored. Most importantly, a mathematical proof confirming our hypothesis of polynomially decaying gradients beyond the classical landscape patch remains an open problem. Bridging these gaps would offer a promising path toward identifying practical quantum advantage in the pre-fault tolerance era.}

\section{Acknowledgments}

We are grateful to Matthis Lehmkühler for his valuable input on the classical simulation of PQCs via Taylor surrogates and for discussions on the phenomenon of barren plateaus. We also thank Christa Zoufal, David Sutter, Stefan Woerner, Zoë Holmes, Ricard Puig and Sacha Lerch for their constructive feedback, Ji\v{r}í \v{C}erný for his insightful comments, and Enea Monzio Compagnoni and Jim Zhao for discussions on optimization. Sabri Meyer, Francesco Scala and Aurelien Lucchi acknowledge the support from SNSF grant No. 214919. Francesco Tacchino acknowledges support from SNSF grant No. 225229 RESQUE.

\bibliographystyle{unsrt}
\bibliography{references}

\resumetoc
\newpage
\onecolumngrid
\appendix

{
  \tableofcontents
}

\section{Comparing Barren-Plateau Mitigation Techniques and Classical Surrogates} \label{sec:comparison}

{We use the following Landau notations to characterize asymptotic behavior of a function $f:\mathbb{R}_{\geq1}\to\mathbb{R}$ in terms of some $g:\mathbb{R}_{\geq1}\to\mathbb{R}$, as $x\to\infty$:}
\begin{align}
    {{f(x)}} &= {{\mathcal{O}(g(x)) \quad:\iff\quad \lim_{x\to\infty}\left|\frac{f(x)}{g(x)}\right|\in[0,\infty)}} \nonumber
    \\ {{f(x)}} &= {{\Omega(g(x)) \quad:\iff\quad \lim_{x\to\infty} \left|\frac{f(x)}{g(x)}\right| \in (0,\infty]}} \nonumber
    \\ {{f(x)}} &= {{\Theta(g(x)) \quad:\iff \quad\lim_{x\to\infty}\left|\frac{f(x)}{g(x)}\right| \in (0,\infty)}}
    \\ {f(x)} &= {o(g(x))\quad:\iff\quad\lim_{x\to\infty}\frac{f(x)}{g(x)}=0,}
    \label{eq:landau-notations}
\end{align}
{{assuming that there exists $x_0\geq 1$ such that $g(x)\neq0$ for all $x\geq x_0$. Furthermore, we say that $f(x)$ and $g(x)$ are \emph{asymptotically equivalent} if the much stronger condition,}}
\begin{equation}
    {{\lim_{x\to\infty}\frac{f(x)}{g(x)} = 1,}} \label{eq:asymptotic-equivalence}
\end{equation}
{{holds true. In that case, we write $f(x)\sim g(x)$ as $x\to\infty$.}}
\bigskip

In order to provide an overview of our results in {gradient scalability} and classical computational complexity, and how they fit into the literature, we compare related barren-plateau mitigation techniques and the performance of classical surrogates with related works, respectively.

\subsection{Comparing Barren-Plateau Mitigation Techniques}

\begin{table*}[h]
    \begin{ruledtabular}
        \renewcommand{\arraystretch}{1.5}
        \begin{tabular}{
            >{\centering\arraybackslash}m{2.5cm}   
            >{\centering\arraybackslash}m{4.5cm} 
            >{\centering\arraybackslash}m{4.5cm} 
            >{\raggedright\arraybackslash}p{4.5cm} 
        }
            & gradient bounds within $\mathcal{O}(D^{-\frac{1}{2}})$ patches 
            & gradient bounds beyond $\mathcal{O}(D^{-\frac{1}{2}})$ patches 
            & model assumptions \\
            \hline
            \textbf{Linear Clifford Encoder (LCE)} 
            & \vspace{0.5cm}$\Omega(1)$ on $\mathcal{O}\left(\|H\|^{-1}D^{-\frac{1}{2}+\delta}\right)$
            & $\Omega(\operatorname{Poly}(N)^{-1})$ ?
            & LCE transformed expectation $C(\theta)=\operatorname{Tr}\big(U(\theta)^\dagger |0\rangle\langle 0| U(\theta) H\big)$ with arbitrary PQC $U(\theta)$, and any observable $H$ with $c_{i}=\Omega(1)$ for some $i\in\mathcal{I}$. \\
            \textbf{Small Angles~\cite{wang2024trainability}} & $\Omega(\operatorname{Poly}(N)^{-1})$ on $\mathcal{O}(\|H\|^{-1}D^{-\frac{1}{2}})$ & none & expectation value of the form $\operatorname{Tr}\big(U(\theta)^\dagger |0\rangle\langle 0| U(\theta) H\big)$, and local observable $H$. \\
            \textbf{Park et al.~\cite{park2024hardware}} & $\exists\gamma>0:\quad\Omega(1)$ on $\gamma D^{-1}$ & none & expectation cost function $\operatorname{Tr}\big(U(\theta)^\dagger |\psi_0\rangle\langle \psi_0| U(\theta) H\big)$ with $|\psi_0\rangle$ s.t. $\|(\nabla_\theta C)(0)\|=\Omega(1)$, imposing single Pauli or local observable $H$. \\
            \textbf{Warm Start~\cite{puig2025variational}}
            & $\Omega(\operatorname{Poly}(N)^{-1})$ on $\mathcal{O}(D^{-\frac{1}{2}})$ 
            & none  
            & warm-started cost function $1-|\langle\psi_0|U(\theta)^\dagger e^{-i\delta t H}U(\theta^*)|\psi_0\rangle|^2$ with $\delta t=\Theta(\|H\|^{-1})$ for an observable $H$, and arbitrary PQC $U(\theta)$.
        \end{tabular}
    \end{ruledtabular}
    \caption{\label{tab:comparing-bp-techniques} 
Comparison of barren-plateau mitigation on small patches. Models within the $\mathcal{O}(D^{-1/2})$ patch are classically simulable (cf. Theorem~\ref{thm:probabilistic-complexity}), while only our method {provides numerical experiments with evidence of a larger landscape patch (cf.~Figs.~\ref{fig:escaping-BP} and~\ref{fig:VQE-experiment})} where no classical surrogate is currently known to exist. To simplify the comparison, we may assume that $\|H\|=\mathcal{O}(1)$.}
\end{table*}

Refer to Table~\ref{tab:comparing-bp-techniques} in order to compare barren-plateau mitigation techniques. In this work, we established the \emph{Linear Clifford Encoder (LCE)} which ensures constant scaling gradient lower bounds on $\mathcal{O}(D^{-\frac{1}{2}})$ patches (cf. Theorem~\ref{thm:general-escaping-barren-plateau}), and is currently the only known mitigation technique {providing empirical evidence of} polynomially decaying gradient lower bounds beyond that patch. Notably, Park et al.~\cite{park2024hardware} also established constant lower gradient bounds, where their assumption is satisfied with the special case where the initial state $|\psi_0\rangle=\tilde{Q}U(0)Q|0\rangle$ matches the LCE transformation. However, they achieve these asymptotics solely within the $\mathcal{O}(D^{-1})$ patch, and they also impose locality of the observable, or $H=P$ being a single Pauli string. Puig et al.~\cite{puig2025variational} use a warm start approach in order to achieve polynomially decaying gradients on the larger $\mathcal{O}(D^{-\frac{1}{2}})$ but do not extend their analysis beyond that patch size. Additionally, we compare with uniform small angle initialization of parameters, where Wang et al.~\cite{wang2024trainability} establish polynomially decaying gradients on patches $\mathcal{O}(D^{-\frac{1}{2}})$. Likewise, they require locality of the observable $H$. A Gaussian small angles initialization strategy is studied by Zhang et al.~\cite{zhang2022escaping}, where they employ gradients evaluated at $\theta=0$ to lower bound the expected gradient norm, thus bearing resemblance to our Taylor approximation technique (cf. Theorem~\ref{thm:gradient-lower-bound}). Similarly, they achieve polynomially decaying asymptotics on $\mathcal{O}(D^{-\frac{1}{2}})$ patches, and require locality of the observable.

\subsection{Comparing Classical Surrogates}

\bigskip
\begin{table*}[h]
    \begin{ruledtabular}
        \renewcommand{\arraystretch}{1.5}
        \begin{tabular}{
            >{\centering\arraybackslash}m{3cm}   
            >{\centering\arraybackslash}m{3.5cm} 
            >{\centering\arraybackslash}m{3.5cm} 
            >{\raggedright\arraybackslash}p{5cm} 
        }
            & worst-case simulation of $\mathcal{O}(D^{-1})$ patches
            & MSE simulation of $\mathcal{O}(D^{-1/2})$ patches 
            & model assumptions \\
            \hline
            \textbf{Taylor Surrogate}
            & \vspace{0.5cm}$\mathcal{O}(|\mathcal{I}|D^{m(\theta,\varepsilon)})$ 
            & $\mathcal{O}\Big(|\mathcal{I}|\,D^{\frac{1}{2\log \big(\frac{1}{q}\big)}\log\big(\frac{c}{\rho\,\varepsilon^2}\big)}\Big)$ 
            & expectation $\operatorname{Tr}\big(U(\theta)^\dagger |0\rangle\langle 0| U(\theta) H\big)$ with arbitrary PQC $U(\theta)$, generated by Clifford and single-qubit Pauli rotation gates. \\
            \textbf{Pauli Path Surrogate~\cite{lerch2024efficient}} 
            & $\mathcal{O}\big(|\mathcal{I}|D^{\log\big(\frac{1}{\varepsilon}\big)}\big)$ 
            & $\mathcal{O}\big(|\mathcal{I}|D^{\log\big(\frac{1}{\rho \varepsilon^2}\big)}\big)$  
            & expectation $\operatorname{Tr}(\mathcal{E}_\theta(\rho) H)$ with density matrix $\rho$, and parameterized channel $\mathcal{E}_\theta$, imposing classically feasible derivatives $D_\theta^{\alpha}C(\theta)$ with certain bounds.
        \end{tabular}
    \end{ruledtabular}
    \caption{\label{tab:comparing-surrogates} 
    Comparing classical surrogates of small landscape patches.}
\end{table*}

Table~\ref{tab:comparing-surrogates} compares our Taylor surrogate performance to that relying on the Pauli path technique in the work of Lerch et al.~\cite{lerch2024efficient}. Both techniques are able to simulate the patch $\mathcal{O}(D^{-1})$ in polynomial time up to a certain approximation error $\varepsilon>0$. In our case, the degree of the polynomial is given by the explicitly determined truncation threshold $m(\theta,\varepsilon)$ of the Taylor surrogate (cf. Lemma~\ref{lemma:truncation-threshold}). When allowing a small probability of failure $\rho>0$, both surrogates are able to extend classical simulability to the larger patch $\mathcal{O}(D^{-\frac{1}{2}})$ (cf. Theorem~\ref{thm:probabilistic-complexity}). This is achieved with the mean-squared error:
\begin{equation}
    \mathbb{E}_\theta[(C(\theta)-C_m(\theta))^2] \leq { c\,\|H\|^2 \frac{(D \sigma^{2})^m}{m^m}},
\end{equation}
for some $c > 0$ (cf. Appendix~\ref{sec:truncation}). When initializing the parameters with $\sigma = q\,\|H\|^{-\frac{1}{m}}\,D^{-\frac{1}{2}}$ for some sufficiently small $0.35< q<1$, we are able to prove that the Taylor surrogate outperforms the Pauli path approach (cf. Corollary~\ref{corr:faster-runtime}), where the efficiency of Gottesman-Knill~\cite{gottesman1998heisenberg, aaronson2004improved} starts to emerge.

\section{Stabilizer Formalism and the Gottesman-Knill Theorem \label{sec:gottesman-knill}}

We presume a familiarity of the reader with basic concepts in group theory. The content of this section may be seen as a summary and a slightly alternative point of view of the introductory works on stabilizer formalism in Ref.~\cite{nielsen2010quantum}. We begin by refreshing the necessary terminology, and conclude with a proof sketch of the version of the Gottesman-Knill theorem~\cite{gottesman1998heisenberg, aaronson2004improved} that we employ in this work. For a more in-depth review of the Pauli and Clifford theory, we encourage the reader to look into Refs.~\cite{mastel2023clifford}.

\subsection{Quantum States in an Exponential Hilbert Space}

Consider a $N$-qubit quantum system on which we intend to perform error-free computation. That system may be described by a Hilbert space
\begin{equation}
    \mathcal{H}_N:=\left\{|\psi\rangle = \sum_{b\in\{0,1\}^N}a_b\,|b\rangle: a_b\in\mathbb{C}\right\},
\end{equation}
spanned by the computational basis $\{|b\rangle:b\in\{0,1\}^N\}$, where valid quantum states satisfy the normalization condition,
\begin{equation}
    \sum_{b\in\{0,1\}^N}|a_b|^2=1
\end{equation}
for all $b\in\{0,1\}^N$.

The Hilbert space $\mathcal{H}_N$ is exponentially large and generally difficult to fully describe classically. Of course there are states that are given by polynomially many amplitudes $a_b$ which remain classically feasible. However, a groundbreaking insight lead to the discovery of \emph{stabilizer states} which are quantum states spanned by exponentially many amplitudes that may be classically described using polynomial resources. These states may be derived with basic concepts from group theory.

\subsection{Stabilizer Group}
Stabilizer formalism is motivated from a concept in group theory called \emph{stabilizer}. In the context of the quantum computing, we say that a unitary operator $U\in\operatorname{U}(2^N)$ \emph{stabilizes} a quantum state $|\psi\rangle\in\mathcal{H}_N$ if $U|\psi\rangle=|\psi\rangle$, i.e., $|\psi\rangle$ is an eigenstate of $U$ with eigenvalue $+1$. From a group-theoretic perspective, the \emph{stabilizer} of a quantum state $|\psi\rangle\in\mathcal{H}_N$ is defined as
\begin{equation}
    \operatorname{Stab}(|\psi\rangle):=\{U\in\operatorname{U}(2^N):U|\psi\rangle = |\psi\rangle\}
\end{equation}
consisting of all unitary operators that stabilize the state $|\psi\rangle$. It can be shown that $\operatorname{Stab}(|\psi\rangle)\subset\mathcal{H}_N$ defines an abelien subgroup for every $|\psi\rangle\in\mathcal{H}_N$. Similarly, we say that $|\psi\rangle$ is a \emph{stabilizer} of a unitary operator $U\in\operatorname{U}(2^N)$ if $U|\psi\rangle=|\psi\rangle$.

\subsection{$N$-qubit Pauli Group}
The Pauli matrices are defined as
\begin{equation}
    I=\left(\begin{matrix}1&0\\0&1\end{matrix}\right),\quad X=\left(\begin{matrix}0&1\\1&0\end{matrix}\right),\quad Y=\left(\begin{matrix}0&-i\\ i&0\end{matrix}\right),\quad\text{and}\quad Z=\left(\begin{matrix}1&0\\0&-1\end{matrix}\right),
\end{equation}
where $i$ denotes the imaginary unit. The multiplicative group dynamics of $\mathcal{P}_1:=\langle X,Y,Z\rangle$ (modulo phases $\pm1,\pm i$) is fully determined by the \emph{Cayley table},
\[
\begin{array}{c|ccc}
    & X & Y & Z \\
    \hline
    \rule{0pt}{10pt}
  X & I & iZ & -iY \\
  Y & -iZ & I & iX \\
  Z & iY & -iX & I
\end{array}
\]
In particular, we have the commutator rules
\begin{equation}
    [P,Q]=0\quad\text{or}\quad\{P,Q\}=0
\end{equation}
for all $P,Q\in\mathcal{P}_1$.

The following table represents the set of stabilizer states for each element in the single-qubit Pauli group $\mathcal{P}_1$ (with $-I$ being the sole pathological case):
\[
\begin{array}{c|cccccccc}
    \text{Operator} & I & X & Y & Z & -I & -X & -Y & -Z \\
    \hline
    \rule{0pt}{10pt}
  \text{Stabilizers} & \mathcal{H}_N & |+\rangle & |i\rangle & |0\rangle & \emptyset & |-\rangle & |-i\,\rangle & |1\rangle \\
\end{array}
\]
where the quantum states
\begin{align}
    |0\rangle:=\left(\begin{matrix}
        1 \\ 0
    \end{matrix}\right),\quad |1\rangle:=\left(\begin{matrix}
        0\\1
    \end{matrix}\right)\quad&\text{\emph{($Z$-axis)}} \nonumber
    \\ |\pm\rangle:=\frac{1}{\sqrt{2}}(|0\rangle\pm|1\rangle)\quad&\text{\emph{($X$-axis)}} \nonumber
    \\ |\pm i\,\rangle:=\frac{1}{\sqrt{2}}(|0\rangle\pm i|1\rangle)\quad&\text{\emph{($Y$-axis)}}
\end{align}
are known as the main axes of the \emph{Bloch sphere}.

A generalization of the group $\mathcal{P}_1$ leads to the definition of the $N$-qubit Pauli group:
\begin{equation}
    \mathcal{P}_N:=\left\{\phi\cdot\bigotimes_{j=1}^N P_j:P_j\in\{I,X,Y,Z\},\,\phi\in\{\pm1,\pm i\}\right\},
\end{equation}
and it can be shown that this defines a finite group with $|\mathcal{P}_N|=4^{N+1}$ elements. The elements of $\mathcal{P}_N$ are known as \emph{Pauli strings} or \emph{Pauli words}.

In the Clifford theory, setting the stage of the Gottesman-Knill theorem discussed further below, we are particularly interested in the stabilizer group consisting of Pauli strings only. For that purpose, we introduce the $N$-qubit \emph{Pauli stabilizer group} of a given quantum state $|\psi\rangle\in\mathcal{H}_N$ which is defined as
\begin{equation}
    \operatorname{Stab}_{\mathcal{P}_N}(|\psi\rangle):=\operatorname{Stab}(|\psi\rangle)\cap\mathcal{P}_N.
\end{equation}
The intersection of two groups forms a group, and $\operatorname{Stab}_{\mathcal{P}_N}(|\psi\rangle)$ is abelian since $\operatorname{Stab}(|\psi\rangle)$ is abelian.

\subsection{The Clifford Group and Stabilizer States}
The $N$-qubit Clifford is defined to be the subgroup of $2^N\times 2^N$ unitary operators that map tensor products of Pauli matrices to tensor products of Pauli matrices through conjugation modulo global phases, $U(1)=\{e^{i\theta}:\theta\in[-\pi,\pi]\}$:
\begin{equation}
    \operatorname{Clifford}(N):=\{Q\in\operatorname{U}(2^N):Q\,\mathcal{P}_N\,Q^\dagger = \mathcal{P}_N\} \,\slash\, U(1).
\end{equation}
In other words, these operators normalize the $N$-qubit Pauli group $\mathcal{P}_N$ modulo global phases. The Clifford group is well-known to be finite, and generated by $H$, $S$, and $\operatorname{CX}$-gates, i.e., $\operatorname{Clifford}(N)=\langle H,S,\operatorname{CX}\rangle_N$ (modulo global phases).

The \emph{stabilizer states} are simply given by the set of all states achieved by applying a Clifford operator to the zero state, i.e.,
\begin{equation}
    \operatorname{Stab}(N):=\{Q|0\rangle:Q\in\operatorname{Clifford}(N)\}
\end{equation}
It can be shown that for each $|\psi\rangle\in\operatorname{Stab}(N)$ there exists a set of basis states $S\equiv\{b\}_{b\in S}\subseteq\{0,1\}^N$ and $k_b\in\mathbb{N}_0$ such that
\begin{equation}
    |\psi\rangle = \frac{1}{\sqrt{|S|}}\sum_{b\in S}i^{k_b}|b\rangle
\end{equation}
defines an equal superposition of basis states with relative phases in $\left\{e^{i\phi}:\phi\in\frac{\pi}{2}\mathbb{Z}\right\}$. Stabilizer states may always be efficiently described classically with the help of the Gottesman-Knill theorem, as elaborated next.

\subsection{The Gottesman-Knill Theorem}

We focus on the following version of the Gottesman-Knill theorem~\cite{gottesman1998heisenberg, aaronson2004improved}. We provide a proof sketch in the sense that we skip minor technical details in order to focus on the main idea behind the argumentation. The goal is to develop an intuitive formal understanding on how the group theory in the stabilizer formalism grants access to exponential savings in the classical description of stabilizer states. For more details, see Ref.~\cite{nielsen2010quantum}.

The version of Gottesman-Knill (cf. Theorem~\ref{thm:gottesman-knill}) that we are going to prove states that every stabilizer state $|\psi\rangle\in\operatorname{Stab}(N)$ is \emph{uniquely} determined by its underlying Pauli stabilizer group $\operatorname{Stab}_{\mathcal{P}_N}(|\psi\rangle)=\langle g_1,\dots,g_N\rangle$, generated by $N$ independent generators $g_1,\dots,g_N\in\mathcal{P}_N$. The generators $g_1,\dots,g_N$ are efficiently described classically with the help of \emph{check matrices}~\cite{nielsen2010quantum, aaronson2004improved}, which establishes that stabilizer states are indeed easy to simulate classically. The proof of Theorem~\ref{thm:gottesman-knill} below breaks down into three main steps:
\begin{itemize}

    \item[I.] \emph{Zero State Description:} First, we establish by induction that the zero state $|0\rangle\in\operatorname{Stab}(N)$ may always be described in terms of group generators $Z_1,\dots,Z_N$ of the corresponding Pauli stabilizer group $\operatorname{Stab}_{\mathcal{P}_N}(|0\rangle)$. That is, $\operatorname{Stab}_{\mathcal{P}_N}(|0\rangle)=\langle Z_1,\dots,Z_N\rangle$.
    
    \item[II.] \emph{Zero State Evolution:} Next, we employ the previous step to prove that any stabilizer state $|\psi\rangle\in\operatorname{Stab}(N)$ may be fully described by evolving the zero state through the underlying Clifford operator. Indeed, by definition there exists $C\in\operatorname{Clifford}(N)$ such that $|\psi\rangle=C|0\rangle$. With the help of the previous step, we then establish that the generators of $\operatorname{Stab}_{\mathcal{P}_N}(|\psi\rangle)=\langle g_1,\dots,g_N\rangle$ are simply obtained by evolving the generators of $\operatorname{Stab}_{\mathcal{P}_N}(|0\rangle)$ through $C$ in the Heisenberg picture, i.e., $g_j=CZ_jC^\dagger$ for all $j\in[N]$. This establishes that each stabilizer state is fully described by $N$ generators.
    
    \item[III.] \emph{Stabilizer State Uniqueness:} Finally, it remains to prove that the generators of $\operatorname{Stab}_{\mathcal{P}_N}(|\psi\rangle)$ are uniquely determined. This is achieved by introducing a certain projector operator $P_{\operatorname{Stab}_{\mathcal{P}_N}(|\psi\rangle)}:\mathcal{H}_N\to V_{\operatorname{Stab}_{\mathcal{P}_N}(|\psi\rangle)}$, that projects the full Hilbert space $\mathcal{H}_N$ onto a certain linear subspace $V_{\operatorname{Stab}_{\mathcal{P}_N}(|\psi\rangle)}$ characterized by the Pauli stabilizer group $\operatorname{Stab}_{\mathcal{P}_N}(|\psi\rangle)$. Afterwards, by using the fact that non-identity Pauli operators are traceless, we establish that the dimension of the subspace $V_{\operatorname{Stab}_{\mathcal{P}_N}(|\psi\rangle)}$ is equal to one. Thus, by normalization of quantum state, this shows that the group generators $g_1,\dots,g_N$ of $\operatorname{Stab}_{\mathcal{P}_N}(|\psi\rangle)$ are sufficient to uniquely determine the stabilizer state $|\psi\rangle$.
    
\end{itemize}

\begin{theorem}[Gottesman-Knill]
    Every stabilizer state $|\psi\rangle\in\operatorname{Stab}(N)$ is uniquely determined by its Pauli stabilizer group, $\operatorname{Stab}_{\mathcal{P}_N}(|\psi\rangle)$. Moreover, the group $\operatorname{Stab}_{\mathcal{P}_N}(|\psi\rangle)$ is spanned by $N$ pair-wise independent generators. \label{thm:gottesman-knill}
\end{theorem}
\noindent
\emph{Proof Sketch.} Let $|\psi\rangle\in\operatorname{Stab}(N)$ be arbitrary. Then, by definition, $|\psi\rangle=C|0\rangle$ for some $C\in\operatorname{Clifford}(N)$. By induction, one establishes that the group
\begin{equation}
    \operatorname{Stab}_{\mathcal{P}_N}(|0\rangle)=\langle Z_1,\dots,Z_N\rangle, \label{eq:base-group}
\end{equation}
is uniquely generated by the pair-wise independent Pauli strings $Z_1,\dots,Z_N$, where we use the conventional notations $V_j:=I_{2^{j-1}}\otimes V \otimes I_{2^{N-j}}$. We use this as a basis to show that
\begin{equation}
    \operatorname{Stab}_{\mathcal{P}_N}(|\psi\rangle) = \langle CZ_1 C^\dagger,\dots,C Z_N C^\dagger \rangle \label{eq:generated-group}
\end{equation}
is uniquely generated by the pair-wise independent Pauli strings $CZ_1C^\dagger,\dots, CZ_NC^\dagger$, where $C\in\operatorname{Clifford}(N)$. To prove Eq.~\eqref{eq:generated-group}, first note that
\begin{align}
    P\in\operatorname{Stab}_{\mathcal{P}_N}(C|0\rangle) \iff PC|0\rangle = C|0\rangle \iff (C^\dagger P C)|0\rangle=|0\rangle \iff \underbrace{C^\dagger P C}_{\in \mathcal{P}_N}\in \operatorname{Stab}_{\mathcal{P}_N}(|0\rangle), \label{eq:prep-equiv}
\end{align}
again using $C\in\operatorname{Clifford}(N)$. Next, we apply Eq.~\eqref{eq:base-group} to establish that there exist exponents $e_1,\dots,e_N\in\{0,1\}$ such that
\begin{align}
    C^\dagger P C\in \operatorname{Stab}_{\mathcal{P}_N}(|0\rangle) = \langle Z_1,\dots, Z_N\rangle &\iff C^\dagger P C= Z_1^{e_1}\cdots Z_N^{e_N} \nonumber
    \\ &\iff P = C\left(\prod_{j=1}^N Z_j^{e_j}\right) C^\dagger = \prod_{j=1}^N(CZ_jC^\dagger)^{e_j} \nonumber
    \\ &\iff P\in\langle CZ_1C^\dagger,\dots, CZ_NC^\dagger\rangle, \label{eq:equivalence-chain}
\end{align}
using $C^\dagger C=CC^\dagger=I$. Combined with Eq.~\eqref{eq:prep-equiv} this proves the equality in Eq.~\eqref{eq:generated-group}. Next, we show that $CZ_1C^\dagger,\dots,CZ_NC^\dagger$ are independent generators. For that purpose, we assume for absurd that the contrary is true. Then there exists an index $i\in[N]$ such that
\begin{equation}
    \operatorname{Stab}_{\mathcal{P}_N}(C|0\rangle) = \langle CZ_jC^\dagger\rangle_{j\in[N]} = \langle CZ_jC^\dagger\rangle_{j\in[N]\setminus\{i\}}. 
\end{equation}
But by the previous equivalence in Eq.~\eqref{eq:equivalence-chain}, this would imply $Z_i\in\langle Z_j\rangle_{j\in[N]}=\langle Z_j\rangle_{j\in[N]\setminus\{i\}}$, which is absurd.

It remains to show that the group $\operatorname{Stab}_{\mathcal{P}_N}(|\psi\rangle)$ (and thus the set of its generators) uniquely determines the stabilizer state $|\psi\rangle$. Consider the space
\begin{equation}
    V_{\operatorname{Stab}_{\mathcal{P}_N}(|\psi\rangle)}:= \{|v\rangle\in\mathcal{H}_N: P|v\rangle=|v\rangle\text{ for all $P\in \operatorname{Stab}_{\mathcal{P}_N}(|\psi\rangle)$}\}. \label{eq:dimension}
\end{equation}
It can be shown that $V_{\operatorname{Stab}_{\mathcal{P}_N}(|\psi\rangle)}\subset\mathcal{H}_N$ defines a linear subspace, and it consists of all states that are fixed by elements of the Pauli stabilizer group $\operatorname{Stab}_{\mathcal{P}_N}(|\psi\rangle)$ (in particular by its generators). We are going to establish that its dimension equals $\dim V_{\operatorname{Stab}_{\mathcal{P}_N}(|\psi\rangle)}=1$. Indeed, this completes the proof because by normalization of quantum states, together with the one-dimensionality, Eq.~\eqref{eq:dimension} implies the existence of a unique quantum state $|\psi\rangle\in\mathcal{H}_N$.

By Eq.~\eqref{eq:generated-group} there exist Pauli generators $g_1,\dots,g_n\in\mathcal{P}_N$ such that $\operatorname{Stab}_{\mathcal{P}_N}(|\psi\rangle)=\langle g_1,\dots,g_N\rangle$. Consider the operator
\begin{equation}
    P_{\operatorname{Stab}_{\mathcal{P}_N}(|\psi\rangle)}:=\frac{1}{2^N}\prod_{j=1}^N(I+g_j).
\end{equation}
Then, we note that $P_{\operatorname{Stab}_{\mathcal{P}_N}(|\psi\rangle)}:\mathcal{H}_N\to V_{\operatorname{Stab}_{\mathcal{P}_N}(|\psi\rangle)}$ defines a projector. Indeed, for each $j\in[N]$ we have
\begin{align}
    \left(\frac{I+g_j}{2}\right)^2 &= \frac{1}{4}(I+2g_j+g_j^2) = \frac{I+g_j}{2} \quad\text{and}\quad \left(\frac{I+g_j}{2}\right)|\psi\rangle = \frac{|\psi\rangle + g_j|\psi\rangle}{2}=|\psi\rangle,
\end{align}
using $g_j^2=I$ and $g_j|\psi\rangle=|\psi\rangle$ since $g_j\in\langle g_j\rangle_{j\in[N]}=\operatorname{Stab}_{\mathcal{P}_N}(|\psi\rangle)$ is a Pauli string that stabilizes $|\psi\rangle$. In other words, for each $j\in[N]$, $\frac{1}{2}(I+g_j)$ is the projector onto the $+1$-eigenspace of $g_j$. Taking the product, this shows that $P_{\operatorname{Stab}_{\mathcal{P}_N}(|\psi\rangle)}$ defines the claimed projector, and it can be shown that it is surjective.

The crucial observation now is that this projector decomposes as
\begin{align}
    P_{\operatorname{Stab}_{\mathcal{P}_N}(|\psi\rangle)}=\frac{1}{2^N}\prod_{j=1}^N(I+g_j)=\frac{1}{2^N}\sum_{e_1,\dots,e_n\in\{0,1\}} g_1^{e_1}\cdots g_N^{e_N} = \frac{1}{2^N}\sum_{g\in\operatorname{Stab}_{\mathcal{P}_N}(|\psi\rangle)}g,
\end{align}
where the second equality is obtained by expanding the product, and the fact that all $g_j$ commute pair-wise (else they must anti-commute which is impossible because of $g_jg_i|\psi\rangle=-g_ig_j|\psi\rangle = -|\psi\rangle$). The last equality simply follows from $\operatorname{Stab}_{\mathcal{P}_N}(|\psi\rangle)=\langle g_1,\dots,g_N\rangle$. This observation immediately implies that taking its trace yields
\begin{equation}
    \operatorname{Tr}P_{\operatorname{Stab}_{\mathcal{P}_N}(|\psi\rangle)} = \frac{1}{2^N}\sum_{g\in\operatorname{Stab}_{\mathcal{P}_N}(|\psi\rangle)}\operatorname{Tr}g = \frac{1}{2^N}\underbrace{\operatorname{Tr}I}_{=2^N} = 1,
\end{equation}
using that $\operatorname{Tr}g = 0$ if and only if $g\neq I$. On the other hand, since the trace of an operator equals the sum of its eigenvalues, we find
\begin{equation}
    1=\operatorname{Tr}P_{\operatorname{Stab}_{\mathcal{P}_N}(|\psi\rangle)} = \sum_{k=1}^{2^N}\underbrace{\lambda(\operatorname{Stab}_{\mathcal{P}_N}(|\psi\rangle))}_{\in\{0,1\}} = \dim\operatorname{Image}(P_{\operatorname{Stab}_{\mathcal{P}_N}(|\psi\rangle)}) = \dim V_{\operatorname{Stab}_{\mathcal{P}_N}(|\psi\rangle)}.
\end{equation}
Here, we used that the eigenvalues of a projector are either 0 or 1. The sum over all eigenvalues of the (linear) projector operator $P_{\operatorname{Stab}_{\mathcal{P}_N}(|\psi\rangle)}:\mathcal{H}_N\to V_{\operatorname{Stab}_{\mathcal{P}_N}(|\psi\rangle)}$ thus counts the dimensions that are not in its kernel, and thus counts the dimension of its image, which precisely equals the dimension of the subspace $V_{\operatorname{Stab}_{\mathcal{P}_N}(|\psi\rangle)}$, proving Eq.~\eqref{eq:dimension}. This completes the proof sketch.
\qed
\bigskip
\\
\indent
This version of Gottesman-Knill proves that there exist states $|\psi\rangle$ (stabilizer states) with $2^N$ non-zero amplitudes, that are classically fully characterized by only $N$ generators of the underlying Pauli stabilizer group $\operatorname{Stab}_{\mathcal{P}_N}(|\psi\rangle)$. Aaronson and Gottesman developed an efficient method, known as the \emph{check matrix algorithm}~\cite{aaronson2004improved}, establishing that a stabilizer state $C|0\rangle$ requires space resources of order $\mathcal{O}(D_C^2)$, where $D_C$ is the number of elementary gates in the circuit $C\in\operatorname{Clifford}(N)$. Furthermore, they showed that the time resources required to obtain $C|0\rangle$ from $|0\rangle$ is of order $\mathcal{O}(D_C)$. In our setting, where we consider a PQC $U(\theta)$ generated by $D_C\geq 1$ elementary Clifford gates and $D\geq1$ parameterized single-qubit (generally non-Clifford) rotation gates, we may generally assume that $D_C = c\,D$ for some constant factor $c>0$, so that simulating $U(0)|0\rangle$ requires $\mathcal{O}(D^2)$ space resources, since $U(0)\in\operatorname{Clifford}(N)$.

Next, we show that simulating the expectation $\langle 0|U(0)^\dagger H U(0)|0\rangle$ yields a runtime of order $\mathcal{O}(D\,|\mathcal{I}|)$, where $|\mathcal{I}|$ is the number of Pauli terms in the observable $H$.

\subsection{Computing Expectations with the Stabilizer Formalism}

We may use the Gottesman-Knill theorem (Theorem~\ref{thm:gottesman-knill}) in order to prove that Hamiltonian expectations of the form $\langle0|C^\dagger H C|0\rangle$ may be computed in polynomial time for each Clifford operator $C\in\operatorname{Clifford}(N)$. The result particularly holds true for PQCs evaluated at Clifford configurations, i.e., $C=U(0)$.
\bigskip
\begin{corollary}\label{corr:stabilizer-expectation}
    Let $|\psi\rangle\in\operatorname{Stab}(N)$ be a stabilizer state, and $P\in\mathcal{P}_N$ a Pauli string. Let $g_1,\dots,g_N\in\mathcal{P}_N$ be the generators of Pauli stabilizer group $\operatorname{Stab}_{\mathcal{P}_N}(|\psi\rangle)$. Then,
    \begin{equation*}
        \langle\psi|P|\psi\rangle = \begin{cases}
            \operatorname{sign}P, & \text{if $[P,g_j]=0$ for all $j\in[N],$} \\ 0,&\text{if there exists $j\in[N]$ such that $\{P,g_j\}=0$,}
        \end{cases}
    \end{equation*}
    where $\operatorname{sign}P\in\{\pm1\}$ denotes the sign of the Pauli string.
\end{corollary}
\noindent
\emph{Proof.} Since $P\in\mathcal{P}_N$ is a Pauli string, it has eigenvalues in $\{\pm1\}$. Therefore, we have
\begin{equation}
    \langle\psi|P|\psi\rangle = p(+1)\cdot 1 + p(-1)\cdot (-1) = p(+1)-p(-1),
\end{equation}
where $p(\pm1)$ is the probability of measuring the outcome $\pm1$ when the system is prepared in the state $|\psi\rangle$. By linearity and the cyclic property of the trace operator, we first establish that
\begin{align}
    p(+1)-p(-1) = \langle\psi|P|\psi\rangle &= \operatorname{Tr}(P|\psi\rangle\langle\psi|) = \operatorname{Tr}\left(\frac{I+P}{2}|\psi\rangle\langle\psi|\right) - \operatorname{Tr}\left(\frac{I-P}{2}|\psi\rangle\langle\psi|\right), \label{eq:stabilizer-expectation}
\end{align}
and by Born's rule, it immediately follows that
\begin{equation}
    p(\pm1)=\operatorname{Tr}\left(\frac{I\pm P}{2}|\psi\rangle\langle\psi|\right).  \label{eq:binary-probability}
\end{equation}
Therefore, the projectors $\frac{1}{2}(I\pm P)$ determine the probability distributions $p(\pm1)$ (of the outcomes $\pm1$) needed for the computation of the expectation value $\langle\psi|P|\psi\rangle$. It turns out that certain commutation rules of the Pauli operators involved suffice to explicitly determine these probabilities. More precisely, if the following holds true:
\begin{equation}
    \begin{cases}
        \text{$[P,g_j]=0$ for all $j\in[N] \implies p(\operatorname{sign} P )=1$ and $p(-\operatorname{sign} P)=0$ \quad\emph{(deterministic outcome)}}
        \\ \text{there exists $j\in[N]$ such that $\{P,g_j\}=0 \implies p(+1)=p(-1)=\frac{1}{2}$ \quad\emph{(coin flip outcome)}}
    \end{cases} \label{eq:stabilizer-expectation2}
\end{equation}
then, combining Eq.~\eqref{eq:stabilizer-expectation2} with Eq.~\eqref{eq:stabilizer-expectation} would yield
\begin{equation}
    \langle\psi|P|\psi\rangle = p(+1) - p(-1) \stackrel{\eqref{eq:stabilizer-expectation2}}{=} \begin{cases}
                \operatorname{sign}P, & \text{if $[P,g_j]=0$ for all $j\in[N],$} \\ 0,&\text{if there exists $j\in[N]$ such that $\{P,g_j\}=0$.}
            \end{cases}
\end{equation}

Therefore, it remains to prove Eq.~\eqref{eq:stabilizer-expectation2}. First, let us handle the case where $[P,g_j]=0$ for all $j\in[N]$. Recall the definition of the linear subspace $V_{\operatorname{Stab}_{\mathcal{P}_N}(|\psi\rangle)}$ in Eq.~\eqref{eq:dimension} introduced in the proof of Theorem~\ref{thm:gottesman-knill}. By assumption, we have for all $j\in[N]$ that
\begin{equation}
    g_jP|\psi\rangle = Pg_j|\psi\rangle = P|\psi\rangle \implies P|\psi\rangle\in V_{\operatorname{Stab}_{\mathcal{P}_N}(|\psi\rangle)}.
\end{equation}
We are going to use this observation in order to establish that $p(\operatorname{sign} P) = 1$. Indeed, since $\dim V_{\operatorname{Stab}_{\mathcal{P}_N}(|\psi\rangle)} = 1$ (as shown in Theorem~\ref{thm:gottesman-knill}), we know that there exists a phase $\phi\in\mathbb{C}$ such that $P|\psi\rangle = \phi|\psi\rangle$. Since $P\in\mathcal{P}_N$, we must have that $\phi\in\{\pm1\}$ (eigenvalues of $P$). As a result, we find that $P|\psi\rangle = \operatorname{sign}(P)|\psi\rangle$. Consequently,
\begin{align}
    p(\operatorname{sign}P) \stackrel{\eqref{eq:binary-probability}}{=} \operatorname{Tr}\left(\frac{I+\operatorname{sign}(P)\,P}{2}|\psi\rangle\langle\psi|\right) = \operatorname{Tr}\left(\frac{I+\operatorname{sign}(P)^2 I}{2}|\psi\rangle\langle\psi|\right) = \operatorname{Tr}(|\psi\rangle\langle\psi|)=1.
\end{align}
In particular, $p(-\operatorname{sign}P) = 1-p(\operatorname{sign}P)=1-1=0$, which proves the deterministic outcome in Eq.~\eqref{eq:stabilizer-expectation2}.

On the other hand, if there exists $j\in[N]$ such that $\{P,g_j\}=0$, then
\begin{equation}
    g_jP|\psi\rangle = -Pg_j|\psi\rangle = -P|\psi\rangle,
\end{equation}
which in turn, once again making use of the cyclic property of the trace operator, implies that
\begin{align}
    p(-1) &\stackrel{\eqref{eq:binary-probability}}{=} \operatorname{Tr}\left(\frac{I-P}{2}|\psi\rangle\langle\psi|\right) = \operatorname{Tr}\left(\frac{I+g_jP}{2}|\psi\rangle\langle\psi|\right) = \frac{1}{2}\langle\psi|\psi\rangle + \frac{1}{2}\underbrace{\langle\psi|g_j}_{\langle\psi|}P|\psi\rangle \nonumber
    \\&\quad= \frac{1}{2}\langle\psi|(I+P)|\psi\rangle = \operatorname{Tr}\left(\frac{I+P}{2}|\psi\rangle\langle\psi|\right) = p(+1).
\end{align}
In particular, $p(+1)=p(-1)=\frac{1}{2}$, as claimed.
\qed
\bigskip
\\
\indent
As previously established by Theorem~\ref{thm:gottesman-knill} we already know that all states involved only require polynomial classical space resources. Together with Corollary~\ref{corr:stabilizer-expectation} this implies that any cost function, $C(0)=\langle0|U(0)^\dagger HU(0)|0\rangle$, evaluated at Clifford configurations (without loss of generality $\theta=0$) may be classically evaluated in polynomial time. Indeed, the time complexity of classically evaluating $\langle\psi|P|\psi\rangle$ for a single Pauli string $P\in\mathcal{P}_N$ scales as $\mathcal{O}(N)$ since at most $N$ of the generators $g_1,\dots,g_N$ need to be compared with $P$ in order to determine the commutator rules in Eq.~\eqref{eq:stabilizer-expectation2}. Doing this for each of the Pauli terms in the observable yields a runtime scaling as $\mathcal{O}(N\,|\mathcal{I}|)$, where $|\mathcal{I}|$ is the number of Pauli terms in the observable $H$.

In this work we make the assumption that the ansatz $U(\theta)$ satisfies that $D=\Theta(N)$ (i.e., the number of parameters is a multiple of the number of qubits) so that the time complexity of evaluating $C(0)=\langle0|U(0)^\dagger HU(0)|0\rangle$ scales as $\mathcal{O}(D\,|\mathcal{I}|)$ as established in our complexity theory results.

\section{PQC Ansatz Architectures \label{sec:ansatz}}
In this work, we consider three PQC model ansätze with $L\geq1$ layers:

\begin{enumerate}
    \item \texttt{mHEA}~\cite{leone2024HEA}: minimalistic Hardware Efficient Ansatz based on Qiskit's \texttt{EfficientSU2}~\cite{efficientsu2} with parameterized $Y$- and $Z$-single-qubit rotation layers, and circular entanglements. The number of parameters satisfies $D=2N(L+1)$.
    \item \texttt{fHEA}: full Hardware Efficient Ansatz based on Qiskit's \texttt{EfficientSU2}~\cite{efficientsu2} with parameterized $X$-, $Y$- and $Z$-single-qubit rotation layers, and full entanglements. The number of parameters satisfies $D=3N(L+1)$.
    \item \texttt{rPQC}: We construct a layer-wise random PQC as follows: independently for each layer, we apply a layer of single-qubit $H$-gates followed by a layer of single-qubit $S$-gates. Each single-qubit gate appears with $0.5$ probability. Then, we apply $N$ uniformly random pairs of $\operatorname{CX}$-gates. The number of parameters satisfies $D=NL$.
\end{enumerate}

Fig.~\ref{fig:PQC-models} illustrates representative examples of the PQC models used in our experiments after applying the LCE transformation $(Q,\tilde{Q})$. The circuit starts with the Clifford $Q$ to diagonalize the Pauli string, and ends with $\tilde{Q}$, used to induce a relative phase between two parameter-shifted terms $C(\pm\pi/2\,e_k)$. Panel (a) shows \texttt{mHEA} with $N=3$ qubits and $D=18$ parameters, LCE transformed using $k=10$ and Pauli string $P=XXZ$. Panel (b) presents \texttt{fHEA} with $N=3$ qubits and $D=27$ parameters, LCE transformed using $k=16$ and $P=IIZ$. Panel (c) depicts \texttt{rPQC} with $N=3$ qubits, $D=6$ parameters, LCE transformed using $k=4$ and $P=IXI$.

\begin{figure}[t]
    \centering
    \begin{subfigure}(a)
        \centering
        \includegraphics[width=\linewidth]{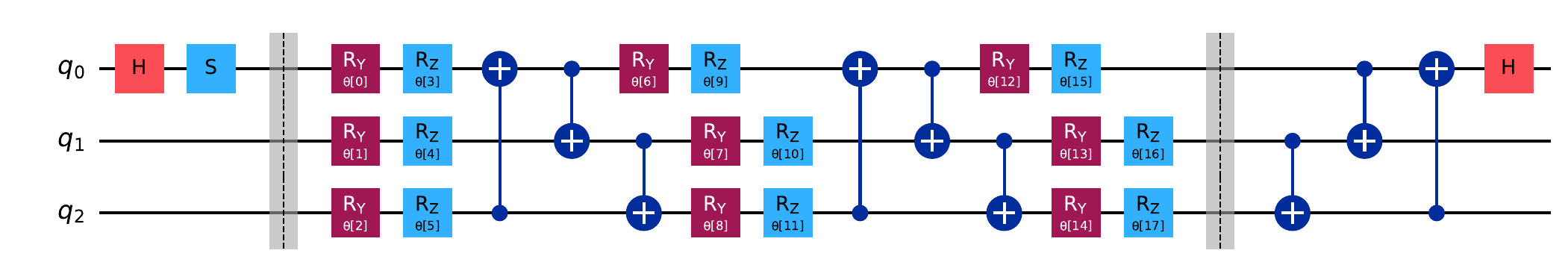}
    \end{subfigure}
    
    \begin{subfigure}(b)
        \centering
        \includegraphics[width=\linewidth]{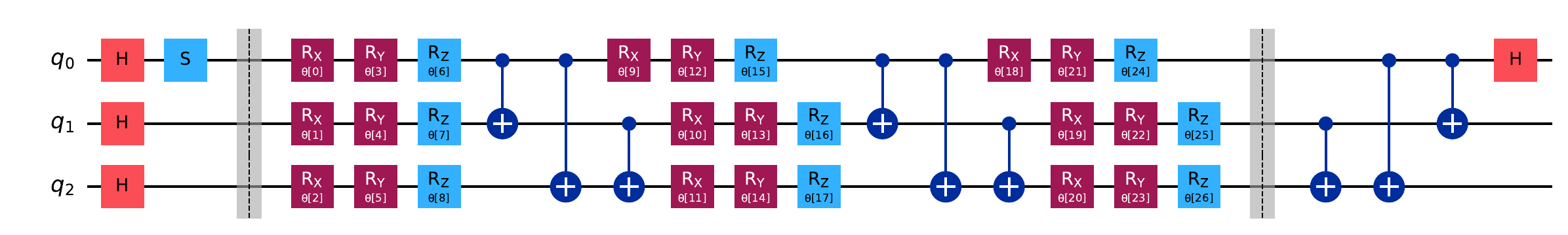}
    \end{subfigure}
    
    \begin{subfigure}(c)
        \centering
        \includegraphics[width=\linewidth]{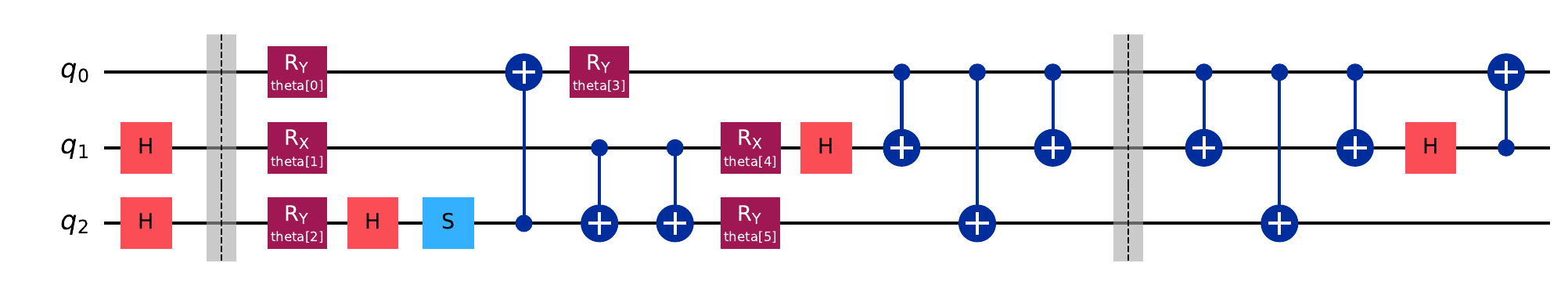}
    \end{subfigure}

    \caption{Examples of two-layer PQCs after LCE transformation. (a) \texttt{mHEA}-circuit with $N=3$, $D=18$, $k=10$, $P=XXZ$. (b) \texttt{fHEA}-circuit with $N=3$, $D=27$, $k=16$, $P=IIZ$. (c) \texttt{rPQC}-circuit with $N=3$, $D=6$, $k=4$, $P=IXI$.}
    \label{fig:PQC-models}
\end{figure}

\section{Comparing State-of-the-Art Simulation Techniques \label{sec:simulation}}
In general, the computational resources needed to evaluate the cost $C(\theta)$ in Eq.~\eqref{eq:expectation} is characterized by the number $\mathcal{O}(|\mathcal{J}|)$ of operations in an estimator of the form
\begin{equation}
    C(\theta)\approx\sum_{j\in \mathcal{J}}f_j(\theta), \label{eq:decomposition-complexity}
\end{equation}
where $f_j:[-\pi,\pi]^D\to\mathbb{R}$ are certain parameter-dependent functions.

\subsection{Naive Evaluation of the Exponential Quadratic Form}
Recall that, if the $n$-qubit system is represented by a Hilbert space $\mathcal{H}_N$ with state vectors of the general form
\begin{equation}
    |\psi\rangle = \sum_{b\in\{0,1\}^N}a_b\,|b\rangle\quad\text{with amplitudes}\quad a_b\in\mathbb{C}:\sum_{b\in\{0,1\}^N}|a_b|^2=1,
\end{equation}
in the computational basis, then the cost is naively expressed as the exponential quadratic form (exponentially large matrix, and exponentially many terms depending on the state vector),
\begin{equation}
    C(\theta) \stackrel{\eqref{eq:expectation}}{=} \langle \psi(\theta)| H |\psi(\theta)\rangle = \sum_{b,b'\in\{0,1\}^N} a_b^*\,a_{b'}\,\langle b |H|b'\rangle = \sum_{b,b'\in\{0,1\}^N}\sum_{i\in\mathcal{I}} a_b^*\,a_{b'}\,\langle b |P_i|b'\rangle. \label{eq:expectation2}
\end{equation}
In that case, we find $|\mathcal{J}|=|\mathcal{I}|\,\#\{0,1\}^{2N}=2^{2N}$, thus implying a worst-case runtime of $\mathcal{O}(|\mathcal{I}|\,2^{2N})$.

The computational cost of evaluating the quadratic form (\ref{eq:expectation}) inherently depends on the specific parameter configuration $\theta$. In general, estimating quantum expectation values is exponentially hard~\cite{bravyi2016improved, bravyi2019simulation, beguvsic2023simulating}, but depending on the values of the parameters we can classically approximate them more efficiently. For example, if $\theta=0$, then we can simulate $C(\theta=0)$ in polynomial time due to the Gottesman-Knill theorem~\cite{gottesman1998heisenberg, aaronson2004improved}.

In what follows, we provide a brief comparison of state-of-the-art simulation techniques for approximating the expectation value $C(\theta)$ as a function of $\theta$. A short overview is found in Table~\ref{tab:simulations} below.

\begin{table*}[h]
    \begin{ruledtabular}
        \renewcommand{\arraystretch}{1.5}
        \begin{tabular}{cccc}
            & exact runtime & circuit restrictions & complexity in terms of $\theta$ \\
            \hline
            \rule{0pt}{3ex}
            Low-Rank Stabilizer Decomposition~\cite{bravyi2016improved, bravyi2019simulation} & $\mathcal{O}(|\mathcal{I}|\,2^{0.48D})$ & 3-locality (cf. Eq. (\ref{eq:lightcone})) & bounding $m(\theta)$ (cf. Fig.~\ref{fig:bravyi}) \\
            Clifford Perturbation Theory~\cite{beguvsic2023simulating, lerch2024efficient} & $\mathcal{O}(|\mathcal{I}|\,2^D)$ & none  & none \\
            Truncated Taylor series & $\mathcal{O}(|\mathcal{I}|\,2^{2D})$ & none & $m = \Theta(\|\theta\|_1)$ (cf. Lemma~\ref{lemma:truncation-threshold})
        \end{tabular}
    \end{ruledtabular}
    \caption{\label{tab:simulations} 
    Comparing state-of-the-art simulation techniques with our classical Taylor surrogate method. These are applied to simulate near-Clifford circuits. The comparison is made in terms of the worst-case runtime required for exactly computing $C(\theta)$, assumptions on the quantum model ansatz, and the analysis performed of the computational complexity as a function of $\theta$.}
\end{table*}

\subsection{Low-Rank Stabilizer Decomposition}
Taking into consideration a parameterized state $|\psi(\theta)\rangle=U(\theta)|0\rangle$, setting the parameters to $0$ results in a stabilizer state~\cite{gottesman1998heisenberg, aaronson2004improved}, $|\psi(0)\rangle\in\operatorname{Stab}(N)$. Such states can be efficiently represented and manipulated using Aaronson-Gottesman’s check matrix algorithm~\cite{ aaronson2004improved} leveraging the Gottesman-Knill theorem~\cite{gottesman1998heisenberg} (cf. Theorem~\ref{thm:gottesman-knill}).

This observation led to the classical simulation technique of Bravyi et al.~\cite{bravyi2016improved, bravyi2019simulation} which is based on a \emph{low-rank stabilizer decomposition} of $|\psi(\theta)\rangle$. Specifically,
\begin{equation}
    |\psi(\theta)\rangle \approx \sum_{\ell=1}^{m} \gamma_\ell(\theta) \, |\phi_\ell\rangle, \label{eq:stabilizer-decomposition} 
\end{equation}
where each $|\phi_\ell\rangle$ defines a stabilizer state, efficiently described within the stabilizer formalism, and $\gamma_\ell(\theta)$ are parameter-dependent complex coefficients.

The computational cost of calculating the expectation value in Eq.~\eqref{eq:expectation2} is directly characterized by the decomposition number $m=\chi(\theta)$ (as a function of $\theta$), also known as the \emph{stabilizer rank}. Also, note that $m=\sqrt{|\mathcal{J}|}$ in Eq.~\eqref{eq:decomposition-complexity} since taking the expectation value requires two copies of the low-rank stabilizer decomposition, hence an extra square root. In particular, starting from a Clifford configuration $\theta=0$, as $U(\theta)$ approaches a circuit containing an increasing number of non-Clifford gates, the number of terms in the decomposition (\ref{eq:stabilizer-decomposition}) grows, ultimately leading to an exponential scaling of the simulation cost (cf. Eq.~\eqref{eq:expectation2}).

Using the stabilizer decomposition, the expectation value can then be approximated as
\begin{equation}
    C(\theta) \approx \sum_{\ell,\ell'=1}^{m}\gamma_\ell(\theta)\,\gamma_{\ell'}(\theta)\,\langle\phi_\ell|H|\phi_{\ell'}\rangle, \label{eq:stabilizer-decomposition-expectation}
\end{equation}
with the approximation error decreasing as $m$ increases. As of this writing, the most efficient algorithm for implementing Eq.~\eqref{eq:stabilizer-decomposition-expectation} is reported by Bravyi et al.~\cite{bravyi2019simulation}, achieving a worst-case computational complexity of $\mathcal{O}(|\mathcal{I}|\,2^{0.48D})$.

Unfortunately, their analysis assumes a circuit architecture limited to $3$-local qubit gates~\cite[Proposition 1]{bravyi2019simulation}, thereby imposing a lightcone constraint that reduces the computational complexity of their method. More precisely, they require that
\begin{equation}
    |\psi(\theta)\rangle = |\psi_1\rangle\otimes\cdots\otimes|\psi_n\rangle \label{eq:lightcone}
\end{equation}
with each state $|\psi_j\rangle$ describing a system of at most three qubits. Consequently, this strongly limits the expressive power of the ansatz model.

In Fig.~\ref{fig:bravyi}, we analyze the stabilizer rank $m = \chi(\theta)$~\cite[Eq.~(14)]{bravyi2019simulation} as a function of the number of qubits $N$, under the same experimental setting as Fig.~\ref{fig:truncation-statistics}. In panel (a), where all parameters are deterministically initialized as $\theta_1 = \cdots = \theta_D$, we observe a clear phase transition in computational complexity, consistent with Fig.~\ref{fig:truncation-statistics}. The red-shaded region indicates the barren-plateau-free regime characterized in Theorem~\ref{thm:general-escaping-barren-plateau}. In contrast, panel (b) shows that for randomly initialized parameters $\theta \sim \mathcal{N}(0, \sigma^2 I_D)$, the decomposition rank $m = \chi(\theta)$ exhibits no meaningful dependence on $N$, since $m \geq 1$ is always a positive integer. This demonstrates that the low-rank stabilizer decomposition technique fails to capture the statistical behavior of $\chi(\theta)$ under random initialization $\mathbb{P}_\theta$. The key issue is that the formula used in~\cite{bravyi2019simulation} assumes parameter correlations not present under i.i.d.\ initializations, thereby violating the conditions required for the law of large numbers. In contrast, our formulation $m = m(\|\theta\|_1)$ satisfies these assumptions asymptotically (cf.~Lemma~\ref{lemma:truncation-threshold}). As a result, the framework of Bravyi et al. is unsuitable for testing initialization strategies based on random sampling.

\begin{figure*}[t]
    \centering
    \includegraphics[width=1\textwidth]{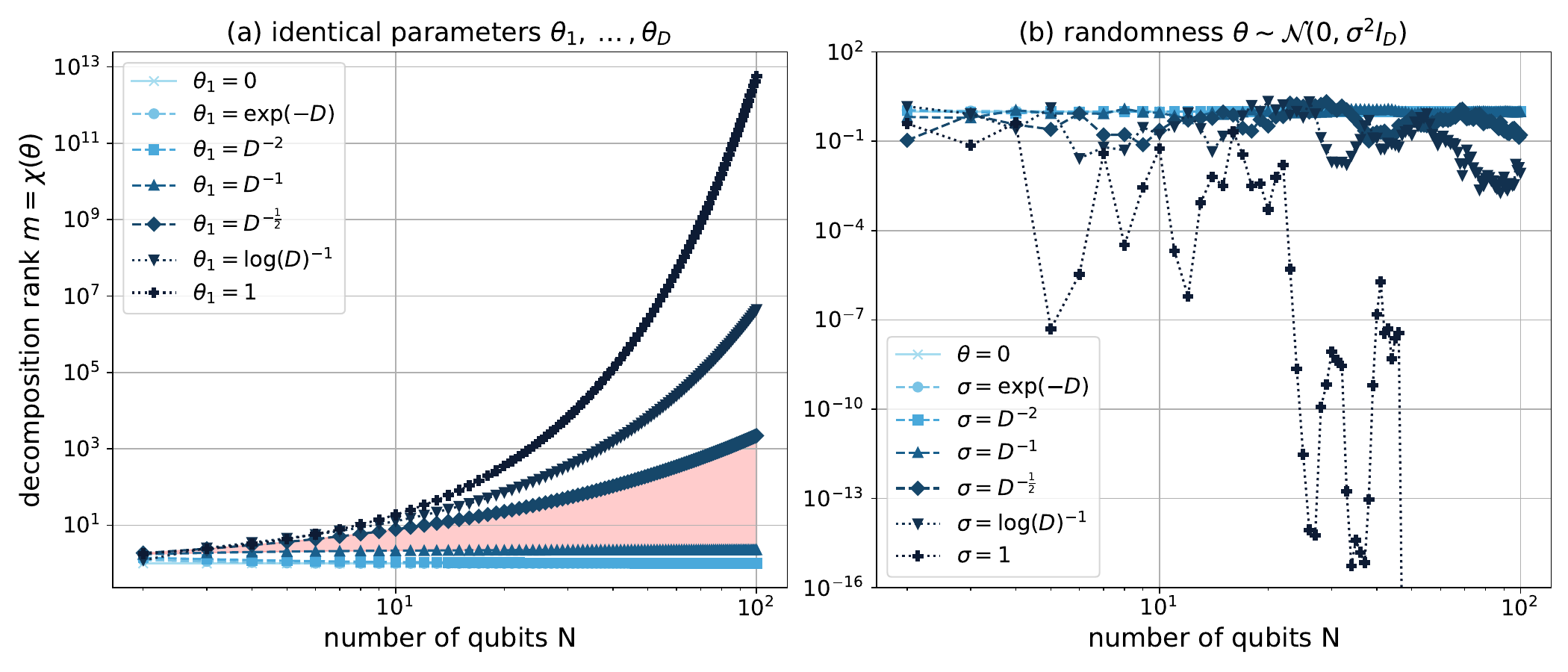}
    \caption{Stabilizer rank $m = \chi(\theta)$ as a function of qubit number $N$ for a model with $D=N$ parameters. (a) Deterministic parameters show a complexity phase transition and a barren-plateau-free region (red). (b) Randomly initialized parameters yield uninformative scaling of $m$ with the number of qubits $N$.}
    \label{fig:bravyi}
\end{figure*}

\begin{figure}[b]
    \centering
    \begin{subfigure}
        \centering
        \includegraphics[width=0.45\linewidth]{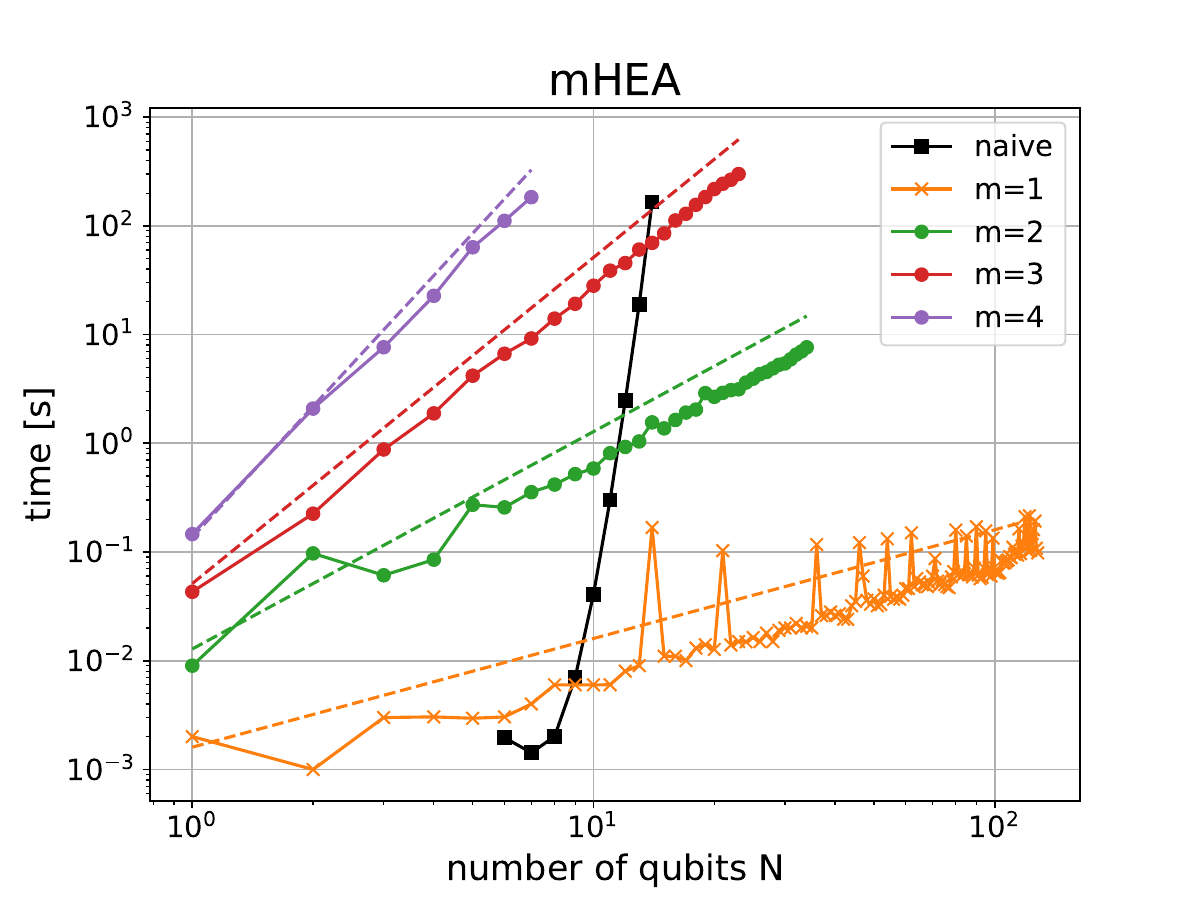}
    \end{subfigure}
    \hspace{0.02\textwidth}
    \begin{subfigure}
        \centering
        \includegraphics[width=0.45\linewidth]{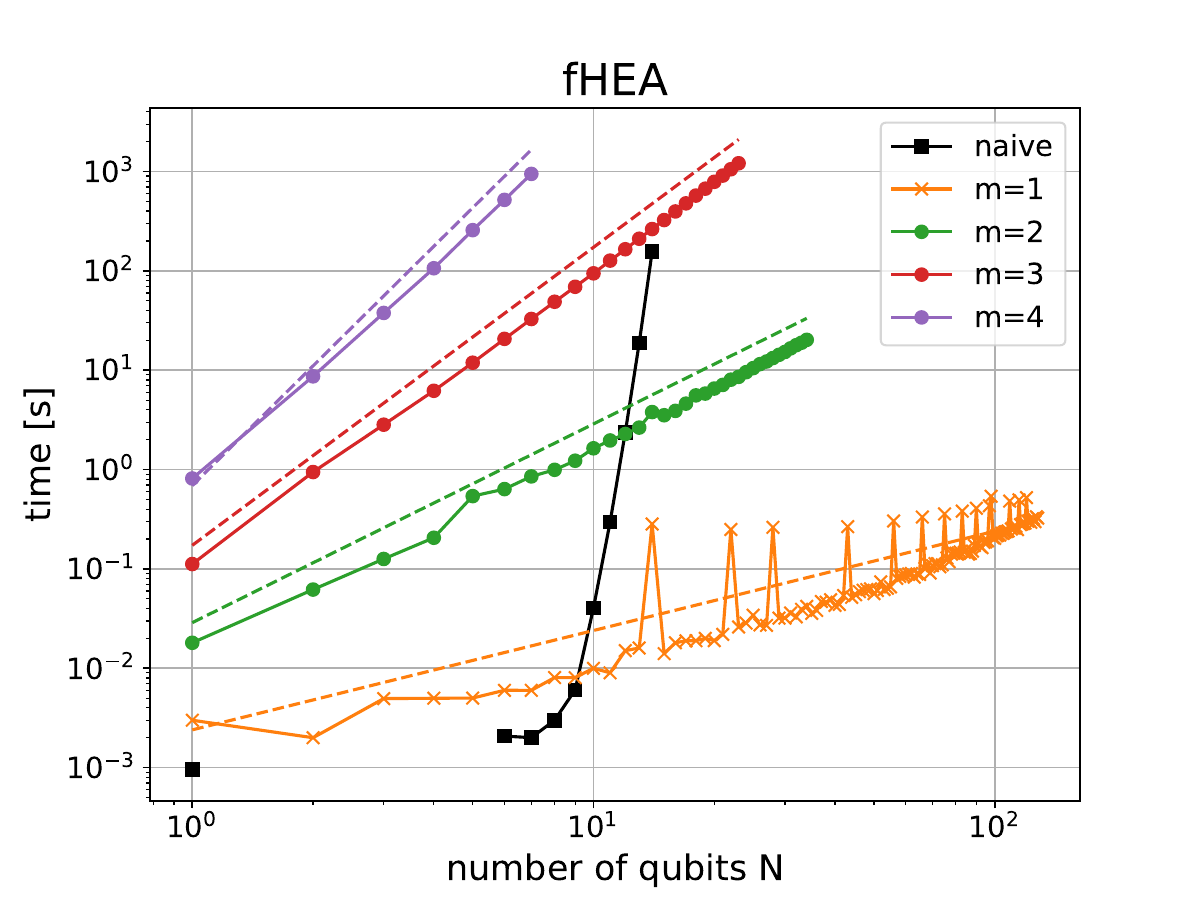}
    \end{subfigure}

    \caption{Simulation time of $C_m(\theta)$ using \texttt{mHEA} (left) and \texttt{fHEA} (right) for increasing qubit number $N$ and Taylor orders $m\leq 4$. Due to the additional number of parameters, \texttt{fHEA} has a computational overhead of $1.5^m$ compared to \texttt{mHEA}.}
    \label{fig:polynomial-complexity}
\end{figure}

\subsection{Pauli Path Formalism}
Begu\v{s}i\'c et al.~\cite{beguvsic2023simulating} take a fundamentally different approach to the problem. Instead of explicitly describing the quantum state $|\psi(\theta)\rangle$ first, they directly expand the expectation value Eq.~\eqref{eq:expectation2} by iteratively applying commutator rules. Specifically, for each Pauli-rotation gate $R_{V_k}$ with $k\in[D]$ in the circuit ansatz, and each Pauli term $P_i$ with $i\in\mathcal{I}$ in the cost function, they use
\begin{align}
    R_{V_k}(\theta_k)^\dagger P_i R_{V_k}(\theta_k) = \begin{cases}
        P_i, & [P_i,V_k]=0, \\
        \cos\frac{\theta_k}{2}P_i+i\sin\frac{\theta_k}{2}V_kP_i,& \{P_i,V_k\}=0,
    \end{cases}
\end{align}
to propagate each Pauli-rotation gate through the observable, where $[\cdot,\cdot]$, and $\{\cdot,\cdot\}$ denote the commutator, and anti-commutator, respectively.

This results in an alternative series expansion of the expectation value:
\begin{equation}
    C(\theta) \approx \sum_{i\in\mathcal{I}}c_i\sum_{\ell=1}^m f_\ell(\theta)\,\langle0|\tilde{P}_\ell|0\rangle, \label{eq:clifford-perturbation}
\end{equation}
where $f_\ell(\theta)$ are sinusoidal functions of $\theta$, and $\tilde{P}_\ell$ are newly propagated Pauli strings. More concretely, the functions $f_\ell(\theta)$ are products of sine and cosine functions so that if $f_\ell(\theta)$ contains many sine factors, it will be negligibly small if $\theta\approx0$ due to $\sin\theta\approx\theta$. This motivates the idea behind the approximation in Eq.~\eqref{eq:clifford-perturbation}, where one truncates the terms containing many sine factors which do not contribute much to the error. This simulation technique is used for surrogating near-Clifford patches in the works of Lerch et al.~\cite{lerch2024efficient}.

In this approach, the decomposition number $m$ depends on how frequently the Pauli-rotation gates in the circuit ansatz fail to commute with the Pauli strings in the observable.
Their simulation strategy has a worst-case complexity of $\mathcal{O}(|\mathcal{I}|\,2^D)$, but if all involved Pauli matrices are uniformly random, the average-case scaling improves to $\mathcal{O}(|\mathcal{I}|\,2^{0.59D})$~\cite{beguvsic2023simulating}.

Although it exhibits worse scaling compared to Bravyi et al.~\cite{bravyi2016improved, bravyi2019simulation} this may be due to their analysis imposing no restrictions on the circuit model ansatz. However, their approach does not provide a closed-form expression for the truncation number $m = K(\theta)$ as a function of $\theta$, making it less convenient for characterizing time complexity in terms of $\theta$. This may be due to the technical challenges of their perturbative expansion in the Heisenberg picture, which makes it difficult to analytically capture the dependency $m=K(\theta)$. In contrast, our approach offers a simple characterization of $m=m(\theta)$ in Lemma~\ref{lemma:truncation-threshold}, though at the expense of increased computational complexity when applied to simulating $C(\theta)$.

\subsection{Truncated Taylor series \label{sec:taylor-complexity}}

Mitarai et al.~\cite{mitarai2022quadratic} first introduced a Taylor approach for approximating the expectation value (\ref{eq:expectation2}) classically. Their method corresponds to the special case of a truncated Taylor expansion with $m=3$:
\begin{equation}
    C(\theta) \approx \sum_{\|\alpha\|_1 < m}(D_\theta^\alpha C)(0)\,\frac{\theta^\alpha}{\alpha!}, \label{eq:taylor2}
\end{equation}
where for each multi-index $\alpha=(\alpha_1,\dots,\alpha_D)$ we use the notations $D_\theta^\alpha:=\partial_{\theta_1}^{\alpha_1}\cdots\partial_{\theta_D}^{\alpha_D}$, $\theta^\alpha:=\theta_1^{\alpha_1}\cdots\theta_D^{\alpha_D}$, $\alpha!:=\alpha_1!\cdots\alpha_D!$, and $\|\alpha\|_1:=|\alpha_1|+\cdots+|\alpha_D|$ denoting the $\ell_1$-norm of $\alpha$.

This expansion provides an explicit polynomial approximation of $C(\theta)$, enabling efficient classical evaluations for small values of $\theta$. Lerch et al.~\cite{lerch2024efficient} also employed a truncated series of this form, not as a simulation method, but rather as part of their analysis in proving their results (cf. their Eq.~(C6), and their Theorems~1 and 2).

One particularly interesting difference regarding the decomposition complexity $|\mathcal{J}|$ is that the Taylor truncation threshold $m$ is used to upper bound a set of multi-indices $\{\alpha\in\mathbb{N}_0^D:\|\alpha\|_1 <m\}$, while the other methods do not iterate over multi-indices. This distinction allows us to characterize complexity more easily as a function of $\theta$ by focusing solely on the multi-index upper bound $m=m(\theta)$ as evident from the proofs of Lemma~\ref{lemma:truncation-threshold} in Appendix~\ref{sec:truncation} and Theorem~\ref{thm:complexity} in the main text, respectively. Additionally, this explains why the scaling of our multi-index threshold $m=m(\theta)$ in Fig.~\ref{fig:truncation-statistics} is simpler compared to the full decomposition threshold used by Bravyi et al.~\cite{bravyi2019simulation} in Fig.~\ref{fig:bravyi}, because an upper bound for multi-indices grows much smaller compared to the direct number of multi-indices. Furthermore, the simplicity of our formula ensures consistency with the law of large numbers, unlike the formula in \cite[Eq. (14)]{bravyi2019simulation} by Bravyi et al., which is not a sum of independent random variables.

Fig.~\ref{fig:polynomial-complexity} illustrates the cost of classically simulating $C_m(\theta)$ under the two PQC models \texttt{mHEA} and \texttt{fHEA} (introduced in Appendix~\ref{sec:ansatz}), for various system sizes $N$ and Taylor expansion orders $m\leq4$. The simulations use uniformly random Pauli strings $H = P_i$ (in particular $|\mathcal{I}|=1$) and initialization $\mathbb{P}_\theta \sim \mathcal{N}(0, \sigma^2 I_D)$ with $\sigma = D^{-1}$. Theorem~\ref{thm:complexity} states that the time complexity of evaluating $C_m(\theta)$ is a polynomial scaling as $\mathcal{O}(D^m)$. Using that \texttt{mHEA} and \texttt{fHEA} each satisfy $D=2N(L+1)$ and $D=3N(L+1)$, respectively, we see that \texttt{fHEA} has a computational overhead of $(3/2)^m=1.5^m$. For large $m$, computing $3^m$ takes significantly more time and space resources than $2^m$. The special case $m=1$ corresponds to the Gottesman-Knill theorem~\cite{gottesman1998heisenberg, aaronson2004improved}. The dashed lines represent the theoretical prediction as per Theorem~\ref{thm:complexity}. As expected, higher-order approximations ($m > 0$) lead to increased computational cost, especially as $N$ grows.

The following sections present the proofs of our theoretical results, organized thematically for clarity. We begin by defining the PQC ansatz employed in our numerical experiments. This is followed by proofs of the high-order parameter-shift rules and cost function regularities. Subsequently, we introduce the Linear Clifford Encoder (LCE) and demonstrate how it can be leveraged to mitigate barren plateaus. We then establish that the LCE transformation preserves both the expressivity of the ansatz and the global structure of the optimization landscape. The subsequent sections address computational complexity. We first derive the minimal truncation threshold required for the Taylor surrogate and use this to characterize the associated computational cost. Finally, we provide additional numerical experiments to support our theoretical findings.

\section{High-Order Parameter-Shift Rules and Cost Regularity \label{sec:parameter-shift}}

In this section, we study the regularity properties of the cost function $C(\cdot)$. Recall from Eq.~\eqref{eq:expectation} that
\begin{equation}
    C(\theta) := \langle\psi(\theta)|H|\psi(\theta)\rangle = \sum_{i\in\mathcal{I}} c_i \langle\psi(\theta)|P_i|\psi(\theta)\rangle. \label{eq:expectation3}
\end{equation}
We first establish smoothness and explicitly demonstrate differentiability using the standard parameter-shift rule~\cite{mitarai2018quantum, schuld2019evaluating}. We then extend this to the higher-order parameter-shift rule~\cite{cerezo2021higher} and leverage this structure to derive strong regularity bounds for the cost function.

For a fixed multi-index $\alpha=(\alpha_1,\dots,\alpha_D)\in\mathbb{N}_0^D$, we define the class $\mathcal{C}^\alpha$ to be the space of functions $f:[-\pi,\pi]^D\to\mathbb{R}$ for which the higher-order derivative $D_\theta^\alpha f$ of $f$ exists and is continuous. Note that the cost function $C(\cdot)$ defines a smooth function, i.e.,
\begin{equation}
    C(\cdot)\in\bigcap_{\alpha\in\mathbb{N}_0^D}\mathcal{C}^{\alpha}=:\mathcal{C}^\infty. \label{eq:smoothnes}
\end{equation}
Indeed, this immediately follows from the fact that the vector-valued mapping $\theta\mapsto |\psi(\theta)\rangle=U(\theta)|0\rangle$ is smooth (since $U(\theta)$ is a product involving smooth factors $\exp(-i\theta_k/2\,V_k)$ as per Eq.~\eqref{eq:PQC}), and that $C(\theta)=\langle\psi(\theta)|H|\psi(\theta)\rangle$ is a quadratic form in the Hilbert space $\mathcal{H}_N$ (that is, a multivariate polynomial).

The standard parameter-shift rule~\cite{mitarai2018quantum, schuld2019evaluating} shows explicitly that $C(\cdot)$ is differentiable:
\bigskip
\begin{lemma}[Vanilla Parameter-Shift Rule]\label{lemma:vanilla-parameter-shift}
    Consider an expectation of the form in Eq.~\eqref{eq:expectation3} with a variational ansatz $U(\theta)$ involving $D\geq1$ parameters. Let $\theta\in[-\pi,\pi]^D$ and $k\in[D]$ be arbitrary. Then, $C(\cdot)$ is continuous, and
    \begin{equation*}
        \partial_{\theta_k}C(\theta)=\frac{1}{2}\left\{C\left(\theta+\frac{\pi}{2}e_k\right)-C\left(\theta-\frac{\pi}{2}e_k\right)\right\},
    \end{equation*}
    where $e_k$ denotes the $k$-th unit vector.
\end{lemma}
\noindent
\emph{Proof.} Let $\theta\in[-\pi,\pi]^D$ and $k\in[D]$ be arbitrary. By definition of $U(\cdot)$, we may decompose
\begin{equation}
    U(\theta)=\prod_{j=D}^1 U_j(\theta_j)W_j=\underbrace{\left(\prod_{j=D}^{k+1} U_j(\theta_j)W_j\right)}_{=:U_{+}}U_k(\theta_k)\underbrace{\left(W_k\prod_{j=k-1}^1U_j(\theta_j)W_j\right)}_{=:U_{-}} = U_{+}U_k(\theta_k)U_{-},
\end{equation}
where $U_j(\theta_j):=\exp(-i\frac{\theta_j}{2}V_{j})$ with each $V_j\in\{X,Y,Z\}$ specifying the Pauli-rotation axis, and $W_j\in\operatorname{Clifford}(N)$, respectively. Note that $U_\pm$ are independent of $\theta_k$. In particular,
\begin{align}
    \partial_{\theta_k}C(\theta_k) &= \partial_{\theta_k}\langle0|U(\theta)^\dagger HU(\theta)|0\rangle = \langle0|U_-^\dagger\partial_{\theta_k}\left\{U_k(\theta_k)^\dagger U_+^\dagger HU_+U_k(\theta_k)\right\}U_-|0\rangle. \label{eq:derivative2}
\end{align}
Mitarai et al.~\cite{mitarai2018quantum} made the crucial observation that (redefining $H_+:=U_+^\dagger\ HU_+$)
\begin{equation}
    [V_k, H_+]=i\left\{U_k\left(\frac{\pi}{2}\right)H_+U_k\left(-\frac{\pi}{2}\right)-U_k\left(-\frac{\pi}{2}\right)H_+U_k\left(\frac{\pi}{2}\right)\right\}, \label{eq:commutator-formula}
\end{equation}
which directly follows from the fact that $U_k(\pm\frac{\pi}{2})=\frac{1}{\sqrt{2}}(I\mp iV_k)$, obtained by applying Euler's identity $U_k(\theta_k)=\exp(-i\frac{\theta_k}{2}V_k)=\cos\frac{\theta_k}{2}I-i\sin\frac{\theta_k}{2}V_k$. Also, note that $V_k=V_k^\dagger$ and $V_k^2=I$. Indeed, Eq.~\eqref{eq:commutator-formula} allows us to compute the derivative
\begin{align}
    \partial_{\theta_k}\left\{U_k(\theta_k)^\dagger U_+^\dagger HU_+U_k(\theta_k)\right\} &= \{\partial_{\theta_k}U_k(\theta_k)\}^\dagger H_+U_k(\theta_k)+U_k(\theta_k)^\dagger H_+\{\partial_{\theta_k}U_k(\theta_k)\} \nonumber
    \\ &= \frac{i}{2}U_k(\theta_k)^\dagger V_k H_+U_k(\theta_k)-\frac{i}{2}U_k(\theta_k)^\dagger H_+V_kU_k(\theta_k)  \nonumber
    \\ & = \frac{i}{2}U_k(\theta_k)^\dagger[V_k,H_+]U_k(\theta_k) \nonumber
    \\ &\stackrel{(\ref{eq:commutator-formula})}{=} -\frac{1}{2}\left\{U_k(\theta_k)^\dagger U_k\left(\frac{\pi}{2}\right) H_+U_k\left(-\frac{\pi}{2}\right)U_k(\theta_k) - U_k(\theta_k)^\dagger U_k\left(-\frac{\pi}{2}\right)H_+U_k\left(\frac{\pi}{2}\right)U_k(\theta_k)\right\}\nonumber
    \\ &= \frac{1}{2}\left\{U_k\left(\theta_k+\frac{\pi}{2}\right)^\dagger H_+U_k\left(\theta_k+\frac{\pi}{2}\right)-U_k\left(\theta_k-\frac{\pi}{2}\right)^\dagger H_+U_k\left(\theta_k-\frac{\pi}{2}\right)\right\},
\end{align}
using $\partial_{\theta_k}U_k(\theta_k) = -\frac{i}{2}V_kU_k(\theta_k)$, and in the last equality, $U_k(\pm\theta_k)^\dagger = U_k(\mp\theta_k)$ and $U_k(\theta_k\pm\frac{\pi}{2})=U_k(\theta_k)U_k(\pm\frac{\pi}{2})$, respectively. Returning to Eq.~\eqref{eq:derivative2}, it now easily follows that
\begin{align}
    \partial_{\theta_k}C(\theta_k) &= \langle0|U_-^\dagger\partial_{\theta_k}\{U_k(\theta_k)^\dagger H_+U_k(\theta_k)\}U_-|0\rangle \nonumber
    \\ &= \frac{1}{2}\langle0|\left\{U_-^\dagger U_k\left(\theta_k+\frac{\pi}{2}\right)^\dagger U_+^\dagger H U_+ U_k\left(\theta_k+\frac{\pi}{2}\right) U_- - U_-^\dagger U_k\left(\theta_k-\frac{\pi}{2}\right)^\dagger U_+^\dagger HU_+U_k\left(\theta_k-\frac{\pi}{2}\right)U_-\right\}|0\rangle \nonumber
    \\ &= \frac{1}{2}\left\{\langle0|U\left(\theta+\frac{\pi}{2}e_k\right)^\dagger HU\left(\theta+\frac{\pi}{2}e_k\right)|0\rangle -\langle0|U\left(\theta-\frac{\pi}{2}e_k\right)^\dagger HU\left(\theta-\frac{\pi}{2}e_k\right)|0\rangle\right\} \nonumber
    \\ &= \frac{1}{2}\left\{C\left(\theta+\frac{\pi}{2}e_k\right)-C\left(\theta-\frac{\pi}{2}e_k\right)\right\},
\end{align}
as desired.
\qed

The next lemma generalizes the previous result to arbitrary high-order derivatives $D_\theta^\alpha C(\cdot)$, and may be considered an explicit proof of the regularity property in Eq.~\eqref{eq:smoothnes}. {The $\alpha$-order parameter-shift rule gives rise to the illustration in Fig.~\ref{fig:Clifford-structure-appendix} which depicts three views of the $\alpha$-order Clifford structure as oper Definition~\ref{def:Clifford structure}. The leftmost picture shows an equidistant grid of points $\xi_{j,\alpha}=\frac{\pi}{2}(\alpha-2j)$ in parameter space stemming from the $\alpha$-order parameter-shift rule in Eq.~\eqref{eq:taylor-coeff2}. The central one depicts the ensemble of Clifford circuits $U(\xi_{j,\alpha})$ obtained by evaluating $U(\cdot)$ at each grid point. Finally, the rightmost one shows the corresponding Heisenberg-evolved list of Pauli strings $P_i^{j,\alpha}$, classically encoding the $\alpha$-order Clifford structure.}

\begin{figure}[h]
    \centering
    \includegraphics[width=.6\linewidth, keepaspectratio, trim={.4cm 2cm 1cm 2.5cm}, clip]{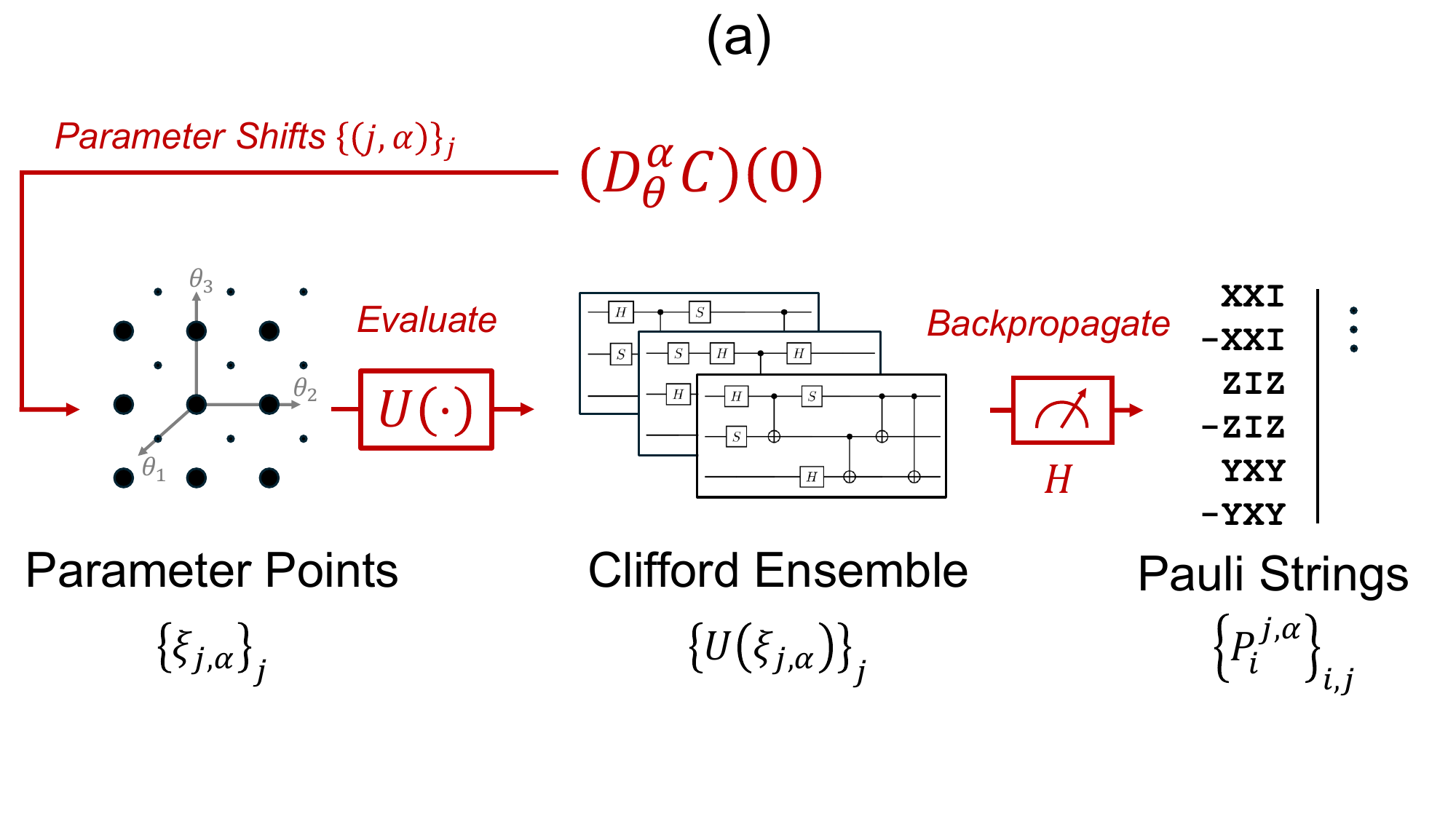}
    \caption{{Three possible interpretations of the $\alpha$-order Clifford structure.}}
    \label{fig:Clifford-structure-appendix}
\end{figure}

\bigskip
\begin{lemma}\label{lemma:parameter-shift}
    Consider an expectation of the form Eq.~\eqref{eq:expectation3} with a variational ansatz $U(\theta)$ involving $D\geq1$ parameters. Let $\theta\in[-\pi,\pi]^D$ and $\alpha\in\mathbb{N}_0^{D}$ be arbitrary but fixed. Then, the higher-order derivative of $C(\theta)$ satisfies
\begin{equation}
    D_\theta^\alpha C(\theta) = \frac{1}{2^{\|\alpha\|_1}}\sum_{j\leq\alpha}(-1)^{\|j\|_1}\binom{\alpha}{j}\,C\left(\theta+\frac{\pi}{2}(\alpha-2j)\right), \label{eq:derivative}
\end{equation}
where ${\binom\alpha j}:= \prod_{k=1}^D{\binom{\alpha_k}{j_k}}$, and the indexing set $\{j\leq\alpha\}$ defined as $\{j\in\mathbb{N}_0^D:j_k\leq\alpha_k\,\forall k\in[D]\}$. The multi-index expressions $\alpha-2j$ are to be understood point-wise. In particular, $(D_\theta^\alpha C)(\cdot)$ is continuous. If $\theta = 0$, the computational cost of evaluating Eq.~\eqref{eq:derivative} is of order $\mathcal{O}(D\,|\mathcal{I}|\,2^{\|\alpha\|_1})$, which is polynomial if $\|\alpha\|_1=\mathcal{O}(\log D)$. Furthermore, $C(\cdot)$ satisfies the strong regularity property that for each $\alpha\in\mathbb{N}_0^D$,  
\begin{equation*}  
    \sup_{\theta\in[-\pi,\pi]^D} |D^\alpha_\theta C(\theta)| \leq \|H\|,
\end{equation*}  
where $\|H\|$ denotes the operator norm of the observable in Eq.~\eqref{eq:observable}.
\end{lemma}
\noindent
\emph{Proof.} Let $\theta\in[-\pi,\pi]^D$ and $\alpha\in\mathbb{N}_0^D$ be arbitrary. By the vanilla parameter-shift rule, Lemma~\ref{lemma:vanilla-parameter-shift}, we know that
\begin{equation}
    \partial_{\theta_k}C(\theta)=\frac{1}{2}\left\{C\left(\theta+\frac{\pi}{2}e_k\right)-C\left(\theta-\frac{\pi}{2}e_k\right)\right\} \label{eq:parameter-shift}
\end{equation}
for all $k\in[D]$, where $e_k$ denotes the $k$-th unit vector. This is precisely the formula for the central finite difference of $C(\cdot)$. Inductively over $\alpha_k\geq1$, one easily establishes the higher-order generalization of the central finite difference formula, i.e.,
\begin{equation}
    \partial_{\theta_k}^{\alpha_k}C(\theta)=\frac{1}{2^{\alpha_k}}\sum_{j_k=0}^{\alpha_k}(-1)^{j_k}\binom{\alpha_k}{j_k}\, C\left(\theta + \pi\left(\frac{\alpha_k}{2}-j_k\right)e_k\right) \label{eq:higher-parameter-shift}
\end{equation}
for all $k\in[D]$. Indeed, the case $\alpha_k=1$ follows directly from Eq.~\eqref{eq:parameter-shift}. Assume that Eq.~\eqref{eq:higher-parameter-shift} is true for some fixed $\alpha_k\geq1$. Then, by induction,
\begin{align}
    \partial_{\theta_k}^{\alpha_k+1}C(\theta) &= \partial_{\theta_k}\{\partial_{\theta_k}^{\alpha_k}C(\theta)\} \nonumber
    \\ &\stackrel{(\ref{eq:higher-parameter-shift})}{=} \frac{1}{2^{\alpha_k}} \sum_{j_k=0}^{\alpha_k}(-1)^{j_k}\binom{\alpha_k}{j_k}\, \partial_{\theta_k}C\left(\theta + \pi\left(\frac{\alpha_k}{2}-j_k\right)e_k\right) \nonumber
    \\ &= \frac{1}{2^{\alpha_k+1}}\sum_{j_k=0}^{\alpha_k}(-1)^{j_k}\binom{\alpha_k}{j_k}\, \left\{C\left(\theta + \pi\left(\frac{\alpha_k+1}{2}-j_k\right)e_k\right)-C\left(\theta - \pi\left(\frac{\alpha_k+1}{2}-j_k\right)e_k\right)\right\} \nonumber
    \\ &= \frac{1}{2^{\alpha_k+1}}\sum_{j_k=0}^{\alpha_k+1}(-1)^{j_k}\binom{\alpha_k+1}{j_k}\, C\left(\theta + \pi\left(\frac{\alpha_k+1}{2}-j_k\right)e_k\right),
\end{align}
where the second equality makes use of Eq.~\eqref{eq:parameter-shift} as follows:
\begin{align}
    \partial_{\theta_k}C\left(\theta + \pi\left(\frac{\alpha_k}{2}-j_k\right)e_k\right) &\stackrel{(\ref{eq:parameter-shift})}{=} \frac{1}{2}\left\{C\left(\theta+\pi\left(\frac{\alpha_k}{2}-j_k\right)e_k+\frac{\pi}{2}e_k\right)-C\left(\theta-\pi\left(\frac{\alpha_k}{2}-j_k\right)e_k-\frac{\pi}{2}e_k\right)\right\} \nonumber
    \\ &= \frac{1}{2}\left\{C\left(\theta+\pi\left(\frac{\alpha_k+1}{2}-j_k\right)e_k\right)-C\left(\theta-\pi\left(\frac{\alpha_k+1}{2}-j_k\right)e_k\right)\right\}.
\end{align}
\par
Next, note that for each pair $k'\neq k$ in $[D]$,
\begin{align}
    \partial_{\theta_{k'}}^{\alpha_{k'}}\partial_{\theta_k}^{\alpha_k}C(\theta) &= \frac{1}{2^{\alpha_k}}\sum_{j_k=0}^{\alpha_k}(-1)^{j_k}\binom{\alpha_k}{j_k}\cdot\partial_{\theta_{k'}}^{\alpha_{k'}}C\left(\theta + \pi\left(\frac{\alpha_k}{2}-j_k\right)e_k\right) \nonumber
    \\ &= \frac{1}{2^{\alpha_k}}\sum_{j_k=0}^{\alpha_k}(-1)^{j_k}\binom{\alpha_k}{j_k}\cdot \frac{1}{2^{\alpha_{k'}}}\sum_{j_{k'}=0}^{\alpha_{k'}}(-1)^{j_{k'}}\binom{\alpha_{k'}}{j_{k'}}\, C\left(\theta + \pi\left(\frac{\alpha_k}{2}-j_k\right)e_k + \pi\left(\frac{\alpha_{k'}}{2}-j_{k'}\right)e_{k'}\right) \nonumber
    \\ &= \frac{1}{2^{\alpha_k+\alpha_{k'}}}\sum_{j_k=0}^{\alpha_j}\sum_{j_{k'}=0}^{\alpha_{k'}} (-1)^{j_k+j_{k'}}\binom{\alpha_k}{j_k}\binom{\alpha_{k'}}{j_{k'}}\, C\left(\theta + \pi\left(\frac{\alpha_k}{2}-j_k\right)e_k + \pi\left(\frac{\alpha_{k'}}{2}-j_{k'}\right)e_{k'}\right).
\end{align}
Applying these calculations successively for each $k\in[D]$, one finally obtains the multi-variate higher-order formula
\begin{align}
    D_\theta^{\alpha}C(\theta) &= \partial_{\theta_1}^{\alpha_1}\cdots\partial_{\theta_D}^{\alpha_D} C(\theta) = \frac{1}{2^{\alpha_1+\cdots+\alpha_D}}\sum_{j_1=0}^{\alpha_1}\cdots\sum_{j_D=0}^{\alpha_D}(-1)^{j_1+\cdots+j_D}\binom{\alpha_1}{j_1}\cdots\binom{\alpha_D}{j_D}C\left(\theta + \pi\sum_{k=1}^D\left(\frac{\alpha_k}{2}-j\right)e_k\right) \nonumber
    \\ &= \frac{1}{2^{\|\alpha\|_1}}\sum_{\substack{j\in\mathbb{N}_0^D \\ j_k\leq\alpha_k\,\forall k}}(-1)^{|j|}\binom{\alpha}{j}\,C\left(\theta+\frac{\pi}{2}\sum_{k=1}^D\left(\alpha_k-2j_k\right)e_k\right),
\end{align}
of the central finite difference, as desired.
\par
In order to show the uniform bound, let $j\in\mathbb{N}_0^D$ be arbitrary with $j_k\leq \alpha_k$ for all $k\in[D]$. Let us define the shifted parameter $\theta_{j,\alpha}:=\theta + \frac{\pi}{2}\sum_{k=1}^D(\alpha_k-2j_k)e_k$. Then, we apply sub-multiplicativity of norms in order to establish the bound
\begin{equation}
    |C(\theta_{j,\alpha})|\,=|\langle\psi_{\theta_{j,\alpha}}|H|\psi_{\theta_{j,\alpha}}\rangle|\,\leq\|H\|\cdot \underbrace{|\langle \psi_{\theta_{j,\alpha}}|\psi_{\theta_{j,\alpha}}\rangle|}_{=1}\, = \|H\|,
\end{equation}
using the notation $|\psi_{\theta_{j,\alpha}}\rangle = U(\theta_{j,\alpha})|0\rangle$. Since this holds for arbitrary $j$ and $\alpha$ as above, by the previous computations, we may conclude with the uniform bound
\begin{align}
    |D_\theta^\alpha C(\theta)|\, &\leq \frac{1}{2^{|\alpha|}}\sum_{\substack{j\in\mathbb{N}_0^D \\ j_k\leq\alpha_k\,\forall k}}\binom{\alpha}{j}\,\underbrace{\left\lvert C\left(\theta+\frac{\pi}{2}\sum_{k=1}^D(\alpha_k-2j_k)e_k\right)\right\rvert}_{=\,|C(\theta_{j,\alpha})|\,\leq\|H\|}\,\leq \|H\|\,\frac{1}{2^{|\alpha|}}\sum_{\substack{j\in\mathbb{N}_0^D \\ j_k\leq\alpha_k\,\forall k}}\binom{\alpha}{j} = \|H\|,
\end{align}
where in the last step follows from the binomial theorem, i.e.,
\begin{align}
    \sum_{\substack{j\in\mathbb{N}_0^D \\ j_k\leq\alpha_k\,\forall k}}\binom{\alpha}{j} &= 
\sum_{j_1=0}^{\alpha_1}\cdots\sum_{j_D=0}^{\alpha_D}\prod_{k=1}^D\binom{\alpha_k}{j_k} =\prod_{k=1}^D \underbrace{\left(\sum_{j_k=0}^{\alpha_k}\binom{\alpha_k}{j_k}\right)}_{=(1+1)^{\alpha_k}} = \prod_{k=1}^D 2^{\alpha_k}=2^{|\alpha|}. \label{eq:binomial-theorem}
\end{align}
Since $\theta\in\mathbb{R}^D$ and $\alpha\in\mathbb{N}_0^D$ were arbitrary, this completes the first part of the lemma.

It remains to show that, if $\theta=0$, then the computational cost of evaluating Eq.~\eqref{eq:derivative} is of order $\mathcal{O}(D\,|\mathcal{I}|\,2^{\|\alpha\|_1})$. Each evaluation of $C(\xi_{j,\alpha})$ using Gottesman-Knill's theorem requires $\mathcal{O}(D\,|\mathcal{I}|)$ operations since all unitaries $U(\xi_{j,\alpha})$ involved are Clifford operators (cf. last part of the proof of Theorem~\ref{thm:complexity} in Appendix~\ref{sec:complexity}). The number of times we need to evaluate $C(\xi_{j,\alpha})$ is given by the cardinality
\begin{align}
    \#\{j\leq\alpha\}&=\#\{j\in\mathbb{N}_0^D:j_k\leq\alpha_k\,\forall k\in[D]\} = \prod_{k=1}^D\underbrace{(\alpha_k+1)}_{\leq 2^{\alpha_k}}\leq 2^{\sum_{k=1}^D\alpha_k} = 2^{\|\alpha\|_1}.
\end{align}
Therefore, the total cost of evaluating Eq.~\eqref{eq:derivative} is indeed $\mathcal{O}(D\,|\mathcal{I}|\,2^{\|\alpha\|_1})$.
\qed

\section{Construction of the Linear Clifford Encoder (LCE) \label{sec:LCE}}

We now construct the LCE transformation used to mitigate the barren plateau problem~\cite{mcclean2018barren}. The approach involves modifying the PQC via a pair $(Q,\tilde{Q})$ of Clifford operators such that the $k$-th gradient component satisfies $(\partial_{\theta_k}C)(0)=\pm1$, ensuring it is non-zero at the origin. This is accomplished by propagating the shifted Clifford operators $U(\xi_{j,\alpha})$ through the Pauli observable $H=P$, yielding diagonal Pauli strings $\tilde{P}_j := U(\xi_{j,\alpha})^\dagger P\, U(\xi_{j,\alpha})$ with a relative phase. Afterwards, we generalize this result to arbitrary, non-identity observables of the general form
\begin{equation}
    H=\sum_{i\in\mathcal{I}}c_iP_i, \label{eq:observable2}
\end{equation}
with Pauli strings $P_i\in\mathcal{P}_N\setminus\{I\}$ and real coefficients $c_i\in\mathbb{R}$.

In what follows, we use $\langle A,B\rangle_N$ to denote the multiplicative subgroup of the unitary group $\operatorname{U}(2^N)$, generated by the matrices $A$ and $B$, modulo the global phase factors $\pm, \pm i$. For example, it is well-known~\cite{nielsen2010quantum} that the Clifford group is equal to $\langle H,S,\operatorname{CX}\rangle_N$. In practice, this means that a Clifford circuit is constructed from elementary gates in the ensemble $\{H,S,\operatorname{CX}\}$. As another example, recall that the $N$-qubit Pauli group is defined as
\begin{equation}
    \mathcal{P}_N:=\langle X,Y,Z\rangle_N=\{\pm P, \pm i P: P = P_1\otimes\cdots\otimes P_N,\, \forall j\in[N]:P_j\in\{I,X,Y,Z\}\}.
\end{equation}
We generally consider
\begin{equation}
    U(\cdot)\in\langle R_X(\cdot),R_Y(\cdot),R_Z(\cdot),\operatorname{Clifford}(N)\rangle_N \label{eq:generated-unitary}
\end{equation}
to be generated by single-qubit rotation gates and Clifford-operators, $\operatorname{Clifford}(N)=\langle H, S, \operatorname{CX}\rangle_N$, acting on a $N$-qubit system. It is well-known that this defines a universal gate set.

\subsection{Explicit Construction of the Linear Clifford Encoder (LCE)}

The following theorem shows how to explicitly construct the LCE pair $(Q,\tilde{Q})$ which plays a central role in our barren plateau mitigation technique. We begin by handling the case where the observable $H=P$ is a single Pauli string. Afterwards, we show that the same construction may be used for arbitrary observables.

\begin{theorem} \label{thm:linear-clifford-encoder-special}
Consider an expectation of the form Eq.~\eqref{eq:expectation3} with a variational ansatz $U(\theta)$ consisting of Clifford-gates and single-qubit Pauli-rotation gates involving $D\geq1$ parameters. Assume that the observable $H=P$ in Eq.~\eqref{eq:observable2} is a single Pauli string. Then, for every $k\in[D]$ there exist two explicitly determined Clifford operators $Q\in\langle H,S,\operatorname{CX}\rangle_N$ and $\tilde{Q}\in \langle H,S\rangle_N$ such that
\begin{equation*}
    (\partial_{\theta_k}\tilde{C})(0)= \frac{1}{2}\left\{\langle0|Q^\dagger \tilde{P}^{0,e_k}\, Q|0\rangle - \langle0|Q^\dagger \tilde{P}^{e_k,e_k}\, Q|0\rangle\right\} = \pm1,
\end{equation*}
after replacing $U(\theta)$ with $\tilde{Q}\,U(\theta)\, Q$, where $\tilde{P}^{j,\alpha}:=U(\xi_{j,\alpha})^\dagger \tilde{Q}^\dagger P\,\tilde{Q} \, U(\xi_{j,\alpha})$, and $\xi_{j,\alpha}:=\frac{\pi}{2}(\alpha-2j)$ for multi-indices $\alpha\in\mathbb{N}_0^D$ and $0\leq j\leq \alpha$ (i.e., $0\leq j_k\leq \alpha_k$ for all $k\in[D]$).
\end{theorem}
\bigskip
\noindent
\emph{Proof.} Let $k\in[D]$ be arbitrary, and the Clifford pair $(Q,\tilde{Q})$ to be determined.

\textbf{Step 1: Higher-Order Parameter-Shift Rule.} First, we apply the higher-order parameter-shift rule, Lemma~\ref{lemma:parameter-shift}, with $\xi_{0,e_k}=\frac{\pi}{2}e_k$ and $\xi_{e_k,e_k}=-\frac{\pi}{2}e_k$, which results in
\begin{align}
    (\partial_{\theta_k}\tilde{C})(0) &= \frac{1}{2}\left\{\tilde{C}(\xi_{0,e_k})-\tilde{C}(\xi_{e_k,e_k})\right\} \nonumber
    \\ &= \frac{1}{2}\left\{\langle0|Q^\dagger U(\xi_{0,e_k})^\dagger \tilde{Q}^\dagger P\,\tilde{Q} \,U(\xi_{0,e_k})\,Q|0\rangle - \langle0|Q^\dagger U(\xi_{e_k,e_k})^\dagger \tilde{Q}^\dagger P\,\tilde{Q} \,U(\xi_{e_k,e_k})\,Q|0\rangle\right\} \nonumber
    \\ &= \frac{1}{2}\left\{\langle 0|Q^\dagger \tilde{P}^{0,e_k}\,Q|0\rangle - \langle 0|Q^\dagger \tilde{P}^{e_k,e_k}\,Q|0\rangle\right\}. \label{eq:linear-pauli-shifts}
\end{align}
\par
The idea is to engineer the pair $(Q,\tilde{Q})$ of Clifford operators as follows: First, we design $\tilde{Q}$ explicitly for inducing a relative phase, i.e.,
\begin{equation}
    \operatorname{sign}\tilde{P}^{0,e_k} \stackrel{!}{\neq} \operatorname{sign}\tilde{P}^{e_k,e_k} \label{eq:unequal-sign}
\end{equation}
Here, the notation ``$!$'' means that we would like to achieve the desired relation by adjusting free parameters (in that case $\tilde{Q}$). Next, after determining $\tilde{Q}$, we explicitly design $Q$ such that it simultaneously diagonalizes $\tilde{P}^{0,e_k}$ and $\tilde{P}^{e_k,e_k}$. That is,
\begin{equation}
    Q^\dagger\tilde{P}^{0,e_k}\,Q,\, Q^\dagger \tilde{P}^{e_k,e_k}\,Q \stackrel{!}{\in} \langle Z\rangle_N. \label{eq:diagonalized}
\end{equation}
Engineering the pair $(Q,\tilde{Q})$ that way thus implies
\begin{equation}
    \langle 0|Q^\dagger \tilde{P}^{0,e_k}\,Q|0\rangle - \langle 0|Q^\dagger \tilde{P}^{e_k,e_k}\,Q|0\rangle \stackrel{\eqref{eq:diagonalized}}{=} \operatorname{sign}\tilde{P}^{0,e_k} - \operatorname{sign}\tilde{P}^{e_k,e_k} \stackrel{\eqref{eq:unequal-sign}}{=} \pm 2,
\end{equation}
which, substituted into Eq.~\eqref{eq:linear-pauli-shifts}, results in the desired equality.
\par
\textbf{Step 2: Relative Phase via $\tilde{Q}$.} Let us begin by designing $\tilde{Q}\in\langle H,S,\operatorname{CX}\rangle_N$. In general, since $U(\cdot)$ is generated according to Eq.~\eqref{eq:generated-unitary}, we may decompose
\begin{equation}
    U(\xi_{j,e_k}) = W_a R_{V_k}\left(\pm\frac{\pi}{2}e_k\right)W_b
\end{equation}
for some Clifford operators $W_a,W_b\in\langle H,S,\operatorname{CX}\rangle_N$. Thus, evolving the input Pauli string $P$ through the Clifford $\tilde{Q}\, U(\xi_{j,e_k})$ in the Heisenberg picture yields the evolved Pauli string,
\begin{equation}
    \tilde{P}^{j,e_k} = U(\xi_{j,\alpha})^\dagger \tilde{Q}^\dagger P\,\tilde{Q} \, U(\xi_{j,\alpha}) = W_b^\dagger R_{V_k}\left(\pm\frac{\pi}{2}e_k\right)^\dagger W_a^\dagger \,\tilde{Q}^\dagger P\,\tilde{Q}\,W_a\,R_{V_k}\left(\pm\frac{\pi}{2}e_k\right)W_b.
\end{equation}
The only way to obtain a relative phase, specifically different signs as per Eq.~\eqref{eq:unequal-sign}, is to design $\tilde{Q}$ systematically such that
\begin{equation}
    R_{V_k}\left(\pm\frac{\pi}{2}e_k\right)^\dagger W_a^\dagger \,\tilde{Q}^\dagger P\,\tilde{Q}\,W_a\,R_{V_k}\left(\pm\frac{\pi}{2}e_k\right) \in \{\pm \Pi,\mp \Pi\} \label{eq:phase-condition}
\end{equation}
for some unsigned Pauli string $\Pi\in\mathcal{P}_N$.

\begin{table}[ht]
\caption{\label{tab:update-rules_rotations} Update rules for conjugating Pauli strings $\Pi$ with Clifford operators $C$.}
    \centering
    \renewcommand{\arraystretch}{1.5}
    \setlength{\tabcolsep}{15pt}
    \begin{tabular}{c | c c c c}
        Clifford $C$ & \multicolumn{4}{c}{Input Pauli $\Pi$ $\rightarrow$ Output Pauli $C^\dagger \Pi\, C$} \\
        \hline
        \hline
        $R_X(\pm\frac{\pi}{2})$    & $X \rightarrow X$ & \hspace{15pt} $Y \rightarrow \mp Z$ & $Z \rightarrow \pm Y$ &  \\
        \hline
        $R_Y(\pm\frac{\pi}{2})$    & $X \rightarrow \pm Z$ & \hspace{15pt} $Y \rightarrow Y$ & $Z \rightarrow \mp X$ &  \\
        \hline
        $R_Z(\pm\frac{\pi}{2})$    & $X \rightarrow \mp Y$ & \hspace{15pt} $Y \rightarrow \pm X$ & $Z \rightarrow Z$ &  \\
    \end{tabular}
\end{table}

With the help of Table~\ref{tab:update-rules_rotations} one establishes that for every pair $(V,\Pi_1)$ of unsigned, single-qubit Pauli-operators, there exists an unsigned Pauli-operator $\Pi_2$ such that $\Pi_1\neq \Pi_2$ and
\begin{equation}
    R_V\left(\pm\frac{\pi}{2}\right)^\dagger \Pi_1\, R_V\left(\pm\frac{\pi}{2}\right) \in\{\pm \Pi_2, \mp \Pi_2\} \quad\text{if and only if}\quad \Pi_1 \notin \{I,V\}.
\end{equation}

We make the ansatz $\tilde{Q}:=\tilde{W}\,W_a^\dagger$ for some Clifford $\tilde{W}$ to be determined, and it will follow that Eq.~\eqref{eq:phase-condition} holds true if and only if
\begin{equation}
    (\tilde{W}^\dagger P \,\tilde{W})_k \notin \{I,V_k\}. \label{eq:reduced-condition}
\end{equation}
Indeed, the factor $W_a^\dagger$ is used to cancel with $W_a$, and the Clifford operator $\tilde{W}$ remains to be designed such that a relative phase is induced as per Eq.~\eqref{eq:unequal-sign}.

We briefly recall the indexing convention for Pauli operators: For a Pauli string $P \in \mathcal{P}_N$, we use the notation $P_k \in \{I, X, Y, Z\}$ to denote the $k$-th component of the tensor product $P = P_1 \otimes \cdots \otimes P_N$. Equivalently, $P_k$ may be interpreted as the $N$-qubit operator $I_{2^{k-1}} \otimes P_k \otimes I_{2^{N-k}}$. For $2\times2$ operators $A$ (like $H$ and $S$), we analogously denote $A_k:=I_{2^{k-1}} \otimes A \otimes I_{2^{N-k}}$ These conventions are used to simplify our construction. The goal is now to construct $\tilde{W}$ such that Eq.~\eqref{eq:reduced-condition} is satisfied:

Recall that we have fixed $k\in[D]$ for the LCE construction. First, consider the case where $P_k\neq I$. If $V_k\neq P_k$, then Eq.~\eqref{eq:reduced-condition} holds true with the identity $\tilde{W}:=I$. If $V_k=P_k$, then define
\begin{equation}
    \tilde{W}:=\begin{cases}
        H_k,& V_k\in\{X,Z\}, \\ S_k,& V_k=Y,
    \end{cases}
\end{equation}
and it follows from Table~\ref{tab:update-rules1_1} below that Eq.~\eqref{eq:reduced-condition} is satisfied.

For the case where $P_k=I$, we are going to decompose $\tilde{W}:=\tilde{W}_1\tilde{W}_2$ into two Clifford operators $\tilde{W}_1\in\langle \operatorname{CX}\rangle_N$, and $\tilde{W}_2\in\langle H,S\rangle_N$, respectively. Due to the fact that $P\neq I$ is non-trivial, there exists $i\in[N]$ such that $P_i\neq I$. Define
\begin{equation}
    \tilde{W}_1 := \begin{cases}
        \operatorname{CX}_{i,k},& P_i\in\{X,Y\}, 
        \\ \operatorname{CX}_{k,i},& P_i=Z,
    \end{cases}
\end{equation}
where the entangling gate $\operatorname{CX}_{i,j}$ has control $i$ and target $j$.
\begin{table}[h]
\caption{\label{tab:update-rules-cx} Update rules for conjugating Pauli strings $R$ with Clifford operators $C$.}
    \centering
    \renewcommand{\arraystretch}{1.5} 
    \setlength{\tabcolsep}{15pt}      
    \begin{tabular}{c | c c c c}
        Clifford $C$ & \multicolumn{4}{c}{Input Pauli $R$ $\rightarrow$ Output Pauli $C^\dagger R\, C$} \\
        \hline
        \hline
        \multirow{2}{*}{$\operatorname{CX}_{i,j}$} & $X_i \rightarrow X_i X_j$ & \hspace{15pt} $Y_i \rightarrow Y_i X_j$ & \hspace{15pt} $Z_i \rightarrow Z_i$ & \\
        & $X_j \rightarrow X_j$ & \hspace{15pt} $Y_j \rightarrow Z_i Y_j$ & \hspace{15pt} $Z_j \rightarrow Z_iZ_j$ & \\
    \end{tabular}
\end{table}

With the help of Table~\ref{tab:update-rules-cx} we see that
\begin{equation}
    (\tilde{W}_1^\dagger P\,\tilde{W}_1)_k = \left.\begin{cases}
        X_k,& P_i\in\{X,Y\}, \\ Z_k,& P_i=Z,
    \end{cases}\right\}\neq I. 
\end{equation}
Hence, we may construct $\tilde{W}_2$ analogous to the previous case, $P_k\neq I$. That is, if $V_k\neq(\tilde{W}_1^\dagger P\,\tilde{W}_1)_k$, then Eq.~\eqref{eq:reduced-condition} holds true with $\tilde{W}_2=I$. Otherwise, if $V_k = (\tilde{W}_1^\dagger P\,\tilde{W}_1)_k$, we may define
\begin{equation}
    \tilde{W}_2:=\begin{cases}
        H_k,& V_k\in\{X,Z\}, \\ S_k,& V_k=Y,
    \end{cases}
\end{equation}
and it follows from Table~\ref{tab:update-rules1_1} below that Eq.~\eqref{eq:reduced-condition} is satisfied.
\par
\textbf{Step 3: Simultaneous Diagonalization via $Q$.} It remains to design $Q\in\langle H, S\rangle_N$. We make the ansatz $Q=Q_1\otimes\cdots\otimes Q_N$ and define
\begin{equation}
    Q_i:=\begin{cases}
        H,& \tilde{P}_i^{0,e_k}=X, \\ SH,& \tilde{P}_i^{0,e_k}=Y, \\ I,&\text{else}.
    \end{cases}
\end{equation}

\begin{table}[ht]
\caption{\label{tab:update-rules1_1}Update rules for conjugating Pauli strings $R$ with Clifford operators $C$.}
    \centering
    \renewcommand{\arraystretch}{1.5} 
    \setlength{\tabcolsep}{15pt}      
    \begin{tabular}{c | c c c c}
        Clifford $C$ & \multicolumn{4}{c}{Input Pauli $R$ $\rightarrow$ Output Pauli $C^\dagger R\, C$} \\
        \hline
        \hline
        $H$    & $X \rightarrow Z$ & \hspace{15pt} $Y \rightarrow -Y$ & $Z \rightarrow X$ &  \\
        \hline
        $S$    & $X \rightarrow -Y$ & \hspace{15pt} $Y \rightarrow X$ & $Z \rightarrow Z$ &  \\
        \hline
        $SH$    & $X \rightarrow Y$ & \hspace{15pt} $Y \rightarrow Z$ & $Z \rightarrow X$ &  \\
    \end{tabular}
\end{table}

With the help of Table~\ref{tab:update-rules1_1}, one obtains that
\begin{equation}
    Q^\dagger \tilde{P}^{0,e_k}\, Q = \bigotimes_{i=1}^N \underbrace{Q_i^\dagger \tilde{P}_i^{0,e_k}\,Q_i}_{\in\{I, \pm Z\}} \in \langle Z\rangle_N
\end{equation}
is diagonal. Since $\tilde{P}^{0,e_k}$ and $\tilde{P}^{e_k,e_k}$ only differ by the sign, we also have that $Q^\dagger\tilde{P}^{e_k,e_k}\,Q\in\langle Z\rangle_N$ is diagonal.

We have thus successfully constructed the pair $(Q,\tilde{Q})$ of Clifford operators such that Eqs.~\eqref{eq:unequal-sign} and~\eqref{eq:diagonalized} are simultaneously satisfied. Therefore, by Eq.~\eqref{eq:linear-pauli-shifts}, we have established that
\begin{equation}
    (\partial_{\theta_k} \tilde{C})(0) = \frac{1}{2}\left\{\langle 0|Q^\dagger \tilde{P}^{0,e_k}\,Q|0\rangle - \langle 0|Q^\dagger \tilde{P}^{e_k,e_k}\,Q|0\rangle\right\} \stackrel{\eqref{eq:diagonalized}}{=} \frac{1}{2}\left\{\operatorname{sign}\tilde{P}^{0,e_k} - \operatorname{sign}\tilde{P}^{e_k,e_k}\right\} \stackrel{\eqref{eq:unequal-sign}}{=} \pm1,
\end{equation}
which completes the proof of the theorem.
\qed

\begin{figure*}[b]
    \centering
    \includegraphics[width=0.9\textwidth]{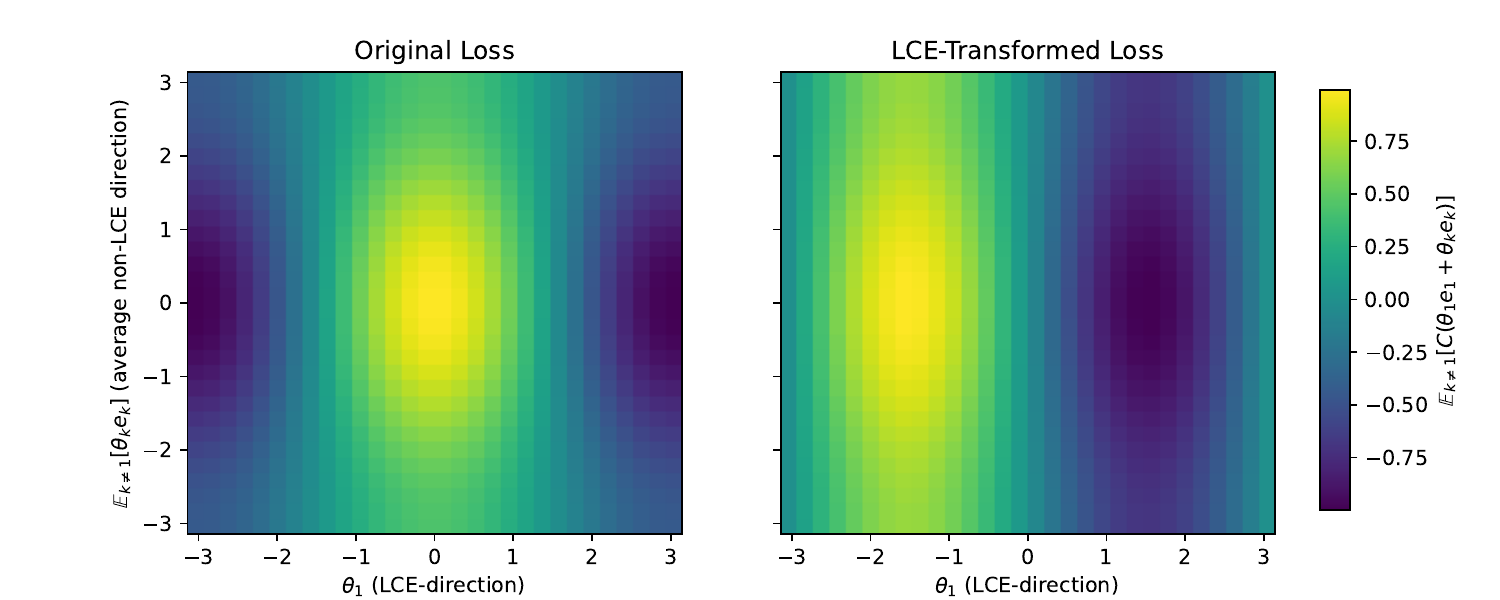}
    \caption{Illustrating the landscape transformation effect of LCE for a \texttt{mHEA}-model with $N=12$ qubits, $L=6$ layers and $H=Z^{\otimes N}$ as the observable. We fix $k=1$ as the LCE direction, and average  the grids $\{\tilde{C}(\theta_1e_1+\theta_{k}e_{k}):\theta_1,\theta_{k}\in[-\pi,\pi]\}$ over non-LCE directions $k\neq1$ (always used as the vertical axis).}
    \label{fig:LCE-heightmap}
\end{figure*}

We refer to the Clifford pair $(Q,\tilde{Q})$ used to transform the circuit $U(\cdot)$ into $\tilde{Q}\,U(\cdot)\,Q$ as the \emph{Linear Clifford Encoder (LCE)}, since it is constructed to render the $k$-th linear Taylor coefficient constant, i.e., $(\partial_{\theta_k} C)(0)=\pm1$.

In Fig.~\ref{fig:LCE-heightmap} we use a numerical experiment to illustrate geometrically how LCE transforms the loss landscape $C(\cdot)\mapsto \tilde{C}(\cdot)$. We consider a $N=12$ qubit \texttt{mHEA}-model with $L=6$ layers, and focus on the global observable $Z^{\otimes N}$. For the LCE construction, we pick the direction $k=1$, and evaluate the cost $\tilde{C}(\theta_1 e_1 + \theta_k e_k)$ for all non-LCE directions $k\neq 1$, where $e_k$ denotes the $k$-th unit vector. This results in an equidistant grid $\{\theta_1 e_1 + \theta_k e_k\}$ (we use $30\times 30$ points) on which we evaluate the cost $C(\cdot)$. Then, we average over all non-LCE directions, i.e., $\mathbb{E}_{k\in[D]\setminus1}[\tilde{C}(\theta_1 e_1+\theta_k e_k)]$, which are always used as the vertical axis in the plot. We can see that LCE ensures a better gradient when initializing close to $\theta=0$, as predicted by Theorem~\ref{thm:general-escaping-barren-plateau}. This is due to initializing at the steepest slope of the landscape (green-blue area where $\tilde{C}(\theta)\approx 0$), which ensures a gradient ``boost'' in the positive LCE-direction (which could be exploited for momentum). The landscape appears to be shifted to the left in the LCE direction by $\pi/2$ (stemming from the operator $Q$ causing a translation), and then slightly stretched in the non-LCE directions (stemming from the operator $\tilde{Q}$ causing a change of basis). In the case of $Z^{\otimes N}$, this transformation guarantees faster convergence to the global minimum value $-1$ when directly employing gradient descent with sufficiently small learning rates (cf. Fig.~\ref{fig:VQE-experiment}). On the other hand, the non-LCE loss landscape has a global maximum at $\theta=0$, showing that using vanilla gradient descent requires more iterations compared to LCE. In fact, using LCE in this situation even results in closer parameter initialization to the solution.

We have explained how to make the gradient constant in one direction. This naturally raises the question of whether multiple gradient components can be controlled simultaneously. In principle, this is possible to some extent: one must solve an over-determined system of operator equations. As the number of components increases, the likelihood of a solution decreases. The following elaborates the details:

Consider a collection $\mathcal{K}\subseteq[D]$ of gradient elements we would like to control. We would like to construct a stronger LCE $(Q,\tilde{Q})$ in the sense that
\begin{equation}
    (\partial_{\theta_k}\tilde{C})(0) = \pm 1\quad\text{for all $k\in\mathcal{K}$}
\end{equation}
after replacing $U(\theta)$ with $\tilde{Q}\,U(\theta)\, Q$. To see how this works, for each $k\in\mathcal{K}$ we decompose
\begin{equation}
    U(\xi_{j,e_k}) = W_{a_k} R_{V_k}\left(\pm\frac{\pi}{2}e_k\right)W_{b_k} \label{eq:first-order-clifford-shifts}
\end{equation}
for some Clifford operators $W_{a_k},W_{b_k}\in\langle H,S,\operatorname{CX}\rangle_N$. The solution (if it exists) is determined by the system of $2\#\mathcal{K}$ operator equations:
\begin{equation}
\left\{
\begin{array}{l}
(W_{a_k}^\dagger \tilde{Q}^\dagger P \,\tilde{Q}\,W_{a_k})_k\notin\{I,V_k\} \\
Q^\dagger W_{b_k}^\dagger R_{V_k}\left(\pm\frac{\pi}{2}e_k\right)^\dagger W_{a_k}^\dagger \,\tilde{Q}^\dagger P\,\tilde{Q}\,W_{a_k}\,R_{V_k}\left(\pm\frac{\pi}{2}e_k\right)W_{b_k}\, Q \in\langle Z\rangle_N
\end{array}
\quad : k\in\mathcal{K} \right\}. \label{eq:LCE-multilinear}
\end{equation}
The difficulty of finding a solution $(Q,\tilde{Q})$ in Eq.~\eqref{eq:LCE-multilinear} is influenced by the circuit ansatz first-order Clifford architecture $\{(W_{a_k},W_{b_k}):k\in\mathcal{K}\}$, which is obtained from the parameter shifts in Eq.~\eqref{eq:first-order-clifford-shifts}. 

If $U(\cdot)$ exhibits additional structure, for example like a hardware-efficient, alternating ansatz (such as \texttt{EfficientSU2}~\cite{efficientsu2}), and if $\mathcal{K}$ chosen such that \emph{all $V_k$ are identical and in the same layer}, then Eq.~\eqref{eq:LCE-multilinear} reduces to a system of two equations (since $W_{a_k}=W_a$ and $W_{b_k}=W_b$ for all $k\in\mathcal{K}$ for some fixed $W_a,W_b\in\operatorname{Clifford}(N)$ independent of $k$), and we may use the algorithm established in Theorem~\ref{thm:linear-clifford-encoder} to construct an explicit solution $(Q,\tilde{Q}$). That is, for alternating ansätze of this form, we have $\#\mathcal{K}\geq N$ since there are $N$-many identical Pauli rotations in a fixed layer.

If we impose more conditions on $U(\cdot)$, it is possible to achieve even larger initial gradients: Suppose the PQC is of the form
\begin{equation}
    U(\theta) = \prod_{\ell = L}^1 W_\ell U_{\ell}(\theta_{\ell}) W_\ell^\dagger
\end{equation}
for some $W_\ell\in\operatorname{Clifford}(N)$ and $U_{\ell}(\theta_{\ell}):=R_{P^{\otimes N}}(\theta_{\ell})R_{S^{\otimes N}}(\theta_{\ell})$ a product of two single-layers of Pauli rotations generated by $P,S\in\{X,Y,Z\}$ with $P\neq S$ (e.g., a layer of $Y$-$Z$-rotations as in Fig.~\ref{fig:PQC-models}(a)), where $\theta_{\ell}\in[-\pi,\pi]^{2N}$ are parameter subvectors of $\theta$ independent ove $\ell\in[L]$. In that case, the number of parameters $D=2NL$ is even, and $\theta=0$ without LCE corresponds to the identity initialization since $W_\ell W_\ell^\dagger=I$ for all $\ell\in[L]$. On the other hand, $\theta=0$ with LCE generally yields a Clifford initialization $U(0)=Q\tilde{Q}$.

As a result, given any gradient component $k\in[D]$, we know that there are exactly $N$ Pauli rotation gates (with generator either $P$ or $S$) living in the same fixed layer $\ell_0\in[L]$, and crucially, we again have $W_{a_k}=W_a$ and $W_{b_k}=W_b$ for all these rotations because of $W_{\ell}W_{\ell}^\dagger=I$ for all $\ell\neq\ell_0$. Therefore, we find that $\#\mathcal{K}\geq NL = \frac{D}{2}$.

\subsection{Generalization to Arbitrary Observables}
We showed how to render the gradient constant along one direction for $H = P$. We now extend this result to an arbitrary linear combinations of Pauli strings.
 Consider a general observable of the form in Eq.~\eqref{eq:observable2}, i.e.,
\begin{equation}
    H = \sum_{i\in\mathcal{I}}c_i\,P_i.
\end{equation}
The linear Clifford encoder constructed in Theorem~\ref{thm:linear-clifford-encoder} may still be used in that case:
\bigskip
\\
\noindent
\textbf{Theorem~\ref{thm:linear-clifford-encoder}}
    \emph{Consider an expectation of the form Eq.~\eqref{eq:expectation} with a variational ansatz $U(\theta)$ consisting of Clifford-gates and single-qubit Pauli-rotation gates involving $D\geq1$ parameters. Let $H$ be an arbitrary observable as in Eq.~\eqref{eq:observable}. Then, for every $k\in[D]$ and every $j\in\mathcal{I}$ there exist two explicitly determined Clifford operators $Q\in\langle H,S,\operatorname{CX}\rangle_N$ and $\tilde{Q}\in \langle H,S\rangle_N$ such that
    \begin{equation}
        (\partial_{\theta_k}\tilde{C})(0) =  \pm c_j+\frac{1}{2}\sum_{i\in\mathcal{I}\setminus\{j\}}c_i\left\{\langle0|Q^\dagger \tilde{P}^{0,e_k}_i\, Q|0\rangle - \langle0|Q^\dagger \tilde{P}^{e_k,e_k}_i\, Q|0\rangle\right\}, \label{eq:general-LCE}
    \end{equation}
    after replacing $U(\theta)$ with $\tilde{Q}\,U(\theta)\, Q$, where, for each $i\in\mathcal{I}$, $\tilde{P}^{j,\alpha}_i:=U(\xi_{j,\alpha})^\dagger \tilde{Q}^\dagger P_i\,\tilde{Q} \, U(\xi_{j,\alpha})$, and $\xi_{j,\alpha}:=\frac{\pi}{2}(\alpha-2j)$ for multi-indices $\alpha\in\mathbb{N}_0^D$ and $0\leq j\leq \alpha$.
}
\bigskip
\\
\noindent
\emph{Proof.} Let $j\in\mathcal{I}$ be arbitrary. By Theorem~\ref{thm:linear-clifford-encoder-special}, there exists a pair $(Q,\tilde{Q})$ of Clifford operators with the desired properties such that,
\begin{equation}
    \frac{c_j}{2}\left\{\langle0|Q^\dagger \tilde{P}^{0,e_k}_j\, Q|0\rangle - \langle0|Q^\dagger \tilde{P}^{e_k,e_k}_j\, Q|0\rangle\right\} = \pm c_j. \label{eq:single-out}
\end{equation}
Hence, after replacing $U(\theta)$ with $\tilde{Q}\,U(\theta)\, Q$, applying Lemma~\ref{lemma:parameter-shift} with $\xi_{0,e_k}=\frac{\pi}{2}e_k$ and $\xi_{e_k,e_k}=-\frac{\pi}{2}e_k$, yields
\begin{align}
    (\partial_{\theta_k}\tilde{C})(0) &= \frac{1}{2}\left\{\tilde{C}(\xi_{0,e_k})-\tilde{C}(\xi_{e_k,e_k})\right\} \nonumber
    \\ &= \frac{1}{2}\sum_{i\in\mathcal{I}}c_i\left\{\langle0|Q^\dagger U(\xi_{0,e_k})^\dagger \tilde{Q}^\dagger P_i\,\tilde{Q} \,U(\xi_{0,e_k})\,Q|0\rangle - \langle0|Q^\dagger U(\xi_{e_k,e_k})^\dagger \tilde{Q}^\dagger P_i\,\tilde{Q} \,U(\xi_{e_k,e_k})\,Q|0\rangle\right\} \nonumber
    \\ &= \frac{c_j}{2}\left\{\langle 0|Q^\dagger \tilde{P}^{0,e_k}_j\,Q|0\rangle - \langle 0|Q^\dagger \tilde{P}^{e_k,e_k}_j\,Q|0\rangle\right\}+\frac{1}{2}\sum_{i\in\mathcal{I}\setminus\{j\}}c_i\left\{\langle 0|Q^\dagger \tilde{P}^{0,e_k}_i\,Q|0\rangle - \langle 0|Q^\dagger \tilde{P}^{e_k,e_k}_i\,Q|0\rangle\right\} \nonumber
    \\ &\stackrel{\eqref{eq:single-out}}{=} \pm c_j+\frac{1}{2}\sum_{i\in\mathcal{I}\setminus\{j\}}c_i\left\{\langle0|Q^\dagger \tilde{P}^{0,e_k}_i\, Q|0\rangle - \langle0|Q^\dagger \tilde{P}^{e_k,e_k}_i\, Q|0\rangle\right\},
\end{align}
as claimed.
\qed

\section{Escaping Barren Plateaus \label{sec:escaping-BP}}
In the previous section, we introduced the Linear Clifford Encoder (LCE) to ensure constant gradient directions at zero initialization. We employ this construction now to mitigate the barren plateau problem~\cite{mcclean2018barren}. In what follows, $\|H\|$ denotes the operator norm of the observable in Eq.~\eqref{eq:observable2}.

\subsection{Taylor Perturbation around Clifford Points}

The proposition below characterizes the perturbation of higher-order derivatives $D_\theta^\alpha C(\theta)$ around their zero evaluation $(D_\theta^\alpha C)(0)$—interpreted here as Taylor coefficients—by an error term $\mathcal{O}(\mathbb{P}_\theta)$ that depends on the initial distribution $\mathbb{P}_\theta$.

\bigskip
\begin{proposition}\label{prop:landscape-statistics}
     Consider an expectation of the form Eq.~\eqref{eq:expectation3} with a variational ansatz $U(\theta)$ involving $D\geq1$ parameters. Then, for each multi-index $\alpha\in\mathbb{N}_0^D$, we have
    \begin{equation*}
        \mathbb{E}_\theta[D_\theta^\alpha C(\theta)] = (D_\theta^\alpha C)(0) + \|H\|\,\mathcal{O}(\mathbb{P}_\theta),
    \end{equation*}
    where the error term $\mathcal{O}(\mathbb{P}_\theta)$ scaling depends on the probability distribution $\mathbb{P}_\theta$.
\end{proposition}
\noindent
\emph{Proof.} Let $\alpha\in\mathbb{N}_0^D$ be an arbitrary multi-index. Since $C(\cdot)$ is analytic, we may use a Taylor expansion, combined with the uniform bound in higher-order parameter-shift rule, Lemma~\ref{lemma:parameter-shift}, finding
\begin{align}
    \mathbb{E}_\theta[D_\theta^\alpha C(\theta)] &= \sum_{\|\beta\|_1\geq 0}\underbrace{(D_\theta^{\alpha+\beta}C)(0)}_{=\,\mathcal{O}(\|H\|)}\,\frac{\mathbb{E}_\theta[\theta^{\alpha+\beta}]}{\alpha!\beta!} = (D_\theta^{\alpha} C)(0) + \mathcal{O}(\|H\|)\sum_{\|\beta\|_1>0}\frac{\mathbb{E}_\theta[\theta^{\alpha+\beta}]}{\alpha!\beta!} = (D_\theta^\alpha C)(0) + \|H\|\,\mathcal{O}(\mathbb{P}_\theta),
\end{align}
where we separated the zero term $(D_\theta^\alpha C)(0)$ (i.e., $\{\|\beta\|_1=0\}$) from the remaining terms in the summation over $\{\|\beta\|_1>0\}$ for which we applied Lemma~\ref{lemma:parameter-shift}. Note that the bound $(D_\theta^{\alpha+\beta}C)(0)=\mathcal{O}(\|H\|)$ also follows from Lemma~\ref{lemma:parameter-shift}. We defined the asymptotic error term
\begin{equation}
    \mathcal{O}(\mathbb{P}_\theta) := \mathcal{O}\left(\sum_{\|\beta\|_1>0}\frac{\mathbb{E}_\theta[\theta^{\alpha+\beta}]}{\alpha!\beta!}\right),
\end{equation}
which has a scaling depending on the probability distribution $\mathbb{P}_\theta$.

Similar to Proposition~\ref{prop:landscape-statistics}, the following theorem lower bounds the scaling of the expected squared gradient norm $\nabla_\theta\,C(\theta)$ under local Gaussian/uniform initialization $\mathbb{P}_\theta$ of the parameters. We generally assume that $\sigma<1$.

\bigskip
\begin{theorem}\label{thm:gradient-lower-bound}
    Consider an expectation of the form Eq.~\eqref{eq:expectation3} with a variational ansatz $U(\theta)$ involving $D\geq1$ parameters. Let $\mathbb{P}_\theta\in\{\mathcal{N}(0,\sigma^2I_D), \operatorname{Unif}([-\sigma,\sigma]^D)\}$ be a probability distribution with i.i.d. components $\theta_1,\dots,\theta_D$. Then, for every $k\in[D]$,
    \begin{equation*}
        \mathbb{E}_\theta[\|\nabla_\theta\,C(\theta)\|^2] \geq \mathbb{E}_{\theta}[(\partial_{\theta_k}C(\theta))^2] = (\partial_{\theta_k} C)(0)^2 + \|H\|^2\,\mathcal{O}(D\sigma^2).
    \end{equation*}
\end{theorem}
\noindent
\emph{Proof.} Let $k\in[D]$ be arbitrary. We may expand the gradient using the Taylor series, Eq.~\eqref{eq:taylor},
\begin{equation}
    \partial_{\theta_k}C(\theta)=(\partial_{\theta_k}C)(0)+\sum_{\|\alpha\|_1>0}\underbrace{(D^{\alpha+e_k}_\theta C)(0)}_{=\,\mathcal{O}(\|H\|)}\frac{\theta^\alpha}{\alpha!} = (\partial_{\theta_k}C)(0) +\mathcal{O}(\|H\|)\sum_{\|\alpha\|_1>0}\frac{\theta^\alpha}{\alpha!}, \label{eq:derivative-expansion}
\end{equation}
where we made use of the strong regularity property in Lemma~\ref{lemma:parameter-shift}, and using the observation that $\partial_{\theta_k} D_\theta^\alpha = D_\theta^{\alpha+e_k}$ for all $\alpha\in\mathbb{N}_0^D$. Taking the expectation then involves the computation of the well-known moments ($\delta$ denoting the Dirac delta function)
\begin{equation}
    \mathbb{E}_\theta[\theta^\alpha]=\delta_{\{\alpha_k\in2\mathbb{N}\,\forall k\}}(\alpha)\cdot\begin{cases} (\alpha-1)!!\,\sigma^\alpha,& \theta\sim\mathcal{N}(0,\sigma^2I_{D}), \\ \sigma^\alpha\prod_{k=1}^{D}(\alpha_k+1)^{-1},&\theta\sim\operatorname{Unif}([-\sigma,\sigma]^{D}), \end{cases} \label{eq:moments-parameters}
\end{equation}
which uses the assumption that $\theta_1,\dots,\theta_{D}$ are independent. Again, we use the canonical notations for $\sigma^\alpha:=\sigma^{\alpha_1}\cdots\sigma^{\alpha_{D}}$ and $(\alpha-1)!!:=\prod_{k\in[D]}(\alpha_k-1)!!$. This particularly implies that
\begin{equation}
    \sum_{\|\alpha\|_1>0}\frac{1}{\alpha!}\mathbb{E}_\theta[\theta^\alpha]=\mathcal{O}(D\sigma^2) = \sum_{\|\alpha\|_1,\|\beta\|_1>0}\frac{1}{\alpha!\,\beta!}\mathbb{E}_\theta[\theta^{\alpha+\beta}], \label{eq:moments-error}
\end{equation}
noting that there are exactly $D$ many multi-indices $\alpha\in\mathbb{N}_0^D$ for which the error bound $\mathbb{E}_\theta[\theta^\alpha]\stackrel{\eqref{eq:moments-parameters}}{=}\mathcal{O}(\sigma^2)$ holds true (namely $\{\alpha=2e_k:k\in[D]\}$). Note that the higher order terms $\mathcal{O}(\sigma^r)$ with $r>2$ are dominated by $\mathcal{O}(\sigma^2)$ since $\sigma<1$. As a result, we find
\begin{align}
    \mathbb{E}_\theta[\|\nabla_\theta\,C(\theta)\|^2] &= \sum_{j=1}^{D}\mathbb{E}_\theta[(\partial_{\theta_j}C(\theta))^2] \nonumber
    \\ &\geq \mathbb{E}_\theta[(\partial_{\theta_k}C(\theta))^2] \nonumber
    \\ &\stackrel{\eqref{eq:derivative-expansion}}{=} (\partial_{\theta_k} C)(0)^2 + 2\underbrace{(\partial_{\theta_k} C)(0)}_{=\,\mathcal{O}(\|H\|)}\,\mathcal{O}(\|H\|)\sum_{\|\alpha\|_1>0}\frac{\mathbb{E}_\theta[\theta^\alpha]}{\alpha!} + \mathcal{O}(\|H\|^2)\sum_{\|\alpha\|_1,\|\beta\|_1>0}\frac{\mathbb{E}_\theta[\theta^{\alpha+\beta}]}{\alpha!\,\beta!} \nonumber
    \\ &\stackrel{\eqref{eq:moments-error}}{=}  (\partial_{\theta_k} C)(0)^2 + \|H\|^2\,\mathcal{O}(D\sigma^2),
\end{align}
as desired.
\qed

\subsection{Constant Gradient Lower Bound in Small Patches}

Having derived a lower bound for the expected squared gradient norm with an error term perturbing a given component $(\partial_{\theta_k}C)(0)$, we combine this result with the LCE (cf. Theorem~\ref{thm:linear-clifford-encoder}) to establish a constant lower bound. In particular, this guarantees a tight escape from barren-plateaus as per Eq.~\eqref{eq:no-barren-plateau}:
\bigskip
\\
\noindent
\begin{corollary} \label{corr:escaping-barren-plateau}
    Consider an expectation of the form Eq.~\eqref{eq:expectation3} with a variational ansatz $U(\theta)$ consisting of Clifford-gates and single-qubit Pauli-rotation gates involving $D\geq1$ parameters. Assume that $H=P_i$ is a single Pauli string. For any arbitrarily small $\delta>0$, initialize $\theta\sim\mathbb{P}_\theta\in\{\mathcal{N}(0,\sigma^2I_D), \operatorname{Unif}([-\sigma,\sigma]^D)\}$ with i.i.d. components $\theta_1,\dots,\theta_D$, where $\sigma=\mathcal{O}(D^{-\frac{1+\delta}{2}})$. Then, there exists a pair of Clifford operators $(Q,\tilde{Q})$ (cf. Theorem~\ref{thm:linear-clifford-encoder}) such that
    \begin{equation*}
        \mathbb{E}_\theta[\|\nabla_\theta\,C(\theta)\|^2] \geq \mathbb{E}_{\theta}[(\partial_{\theta_k}C(\theta))^2] = 1 + \mathcal{O}(D^{-\delta}),
    \end{equation*}
    after replacing $U(\theta)$ by $\tilde{Q}\,U(\theta)\,Q$.
\end{corollary}
\bigskip
\noindent
\emph{Proof.} Let $k\in[D]$ be arbitrary. By Theorem~\ref{thm:gradient-lower-bound} we know that
\begin{equation}
    \mathbb{E}_\theta[\|\nabla_\theta\,C(\theta)\|^2] \geq \mathbb{E}_{\theta}[(\partial_{\theta_k}C(\theta))^2] = (\partial_{\theta_k} C)(0)^2 + \mathcal{O}(D\sigma^2),
\end{equation}
noting that $\|H\|=1$ if $H=P$. On the other hand, Theorem~\ref{thm:linear-clifford-encoder} shows how to find the desired pair $Q$, $\tilde{Q}$ of Clifford operators such that
\begin{equation}
    (\partial_{\theta_k}C)(0) = \pm1
\end{equation}
after replacing $U(\theta)$ with $\tilde{Q}\,U(\theta)\, Q$. Therefore, observing that
\begin{align}
    \sigma=\mathcal{O}\left(D^{-\frac{1+\delta}{2}}\right) \implies \mathcal{O}(D\sigma^2) = \mathcal{O}(D^{-\delta})
\end{align}
concludes the proof.
\qed
\bigskip
\\
\indent
As in Theorem~\ref{thm:linear-clifford-encoder}, this result may be generalized to arbitrary observables $H$ as defined in Eq.~\eqref{eq:observable2} with a lower bound depending on the distribution of the Pauli coefficients $\{c_i\}_{i\in\mathcal{I}}$.
\bigskip
\\
\noindent
\textbf{Theorem \ref{thm:general-escaping-barren-plateau}.}
    \emph{Let $H$ be an arbitrary observable as in Eq.~\eqref{eq:observable} with Pauli terms indexed by $\mathcal{I}$. Let $\delta>0$ and initialize $\theta \sim \mathbb{P}_\theta \in \{\mathcal{N}(0,\sigma^2 I_D), \operatorname{Unif}([-\sigma,\sigma]^D)\}$ with i.i.d. components and $\sigma = \mathcal{O}(\|H\|^{-1}D^{-\frac{1+\delta}{2}})$. Then, for each $i_0\in\mathcal{I}$, there exists a pair of Clifford operators $(Q,\tilde{Q})$ such that
    \begin{equation*}
        \mathbb{E}_\theta[\|\nabla_\theta\,\tilde{C}(\theta)\|^2] \geq \mathbb{E}_{\theta}[(\partial_{\theta_k}\tilde{C}(\theta))^2] = \beta_{i_0}(H)^2 + \mathcal{O}(D^{-\delta}),
    \end{equation*}
    where 
    \begin{align*}
        \beta_{i_0}(H) := \pm c_{i_0} + \frac{1}{2} \sum_{i \in \mathcal{I} \setminus \{i_0\}} c_i \big\{ 
        \langle 0|\tilde{P}^{0,e_k}_i|0\rangle - \langle 0| \tilde{P}^{e_k,e_k}_i |0\rangle \big\},
    \end{align*}
    defining $\tilde{P}^{j,\alpha}_i:=Q^\dagger U(\xi_{j,\alpha})^\dagger \tilde{Q}^\dagger P_i \tilde{Q} U(\xi_{j,\alpha})Q$ and $\xi_{j,\alpha}:=\frac{\pi}{2}(\alpha - 2j)$ for $0\leq j\leq \alpha$, and $\tilde{C}(\cdot)$ denoting the LCE transformed cost. In particular, $\beta_{i_0}(H)=1$ if $H=P$ is a single Pauli string. {In particular, $\mathbb{E}_\theta[\|\nabla_\theta\,\tilde{C}(\theta)\|^2] = \Omega(1)$ with probability at least $1-\mathcal{O}(|\mathcal{I}|2^{-N})$ if $c_{i_0}=\Omega(1)$.}}
\bigskip
\\
\noindent
\emph{Proof.}  Let $k\in[D]$ and $i_0\in\mathcal{I}$ be arbitrary. By Theorem~\ref{thm:gradient-lower-bound} we know that
\begin{equation}
    \mathbb{E}_\theta[\|\nabla_\theta\,C(\theta)\|^2] \geq \mathbb{E}_{\theta}[(\partial_{\theta_k}C(\theta))^2] = (\partial_{\theta_k} C)(0)^2 + \|H\|^2\,\mathcal{O}(D\sigma^2),
\end{equation}
On the other hand, Theorem~\ref{thm:linear-clifford-encoder} yields the desired pair $(Q,\tilde{Q})$ of Clifford operators such that
\begin{equation}
    (\partial_{\theta_k}\tilde{C})(0) = \pm c_{i_0} + \frac{1}{2} \sum_{i \in \mathcal{I} \setminus \{i_0\}} c_i \big\{ 
        \langle 0|\tilde{P}^{0,e_k}_i|0\rangle - \langle 0| \tilde{P}^{e_k,e_k}_i |0\rangle \big\} = \beta_{i_0}(H),
\end{equation}
after replacing $U(\theta)$ with $\tilde{Q}\,U(\theta)\, Q$ via the LCE transformation. Therefore, observing that
\begin{align}
    \sigma=\mathcal{O}\left(\|H\|^{-1}D^{-\frac{1+\delta}{2}}\right) \implies \|H\|^2\,\mathcal{O}(D\sigma^2) = \mathcal{O}(D^{-\delta}).
\end{align}
{The second statement follows from the assumption that $c_{i_0}=\Omega(1)$, which dominates the decaying term $\mathcal{O}(D^{-\delta})$, and Lemma~\ref{lemma:observable-prob} shown independently below.} This concludes the proof.
\qed
\bigskip
\par
The observable-dependent lower bound $\beta_{i_0}(H)^2$ in Theorem~\ref{thm:general-escaping-barren-plateau} is only meaningful if it behaves like $\Omega(1)$, i.e., when it dominates error term $\mathcal{O}(D^{-\delta})$. The scaling behavior of $\beta_{i_0}(H)$ is less straightforward to analyze due to the cancellations between Pauli terms. The following lemma shows that the probability of total cancellation (i.e., $\beta_{i_0}(H)=0$), conditioned on $P_{i_0}$ which is given for the LCE construction, is vanishing exponentially in the number of qubits $N$:
\bigskip
\\
\noindent
\textbf{Lemma \ref{lemma:observable-prob}.}
    \emph{Let $H=\sum_{i\in\mathcal{I}}c_iP_i$ be a random observable with $P_i\stackrel{i.i.d.}{\sim}\operatorname{Unif}(\mathcal{P}_N\setminus\{I\})$. Then, for each $i_0\in\mathcal{I}$, we have the conditional probability that
    \begin{equation*}
    \mathbb{P}\left(\beta_{i_0}(H)^2 = c_{i_0}^2 \mid P_{i_0}) \geq 1-\mathcal{O}(|\mathcal{I}|\,2^{-N}\right),
    \end{equation*}
    as $N\to\infty$.}
\bigskip
\\
\noindent
\emph{Proof.} Recall the definition of the $N$-qubit Pauli group,
\begin{equation}
    \mathcal{P}_N:=\{\pm P,\pm iP: P\in\{I,X,Y,Z\}^{\otimes N}\},
\end{equation}
which contains $4\cdot 4^N=4^{N+1}$ elements. In the construction of LCE, we were particularly interested in the subgroup
\begin{equation}
    \langle I,Z\rangle_N = \{\pm P, \pm i P: P\in\{I,Z\}^{\otimes N}\}
\end{equation}
of diagonal Pauli strings, containing $4\cdot 2^{N}=2^{N+2}$ elements.
\par
First, we note that a uniformly random (non-identity) Pauli string $P$ satisfies that
\begin{equation}
    \mathbb{P}(\langle0|P|0\rangle = 1)=\frac{\#\langle I,Z\rangle_N\setminus\{I\}}{\#\mathcal{P}_N\setminus \{I\}} = \frac{2^{N+2}-1}{4^{N+1}-1} = \mathcal{O}(2^{-N}),
\end{equation}
since the only non-zero Pauli expectations are achieved by diagonal operators in $\langle I,Z\rangle_N$. In other words, the probability of having a non-zero expectation is determined by the total number of non-identity diagonal Pauli matrices divided by the total number of non-identity Pauli matrices. That is, for any uniformly random (non-identity) Pauli string $P$ we have
\begin{equation}
    \mathbb{P}(\langle0|P|0\rangle=0) = 1-\mathcal{O}(2^{-N}), \label{eq:pauli-prob}
\end{equation}
which we will use to conclude with the lower bound.
\par
Recall from the proof of Theorem~\ref{thm:linear-clifford-encoder-special} that for each $(j,\alpha)$, and $i\in\mathcal{I}$,
\begin{equation}
    \tilde{P}_i^{j,\alpha}:=Q^\dagger U_{j,\alpha}^\dagger \tilde{Q}^\dagger P_i\,\tilde{Q} \, U_{j,\alpha} Q
\end{equation}
for some deterministic $U_{j,\alpha}\in\operatorname{Clifford}(N)$ (i.e., $U(\cdot)$ evaluated at certain parameter points) depending on the LCE direction $k\in[D]$ (since $\alpha=e_k$ and $j\in\{0,e_k\}$) which is deterministically fixed by assumption. On the other hand, the LCE pair $(Q,\tilde{Q})$ depends on both the (deterministic) LCE direction $k\in[D]$, and Pauli term $i_0\in\mathcal{I}$.

Crucially, the LCE pair $(Q,\tilde{Q})$ is explicitly constructed from the random Pauli string $P_{i_0}$ in order to isolate the Pauli coefficient $c_{i_0}$, thus causing $(Q,\tilde{Q})$ to be random as well. Hence, given the choice of the Pauli string $P_{i_0}$ for the LCE construction, the pair $(Q,\tilde{Q})$ is also given, which in turn results in $\tilde{P}_i^{j,\alpha}$ being pair-wise independent in $i\in\mathcal{I}$ with respect to the conditional probability $\mathbb{P}(\cdot \mid P_{i_0})$.

Defining $d_i:=\langle 0|\tilde{P}^{0,e_k}_i|0\rangle - \langle 0| \tilde{P}^{e_k,e_k}_i |0\rangle$, we thus obtain
\begin{align}
    \mathbb{P}(d_i=0\,\forall i\in\mathcal{I}\setminus\{i_0\} \mid P_{i_0}) \stackrel{i.i.d.}{=} \prod_{i\in\mathcal{I}\setminus\{i_0\}} \mathbb{P}(d_i=0 \mid P_{i_0}) \stackrel{\eqref{eq:pauli-prob}}{\geq} (1-\mathcal{O}(2^{-N}))^{|
    \mathcal{I}|} = 1-\mathcal{O}(|\mathcal{I}|\,2^{-N}).
\end{align}
Consequently, by the previous estimation,
\begin{equation}
    \mathbb{P}\left(\sum_{i\in\mathcal{I}\setminus\{i_0\}}c_id_i=0 \Bigm\vert P_{i_0}\right) \geq \mathbb{P}(d_i=0\,\forall i\in\mathcal{I}\setminus\{i_0\} \mid P_{i_0}) \geq 1 - \mathcal{O}(|\mathcal{I}|\,2^{-N}),
\end{equation}
which directly implies the desired claim since by definition,
\begin{equation}
    \beta_{i_0}(H) = \pm c_{i_0} + \frac{1}{2} \sum_{i \in \mathcal{I} \setminus \{i_0\}} c_id_i.
\end{equation}
\qed

\section{State Expressivity of Parameterized Quantum Circuits \label{sec:state-expressivity}}
It is essential to verify that the LCE transformation $C(\cdot)\mapsto\tilde{C}(\cdot)$ preserves desirable properties of the optimization landscape.  Recall that $\tilde{C}(\cdot)$ denotes the LCE transformed cost obtained by replacing $U(\cdot)$ with $\tilde{U}(\cdot) := \tilde{Q} U(\cdot) Q$ as in Theorem~\ref{thm:linear-clifford-encoder}. We show that both expressivity and global optimality remain intact, using the following geometric notion of expressivity:

\begin{definition}[$\varepsilon$-State-Expressivity]\label{def:expressivity}
    Let $\varepsilon > 0$. A PQC $U(\cdot)$ is said to be $\varepsilon$-state-expressive if for every state $|\psi\rangle \in \mathcal{H}_N$ and every stabilizer state $|\psi_0\rangle \in \operatorname{Stab}(N)$, there exists a parameter configuration $\theta \in [-\pi, \pi]^D$ such that
    \begin{equation}
        \|U(\theta)|\psi_0\rangle - |\psi\rangle\|_2 < \varepsilon.
    \end{equation}
\end{definition}

Here, it suffices to consider the metric
\begin{equation}
    \||\psi\rangle - |\phi\rangle\|_2 := \sqrt{2(1 - \operatorname{Re}\langle \psi | \phi \rangle)},
\end{equation}
induced by the $\ell_2$-norm, to measure the distance between quantum states $|\psi\rangle, |\phi\rangle \in \mathcal{H}_N$. The reason for considering arbitrary stabilizer inputs $|\psi_0\rangle \in \operatorname{Stab}(N)$ is that the expressivity of the model $U(\cdot)$ should not depend on a finite set of classically tractable input states.

This definition captures the geometric capacity of the ansatz $U(\cdot)$ to explore the Hilbert space $\mathcal{H}_N$. Intuitively, if $U(\cdot)$ explores $\mathcal{H}_N$ in an unbiased fashion, then the KL-divergence $D_{\mathrm{KL}}(\hat{P}_{\mathrm{PQC}}(F;\theta)\,\|\,P_{\mathrm{Haar}}(F))$ in Ref.~\cite{sim2019expressibility} is small if and only if $U(\cdot)$ is $\varepsilon$-state-expressive for sufficiently small $\varepsilon > 0$.

Recall that we consider a Variational Quantum Eigensolver (VQE)~\cite{Peruzzo2014VQE} problem of the form
\begin{equation}
    \theta_*=\arg\min_{\theta\in[-\pi,\pi]^D}C(\theta), \label{eq:minimizer2}
\end{equation}
where $C(\theta_*)$ approximates the ground state energy $\lambda_{\min}(H)$ of the observable in Eq.~\eqref{eq:observable2}. We first show that sufficient $\varepsilon$-state-expressivity guarantees the existence of a good solution to Eq.~\eqref{eq:minimizer2}:

\bigskip
\begin{lemma}\label{lemma:vqe-solution-existence}
    Assume that the quantum model $U(\cdot)$ is $\frac{\varepsilon}{2\|H\|}$-state-expressive. Then, there exists a configuration $\theta_*\in[-\pi,\pi]^D$ such that
    \begin{equation*}
        |C(\theta_*)-\lambda_{\min}(H)| < \varepsilon,
    \end{equation*}
    where $\lambda_{\min}(H)$ denotes the smallest eigenvalue of the observable in Eq.~\eqref{eq:observable2}.
\end{lemma}
\noindent
\emph{Proof.} Let $|\psi\rangle\in\mathcal{H}_N$ be the eigenstate corresponding to the smallest eigenvalue $\lambda_{\min}(H)$ of $H$. That is, $H|\psi\rangle=\lambda_{\min}(H)|\psi\rangle$. Then, since $U(\cdot)$ is $\frac{\varepsilon}{2\|H\|}$-state-expressive, there exists a configuration $\theta_*\in[-\pi,\pi]^D$ such that
\begin{equation}
    \|U(\theta_*)|0\rangle - |\psi\rangle\| < \frac{\varepsilon}{2\|H\|}. \label{eq:state-bound}
\end{equation}
As a consequence, defining $|\phi\rangle:=U(\theta_*)|0\rangle$,
\begin{align}
    |C(\theta_*) - \lambda_{\min}(H)| &= |\langle\phi|H|\phi\rangle - \langle\psi|H|\psi\rangle| \nonumber
    \\ &= |(\langle\phi|-\langle\psi|)H|\phi\rangle + \langle\psi|H(|\phi\rangle-|\psi\rangle)| \nonumber
    \\ &\leq |(\langle\phi|-\langle\psi|)H|\phi\rangle|+|\langle\psi|H(|\phi\rangle-|\psi\rangle)| \nonumber
    \\ &\leq \|H\|\,\underbrace{\||\phi\rangle\|}_{=1}\,\||\phi\rangle - |\psi\rangle\| + \|H\|\,\underbrace{\||\psi\rangle\|}_{=1}\,\||\phi\rangle - |\psi\rangle\| = 2\|H\|\,\||\phi\rangle - |\psi\rangle\| \stackrel{\eqref{eq:state-bound}}{<} \varepsilon,
\end{align}
where we employed the triangle inequality and Cauchy Schwarz, respectively.
\qed
\bigskip
\\
\noindent
Lemma~\ref{lemma:vqe-solution-existence} proves that, under its assumptions, $C(\theta_*)$ provides a good approximation of the ground energy of $H$, and thus solves the VQE problem in Eq.~\eqref{eq:minimizer2}. The next step is to establish that the LCE transformation $C(\cdot)\mapsto\tilde{C}(\cdot)$ does not compromise $\varepsilon$-state-expressivity. That is, the quantum model is just as $\varepsilon$-state-expressive after replacing $U(\cdot)$ with $\tilde{Q}\,U(\cdot)\,Q$ as per Theorem~\ref{thm:linear-clifford-encoder}:
\bigskip
\begin{lemma}\label{lemma:invariant-expressivity}
    If the quantum model $U(\cdot)$ is $\varepsilon$-state-expressive, then so is $\tilde{Q}\,U(\cdot)\,Q$ for each $Q,\tilde{Q}\in\operatorname{Clifford}(N)$. In particular, the LCE transformation $C(\cdot)\mapsto\tilde{C}(\cdot)$ leaves the $\varepsilon$-state-expressivity of the quantum model $U(\cdot)$ invariant.
\end{lemma}
\noindent
\emph{Proof.} Let $|\psi_0\rangle\in\operatorname{Stab}(N)$, and $|\psi\rangle\in\mathcal{H}_N$ be arbitrary. By assumption that $U(\cdot)$ is $\varepsilon$-state-expressive, we find
\begin{align}
    \|\tilde{Q}U(\theta)Q|\psi_0\rangle- |\psi\rangle\| = \|\tilde{Q}U(\theta)|\tilde{\psi}_0\rangle-|\psi\rangle\| = \|U(\theta)|\tilde{\psi}_0\rangle-|\tilde{\psi}\rangle\| < \varepsilon,
\end{align}
using that $|\tilde{\psi}_0\rangle:=Q|\psi_0\rangle\in\operatorname{Stab}(N)$ defines a stabilizer state, and that $\tilde{Q}$ is unitary, leaving the norm invariant in the second equality with $|\tilde{\psi}\rangle:=\tilde{Q}^\dagger|\psi\rangle$.
\qed
\bigskip
\\
\indent
It remains to show that the global quality of the optimization landscape is preserved under the LCE transformation $C(\cdot)\mapsto\tilde{C}(\cdot)$. Specifically, we prove that the global minimum value satisfies $C(\theta_*)\approx\tilde{C}(\theta_*)$, with an error controllable by assuming stronger $\varepsilon$-state-expressivity. The proof relies on ensuring that a change of basis leaves the eigenspectrum invariant:
\bigskip
\begin{lemma}\label{lemma:invariant-eigenvalues}
    Let $\tilde{Q}$ be an arbitrary unitary operator, and define $\tilde{H}:=\tilde{Q}^\dagger H\,\tilde{Q}$, where $H$ denotes the observable in Eq.~\eqref{eq:observable}. Then, $H$ and $\tilde{H}$ share the same eigenspectrum.
\end{lemma}
\noindent
\emph{Proof.} Let $\lambda(H)$ be an arbitrary eigenvalue of $H$, and take the corresponding eigenstate $|\psi\rangle\in\mathcal{H}_N$. In other words, $H|\psi\rangle = \lambda(H)|\psi\rangle$. Define the state $|\tilde{\psi}\rangle:=\tilde{Q}^\dagger|\psi\rangle$. Then,
\begin{equation}
    \tilde{H}|\tilde{\psi}\rangle = \tilde{Q}^\dagger H\underbrace{\tilde{Q}\tilde{Q}^\dagger}_{=I}|\psi\rangle = \tilde{Q}^\dagger \underbrace{H|\psi\rangle}_{=\lambda(H)|\psi\rangle} = \lambda(H)\tilde{Q}^\dagger|\psi\rangle = \lambda(H)|\tilde{\psi}\rangle,
\end{equation}
implying that $\lambda(H)=\lambda(\tilde{H})$. The claim follows since $\lambda(H)$ was arbitrary.
\qed
\bigskip
\\
\indent
We are now ready to prove the remaining statement:
\bigskip
\begin{proposition}\label{prop:invariant-minimum}
    Assume that the quantum model $U(\cdot)$ is $\frac{\varepsilon}{4\|H\|}$-state-expressive. Then, we have that
    \begin{equation*}
        \left\lvert \min_{\theta\in[-\pi,\pi]^D} C(\theta) - \min_{\theta\in[-\pi,\pi]^D}\tilde{C}(\theta) \right\rvert < \varepsilon,
    \end{equation*}
    where $\tilde{C}(\cdot)$ denotes the cost after replacing $U(\cdot)$ by $\tilde{Q}\,U(\cdot)\,Q$. This ensures that the LCE does not compromise the quality of the global minimum.
\end{proposition}
\noindent
\emph{Proof.} By definition, the LCE transformed cost equals
\begin{equation}
    \tilde{C}(\theta) := \langle 0| Q^\dagger U(\theta)^\dagger\tilde{H}U(\theta) Q|0\rangle,
\end{equation}
where we introduced the transformed observable $\tilde{H}=\tilde{Q}^\dagger H \tilde{Q}$. The goal is to prove that $C(\cdot)$ and $\tilde{C}(\cdot)$ have approximately the same global minimum. The idea is to first make use of state expressivity in order ensure that $C(\cdot)$ is able to approximate the smallest eigenvalue $\lambda_{\min}(H)$. This step is achieved with the help of Lemma~\ref{lemma:vqe-solution-existence}. Next, we use Lemma~\ref{lemma:invariant-expressivity} to establish an identical bound for the LCE transformed cost $\tilde{C}(\cdot)$ approximating $\lambda_{\min}(\tilde{H})$. To conclude with the proof, we employ Lemma~\ref{lemma:invariant-eigenvalues} stating that the change of basis $H\mapsto\tilde{Q}^\dagger H \tilde{Q}=\tilde{H}$ leaves the eigenspectrum invariant, i.e., $\lambda_{\min}(H)=\lambda_{\min}(\tilde{H})$, allowing us to combine both previous bounds for the desired result.

We now elaborate the above proof strategy more formally: Since $U(\cdot)$ is $\frac{\varepsilon}{4\|H\|}$-state-expressive by assumption, it follows from Lemma~\ref{lemma:vqe-solution-existence} that we can approximate the lowest eigenvalue for the original cost:
\begin{equation}
    \left\vert\min_{\theta\in[-\pi,\pi]^D}C(\theta) - \lambda_{\min}(H)\right\vert<\frac{\varepsilon}{2}.
\end{equation}
By Lemma~\ref{lemma:invariant-expressivity} we know that the LCE transformation leaves $\varepsilon$-expressivity invariant, thus establishing the exact same bound for the LCE transformed cost:
\begin{equation}
    \left\vert\min_{\theta\in[-\pi,\pi]^D}\tilde{C}(\theta)-\lambda_{\min}(\tilde{H})\right\vert < \frac{\varepsilon}{2}.
\end{equation}
As a result, by the triangle inequality,
\begin{align}
    \left\lvert \min_{\theta\in[-\pi,\pi]^D} C(\theta) - \min_{\theta\in[-\pi,\pi]^D}\tilde{C}(\theta) \right\rvert  &\leq \left\vert \min_{\theta\in[-\pi,\pi]^D} C(\theta) - \lambda_{\min}(H) \right\vert + \left\vert \min_{\theta\in[-\pi,\pi]^D} \tilde{C}(\theta) - \lambda_{\min}(H)\right\vert \nonumber
    \\ &= \left\vert \min_{\theta\in[-\pi,\pi]^D} C(\theta) - \lambda_{\min}(H) \right\vert + \left\vert \min_{\theta\in[-\pi,\pi]^D} \tilde{C}(\theta) - \lambda_{\min}(\tilde{H})\right\vert < \frac{\varepsilon}{2}+\frac{\varepsilon}{2}=\varepsilon,
\end{align}
where we made use of Lemma~\ref{lemma:invariant-eigenvalues}, which states that $\lambda_{\min}(H)=\lambda_{\min}(\tilde{H})$.
\qed
\bigskip
\\
\indent
Proposition~\ref{prop:invariant-minimum} confirms that the LCE transformation preserves the quality of the global minimum. Notably, we make no structural assumptions on the PQC ansatz $U(\cdot)$ (e.g., no lightcone or locality constraints), allowing us to assume $\varepsilon$-state-expressivity for arbitrarily small $\varepsilon > 0$.

\section{Minimal Truncation Threshold of the Taylor Surrogate \label{sec:truncation}}
We approximate the PQC using the Taylor surrogate $C_m(\theta)$ from Eq.~\eqref{eq:taylor}, where the truncation threshold $m = m(\theta)$ determines computational complexity. To meet a prescribed surrogate error tolerance, we analyze the minimal necessary $m$ via two approaches:
\begin{enumerate}
    \item[\textbf{1.}] \textbf{Worst-Case Error:} $m = m(\theta)$ is a deterministic function of a fixed configuration $\theta \in [-\pi, \pi]^D$;
    \item[\textbf{2.}] \textbf{Mean-Squared Error:} $m = m(\mathbb{P}_\theta)$ is defined probabilistically based on an initialization distribution $\mathbb{P}_\theta$.
\end{enumerate}

\subsection{Worst-Case Error Analysis}
The following lemma quantifies the (deterministic) worst-case error of the approximation $C_m(\theta)\approx C(\theta)$ as a function of $\theta$.

\begin{lemma} \label{lemma:taylor-remainder}
    Consider an expectation of the form Eq.~\eqref{eq:expectation3} with a variational ansatz $U(\theta)$ involving $D\geq1$ parameters, and suppose we simulate it classically using the Taylor series $C_m(\theta)$ defined in Eq.~\eqref{eq:taylor2}. Let $\theta\in[-\pi,\pi]^D$ and $m\in\mathbb{N}$ be arbitrary. Then, for every $\theta\in[-\pi,\pi]^D$ and $m\in\mathbb{N}$, we have
    \begin{equation}
        |C(\theta)-C_m(\theta)|\leq \frac{\|H\|}{m!}\|\theta\|_1^{m}, \label{eq:taylor-error}
    \end{equation}
    where $\|H\|$ is the operator norm of the observable in Eq.~\eqref{eq:observable2}.
\end{lemma}
\noindent
\emph{Proof.} Let $\theta\in[-\pi,\pi]^D$ and $m\in\mathbb{N}$ be arbitrary. Using Lagrange's form of the remainder in the Taylor series, we find an element $\xi\in[0,\theta]$ in the line segment from $0$ to $\theta$, such that
\begin{align}
    C(\theta) - C_m(\theta) &= \sum_{\|\alpha\|_1 \geq m}(D_\theta^\alpha C)(0)\,\frac{\theta^\alpha}{\alpha!} = \sum_{\|\alpha\|_1=m}\underbrace{(D_\theta^\alpha C)(\xi)}_{\leq\|H\|}\,\frac{\theta^\alpha}{\alpha!} \nonumber
    \\ &\leq \|H\|\sum_{\|\alpha\|_1=m}\left\lvert \frac{\theta^\alpha}{\alpha!}\right\rvert \leq \|H\|\,\frac{1}{m!}\left(\sum_{k=1}^D|\theta_k|\right)^{m} = \frac{\|H\|}{m!}\|\theta\|_1^{m},
\end{align}
where, in the first estimate we made use of the uniform bound in the higher-order parameter-shift rule, Lemma~\ref{lemma:parameter-shift}, and the multi-nomial theorem in the second, i.e.,
\begin{equation}
    \sum_{\|\alpha\|_1=m}\binom{m}{\alpha}\theta^\alpha = \left(\sum_{k=1}^D\theta_k\right)^m, \quad \binom{m}{\alpha} := \frac{m!}{\alpha_1!\cdots\alpha_D!}.
\end{equation}
\qed
\bigskip
\\
\indent
In order to solve for the truncation threshold $m$, we make use of asymptotic properties of the Lambert-$W$-function~\cite{corless1996lambert}:
\bigskip
\begin{lemma}\label{lemma:lambert-asymptotics}
    Let $W(\cdot)$ denote the main branch of the Lambert-$W$-function~\cite{corless1996lambert}. Then,
    \begin{equation*}
        W(x)=x+\mathcal{O}(x^2)
    \end{equation*}
    as $x\to0$.
\end{lemma}
\noindent
\emph{Proof.} By definition, $W(x)\,e^{W(x)}=x$. In particular, $W(0)=0$. We shall use the implicit function theorem to prove the lemma statement. For that purpose, let us define the analytic function $F(w,x):=we^w-x$. By the previous observation, we have $F(0,0)=0$. Next, we note that $\partial_w F(w,x) = (w+1)e^w$ which implies $(\partial_w F)(0,0)=1\neq0$. By the implicit function theorem, there exists a unique analytic function $W(\cdot)$ satisfying $W(0)=0$ and $F(W(x),x)=0$ in a neighborhood of $x=0$. By uniqueness, $W(\cdot)$ has to coincide with the principal branch of the lambert-$W$-function~\cite{corless1996lambert}. Since $W(\cdot)$ is analytic (in a neighborhood of $x=0$), we may consider the Taylor expansion,
\begin{equation}
    W(x)=W(0)+W'(0)\,x+\mathcal{O}(x^2). \label{eq:taylor-lambert}
\end{equation}
Taking the derivative on both sides of $W(x)e^{W(x)}=x$ yields
\begin{equation}
    W'(x)e^{W(x)}(1+W(x))=1 \implies W'(x)=\frac{e^{-W(x)}}{1+W(x)} \implies W'(0)=1
\end{equation}
making use of $W(0)=0$. Thus, Eq.~\eqref{eq:taylor-lambert} reduces to the desired asymptotics $W(x)=x+\mathcal{O}(x^2)$ as $x\to0$.
\qed
\bigskip
\\
\indent
We are now ready to deterministically lower bound the truncation threshold $m(\theta)$:
\bigskip
\begin{lemma}\label{lemma:truncation-threshold}
    Consider an expectation in Eq.~\eqref{eq:expectation3} with a variational ansatz $U(\theta)$ involving $D\geq1$ parameters, simulated classically via the Taylor series $C_m(\theta)$ from Eq.~\eqref{eq:taylor2}. Assume the Taylor remainder in Eq.~\eqref{eq:taylor-error} is bounded by a tolerance $\varepsilon>0$, and let $m=m(\theta)$ be the truncation order in Eq.~\eqref{eq:taylor2}. Then, the truncation threshold $m=m(\theta)$ satisfies the lower bound
    \begin{align*}
     m(\theta)\geq   e\|\theta\|_1\exp\left(W\left(\frac{1}{e\|\theta\|_1}\log\left(\frac{\|H\|}{\varepsilon\sqrt{e^3\|\theta\|_1}}\right)\right)\right)-\frac{1}{2},
    \end{align*}
    where $e\approx 2.71828$ is the natural constant, $W(\cdot)$ denotes the main branch of the Lambert-$W$ function~\cite{corless1996lambert}, and $\|H\|$ is the operator norm of the observable in Eq.~\eqref{eq:observable2}. In particular, if $\|\theta\|_1\to\infty$ increases as $D\to\infty$, then the asymptotic equivalence $m(\theta)\sim e\|\theta\|_1$ is sufficient for the truncation threshold to ensure the error tolerance $\varepsilon$ (where the contributions of $\|H\|$ and $\varepsilon$ become negligible).
\end{lemma}
\noindent
\emph{Proof.} Let $\varepsilon>0$ be a given error tolerance. Without loss of generality $\theta\neq0$, so that $m(\theta)\geq2$. Otherwise if $\theta=0$, the Taylor approximation is exact with $m(0)=1$. Utilizing a Stirling approximation yields the lower bound
\begin{equation}
    m! \leq e\sqrt{m}\left(\frac{m}{e}\right)^m \label{eq:stirling}
\end{equation}
for all $m\geq 2$. This bound is derived from the remark of Robbins~\cite{robbins1955remark} that for all $m\geq1$,
\begin{equation}
    m!\leq \sqrt{2\pi m}\left(\frac{m}{e}\right)^me^{\frac{1}{12m}},
\end{equation}
and verifying that for all $m\geq 2$,
\begin{equation}
    \sqrt{2\pi}e^{\frac{1}{12m}} \leq e.
\end{equation}

As a result, we require by Lemma~\ref{lemma:taylor-remainder} that, that the Taylor approximation error is bounded by
\begin{equation}
    \log\varepsilon\stackrel{!}{\geq}\log\left(\frac{\|H\|}{m!}\|\theta\|_1^m\right) \stackrel{\eqref{eq:stirling}}{\geq} \log\left(\frac{\|H\|}{e\sqrt{m}}\frac{(e\|\theta\|_1)^m}{m^m}\right) = \log\|H\|\,-\, \frac{1}{2}\log(e^2m) - m\log\left(\frac{m}{e\|\theta\|_1}\right). \label{eq:error-bound-log}
\end{equation}
Noting that $\log\varepsilon=-\log(1/\varepsilon)$, we may manipulate the following expression:
\begin{align}
    \gamma:=\frac{1}{e\|\theta\|_1}\log\left(\frac{\|H\|}{\varepsilon\sqrt{e^3\|\theta\|_1}}\right) &= -\frac{1}{e\|\theta\|_1}\log\varepsilon + \frac{1}{e\|\theta\|_1}\log\|H\|  \,-\, \frac{1}{2e\|\theta\|_1}\log(e^3\|\theta\|_1) \nonumber
    \\ &\stackrel{\eqref{eq:error-bound-log}}{\leq} \frac{1}{2e\|\theta\|_1}\log(e^2 m) + \frac{m}{e\|\theta\|_1}\log\left(\frac{m}{e\|\theta\|_1}\right) - \frac{1}{2e\|\theta\|_1}\log(e^3\|\theta\|_1) \nonumber
    \\ &= \frac{1}{2e\|\theta\|_1}\underbrace{\left\{\log(e^2m)-\log(e^3\|\theta\|_1)\right\}}_{=\log\left(\frac{m}{e\|\theta\|_1}\right)} + \frac{m}{e\|\theta\|_1}\log\left(\frac{m}{e\|\theta\|_1}\right)
    \\ &= \frac{2m+1}{2e\|\theta\|_1}\log\left(\frac{m}{e\|\theta\|_1}\right) \leq \frac{2m+1}{2e\|\theta\|_1}\log\left(\frac{2m+1}{2e\|\theta\|_1}\right).\label{eq:upper-bound2}
\end{align}
We have therefore established the bound
\begin{equation}
    \gamma \stackrel{\eqref{eq:upper-bound2}}{\leq} y\log y=:f(y),\quad\text{substituting $y:=\frac{2m+1}{2e\|\theta\|_1}$}. \label{eq:upper-bound3}
\end{equation}
The equation $f(y)=\gamma$ is known to be solved by the main branch of the Lambert-$W$-function~\cite{corless1996lambert} with solution $y = e^{W(\gamma)}$. Moreover, the function $f$ is invertible on $y>1$. In that case,
\begin{equation}
    \frac{2m+1}{2e\|\theta\|_1}=y \stackrel{\eqref{eq:upper-bound3}}{\geq} f^{-1}(\gamma) = e^{W(\gamma)} = \frac{1}{e\|\theta\|_1}\log\left(\frac{\|H\|}{\varepsilon\sqrt{e^3\|\theta\|_1}}\right),
\end{equation}
or equivalently,
\begin{equation}
    m \geq   e\|\theta\|_1\exp\left(W\left(\frac{1}{e\|\theta\|_1}\log\left(\frac{\|H\|}{\varepsilon\sqrt{e^3\|\theta\|_1}}\right)\right)\right)-\frac{1}{2} =:b(\theta). \label{eq:lower-bound-m}
\end{equation}

In order to prove that $y>1$ always holds true in our case, we prove the following: If $\|\theta\|_1\to\infty$ as $D\to\infty$, then there exists a lower bound $b(\theta) \sim e\|\theta\|_1$ as $D\to\infty$, where $b(\theta)$ is the lower bound defined in Eq.~\eqref{eq:lower-bound-m}. Let us substitute
\begin{equation}
    x:=\frac{1}{e\|\theta\|_1}\log\left(\frac{\|H\|}{\varepsilon\sqrt{e^3\|\theta\|_1}}\right).
\end{equation}
Observe that $x\to0$ as $D\to\infty$ by assumption on the asymptotics of $\|\theta\|_1$ (even if $\|H\|$ and/or $\varepsilon>0$ depend on $D$, provided that their ratio is not exponential in $D$ which is insure by $|\mathcal{I}|=\mathcal{O}(\operatorname{Poly}(D))$). Moreover, by definition of the Lambert-$W$-function~\cite{corless1996lambert}, we have $W(x)e^{W(x)}=x$. Therefore, applying Lemma~\ref{lemma:lambert-asymptotics} (which tells us that $x/W(x)\sim 1$ as $x\to 0$),
\begin{align}
    m(x) \stackrel{\eqref{eq:lower-bound-m}}{\geq} b(\theta) = e\|\theta\|_1\, \exp(W(x))-\frac{1}{2} = e\|\theta\|_1\, \frac{x}{W(x)}-\frac{1}{2} \sim e\|\theta\|_1 \label{eq:final-bound-m}
\end{align}
as $D\to\infty$ (since $x\to0$). In particular, recalling Eq.~\eqref{eq:upper-bound3}, $y>1$ equivalently means that
\begin{equation}
    m > e\|\theta\|_1-1,
\end{equation}
which always holds true by Eq.~\eqref{eq:final-bound-m}. Moreover, Eq.~\eqref{eq:final-bound-m} also shows that $m(\theta)\sim e\|\theta\|_1$ as $D\to\infty$ is sufficient for the truncation threshold to ensure the error tolerance $\varepsilon>0$.
\qed
\bigskip
\\
\indent
The asymptotics of the lower bound in Lemma~\ref{lemma:truncation-threshold} may be used to establish useful statistical insights about the expected truncation threshold, given a certain initial distribution $\mathbb{P}_\theta$:
\bigskip
\begin{proposition}\label{prop:truncation-statistics} 
    Consider an expectation of the form in Eq.~\eqref{eq:expectation3} with a variational ansatz $U(\theta)$ involving $D\geq1$ parameters, and suppose we simulate it classically using the Taylor series $C_m(\theta)$ defined in Eq.~\eqref{eq:taylor2} within some error tolerance $\varepsilon$. Let $m=m(\theta)$ be the truncation threshold in the Taylor-approximation in Eq.~\eqref{eq:taylor2}. Initialize $\theta\sim\mathbb{P}_\theta\in\{\mathcal{N}(0,\sigma^2I_D), \operatorname{Unif}([-\sigma,\sigma]^D)\}$ with i.i.d. components $\theta_1,\dots,\theta_D$, where $\sigma=D^{-r}$ for some $r\in[0,1)$. Then, as $D\to\infty$, the expected truncation threshold behaves according to
    \begin{equation*}
        \mathbb{E}_\theta[m(\theta)] = \Theta(D^{1-r}).
    \end{equation*}
\end{proposition}
\noindent
\emph{Proof.} Let $r\in[0,1)$ be arbitrary. By Lemma~\ref{lemma:truncation-threshold} we know that $m(\theta)\sim e\|\theta\|_1$ as $D\to\infty$ is sufficient to simulate $C_m(\theta)$ within some error tolerance $\varepsilon>0$. By the law of large numbers we may simply take the expectation value of the $\ell_1$-norm which gives
\begin{align}
\mathbb{E}_\theta[\|\theta\|_1]=\sum_{k=1}^D\mathbb{E}_\theta[|\theta_k|]=D\,\underbrace{\mathbb{E}_\theta[|\theta_1|]}_{=\,\Theta(\sigma)} = D\,\Theta(D^{-r}) = \Theta(D^{1-r}),
\end{align}
where the second equality uses that the components of $\theta$ are i.i.d.
\qed

\subsection{Mean-Squared Error Analysis}
Next, we quantify the mean squared error of the approximation $C_m(\theta)\approx C(\theta)$ with respect to the local distribution $\mathbb{P}_\theta\in\{\mathcal{N}(0,\sigma^2I_D),\operatorname{Unif}([-\sigma,\sigma]^D)\}$ for some $\sigma>0$. However, we first require the following counting argument to conclude with the result:
\bigskip
\begin{lemma}\label{lemma:stars-and-bars}
    Let $m=\mathcal{O}(1)$ be of constant order, i.e., independent of $D$. Then,
    \begin{equation}
        \sum_{\|\alpha\|_1<m}1 = \mathcal{O}(D^{m-1}), \label{eq:stars-and-bars}
    \end{equation}
    where $\{\|\alpha\|_1<m\}:=\{\alpha\in\mathbb{N}_0^D:|\alpha_1|+\cdots+|\alpha_D|<m\}$. {On the other hand, if $m=o(D)$ strictly increases sub-linearly as $D\to\infty$, then}
    \begin{equation}
        {\sum_{\|\alpha\|_1<m}1 = \Omega\left(\operatorname{Poly}(D)^{-1}\left(\frac{eD}{m}\right)^{m}\right)}
    \end{equation}
    {is super-polynomially lower bounded. In particular, if $m=\Theta(D^\delta)$ for some $\delta\in(0,1)$, then $\#\{\|\alpha\|_1<m\}$ is at least super-polynomial in $D$.}
\end{lemma}
\noindent
\emph{Proof.} We begin by proving Eq.~\eqref{eq:stars-and-bars}. By the ``balls and cells'' formula~\cite[Eq.~(5.2)]{feller1991introduction}, we count
\begin{align}
    \sum_{\|\alpha\|_1<m}1 &= \sum_{i=0}^{m-1}\left(\sum_{\|\alpha\|_1=i}1\right) = \sum_{i=0}^{m-1}\binom{D+i-1}{i} = {\binom{D+m-1}{m-1}}, \label{eq:counting}
\end{align}
{where the last identity follows from the hockey stick identity~\cite{jones1996generalized} from combinatorics, i.e.,}
\begin{equation}
    {\binom{n+1}{k+1} = \sum_{j=k}^n\binom{j}{k}\quad\text{for all $n,k\in\mathbb{N}$ with $n\geq k$.}} \label{eq:hockeystick}
\end{equation}
{Indeed, by symmetry of the binomial coefficient, that is $\binom{n}{k}=\binom{n}{n-k}$ for all $n,k\in\mathbb{N}_0$, we find}
\begin{equation}
    {\sum_{i=0}^{m-1}\binom{D+i-1}{i} = \sum_{i=0}^{m-1}\binom{D+i-1}{D-1} = \sum_{j=D-1}^{D+m-2}\binom{j}{D-1} \stackrel{\eqref{eq:hockeystick}}{=} \binom{D+m-1}{D} = \binom{D+m-1}{m-1},}
\end{equation}
{where we applied symmetric in the first and last identity, respectively.}

{We are now ready to prove the polynomial upper bound in the case where $m=\mathcal{O}(1)$. For that purpose, note that if $D\geq m-1$ (which holds true as $D\to\infty$ the assumption that $m=\mathcal{O}(1)$), then}
\begin{equation}
    {\binom{D+m-1}{m-1} = \frac{(D+m-1)(D+m-2)\cdots(D+1)D}{(m-1)!} \leq \frac{(D+m-1)^{m-1}}{(m-1)!} \leq \frac{(2D)^{m-1}}{(m-1)!} = \mathcal{O}(D^{m-1}),}
\end{equation}
{where we used $m=\mathcal{O}(1)$ in the last step.}

{It remains to prove the super-polynomial lower bound if $m\to\infty$ as $D\to\infty$. With the help of Stirling's inequality, $n! \sim \sqrt{2\pi n}\left(\frac{n}{e}\right)^n$, we may lower bound}
\begin{align}
    {\binom{D+m-1}{m-1}} &= {\frac{(D+m-1)(D+m-2)\cdots(D+1)\cdot D}{(m-1)!}} \nonumber
    \\ &\geq {\frac{D^{m-1}}{(m-1)!} \sim \frac{(eD)^{m-1}}{e\sqrt{m-1}(m-1)^{m-1}} \geq \underbrace{\frac{1}{e^2D\sqrt{m-1}}}_{\Theta(\operatorname{Poly}(D)^{-1})}\left(\frac{eD}{m}\right)^{m},} \label{eq:asymbound1}
\end{align}
{using that $m=o(D)$ strictly increases sub-linearly as $D\to\infty$, which directly proves that \eqref{eq:asymbound1} is super-polynomial and sub-exponential.}
\qed
\bigskip
\\
\indent
We are now equipped with all necessary tools to bound the mean-squared error:
\bigskip
\begin{lemma}{\label{lemma:mean-square-error}}
    Consider an expectation of the form in Eq.~\eqref{eq:expectation3} with a variational ansatz $U(\theta)$ involving $D\geq1$ parameters, and suppose we simulate it classically using the Taylor series $C_m(\theta)$ defined in Eq.~\eqref{eq:taylor2}.  Initialize $\theta\sim\mathbb{P}_\theta\in\{\mathcal{N}(0,\sigma^2I_D), \operatorname{Unif}([-\sigma,\sigma]^D)\}$ with i.i.d. components $\theta_1,\dots,\theta_D$. Then, if $m=\mathcal{O}(1)$, the mean squared error satisfies the bound
    \begin{equation}
        \mathbb{E}_\theta[(C(\theta)-C_m(\theta))^2] = {\mathcal{O}\left(\|H\|^2\frac{(D\sigma^2)^m}{m^m}\right)}, \label{eq:mean-squared-error}
    \end{equation}
    where $\|H\|$ is the operator norm of the observable in Eq.~\eqref{eq:observable2}. In particular, if {$\sigma \leq q\sqrt{m} \,\|H\|^{-\frac{1}{m}}D^{-\frac{1}{2}}$} for some $0<q<1$, then the mean squared error decays like $\mathcal{O}(q^{2m})$.
\end{lemma}
\noindent
\emph{Proof.} We use the Lagrange-remainder to quantify the error,
\begin{align}
    C(\theta) - C_m(\theta) &= \sum_{\|\alpha\|_1 \geq m}(D_\theta^\alpha C)(0)\,\frac{\theta^\alpha}{\alpha!} = \sum_{\|\alpha\|_1=m}\underbrace{(D_\theta^\alpha C)(\xi)}_{\leq\mathcal{O}(\|H\|)}\,\frac{\theta^\alpha}{\alpha!}=\mathcal{O}(\|H\|)\sum_{\|\alpha\|_1=m}\frac{\theta^\alpha}{\alpha!}, \label{eq:lagrange-remainder}
\end{align}
employing the uniform bound $(D_\theta^\alpha C)(\xi)\leq\mathcal{O}(\|H\|)$ in the higher-order parameter-shift rule, Lemma~\ref{lemma:parameter-shift}, in the last estimation. {As a result,}

\begin{align}
    {\mathbb{E}_{\theta}[(C(\theta)-C_m(\theta))^2]} &= {\mathcal{O}(\|H\|^2)\,\mathbb{E}_\theta\left[\left(\sum_{\|\alpha\|_1=m}\frac{\theta^\alpha}{\alpha!}\right)^2\right] = \mathcal{O}(\|H\|^2)\frac{1}{m!^2}\mathbb{E}_\theta\left[\left(\sum_{k=1}^D\theta_k\right)^{2m}\right],} \label{eq:MSE1}
\end{align}
{where the last identity follows from the multinomial theorem. In order to compute the last expectation, we first focus on the case $\mathbb{P}_\theta=\mathcal{N}(0,\sigma^2)$. In that case, we recall from basic probability theory that the sum $\sum_{k=1}^D\theta_k\sim\mathcal{N}(0,D\sigma^2)$ of independent Gaussian random variables remains a Gaussian with well-known moments explicitly given by}
\begin{equation}
    {\mathbb{E}_\theta\left[\left(\sum_{k=1}^D\theta_k\right)^{2m}\right] = D^m\sigma^{2m} (2m-1)!! = D^m\sigma^{2m}\frac{(2m)!}{2^mm!},} \label{eq:MSE2}
\end{equation}
{noting that the double factorial $(2m-1)!! := (2m-1)(2m-3)\cdots 3\cdot 1$ satisfies the identity $(2m-1)!!=\frac{(2m)!}{2^mm!}$. Indeed,}
\begin{align}
    {(2m)! = \underbrace{(2m)(2m-2)\cdots 4\cdot 2}_{=(2m)!!} \cdot \underbrace{(2m-1)(2m-2)\cdots 3\cdot 1}_{(2m-1)!!},}
\end{align}
{and}
\begin{align}
    {(2m)!! = (2m)(2m-2)\cdots 4\cdot 2 = 2^m m\cdot(m-1)\cdots 2\cdot 1 = 2^m m!}
\end{align}
{so that $(2m)! = 2^mm!\cdot (2m-1)!!$ as claimed. This shows that}
\begin{equation}
    {\mathbb{E}_{\theta}[(C(\theta)-C_m(\theta))^2] \stackrel{\eqref{eq:MSE1}}{=} \mathcal{O}(\|H\|^2)\frac{1}{m!^2}\mathbb{E}_\theta\left[\left(\sum_{k=1}^D\theta_k\right)^{2m}\right] \stackrel{\eqref{eq:MSE2}}{=} \mathcal{O}(\|H\|^2)\frac{(2m)!}{2^mm!^3}(D\sigma^2)^m.}
\end{equation}
{Next, by utilizing Stirling's formula, $m!\sim \sqrt{2\pi m}\left(\frac{m}{e}\right)^m$, we asymptotically establish}
\begin{align}
    {\frac{(2m)!}{2^mm!^3} \sim \frac{\sqrt{2\pi(2m)}\left(\frac{2m}{e}\right)^{2m}}{2^m(\sqrt{2\pi m})^3\left(\frac{m}{e}\right)^{3m}} = \frac{\sqrt{4\pi \cancel{m}}\,2^{\cancel{2}m}\cancel{\left(\frac{m}{e}\right)^{2m}}}{2\pi m\sqrt{2\pi \cancel{m}}\,\cancel{2^m}\left(\frac{m}{e}\right)^m\cancel{\left(\frac{m}{e}\right)^{2m}}} = \mathcal{O}\left(\frac{(2e)^m}{m^{m+1}}\right),}
\end{align}
{which results in the desired bound,}
\begin{equation}
    {\mathbb{E}_{\theta}[(C(\theta)-C_m(\theta))^2] = \mathcal{O}(\|H\|^2)\frac{(2m)!}{2^mm!^3}(D\sigma^2)^m = \mathcal{O}\left(\|H\|^2 \frac{(2eD\sigma^2)^m}{m^{m+1}}\right) = \mathcal{O}\left(\|H\|^2\frac{(D\sigma^2)^m}{m^m}\right),}
\end{equation}
{using $m=\mathcal{O}(1)$ in the last identity.}

{In the case where $\mathbb{P}_\theta=\operatorname{Unif}([-\sigma,\sigma]^D)$ is i.i.d. uniform in each component, then $\mathbb{E}_\theta[\theta_k]=0$ and $\operatorname{Var}_{\theta}[\theta_k]=\frac{\sigma^2}{3}$ for all $k\in[D]$. By the central limit theorem, the sum}
\begin{equation}
    {\sum_{k=1}^D\theta_k \stackrel{d}{\to} \mathcal{N}\left(0,\frac{D\sigma^2}{3}\right)}
\end{equation}
{converges in distribution to a Gaussian as $D\to\infty$. This allows us to reduce this case asymptotically to the previous from Eq.~\eqref{eq:MSE1}.}

The second statement is immediately obtained by solving for $\sigma$ in {$\|H\|^2\frac{D^m\sigma^{2m}}{m^m} \stackrel{!}{\leq} c_0q^{2m}$, where $c_0>0$ is some constant. Indeed, rearranging the inequality gives}
\begin{equation}
    {\sigma^{2m}\leq\frac{c_0q^{2m}\,m^m}{\|H\|^2D^m} \iff \sigma \leq \underbrace{c_0^{-1/2m}}_{=\Theta_m(1)}\frac{q\sqrt{m}}{\|H\|^{1/m}\sqrt{D}} = \mathcal{O}(q\sqrt{m}\,\|H\|^{-1/m}D^{-\frac{1}{2}}).}
\end{equation}
\qed
\bigskip
\\
\indent
The next step is to establish a lower bound for the necessary truncation threshold $m=m(\mathbb{P}_\theta)$ in order to Taylor surrogate the cost in Eq.~\eqref{eq:expectation3}, i.e., evaluate $C_m(\theta)$ with high probability and error of at most $\varepsilon>0$:
\bigskip
\begin{lemma}\label{lemma:probabilistic-truncation}
    Consider an expectation of the form in Eq.~\eqref{eq:expectation3} with a variational ansatz $U(\theta)$ involving $D\geq1$ parameters, and suppose we simulate it classically using the Taylor series $C_m(\theta)$ defined in Eq.~\eqref{eq:taylor2}. Assume {$\|H\|=\mathcal{O}(1)$ and} that the mean squared error in Eq.~\eqref{eq:mean-squared-error} is bounded by a given tolerance $\varepsilon>0$. Initialize $\theta\sim\mathbb{P}_\theta\in\{\mathcal{N}(0,\sigma^2I_D), \operatorname{Unif}([-\sigma,\sigma]^D)\}$ with i.i.d. components $\theta_1,\dots,\theta_D$, where {$\sigma = \mathcal{O}(q\,D^{-\frac{1}{2}})$ for some $0<q<1$}. Then, there exists a constant $c>0$ such that, with probability at least $1-\rho$, a constant truncation threshold $m=m(\mathbb{P}_\theta)$ satisfying
    \begin{equation*}
        {m(\mathbb{P}_\theta)\geq \frac{1}{2}\,\frac{\log\left(\frac{c}{\rho\,\varepsilon^2}\right)}{\log \left(\frac{1}{q}\right)}}
    \end{equation*}
    is sufficient to simulate $C_m(\theta)$ with error at most $\varepsilon>0$. 
\end{lemma}
\noindent
\emph{Proof.} Let $\delta,\varepsilon,\rho>0$ be arbitrary. {Assume that $\|H\|=\mathcal{O}(1)$. If} $\sigma = {\mathcal{O}(}q\,D^{-\frac{1}{2}}{)}$, Lemma~\ref{lemma:mean-square-error} yields
\begin{equation}
    \mathbb{E}_\theta[(C(\theta)-C_m(\theta))^2] \leq c\,q^{2m} \label{eq:mse-bound3}
\end{equation}
for some constant $c>0$. In that case, {Markov's inequality bounds} the probability of failure,
\begin{equation}
    \mathbb{P}_\theta(|C(\theta)-C_m(\theta)|\geq\varepsilon) \leq \frac{\mathbb{E}_\theta[(C(\theta)-C_m(\theta))^2]}{\varepsilon^2} \stackrel{\eqref{eq:mse-bound3}}{\leq} \frac{c\,q^{2m}}{\varepsilon^2}\stackrel{!}{\leq} \rho.
\end{equation}
Solving for the truncation threshold $m$ in $\varepsilon^{-2}c\,q^{2m}\leq\rho$ implies
\begin{equation}
    2m\log q \leq \log\left(\frac{\rho\,\varepsilon^2}{c}\right),
\end{equation}
or equivalently, since $\log q = -\log\left(\frac{1}{q}\right)$ because of $0<q<1$,
\begin{equation}
    m\geq \frac{1}{2}\,\frac{\log\left(\frac{c}{\rho\,\varepsilon^2}\right)}{\log \left(\frac{1}{q}\right)}.
\end{equation}
\qed

\section{Computational Complexity \label{sec:complexity}}
The first part of this section is dedicated to results in complexity theory showing that the Hamiltonian ground energy decision problem for observables of the form in Eq~\eqref{eq:condition-observable} are classically infeasible in general, unless $\texttt{BPP}=\texttt{QMA}$.

\subsection{The Complexity Class of LCE Hamiltonians} \label{appendix:Hamiltonian}

In order to successfully mitigate barren plateaus of arbitrarily expressive PQC unitaries $U(\theta)$ on parameter landscape patches with LCE, we are restricted to the class of observables,
\begin{equation}
    H=\sum_{i\in\mathcal{I}}c_iP_i\quad\text{with}\quad \|H\|\,=\mathcal{O}(1)\quad\text{and}\quad c_{i_0}=\Omega(1)\quad\text{for some $i_0\in \mathcal{I}$}, \label{eq:LCE-observale}
\end{equation}
where $\mathcal{I}\subset\{I,X,Y,Z\}^{\otimes N}\setminus\{I_{2^N}\}$ is an arbitrary Pauli indexing set with $|\mathcal{I}|\,=\mathcal{O}(\operatorname{Poly}(N))$ and $c_i=\mathcal{O}(1)$ for all $i\in\mathcal{I}$. Here, we rigorously prove that, in general, finding the ground energy of such observables is classically hard in the sense that there exists no classical algorithm able to solve the problem under standard complexity theory assumptions. Since the Pauli strings define a $\mathbb{R}$-linear basis of the $4^N$-dimensional Hilbert-Schmidt vector space $\operatorname{Herm}(2^N)$ of Hermitian matrices, we are particularly interested in the following class

\begin{equation}
    \mathscr{H}_{\text{LCE}}:= \left\{H + \sum_{j\in\mathcal{J}}c_jP_j \,\text{ traceless}\mathrel{\bigg|} \text{$H\in\operatorname{Herm}(2^N):\|H\|\,=\mathcal{O}(1)$,\,\,$\mathcal{J}\neq\emptyset:|\mathcal{J}|\,=\mathcal{O}(1),\, \forall j\in\mathcal{J}: c_j=\Theta(1)$}\right\} \label{eq:hamiltonian-class}
\end{equation}
of (arbitrarily local or global) traceless, normalized $N$-qubit Hamiltonians perturbed by an $\mathcal{O}(1)$-sparse $\Theta(1)$-linear combination of Pauli strings, or for short; the class $\mathscr{H}_{\text{LCE}}$ of LCE Hamiltonians. First, we note that every Hamiltonian $H\in\mathscr{H}_{\text{LCE}}$ automatically satisfies Eq.~\eqref{eq:LCE-observale}. Indeed, by the triangle inequality,
\begin{align}
    \left\|H+\sum_{j\in\mathcal{J}}c_jP_j\right\|\,\leq\underbrace{\|H\|}_{=\mathcal{O}(1)}\, +\sum_{j\in\mathcal{J}}\,\underbrace{|c_j|}_{\Theta(1)}\cdot\underbrace{\|P_j\|}_{=1} = \mathcal{O}(1) + \Theta(1)\cdot\underbrace{\sum_{j\in\mathcal{J}}1}_{=|\mathcal{J}|\,=\mathcal{O}(1)} = \mathcal{O}(1),
\end{align}
noting that $\|P_i\|\,=1$ for every Pauli string $P_i$. This shows that LCE Hamiltonians are normalized. By definition, the LCE perturbation itself has constant scaling coefficients $c_{j}=\Omega(1)$ for all $j\in\mathcal{J}$ thus addressing that condition in Eq.~\eqref{eq:LCE-observale}. Finally, it remains to see (noting $H\neq0$) that $\operatorname{Tr}H=0$ in Eq.~\eqref{eq:LCE-observale} if and only if $I_{2^N}\notin\mathcal{I}$ since all Pauli strings $P_i$ are traceless with $\operatorname{Tr}I_{2^N}=2^N$ being the only exception. This shows that every LCE Hamiltonian in Eq.~\eqref{eq:hamiltonian-class} indeed satisfies the LCE condition in Eq.~\eqref{eq:LCE-observale}.

In the following, we present two proofs showing that there exists no ``one-size-fits-all" (deterministic or probabilistic) classical algorithm that can find the ground energy of an observable in the class $\mathscr{H}_{\text{LCE}}$. We do this by reducing the problem to well-known results from complexity theory: The complexity class $\texttt{QMA}$ is the quantum analogue of $\texttt{NP}$ which contains both $\texttt{BPP}$ and $\texttt{P}$~\cite{aharonov2002quantum}. It is well-known that the $k$-local Hamiltonian problem defines a \texttt{QMA}-complete problem, even for $k=2$~\cite{kempe2006complexity}. In that case, ``\texttt{QMA}-completeness" means that any \texttt{QMA} problem can be reduced to the $k$-local Hamiltonian problem. Note that $\texttt{BPP}\subseteq\texttt{QMA}$~\cite{aharonov2002quantum}. In fact, it is a standard assumtion in complexity theory that $\texttt{BPP}\neq\texttt{QMA}$. Otherwise, any \texttt{QMA} problem (in particular every $k$-local Hamiltonian problem) is guaranteed to be solvable by some probabilistic classical algorithm~\cite{kempe2006complexity}. As a result, if $\texttt{BPP}\neq\texttt{QMA}$, then there exists no classical algorithm that generally finds the ground energy of any given $2$-local Hamiltonian. Increasing the locality $k$ only makes the Hamiltonian problem more difficult.

Here, we mainly focus on the complexity of the $\mathscr{H}_{\text{LCE}}$-class Hamiltonian problem. As a disclaimer, we do not claim that there always exists a LCE initialization strategy leading to training dynamics that converges to the unperturbed solution $\lambda_{\min}(H)$. We further emphasize that this work focuses solely on the theory of initialization and we thus leave convergence results as open problems for future efforts.

Let us briefly motivate the strategy behind our probabilistic proof first: Given an arbitrary (local or global) Hamiltonian $H\in\operatorname{Herm}(2^N)$ with $\|H\|\,=\mathcal{O}(1)$, we assume by contradiction that the ground energy of any LCE perturbation $H+ V_{\mathcal{J}}\in\mathscr{H}_{\text{LCE}}$ of $H$ can be solved in \texttt{BPP}. We are then going to use the output $\lambda\approx \lambda_{\min}(\tilde{H})$ of such a classical \texttt{BPP} algorithm in combination with a  certain, sufficiently small and randomly sampled LCE perturbation $V_{\mathcal{J}}$ to find the original solution $\lambda_{\min}(H)$. In particular, if $H$ is chosen to be an arbitrary $2$-local Hamiltonian, this would thus imply that $\texttt{QMA}=\texttt{BPP}$ by \texttt{QMA}-completeness of the $2$-local Hamiltonian problem, which is absurd under standard complexity assumptions. Therefore, if $\texttt{BPP}\neq\texttt{QMA}$, no such algorithm exists, hence implying non-classicality of the class $\mathscr{H}_{\text{LCE}}$ of LCE Hamiltonians. That is, in practice we are completely free to choose \emph{any} LCE perturbation $V_{\mathcal{J}}$ strategy with the goal of finding the ground energy $\lambda_{\min}(H)$ of an arbitrary normalized Hamiltonian $H$. Finding the ground energy of the corresponding LCE perturbation $H+V_{\mathcal{J}}\in\mathscr{H}_{\text{LCE}}$ remains classically hard in general.

We first focus on a simpler, deterministic proof to illustrate the idea behind our proof. In doing so, we make use of a major conjecture in complexity theory, namely the \emph{Quantum Probabilistically Checkable Proofs}  (\texttt{QPCP}) conjecture~\cite{aharonov2013guest, natarajan2018low} and the standard assumption that $\texttt{P}\neq\texttt{QMA}$. The deterministic proof only guarantees non-existence of deterministic algorithms that can solve the $\mathscr{H}_{\text{LCE}}$-class Hamiltonian problems. Indeed, at the time of writing, it remains an open question whether randomized algorithms are strictly more powerful than deterministic ones, i.e., whether $\texttt{P}\neq\texttt{BPP}$. Note that this is insufficient to prove total non-existence of classical algorithms since there could still exist a randomized algorithm solving the problem. Nevertheless, we provide a slightly more sophisticated probabilistic proof (inspired by the technique of the deterministic proof) that does not make use of the $\texttt{QPCP}$ conjecture, and only requires the standard assumption that $\texttt{BPP}\neq\texttt{QMA}$.

The classical \texttt{PCP} theorem states that every problem in \texttt{NP} has a probabilistically verifiable, classically feasible proof requiring $\mathcal{O}(\log n)$ random bits and only $\mathcal{O}(1)$ bits of the proof to be read ~\cite{dinur2007pcp}. The quantum analogue, i.e., \texttt{QPCP}, is believed to hold true but no proof has been found at the time of writing. Most notably, the \texttt{QPCP} conjecture is mathematically equivalent to the question of whether the local Hamiltonian problem with \emph{at least constant} promise gaps (rather than at least inversely polynomial gaps~\cite{kempe2006complexity}) is \texttt{QMA}-hard~\cite{aharonov2013guest} or not.

\subsubsection{The Local Hamiltonian Problem and Classical Simulability}

Before we prove these claims, we set the context by formally introducing the necessary concepts. First, we clarify what we mean by the \emph{local Hamiltonian problem} in the sense of Ref.~\cite{kempe2006complexity}:

\begin{definition}[$k$-local Hamiltonian Problem]
    We are given a decomposition
    \begin{equation*}
        H = \sum_{j=1}^r H_j
    \end{equation*}
    of a $N$-qubit Hamiltonian $H\in\operatorname{Herm}(2^N)$ with $k$-local Hermitian terms $H_j\in\operatorname{Herm}(2^N)$ (i.e., each $H_j$ acts on at most $k$ qubits) with $\Theta(\operatorname{Poly}(N))$-bit description of its entries such that $\|H_j\|\,=\mathcal{O}(\operatorname{Poly}(N))$ for all $j\in[r]$, where $r=\Theta(\operatorname{Poly}(N))$. Let $\lambda_{\min}(H)\in\mathbb{R}$ denote the smallest eigenvalue of $H$. The $k$-local Hamiltonian problem is the task of deciding whether
    \begin{equation}
        \texttt{YES:}\,\,\,\lambda_{\min}(H) < a\quad\text{or}\quad \texttt{NO:}\,\,\,\lambda_{\min}(H)> b \label{eq:k-local-hamiltonian-problem}
    \end{equation}
    for some given rational numbers $a,b\in\mathbb{Q}$ of $\Theta(\operatorname{Poly}(N))$-bit description with $a<b$ and promise gap satisfying
    \begin{equation}
        b-a=\Omega(\operatorname{Poly}(N)^{-1}).
    \end{equation}
\end{definition}

The most interesting condition of the $k$-local Hamiltonian problem is the (asymptotic) size of the promise gap $b-a>0$ as $N\to\infty$. For larger gaps the problem of deciding \texttt{YES} or \texttt{NO} generally becomes easier, which is the main difficulty in the \texttt{QPCP} conjecture, namely, whether a constant promise gap $b-a=\Omega(1)$ is already sufficient for \texttt{QMA}-hardness~\cite{aharonov2013guest}. On the other hand, decreasing the promise gap $b-a$ only makes the decision problem harder. For the purpose of our analysis, since the promise gap is at least inversely polynomial in $N$, we may assume without loss of generality that $\|H\|\,=\mathcal{O}(1)$ since one may rescale $H\mapsto\frac{1}{r}H$, and due to $r=\Theta(\operatorname{Poly}(N))$, this transformation asymptotically leaves the promise gap invariant as $N\to\infty$.

Note that a sufficiently good approximation of $\lambda_{\min}(H)$ can be used to decide Eq.~\eqref{eq:k-local-hamiltonian-problem}. This specifically motivates what we mean by a \texttt{BPP} and \texttt{P} algorithm, respectively, that can solve the $\mathscr{H}_{\text{LCE}}$-class Hamiltonian problem:

\begin{definition} \label{def:BPP-algorithm}
    A \texttt{BPP} algorithm $\mathcal{A}$ is said to solve the $\mathscr{H}_{\text{LCE}}$-class Hamiltonian problem if for all $H\in\mathscr{H}_{\text{LCE}}$ and all $\varepsilon>0$, the algorithm $\mathcal{A}$ running on a probabilistic Turing machine outputs $\lambda\in\mathbb{R}$ such that
    \begin{equation*}
        |\lambda-\lambda_{\min}(H)|\,<\varepsilon
    \end{equation*}
    with success probability strictly larger than $\frac{1}{2}$, and does so in polynomial time, $\mathcal{O}\left(\operatorname{Poly}(N,\frac{1}{\varepsilon})\right)$. If $\mathcal{A}$ solves the $\mathscr{H}_{\texttt{LCE}}$-class Hamiltonian problem on a deterministic Turing machine (i.e., with success probability 1), then $\mathcal{A}$ is called a \texttt{P} algorithm that solves the $\mathscr{H}_{\texttt{LCE}}$-class Hamiltonian problem.
\end{definition}

By definition it follows that if such an algorithm $\mathcal{A}$ exists, then the $\mathscr{H}_{\text{LCE}}$-class Hamiltonian problem lives in the complexity class \texttt{BPP}, and analogously in \texttt{P} if $\mathcal{A}$ defines a \texttt{P} algorithm.

\subsubsection{Deterministic Proof using $\texttt{P}\neq\texttt{QMA}$ and the \texttt{QPCP} Conjecture}

We begin with the more straightforward, deterministic proof that relies on the the standard assumption that $\texttt{P}\neq\texttt{QMA}$ and additionally assumes the \texttt{QPCP} conjecture to be true. First, we establish the following approximation argument, which intuitively establishes that any normalized Hamiltonian can be perturbed with small deviation from the unperturbed ground energy:

\begin{lemma} \label{prop:reduction-normalized-hamiltonian}
    Let $H\in\operatorname{Herm}(2^N)$ be an arbitrary Hamiltonian with $\|H\|\,=\mathcal{O}(1)$. Then, for all $\varepsilon>0$ independent of $N$ (i.e., $\varepsilon=\Theta(1)$) there exists $\tilde{H}\in\mathscr{H}_{\text{LCE}}$ such that
    \begin{equation*}
        |\lambda_{\min}(\tilde{H}) - \lambda_{\min}(H)| \,< \varepsilon.
    \end{equation*}
\end{lemma}
\smallskip
\noindent
\emph{Proof.} Let $\varepsilon>0$ be arbitrary but fixed, that is $\varepsilon=\Theta(1)$. Let $H\in\operatorname{Herm}(2^N)$ be arbitrary with $\|H\|\,=\mathcal{O}(1)$. The central idea is to perturb $H$ with a deterministic LCE perturbation $V_{\mathcal{J}}$ with operator norm smaller than $\varepsilon>0$. More precisely, for some non-empty $\mathcal{J}\subseteq\{0,1,2,3\}^N\setminus\{0\}$ with $|\mathcal{J}|\,=\mathcal{O}(1)$ we define
\begin{equation}
    \tilde{H} := H + V_{\mathcal{J}}\quad\text{where}\quad V_{\mathcal{J}}:= \frac{1}{|\mathcal{J}|}\sum_{j\in \mathcal{J}}a_jQ_j, \label{eq:hamiltonian-construction}
\end{equation}
with $a_j\in(-\varepsilon,\varepsilon)\setminus\{0\}$ (in particular $a_j=\Theta(1)$ since $\varepsilon=\Theta(1)$ by assumption), and arbitrary non-identity Pauli strings $Q_j$ indexed by $j\in \mathcal{J}$. By construction, we have that $V_{\mathcal{J}}$ defines a LCE perturbation in the sense of Eq.~\eqref{eq:LCE-observale}. Indeed, this follows from $a_j\in\Theta(1)$ for all $j\in\mathcal{J}$, and $|\mathcal{J}|\,=\Theta(1)$ due to $\mathcal{J}\neq\emptyset$ being a discrete set. In particular, $\tilde{H}\in\mathscr{H}_{\text{LCE}}$ by using the assumption that $\|H\|\,=\mathcal{O}(1)$. Furthermore, by the triangle inequality,
\begin{align}
    \sup_{|\psi\rangle}|\langle\psi|V_{\mathcal{J}}|\psi\rangle| \,\leq \frac{1}{|\mathcal{J}|}\sum_{j\in\mathcal{J}}\underbrace{|a_j|}_{<\varepsilon}\,\underbrace{\sup_{|\psi\rangle}|\langle\psi|Q_j|\psi\rangle|}_{\leq 1} \,< \frac{\varepsilon}{|\mathcal{J}|}\sum_{j\in\mathcal{J}} 1 = \varepsilon, \label{eq:term-bound}
\end{align}
using the fact that Pauli strings have eigenvalues in $\pm1$, and thus uniformly bounded operator norm $\|Q_j\|\,=1$ for all $j\in\mathcal{J}$. As a result, by the Weyl inequality,
\begin{align*}
    |\lambda_{\min}(\tilde{H}) - \lambda_{\min}(H)|\, \stackrel{\eqref{eq:hamiltonian-construction}}{=} |\lambda_{\min}(H + V_{\mathcal{J}}) - \lambda_{\min}(H)| \, \stackrel{\text{\tiny Weyl}}{\leq} \sup_{|\psi\rangle}|\langle\psi|V_{\mathcal{J}}|\psi\rangle|\,\stackrel{\eqref{eq:term-bound}}{<} \varepsilon.
\end{align*}
\qed
\bigskip
\par
We are now going to show that there exists no \texttt{P} algorithm that can solve the $\mathscr{H}_{\text{LCE}}$-class problem, unless the \texttt{QPCP} conjecture is false or $\texttt{P}=\texttt{QMA}$:

\begin{proposition} \label{prop:classical-infeasbility}
    If the \texttt{QPCP} conjecture holds true and if $\texttt{P}\neq\texttt{QMA}$, then there exists no deterministic classical algorithm that can solve the $\mathscr{H}_{\text{LCE}}$-class Hamiltonian problem. In particular, the problem of finding the ground energy of an arbitrary Hamiltonian $H\in\mathscr{H}_{\text{LCE}}$ is guaranteed to be classically infeasible on deterministic machines in general.
\end{proposition}
\bigskip
\noindent
\emph{Proof.} Assume that $\texttt{P}\neq\texttt{QMA}$ and the \texttt{QPCP} conjecture to be true. Suppose by contradiction that there exists a \texttt{P} algorithm $\mathcal{A}$ that can solve the $\mathscr{H}_{\text{LCE}}$-class Hamiltonian problem. Let $\varepsilon>0$ to be determined and $H\in\operatorname{Herm}(2^N)$ an arbitrary Hamiltonian with $\|H\|\,=\mathcal{O}(1)$. Let $a,b\in\mathbb{Q}$ be given boundaries of a promise gap for $\lambda_{\min}(H)$ according to Eq.~\eqref{eq:k-local-hamiltonian-problem}. Crucially, the \texttt{QPCP} conjecture allows us to assume that $b-a=\Theta(1)$. By Proposition~\ref{prop:reduction-normalized-hamiltonian} we know that there exists $\tilde{H}\in\mathscr{H}_{\text{LCE}}$ such that
\begin{equation}
    |\lambda_{\min}(\tilde{H})-\lambda_{\min}(H)|\,<\frac{\varepsilon}{2}. \label{eq:first-ineq}
\end{equation}
Furthermore, by the assumption of classicality of $\mathscr{H}_{\text{LCE}}$, we know that the \texttt{P} algorithm $\mathcal{A}$ outputs $\lambda\in\mathbb{R}$ such that
\begin{equation}
    |\lambda_{\min}(\tilde{H})-\lambda|\, <\frac{\varepsilon}{2}. \label{eq:second-ineq}
\end{equation}
Therefore, by combining the two inequalities and choosing $\varepsilon:=\frac{b-a}{2}=\Theta(1)$, we find that
\begin{equation}
    \begin{cases} \lambda \, < \,\,\lambda_{\min}(\tilde{H}) + \frac{\varepsilon}{2} \, < \,\,\lambda_{\min}(H) + \varepsilon  < a + \varepsilon = \frac{a+b}{2},&\text{if $\lambda_{\min}(H) < a$} \\
    \lambda \stackrel{\eqref{eq:second-ineq}}{>} \lambda_{\min}(\tilde{H}) - \frac{\varepsilon}{2} \stackrel{\eqref{eq:first-ineq}}{>} \lambda_{\min}(H) - \varepsilon  > b - \varepsilon = \frac{a+b}{2},&\text{if $\lambda_{\min}(H) > b$}
    \end{cases}
\end{equation}
As a result, the \texttt{P} algorithm $\mathcal{A}$ successfully solves the decision problem in Eq.~\eqref{eq:k-local-hamiltonian-problem} by deciding
\begin{equation}
    \begin{cases}
        \texttt{YES,}& \text{if }\lambda<\frac{a+b}{2} \\
        \texttt{NO,}& \text{if }\lambda > \frac{a+b}{2} \label{eq:decision}
    \end{cases}
\end{equation}

Since $H$ with $\|H\|\,=\mathcal{O}(1)$ and $a,b\in\mathbb{Q}$ with $b-a=\Theta(1)$ were both arbitrary, it especially follows that the $k$-local Hamiltonian problem with constant promise gaps is in \texttt{P} for every $k\geq 2$. On the other hand, the point of this argument is using the \texttt{QPCP} conjecture which states that the local Hamiltonian problem with constant promise gaps $b-a=\Omega(1)$ is \texttt{QMA}-hard~\cite{aharonov2013guest, natarajan2018low}, i.e., it is at least as hard as all \texttt{QMA} problems. Since $\texttt{P}\subseteq\texttt{QMA}$~\cite{aharonov2002quantum}, we thus must have that $\texttt{P}=\texttt{QMA}$. This is the desired contradiction to our assumption that $\texttt{P}\neq\texttt{QMA}$. Consequently, no such classical algorithm $\mathcal{A}$ can exist.
\qed

\subsubsection{Probabilistic Proof using $\texttt{BPP}\subsetneq\texttt{QMA}$ only}

Here, we prove that one does not require the \texttt{QPCP} conjecture in order to show that there exists no classical (probabilistic) \texttt{BPP} algorithm that can solve the $\mathscr{H}_{\text{LCE}}$-class Hamiltonian problem, provided the standard assumption that $\texttt{BPP}\neq\texttt{QMA}$. First, we need to make the following technical reduction for the probabilistic proof to work flawlessly:

\begin{lemma} \label{lemma:decision-reduction}
    Let $\mathscr{H}\subseteq\operatorname{Herm}(2^N)$ be an arbitrary class of Hamiltonians and $\eta=\Theta(1)$ a constant. Suppose that solving the Hamiltonian decision problem of the form in Eq.~\eqref{eq:k-local-hamiltonian-problem} with promise gap $b-a>0$ for the class $\mathscr{H}$ has some complexity $\mathcal{C}\in\{\texttt{P}, \texttt{BPP}, \texttt{NP}, \texttt{BQP}, \texttt{QMA},\dots\}$. Then, solving the decision problem for the affine class $\mathscr{H}':=\{H+\eta I_{2^N}:H\in\mathscr{H}\}$ leaves the promise gap invariant and thus has exactly the same complexity $\mathcal{C}$, where $I_{2^N}$ denotes the $2^N\times 2^N$ identity matrix. In particular, the $2$-local Hamiltonian problem remains \texttt{QMA}-complete after such an affine transformation.
\end{lemma}
\smallskip
\noindent
\emph{Proof.} Let $\mathscr{H}\subseteq\operatorname{Herm}(2^N)$ be an arbitrary Hamiltonian class and let $\mathcal{C}$ denote the complexity of the Hamiltonian decision problem with promise gap $b-a>0$ as stated in the lemma above. Let $\eta=\Theta(1)$ and $H\in\mathscr{H}$ be arbitrary and define $\tilde{H}:=H+\eta I_{2^N}$. Recall from linear algebra that the eigenvalues of $H$ are defined as the roots of the characteristic polynomial $P_H(X)=\det(H-XI_{2^N})\in\mathbb{C}[X]$ in the variable $X$. Therefore, the affine transformation $H\mapsto \tilde{H}=H+\eta I_{2^N}$ has the effect of shifting the eigenspectrum by $\eta$. In particular,
\begin{equation}
    \lambda_{\min}(\tilde{H}) = \lambda_{\min}(H) + \eta
\end{equation}
with an accordingly transformed promise gap for the transformed decision problem given by $\tilde{a}:=a+\eta$ and $\tilde{b}:=b+\eta$. Notably, this affine transformation leaves the promise gap invariant:
\begin{equation}
    \tilde{b}-\tilde{a} = (b+\eta) - (a+\eta) = b-a.
\end{equation}
Therefore, the transformed problem, i.e., the Hamiltonian problem for the class $\mathscr{H}'$ has exactly the same complexity as the original problem, namely $\mathcal{C}$.
\qed
\bigskip
\par
Lemma~\ref{lemma:decision-reduction} allows us to assume without loss of generality that either $a=\Theta(1)$ or $b=\Theta(1)$ is constant in the formulation of the $2$-local Hamiltonian problem in Eq.~\eqref{eq:k-local-hamiltonian-problem}. Now, the idea of our probabilistic proof is to construct a randomly sampled LCE perturbation $V_{\mathcal{J}}$ by using the promise gap boundaries $a$ and $b$ which are a priori given by the problem statement. As detailed below, since at least one of them is constant, this ensures that $V_{\mathcal{J}}$ indeed defines a LCE perturbation. We are now ready to prove non-existence of \texttt{BPP} algorithms that can solve the $\mathscr{H}_{\text{LCE}}$-class Hamiltonian problem:

\begin{theorem}
    If $\texttt{BPP}\neq\texttt{QMA}$, then there exists no \texttt{BPP} algorithm that can solve the $\mathscr{H}_{\text{LCE}}$-class Hamiltonian problem. In particular, the problem of finding the ground energy of an arbitrary Hamiltonian $H\in\mathscr{H}_{\text{LCE}}$ is guaranteed to be classically infeasible in general.
\end{theorem}
\smallskip
\noindent
\emph{Proof.} Assume that $\texttt{BPP}\neq\texttt{QMA}$ and suppose by contradiction that there exists a \texttt{BPP} algorithm $\mathcal{A}$ that can solve the $\mathscr{H}_{\text{LCE}}$-class Hamiltonian problem. Let $\varepsilon>0$ to be determined and let $H\in\operatorname{Herm}(2^N)$ be an arbitrary Hamiltonian with $\|H\|\,=\mathcal{O}(1)$. Let $a,b\in\mathbb{Q}$ be the given rational boundaries of the promise gap $b-a=\Omega(\operatorname{Poly}(N)^{-1})$. Without loss of generality we may assume that $b-a\to0$ as $N\to\infty$, otherwise, $b-a=\Theta(1)$ is constant and the theorem follows directly from $\texttt{P}\subseteq\texttt{BPP}$ and Proposition~\ref{prop:classical-infeasbility} by an analogous deterministic LCE perturbation $V_{\mathcal{J}}$ without having to assume the \texttt{QPCP} conjecture (since the promise gap is already constant in that case). Thus, it suffices to handle the case where the promise gap $b-a$ vanishes as $N\to\infty$.

First, we construct a randomly sampled LCE perturbation $V_{\mathcal{J}}$ to establish a probabilistic analogue of Lemma~\ref{prop:reduction-normalized-hamiltonian} with sufficiently high probability for compatibility with our \texttt{BPP} algorithm $\mathcal{A}$. Here, we employ the reduction argument of Lemma~\ref{lemma:decision-reduction}, allowing us to assume that $a=\Theta(1)$
 or $b=\Theta(1)$. Therefore,
 \begin{equation*}
     V_{\mathcal{J}}:=c_0\sum_{j\in\mathcal{J}}\underbrace{\left(a_j-\frac{a+b}{2}\right)}_{=\Theta(1)}Q_j
 \end{equation*}
 almost surely defines a LCE perturbation for $a_j\stackrel{\text{i.i.d.}}{\sim}\operatorname{Unif}([a,b])$, where $Q_j\in\mathcal{P}_N\setminus\{I\}$ are arbitrary non-identity Pauli strings indexed by $\mathcal{J}$ such that $|\mathcal{J}|\,=\mathcal{O}(1)$, where $c_0=\Theta(1)$ is an arbitrarily small constant that is chosen towards the end of the proof to control the probability. In particular
 \begin{equation}
    \tilde{H}:=H+V_{\mathcal{J}}\in\mathscr{H}_{\text{LCE}}. \label{eq:LCE-perturbation}
 \end{equation}
 
 The next step is to compute the mean and variance of the largest quantum measurement outcome of the LCE perturbation $V_{\mathcal{J}}$ with respect to the randomness of $a_j$ in order concentrate the probability of the desired event that $|\lambda_{\min}(\tilde{H})-\lambda_{\min}(H)|$ is sufficiently small. For that purpose, let $|\psi\rangle\in\mathcal{H}_N$ be an arbitrary $N$-qubit state. Then, the expected measurement outcome of the LCE perturbation $V_{\mathcal{J}}$ has mean
 \begin{align}
     \mathbb{E}[\langle\psi| V_{\mathcal{J}}|\psi\rangle] &= c_0\sum_{j\in\mathcal{J}}\underbrace{\left(\mathbb{E}[a_j]-\frac{a+b}{2}\right)}_{=0}Q_j = 0
 \end{align}
 by construction since
 \begin{equation}
     \mathbb{E}[a_j] = \frac{1}{b-a}\int_a^b x\,dx = \frac{1}{b-a}\cdot\frac{b^2-a^2}{2} = \frac{1}{b-a}\cdot\frac{(b-a)(b+a)}{2} = \frac{a+b}{2} \label{eq:LCE-mean}
 \end{equation}
for all $j\in\mathcal{J}$. Therefore, since the operator norm of Pauli matrices is equal to $1$, that is $\langle\psi|Q_j|\psi\rangle\leq 1$ for all $j\in\mathcal{J}$, the expected measurement outcome of the LCE perturbation $V_{\mathcal{J}}$ has variance
\begin{align}
    \operatorname{Var}&[\langle\psi|V_{\mathcal{J}}|\psi\rangle] = \mathbb{E}[\langle\psi|V_{\mathcal{J}}|\psi\rangle^2] = c_0^2\sum_{j,j'\in\mathcal{J}} \mathbb{E}\left[\left(a_j-\frac{a+b}{2}\right)\left(a_{j'}-\frac{a+b}{2}\right)\right]\langle\psi|Q_jQ_{j'}|\psi\rangle \nonumber
    \\ &= c_0^2\sum_{j\in\mathcal{J}}\mathbb{E}\left[\left(a_j-\frac{a+b}{2}\right)^2\right]\underbrace{\langle\psi|Q_j|\psi\rangle^2}_{\leq 1} + \,c_0^2\sum_{\substack{j,j'\in\mathcal{J} \\ j\neq j'}} \underbrace{\left(\mathbb{E}[a_j]-\frac{a+b}{2}\right)}_{\stackrel{\eqref{eq:LCE-mean}}{=}0}\underbrace{\left(\mathbb{E}[a_{j'}]-\frac{a+b}{2}\right)}_{\stackrel{\eqref{eq:LCE-mean}}{=}0}\langle\psi|Q_j|\psi\rangle\langle\psi| Q_{j'}|\psi\rangle \nonumber
    \\ &\leq c_0^2\sum_{j\in\mathcal{J}}\mathbb{E}\left[\left(a_j-\frac{a+b}{2}\right)^2\right] = \frac{c_0^2(b-a)^2}{12}\sum_{j\in\mathcal{J}}1 = \frac{c_0^2|\mathcal{J}|}{12}(b-a)^2\stackrel{N\to\infty}{\to}0, \label{eq:LCE-variance}
\end{align}
recalling that the promise gap $b-a>0$ vanishes as $N\to\infty$, where in the second-to-last equation we computed
\begin{align}
    \mathbb{E}\left[\left(a_j-\frac{a+b}{2}\right)^2\right] &= \mathbb{E}[a_j^2]-2\underbrace{\mathbb{E}[a_j]}_{\stackrel{\eqref{eq:LCE-mean}}{=}\frac{a+b}{2}}\frac{a+b}{2}+\frac{(a+b)^2}{4} = \mathbb{E}[a_j^2] - \frac{(a+b)^2}{4} \nonumber
    \\ &= \underbrace{\frac{a^2+ab+b^2}{3}}_{=\mathbb{E}[a_j^2]} +\, \frac{a^2+2ab+b^2}{4} = \frac{1}{12}(4a^2+4ab+4b^2-3a^2-6ab-3b^2) \nonumber
    \\ &= \frac{1}{12}(a^2-2ab+b^2) = \frac{(b-a)^2}{12}
\end{align}
with second moments
\begin{align}
    \mathbb{E}[a_j^2] &= \frac{1}{b-a}\int_a^b x^2\,dx = \frac{1}{b-a}\cdot\frac{b^3-a^3}{3} = \frac{1}{b-a}\cdot\frac{(b-a)(b^2+ab+a^2)}{3} = \frac{a^2+ab+b^2}{3}
\end{align}
for all $j\in\mathcal{J}$.

This means that both the mean and variance with respect to $a_j$ of the expected quantum measurement outcome of $V_{\mathcal{J}}$ are independent of $|\psi\rangle$ by construction. Now, let us again take $\varepsilon:=\frac{b-a}{2}\stackrel{N\to\infty}{\to}0$ so that by the Weyl and Chebyshev inequalities,

\begin{align}
    \mathbb{P}_a\left(|\lambda_{\min}(\tilde{H}) - \lambda_{\min}(H)|\,<\frac{\varepsilon}{2}\right) &\stackrel{\eqref{eq:LCE-perturbation}}{=} \mathbb{P}_a\left(|\lambda_{\min}(H+V_{\mathcal{J}}) - \lambda_{\min}(H)|\,<\frac{\varepsilon}{2}\right) \stackrel{\text{\tiny Weyl}}{\geq} \mathbb{P}_a\left(\sup_{|\psi\rangle}|\langle\psi|V_{\mathcal{J}}|\psi\rangle|\, < \frac{\varepsilon}{2}\right) \nonumber
    \\ &\geq 1 - \frac{4}{\varepsilon^2} \operatorname{Var}\left[\sup_{|\psi\rangle}\langle\psi|V_{\mathcal{J}}|\psi\rangle\right] \stackrel{\eqref{eq:LCE-variance}}{=} 1-\frac{(b-a)^2}{3\varepsilon^2}\,c_0^2|\mathcal{J}|\,=1-\underbrace{\frac{4}{3}\,c_0^2|\mathcal{J}|}_{=:\delta_0}\label{eq:first-ineq2}
\end{align}
by our choice of $\varepsilon>0$. In particular, since $|\mathcal{J}|\,=\mathcal{O}(1)$, the error between the perturbed ground energy $\lambda_{\min}(\tilde{H})$ and the exact solution $\lambda_{\min}(H)$ is bounded by a fraction of the vanishing promise gap $\mathcal{O}(b-a)$ with arbitrarily high probability by choosing the constant $c_0>0$ to be sufficiently small. Note that this argument only works due to the fact that $|\mathcal{J}|\,=\mathcal{O}(1)$ since we require that $c_0=\Theta(1)$ for a valid perturbation $V_{\mathcal{J}}$.

The final step is to show that the \texttt{BPP} algorithm $\mathcal{A}$ successfully solves the decision problem in Eq.~\eqref{eq:k-local-hamiltonian-problem} and conclude with a contradiction. By the assumption on the \texttt{BPP} algorithm $\mathcal{A}$, we know that
\begin{equation}
    |\lambda_{\min}(\tilde{H}) - \lambda|\,<\frac{\varepsilon}{2} \label{eq:second-ineq2}
\end{equation}
with probability $p$ strictly larger than $\frac{1}{2}$ in feasible time, using $\varepsilon=\frac{b-a}{2}=\Omega(\operatorname{Poly}(N)^{-1})$. That is, Eq.~\eqref{eq:second-ineq2} occurs with probability at least $p=\frac{1}{2}+\delta$ for some fixed $\delta\in\left(0,\frac{1}{2}\right]$. Hence, combining both inequalities results into 
\begin{equation}
    \begin{cases} \lambda \,\, < \,\,\lambda_{\min}(\tilde{H}) + \frac{\varepsilon}{2} \,\,\,\, < \,\,\,\lambda_{\min}(H) + \varepsilon  < a + \varepsilon = \frac{a+b}{2},&\text{if $\lambda_{\min}(H) < a$} \\
    \lambda \stackrel{\eqref{eq:second-ineq2}}{>} \lambda_{\min}(\tilde{H}) - \frac{\varepsilon}{2} \stackrel{\eqref{eq:first-ineq2}}{>} \lambda_{\min}(H) - \varepsilon  > b - \varepsilon = \frac{a+b}{2},&\text{if $\lambda_{\min}(H) > b$} \label{eq:decision2}
    \end{cases}
\end{equation}
with probability at least $\frac{1}{2}+(\delta-\delta_0)$ which is arbitrarily close to $\frac{1}{2}$ by choosing $c_0>0$ to be sufficiently small as shown below. Indeed, to justify the lower bound of the probability of Eq.~\eqref{eq:decision2} occurring, let us denote the probabilistic events
\begin{equation}
    \mathcal{E}_1 := \left\{|\lambda_{\min}(\tilde{H})-\lambda|\,<\frac{\varepsilon}{2}\right\}\quad\text{and}\quad \mathcal{E}_2 := \left\{|\lambda_{\min}(\tilde{\mathcal{H}})-\lambda_{\min}(H)|\,<\frac{\varepsilon}{2}\right\}.
\end{equation}
Then, by elementary probability theory,
\begin{equation*}
    \mathbb{P}(\text{$\mathcal{E}_1$ and $\mathcal{E}_2$}) =\quad \underbrace{\mathbb{P}(\mathcal{E}_1)}_{\geq\frac{1}{2}+\delta} \quad + \underbrace{\mathbb{P}(\mathcal{E}_2)}_{\stackrel{\eqref{eq:first-ineq2}}{\geq} 1-\delta_0} -\quad  \underbrace{\mathbb{P}(\text{$\mathcal{E}_1$ or $\mathcal{E}_2$})}_{\leq 1} \quad\geq  \frac{1}{2}+\underbrace{(\delta-\delta_0)}_{\stackrel{!}{>}0}
\end{equation*}
with respect to the joint randomness of the \texttt{BPP} algorithm $\mathcal{A}$ and the uniformly sampled coefficients $a_j$, as claimed. Crucially, the difference $\delta-\delta_0>0$ can be made arbitrarily small by choosing a suitable constant $c_0>0$ in the definition of $\delta_0$. This ensures that the probability of Eq.~\eqref{eq:decision2} occurs with a probability strictly larger than $\frac{1}{2}$ in accordance with Definition~\ref{def:BPP-algorithm}.

As a consequence of Eq.~\eqref{eq:decision2}, the \texttt{BPP} algorithm $\mathcal{A}$ successfully solves the decision problem in Eq.~\eqref{eq:k-local-hamiltonian-problem} by deciding
\begin{equation}
    \begin{cases}
        \texttt{YES,}& \text{if }\lambda<\frac{a+b}{2} \\
        \texttt{NO,}& \text{if }\lambda > \frac{a+b}{2}
    \end{cases}
\end{equation}
with probability at least $\frac{1}{2}+\delta'$ for $\delta':=\delta-\delta_0>0$ which can be made arbitrarily small by choosing the constant $c_0>0$ accordingly.

Since $H$ with $\|H\|\,=\mathcal{O}(1)$ and $a,b\in\mathbb{Q}$ with $b-a=\Omega(\operatorname{Poly}(N)^{-1})$ were both arbitrary, it especially follows that the $2$-local Hamiltonian problem with inversely polynomial promise gaps is in \texttt{BPP}. Due to the fact that the $2$-local Hamiltonian problem is \texttt{QMA}-complete~\cite{kempe2006complexity}, and that $\texttt{BPP}\subseteq\texttt{QMA}$~\cite{aharonov2002quantum}, we thus find $\texttt{BPP}=\texttt{QMA}$ which is the desired contradiction to our assumption that $\texttt{BPP}\neq\texttt{QMA}$. Therefore, there cannot exist such a classical algorithm $\mathcal{A}$.
\qed

We have thus rigorously proven that there exists no classical algorithm that can find the ground energy of observables in the class $\mathscr{H}_{\texttt{LCE}}$ of LCE Hamiltonians, either provided that $\texttt{BPP}\neq\texttt{QMA}$. Note that we only made a statement about non-classicality and gradient scalability at initialization via LCE perturbations. In particular, there is no guarantee that a VQA can generally find the ground energy of a Hamiltonian in $\mathscr{H}_{\text{LCE}}$ due to the possibility of solutions $\theta_*$ existing outside landscape patches on which LCE fails to exhibit scalable gradient signals. It remains an open question of whether there exist Hamiltonians in $\mathscr{H}_{\text{LCE}}$ with VQA solutions $\theta_*$ lying inside LCE-effective landscape patches for which no classical algorithm exists. Since this work focuses on the theory of initialization only, we leave this question as an open problem.

\subsection{Deterministic Worst-Case Simulation Complexity}

The next parts leverage the results of the previous section, and focuses on the computational complexity of simulating the cost function $C(\theta)$ in Eq.~\eqref{eq:expectation3} with the help of the Taylor surrogate $C_m(\theta)$ in Eq.~\eqref{eq:taylor2}. We shall use $\operatorname{Time}C_m(\theta)$ to denote the time complexity of evaluating $C_m(\theta)$, which is, by definition, the number of operations required for evaluating the expression for $C_m(\theta)$.

To quantify the time $\operatorname{Time}C_m(\theta)$ required to evaluate $C_m(\theta)$ classically, we begin with the deterministic error analysis, where we are able to highlight a computational phase transition from polynomial to super-polynomial complexity, which is characterized by the $\ell_1$-norm $\|\theta\|_1$ of the parameter configuration:
\bigskip
\\
\noindent
\textbf{Theorem~\ref{thm:complexity}.} (Deterministic Complexity) \emph{Consider an expectation of the form Eq.~\eqref{eq:expectation3} with a variational ansatz $U(\theta)$ involving $D\geq1$ parameters, and suppose we simulate it classically using the Taylor series $C_m(\theta)$ defined in Eq.~\eqref{eq:taylor2} with truncation threshold $m = m(\theta)$ given by Lemma~\ref{lemma:truncation-threshold}. Then, if $\|\theta\|_1=\mathcal{O}(1)$, the runtime of computing $C_m(\theta)$ is polynomial, namely $\mathcal{O}\left(|\mathcal{I}|D^{m}\right)$. On the other hand, if { $\|\theta\|_1 = o(D)$ strictly increases sub-linearly} as $D\to\infty$, {and $\#\{k\in[D]:\exists \xi\in\frac{\pi}{2}\mathbb{Z}:|\theta_k-\xi|> \Omega(D^{-1})\}=\Omega(D)$, then the} runtime becomes super-polynomial,
\begin{align*}
    \operatorname{Time}C_m(\theta) = {\Omega\left(\operatorname{Poly}(D)^{-1}\left(\frac{D}{\|\theta\|_1}\right)^{e\|\theta\|_1}\right),}
\end{align*}
{and is at most $\mathcal{O}(|\mathcal{I}|D\,2^{2D})$ in the worst case}}.
\bigskip
\\
\noindent
\emph{Proof.} By the higher-order parameter-shift rule, Lemma~\ref{lemma:parameter-shift}, we may write
\begin{equation}
    C_m(\theta) = \sum_{{\|\alpha\|_1}<m}(D_\theta^\alpha C)(0)\,\frac{\theta^\alpha}{\alpha!} = \sum_{{\|\alpha\|_1}<m}\frac{\theta^\alpha}{\alpha!}\sum_{\substack{j\in\mathbb{N}_0^D \\ j_k\leq\alpha_k\,\forall k}}(-1)^{|j|}\binom{\alpha}{j}\,C(\xi_{j,\alpha}) = \sum_{(j,\alpha)\in\Lambda_m}\frac{\theta^\alpha}{\alpha!}(-1)^{|j|}\binom{\alpha}{j}\,C(\xi_{j,\alpha}), \label{eq:taylorsurrogate}
\end{equation}
where $\xi_{j,\alpha}:=\frac{\pi}{2}\sum_{k=1}^D(\alpha_k-2j_k)e_k$, with $e_k$ denoting the $k$-th unit vector, and indexing set
\begin{equation}
    \Lambda_m := \{(j,\alpha)\in\mathbb{N}_0^D\times\mathbb{N}_0^D: {\|\alpha\|_1}\,< m,\,\forall k\in[D]:0\leq j_k\leq\alpha_k\}.
\end{equation}
The computational complexity for each of the evaluation of $C(\xi_{j,\alpha})$ using Gottesman-Knill is of order $\mathcal{O}(D\,|\mathcal{I}|)$~\cite{gottesman1998heisenberg, aaronson2004improved}, which is the most expensive computation in our simulation strategy. At the end of the proof, we will justify why Gottesman-Knill may be applied for each $\xi_{j,\alpha}$. As a result, the time resources required to compute $C_m(\theta)$ scales like
\begin{align}
    \operatorname{Time}C_m(\theta) &= \mathcal{O}(D\,|\mathcal{I}|)\cdot \#\{C(\xi_{j,\alpha}):(j,\alpha)\in\Lambda_m\} \nonumber
    \\&\leq\mathcal{O}(D\,|\mathcal{I}|)\,\#\Lambda_m = \mathcal{O}(D\,|\mathcal{I}|)\sum_{{\|\alpha\|_1}< m}\underbrace{\prod_{k=1}^D(\alpha_k+1)}_{\leq 2^{\|\alpha\|_1}\leq 2^{m-1}} \leq \mathcal{O}\left(D\,|\mathcal{I}|\,2^{m}\sum_{\|\alpha\|_1<m}1\right). \label{eq:time}
\end{align}
If $\|\theta\|_1=\mathcal{O}(1)$, then we know that $m=m(\theta)=\mathcal{O}(1)$ by Lemma~\ref{lemma:truncation-threshold}. Therefore, we may employ Eq.~\eqref{eq:stars-and-bars} in Lemma~\ref{lemma:stars-and-bars} to establish the polynomial complexity,
\begin{equation}
    \operatorname{Time}C_m(\theta) = \mathcal{O}\left(D\,|\mathcal{I}|\,\sum_{\|\alpha\|_1<m}1\right) = \mathcal{O}(|\mathcal{I}|D^m), \label{eq:polynomial-time-equation}
\end{equation}
as claimed.

Next, we will prove the super-polynomial case where {$\|\theta\|_1 = o(D)$ strictly increases sub-linearly as $D\to\infty$. Here is where we will use the assumption that $\#\{k\in[D]:\exists \xi\in\frac{\pi}{2}\mathbb{Z}:|\theta_k-\xi|> \Omega(D^{-1})\}=\Omega(D)$, otherwise since $U(\xi)\in\mathcal{C}_N$ is Clifford for all $\frac{\pi}{2}\mathbb{Z}$ (as shown at the end of the proof), one may adapt the Clifford architecture of $U(\cdot)$ to obtain a new PQC $\tilde{U}(\cdot)$ such that $\tilde{U}(0)=U(\xi)$, and as a result, find a translational change of coordinates $\theta\mapsto\tilde{\theta}$ such that $\|\tilde{\theta}\|_1 = \mathcal{O}(1)$, giving rise to a polynomial time Taylor surrogate by the previous case. Therefore, this assumption is made in order to rule out classical simulability beyond the case $\|\theta\|_1=\mathcal{O}(1)$ by ensuring that most of the rotation gates $R_{V_k}(\theta_k)$ are sufficiently far from a single-qubit Clifford gate.} 

{We continue by proving the super-polynomial lower bound. The computational cost of evaluating the Taylor surrogate in Eq.~\eqref{eq:taylorsurrogate} is lower bounded by}
\begin{align}
    {\operatorname{Time}C_m(\theta)} &\geq{\min_{(j,\alpha)\in\Lambda_m}\underbrace{\operatorname{Time}C(\xi_{j,\alpha})}_{\geq1}\cdot\sum_{(j,\alpha)\in\Lambda_m}1} \nonumber
    \\ &\geq {\sum_{\|\alpha\|_1<m}\#\{k\in[D]:0\leq j_k\leq \alpha_k\} = \sum_{\|\alpha\|_1<m}\prod_{k=1}^D\underbrace{(\alpha_k+1)}_{\geq 1} \geq \sum_{\|\alpha\|_1<m} 1 = \Omega\left(\operatorname{Poly}(D)^{-1}\left(\frac{D}{\|\theta\|_1}\right)^{e\|\theta\|_1}\right),}
\end{align}
{where the last equality follows from Lemma~\ref{lemma:stars-and-bars} and applying Lemma~\ref{lemma:truncation-threshold} which tells us that} {$m=m(\theta)\sim e\|\theta\|_1$ as $D\to\infty$.}

What one could think about is whether it is possible to significantly reduce the cardinality of $\Lambda_m$ by using the fact that $C(\cdot)$ is $2\pi$-periodic in each direction in parameter space.
\par
Indeed, note that $(\xi_{j,\alpha})_k$ corresponds to a shift of $2\pi$ in the $k$-th coordinate if and only if $\alpha_k-2j_k = 4$. Therefore, $C(\xi_{j,\alpha})$ depends only on $\alpha_k-2j_k\bmod 4$ for all $k\in[D]$. In other words:
\begin{equation}
    \text{for each $(j,\alpha)\in\Lambda_m$ there exists $\beta_{j,\alpha}\in\{0,1,2,3\}^D$ such that $C(\xi_{j,\alpha})=C\left(\frac{\pi}{2}\beta_{j,\alpha}\right)$}.
\end{equation}
This particularly means that in the worst case scenario, we require at most $4^D=2^{2D}$ different evaluations of $C(\xi_{j,\alpha})$. As a result, it is possible to optimize the worst-case runtime to achieve
\begin{equation}
    \operatorname{Time}C_m(\theta) = \mathcal{O}(|\mathcal{I}|\,D\,2^{2D}),
\end{equation}
as claimed. Note that this scales worse compared to the computational complexity $\mathcal{O}(|\mathcal{I}|\,2^{2N})$ of naively evaluating the exponential quadratic form in Eq.~\eqref{eq:expectation}, since high expressivity requires that $N\ll D$. 

It remains to justify the application of Gottesman-Knill's theorem for each $\xi_{j,\alpha}$. It suffices to show that $U(\xi_{j,\alpha})\in\operatorname{Clifford}(N)$ because of $C(\xi_{j,\alpha})=\langle 0 |U(\xi_{j,\alpha})^\dagger \hat{O} U(\xi_{j,\alpha})|0\rangle$ {(where $\hat{O}$ denotes the observable in Eq.~\eqref{eq:observable} here in order to avoid confusion with the Hadamard gate)}. Indeed, note that $e^{-i\frac{\pi}{4}}R_X\left(\frac{\pi}{2}\right)=HSH\in\operatorname{Clifford}(N)$, $R_Y(\frac{\pi}{2})=XH\in\operatorname{Clifford}(N)$, and $e^{-i\frac{\pi}{4}}R_Z(\frac{\pi}{2})= S\in\operatorname{Clifford}(N)$ (global phases are irrelevant). This particularly means for multiples of $\frac{\pi}{2}$ that $R_X(\frac{\pi}{2}\mathbb{Z}),R_Y(\frac{\pi}{2}\mathbb{Z}),R_Z(\frac{\pi}{2}\mathbb{Z})\subset\operatorname{Clifford}(N)$, respectively. This follows from the fact that $\operatorname{Clifford}(N)$ defines a group. As a result, since $\frac{\pi}{2}\left(\alpha_k-2j_k\right)\in \frac{\pi}{2}\mathbb{Z}$ is also multiple of $\frac{\pi}{2}$, we find
\begin{equation}
    U(\xi_{j,\alpha}) \in \left\langle R_X\left(\frac{\pi}{2}\mathbb{Z}\right),R_Y\left(\frac{\pi}{2}\mathbb{Z}\right), R_Z\left(\frac{\pi}{2}\mathbb{Z}\right), \operatorname{Clifford}(N)\right\rangle_N\subset\operatorname{Clifford}(N). \label{eq:clifford-unitary}
\end{equation}
By the Gottesman-Knill theorem, we thus know that the cost evaluation $C(\xi_{j,\alpha})$ is easy to simulate classically, i.e., it can be computed in polynomial time $\mathcal{O}(D\,|\mathcal{I}|)$ on a conventional computer~\cite{gottesman1998heisenberg, aaronson2004improved}.
\qed

\subsection{Probabilistic Mean-Squared Simulation Complexity}
Next, we use the mean-squared error analysis in order to determine the Gaussian/uniform initialization strategy $\mathbb{P}_\theta$ that covers the largest landscape patch where Taylor surrogating the cost $C(\cdot)$ remains polynomially efficient.
\bigskip
\\
\noindent
\textbf{Theorem~\ref{thm:probabilistic-complexity}.} \emph{Consider an expectation of the form in Eq.~\eqref{eq:expectation3} with a variational ansatz $U(\theta)$ involving $D\geq1$ parameters, and suppose we simulate it classically using the Taylor series $C_m(\theta)$ defined in Eq.~\eqref{eq:taylor2}. Initialize the parameters $\theta\sim\mathbb{P}_\theta\in\{\mathcal{N}(0,\sigma^2I_D), \operatorname{Unif}([-\sigma,\sigma]^D)\}$ with i.i.d. components $\theta_1,\dots,\theta_D$ according to Eq.~\eqref{eq:mean-squared-error2} with constant $c>0$. {Assume further that $\|H\|=\mathcal{O}(1)$}. Then, with probability at least $1-\rho$, the computational cost $\operatorname{Time}C_m(\theta\sim\mathbb{P}_\theta)$ of simulating $C(\theta)$ up to an error $\varepsilon>0$, is of polynomial order,
\begin{equation*}
    {\mathcal{O}\left(|\mathcal{I}|\,D^{\frac{1}{2\log \left(\frac{1}{q}\right)}\log\left(\frac{c}{\rho\,\varepsilon^2}\right)}\right)}
\end{equation*}
for non-asymptotic local initialization, i.e., {$\sigma = \mathcal{O}(qD^{-\frac{1}{2}})$} for some $0<q<1$}
\bigskip
\\
\noindent
\emph{Proof.} By the higher-order parameter-shift rule, Lemma~\ref{lemma:parameter-shift}, we may write
\begin{equation}
    C_m(\theta) = \sum_{|\alpha|<m}(D_\theta^\alpha C)(0)\,\frac{\theta^\alpha}{\alpha!} = \sum_{|\alpha|<m}\frac{\theta^\alpha}{\alpha!}\sum_{\substack{j\in\mathbb{N}_0^D \\ j_k\leq\alpha_k\,\forall k}}(-1)^{|j|}\binom{\alpha}{j}\,C(\xi_{j,\alpha}) = \sum_{(j,\alpha)\in\Lambda_m}\frac{\theta^\alpha}{\alpha!}(-1)^{|j|}\binom{\alpha}{j}\,C(\xi_{j,\alpha}), \label{eq:taylor-formula}
\end{equation}
where $\xi_{j,\alpha}:=\frac{\pi}{2}\sum_{k=1}^D(\alpha_k-2j_k)e_k$, with $e_k$ denoting the $k$-th unit vector, and indexing set
\begin{equation}
    \Lambda_m := \{(j,\alpha)\in\mathbb{N}_0^D\times\mathbb{N}_0^D: \|\alpha\|_1\,< m,\,\forall k\in[D]:0\leq j_k\leq\alpha_k\}.
\end{equation}
The computational complexity for each of the evaluation of $C(\xi_{j,\alpha})$ using Gottesman-Knill is of order $\mathcal{O}(D\,|\mathcal{I}|)$, which is the most expensive computation in our simulation strategy. We have already justified in the proof of Theorem~\ref{thm:complexity}
 why Gottesman-Knill may be applied for each $\xi_{j,\alpha}$. As a result, the time resources requires to compue $C_m(\theta)$ scales like
\begin{align}
    \operatorname{Time}C_m(\theta) &= \mathcal{O}(D\,|\mathcal{I}|)\cdot \#\{C(\xi_{j,\alpha}):(j,\alpha)\in\Lambda_m\} \nonumber
    \\&\leq\mathcal{O}(D\,|\mathcal{I}|)\,\#\Lambda_m = \mathcal{O}(D\,|\mathcal{I}|)\sum_{|\alpha|< m}\underbrace{\prod_{k=1}^D(\alpha_k+1)}_{\leq 2^{\|\alpha\|_1}\leq 2^{m-1}} \leq \mathcal{O}\left(D\,|\mathcal{I}|\,2^{m}\sum_{\|\alpha\|_1<m}1\right).
\end{align}
For $\|\theta\|_1=\mathcal{O}(1)$, we have already establish polynomial complexity in Eq.~\eqref{eq:polynomial-time-equation}, i.e.,
\begin{equation}
    \operatorname{Time}C_m(\theta) = \mathcal{O}\left(D\,|\mathcal{I}|\,\sum_{\|\alpha\|_1<m}1\right) = \mathcal{O}(|\mathcal{I}|D^m),
\end{equation}
By Lemma~\ref{lemma:probabilistic-truncation} we know that, with probability at least $1-\rho$, a constant truncation threshold satisfying
\begin{equation}
    {m\geq \frac{1}{2}\,\frac{\log\left(\frac{c}{\rho\,\varepsilon^2}\right)}{\log \left(\frac{1}{q}\right)}}
\end{equation}
is sufficient to simulate $C_m(\theta)$ with error at most $\varepsilon>0$.
\qed

\subsection{Complexity Comparison with the Pauli Path Surrogate}

We conclude our complexity analysis by showing that our Taylor surrogate outperforms the Pauli path surrogate employed by Lerch et al.~\cite{lerch2024efficient}, as long as the patch is sufficiently small:
\bigskip
\begin{corollary}\label{corr:faster-runtime}
    Consider an expectation of the form in Eq.~\eqref{eq:expectation3} with a variational ansatz $U(\theta)$ involving $D\geq1$ parameters, and suppose we simulate it classically using the Taylor series $C_m(\theta)$ defined in Eq.~\eqref{eq:taylor2}. {Assume further that $\|H\|=\mathcal{O}(1)$.} Initialize $\theta\sim\mathbb{P}_\theta\in\{\mathcal{N}(0,\sigma^2I_D), \operatorname{Unif}([-\sigma,\sigma]^D)\}$ with i.i.d. components $\theta_1,\dots,\theta_D$, where {$\sigma = \mathcal{O}(qD^{-1/2})$} for some $0<q<1$, so that
    \begin{equation*}
        \mathbb{E}_\theta[(C(\theta)-C_m(\theta))^2] \leq c\,q^{2m}
    \end{equation*}
    for some constant $c>0$ as per Lemma~\ref{lemma:mean-square-error}.

    As shown by Lerch et al. in Ref.~\cite[Theorem 2, Eq. (13)]{lerch2024efficient}, with probability at least $1-\rho$, the Pauli-propagation algorithm $\hat{C}(\theta)$ can simulate the cost $C(\theta)$ up to error $\varepsilon>0$ using
    \begin{equation*}
        \operatorname{Time}\hat{C}(\theta\sim\mathbb{P}_\theta)=\gamma_1\,|\mathcal{I}|\,D^{\log\left(\frac{1}{\rho\,\varepsilon^2}\right)}
    \end{equation*}
    many operations for some constant $\gamma_1>0$. On the other hand, by Theorem~\ref{thm:probabilistic-complexity}, we know that, with probability at least $1-\rho$, the Taylor-simulation algorithm $C_m(\theta)$ can simulate the cost $C(\theta)$ up to error $\varepsilon>0$ using
    \begin{equation*}
        \operatorname{Time}C_m(\theta\sim\mathbb{P}_\theta) = \gamma_2\,|\mathcal{I}|\,D^{\frac{1}{2\log \left(\frac{1}{q}\right)}\log\left(\frac{c}{\rho\,\varepsilon^2}\right)}
    \end{equation*}
    many operations for some constant $\gamma_2>0$.

    Then, setting $\gamma:=\frac{\gamma_1}{\gamma_2}>0$, our algorithm outperforms the Pauli-propagation technique, i.e., $\operatorname{Time}C_m(\theta)<\operatorname{Time}\hat{C}(\theta)$, if and only if
    \begin{equation}
    q < \exp\left(-\frac{\log\left(\frac{c}{\rho\varepsilon^2}\right)}{2\log\left(\frac{1}{\rho\varepsilon^2}\right)+\frac{2\log\gamma}{\log D}}\right), \label{eq:q-bound}
    \end{equation}
    provided that
    \begin{equation}
        D > \exp\left(\frac{\log\gamma}{\log\rho+2\log\varepsilon}\right) \label{eq:D-bound}
    \end{equation}
    in the case where $\gamma<1$.
\end{corollary}
\noindent
\emph{Proof.} As shown by Lerch et al. in Ref.~\cite[Theorem 2, Eq. (13)]{lerch2024efficient}, with probability at least $1-\rho$, the Pauli-propagation algorithm $\hat{C}(\theta)$ can simulate the cost $C(\theta)$ up to error $\varepsilon>0$ using
\begin{equation}
    \operatorname{Time}\hat{C}(\theta)=\gamma_1\,|\mathcal{I}|\,D^{\log\left(\frac{1}{\rho\,\varepsilon^2}\right)}
\end{equation}
many operations for some constant $\gamma_1$. On the other hand, by Theorem~\ref{thm:probabilistic-complexity}, we know that, with probability at least $1-\rho$, the Taylor surrogate $C_m(\theta)$ can simulate the cost $C(\theta)$ up to error $\varepsilon>0$ using
\begin{equation}
    \operatorname{Time}C_m(\theta) = \gamma_2\,|\mathcal{I}|\,D^{\frac{1}{2\log \left(\frac{1}{q}\right)}\log\left(\frac{c}{\rho\,\varepsilon^2}\right)}
\end{equation}
many operations for some constant $\gamma_2>0$. Setting $\gamma:=\frac{\gamma_1}{\gamma_2}>0$, we see that $\operatorname{Time}C_m(\theta)<\operatorname{Time}\hat{C}(\theta)$ if and only if
\begin{equation}
     \gamma\,D^{\log\left(\frac{1}{\rho\,\varepsilon^2}\right)} = D^{\log\left(\frac{1}{\rho\,\varepsilon^2}\right)+\frac{\log\gamma}{\log D}} > D^{\frac{1}{2\log\left(\frac{1}{q}\right)}\log\left(\frac{c}{\rho\,\varepsilon^2}\right)}.
\end{equation}
Taking the logarithm on both sides yields
\begin{equation}
    \log\left(\frac{1}{\rho\varepsilon^2}\right)+\frac{\log\gamma}{\log D} > \frac{1}{2\log\left(\frac{1}{q}\right)}\log\left(\frac{c}{\rho\varepsilon^2}\right),
\end{equation}
which may equivalently be rewritten as
\begin{equation}
    \log\left(\frac{1}{q}\right) > \frac{\log\left(\frac{c}{\rho\varepsilon^2}\right)}{2\log\left(\frac{1}{\rho\varepsilon^2}\right)+\frac{2\log\gamma}{\log D}}, \label{eq:comparing-times}
\end{equation}
if and only if the inequality
\begin{equation}
    \log\left(\frac{1}{\rho\varepsilon^2}\right)+\frac{\log\gamma}{\log D} > 0 \label{eq:inequality}
\end{equation}
holds true. Eq.~\eqref{eq:inequality} holds true trivially if $\gamma>1$ because of $\rho,\varepsilon<1$. On the other hand, if $\gamma<1$, then Eq.~\eqref{eq:inequality} is equivalent to
\begin{equation}
    \log\left(\frac{1}{\rho\varepsilon^2}\right) > \frac{\log\left(\frac{1}{\gamma}\right)}{\log D} \iff D > \exp\left(\frac{\log\left(\frac{1}{\gamma}\right)}{\log\left(\frac{1}{\rho\varepsilon^2}\right)}\right) = \exp\left(\frac{\log\gamma}{\log\rho+2\log\varepsilon}\right),
\end{equation}
as assumed by the statement of the corollary.
\par
Returning to Eq.~\eqref{eq:comparing-times}, we find that it equivalently reads
\begin{equation}
    q < \exp\left(-\frac{\log\left(\frac{c}{\rho\varepsilon^2}\right)}{2\log\left(\frac{1}{\rho\varepsilon^2}\right)+\frac{2\log\gamma}{\log D}}\right),
\end{equation}
which concludes the proof.
\qed
\bigskip
\\
\indent
To better understand the practical implications of Corollary~\ref{corr:faster-runtime}, we numerically evaluate the bound on the surrogate patch size parameter $q$ and, if $\gamma<1$, the number of parameters $D$ required for outperforming the Pauli-propagation technique. Fig.~\ref{fig:patchsizes} panels (a) to (c) shows the upper bounds on $q$ (cf. Eq.~\eqref{eq:q-bound}) across various values of the constants $c,\gamma$ and parameter dimensions $D$. Panel (d) separately shows the behavior of the lower bound for $D$ (cf. Eq.~\eqref{eq:D-bound}) when $\gamma<1$. The plots clarify that our Taylor-based surrogate becomes advantageous in practically relevant scenarios, especially for large-scale models with high precision and low probability of failure (i.e., small $\varepsilon,\rho>0$). For constants $c,\gamma\in[10^{-6},10^6]$ and arbitrary model sizes $D\geq1$, the worst case scenario requires $q\geq 0.35$, which does not limit practicality.

\begin{figure*}[htbp]
    \centering
    \includegraphics[width=1\linewidth]{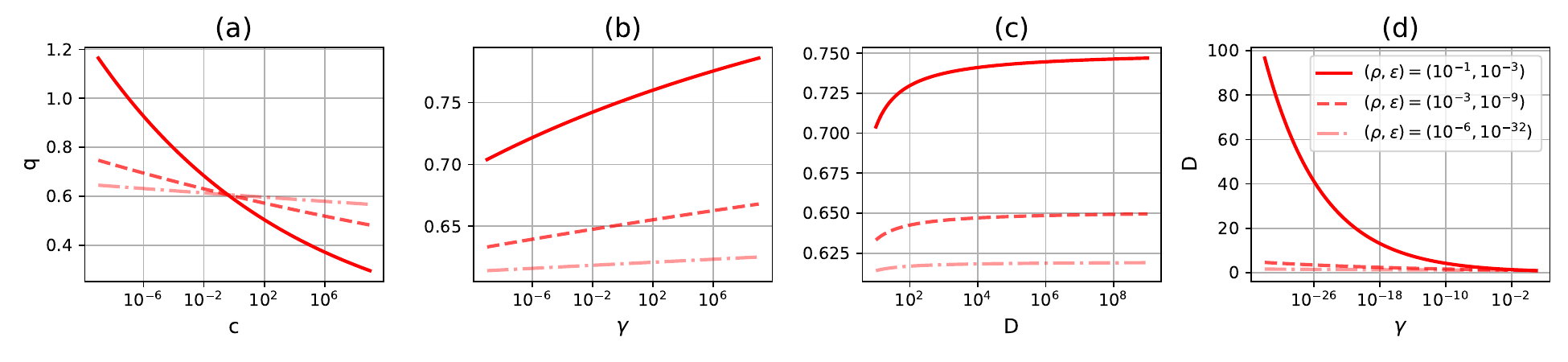}
    \caption{Behavior of the required quantities $q$ and $D$, for a polynomial speedup compared to the Pauli path simulation technique~\cite{lerch2024efficient}, as established by Corollary~\ref{corr:faster-runtime}. (a) - (b): Upper bound of $q$. (d): Lower bound of $D$.}
    \label{fig:patchsizes}
\end{figure*}

\subsection{Efficient Recursive Classical Implementation}
In a practical implementation of our simulation strategy, one has to iterate over the summation $\{\|\alpha\|_1<m\}$ in Eq.~\eqref{eq:taylor2}. Concretely,
we are interested in finding all multi-index combinations in the set:
\begin{equation}
    \mathcal{A}_{m,D}:=\{\alpha=(\alpha_1,\dots,\alpha_D)\in\mathbb{N}_0^D:\|\alpha\|_1<m\}.
\end{equation}
When using a \texttt{for}-loop naively, one will be confronted with unnecessary computational overheads. Alternatively, we propose a recursive method that generates all combinations in the set $\mathcal{A}_{m,D}$, which indeed turns out to avoid the aforementioned overheads (cf. Fig.~\ref{fig:polynomial-complexity}).
\par
For that purpose, let us define $\mathbb{N}_0^0:=\{\emptyset\}$, $\|\emptyset\|_1:=0$ and $\dim(\emptyset):=-\infty$. Consider the recursion
\begin{equation}
    G_{m,D}(\alpha,\sigma):=\begin{cases} \bigcup_{s=0}^{m-\sigma-1}\{G_{m,D}(\alpha\cup(s),s)\},& \dim(\alpha)<D, \\ (\alpha),&\dim(\alpha)=D, \end{cases}
\end{equation}
where for tuples we use $\alpha\cup(s):=(\alpha,s)$ to denote to operation of appending elements to the tuple. Also, $\dim(\alpha)\geq0$ denotes the number of components of non-empty tuples $\alpha$.
\bigskip
\begin{lemma}
    We have that
    \begin{equation*}
        G_{m,D}(\emptyset,0)=\mathcal{A}_{m,D}
    \end{equation*}
    for all $m,D\in\mathbb{N}$.
\end{lemma}
\noindent
\emph{Proof.} Let $m\in\mathbb{N}$ be arbitrary. First, we are going to establish for all $D\geq1$ that
\begin{equation}
    G_{m,D}(\emptyset,0)=\bigcup_{(s_1,\dots,s_{D-1})\in\mathcal{A}_{m,D-1}}\,\bigcup_{s_D=0}^{m-\left(\sum_{j=1}^{D-1}s_j\right)-1}\{(s_1,\dots,s_{D-1},s_D)\}. \label{eq:union-identity}
\end{equation}
We do this by showing $G_{m,D-1}(\emptyset,0)=\mathcal{A}_{m,D-1}$ by induction. Indeed, if $D=1$, we find
\begin{align}
    G_{m,1}(\emptyset,0)=\bigcup_{s_1=0}^{m-1}\{G_{m,1}(\underbrace{(s_1)}_{\dim(\cdot)=1=D},s_1)\} = \bigcup_{s_1=0}^{m-1}\{(s_1)\} = \{(0),(1),\dots,(m-1)\}=\mathcal{A}_{m,1}.
\end{align}
Assume by induction that $G_{m,D-1}(\emptyset,0)=\mathcal{A}_{m,D-1}$ holds true for some fixed $D$. In that case,
\begin{align}
    G_{m,D}(\emptyset,0) &= \bigcup_{s_1=0}^{m-1}\{G_{m,D}((s_1),s_1)\} = \bigcup_{s_1=0}^{m-1} \, \bigcup_{s_2=0}^{m-s_1-1}\{G_{m,D}((s_1,s_2),s_1+s_2)\} \nonumber
    \\ &= \cdots = \bigcup_{s_1=0}^{m-1} \, \bigcup_{s_2=0}^{m-s_1-1} \cdots \bigcup_{s_{D-1}=0}^{m-\left(\sum_{j=1}^{D-2}s_j\right)-1} \, \bigcup_{s_{D}=0}^{m-\left(\sum_{j=1}^{D-1}s_j\right)-1}\{G_{m,D}(\underbrace{(s_1,\dots,s_{D-1},s_D)}_{\dim(\cdot)=D},\sum_{j=1}^D s_j\} \nonumber
    \\ &= \bigcup_{s_1=0}^{m-1} \, \bigcup_{s_2=0}^{m-s_1-1} \cdots \bigcup_{s_{D-1}=0}^{m-\left(\sum_{j=1}^{D-2}s_j\right)-1} \, \left(\bigcup_{s_{D}=0}^{m-\left(\sum_{j=1}^{D-1}s_j\right)-1} \{(s_1,\dots,s_{D-1},s_D)\}\right) \nonumber
    \\ &= \bigcup_{(s_1,\dots,s_{D-1})\in \underbrace{G_{m,D-1}(\emptyset,0)}_{=\mathcal{A}_{m,D-1}}} \left(\bigcup_{s_{D}=0}^{m-\left(\sum_{j=1}^{D-1}s_j\right)-1} \{(s_1,\dots,s_{D-1},s_D)\}\right),
\end{align}
where in the last equality we simply used that, by definition,
\begin{align}
    \bigcup_{(s_1,\dots,s_{D-1})\in G_{m,D-1}(\emptyset,0)}\{(s_1,\dots,s_{D-1})\} = \underbrace{G_{m,D-1}(\emptyset,0)}_{\mathcal{A}_{m,D-1}}=\bigcup_{s_1=0}^{m-1}\,\bigcup_{s_2=0}^{m-s_1-1}\cdots\bigcup_{s_{D-1}=0}^{m-\left(\sum_{j=1}^{D-2}s_j\right)-1}\{(s_1,\dots,s_{D-1})\},
\end{align}
Therefore, by induction, we know that (\ref{eq:union-identity}) holds true. Let us now use this identity to conclude with the claim $G_{m,D}(\emptyset,0)=\mathcal{A}_{m,D}$.
\smallskip
\par
``$\subseteq$'': If $\alpha=(\alpha_1,\dots,\alpha_D)\in G_{m,D}(\emptyset,0)$, then (\ref{eq:union-identity}) implies $\alpha_D\leq m-\left(\sum_{j=1}^{D-1}s_j\right)-1$. This immediately results in the bound
\begin{equation}
    \|\alpha\|_1=\alpha_D+\sum_{j=1}^{D-1}\alpha_j \leq m-1.
\end{equation}
In particular, $\alpha\in\mathcal{A}_{m,D}$.
\par
\smallskip
``$\supseteq$'': On the other hand, if $\alpha= (\alpha_1,\dots,\alpha_D)\in\mathcal{A}_{m,D}$, then $\|\alpha\|_1=\alpha_D+\sum_{j=1}^{D-1}<m$ yields
\begin{equation}
    \alpha_D\leq m-\left(\sum_{j=1}^{D-1}\alpha_j\right)-1.
\end{equation}
In particular
\begin{equation}
    \alpha=(\alpha_1,\dots,\alpha_{D-1},\alpha_D)\in\bigcup_{s_D=0}^{m-\left(\sum_{j=1}^{D-1}\alpha_j\right)-1}\{(\alpha_1,\dots,\alpha_{D-1},s_D)\}.
\end{equation}
Additionally, $\|\alpha\|_1<m$ means that one necessarily must have $(\alpha_1,\dots,\alpha_{D-1})\in\mathcal{A}_{m,D-1}$. Indeed, otherwise we have the contradiction
\begin{equation}
    m\leq\sum_{j=1}^{D-1}\alpha_j\leq \sum_{j=1}^D\alpha_j=\|\alpha\|_1<m.
\end{equation}
Therefore,
\begin{equation}
    \alpha\in\bigcup_{(s_1,\dots,s_{D-1})\in\mathcal{A}_{m,D-1}} \bigcup_{s_D=0}^{m-\left(\sum_{j=1}^{D-1}\alpha_j\right)-1}\{(s_1,\dots,s_{D-1},s_D)\} = G_{m,D}(\emptyset,0),
\end{equation}
where in the last equality we made use of our previous claim (\ref{eq:union-identity}).
\qed

\section{Additional Numerical Experiments \label{sec:appendix-experiments}}
This section expands upon the numerical experiments in Sec.~\ref{sec:experiments}. In particular, we show the scaling of initial gradients for all circuit designs considered in this work, VQE experiments with finite sampling and different learning rates, and ultimately VQE experiments for the Heisenberg model.

\subsection{Gradient Scaling Experiments \label{sec:extra-benchmarking}}

We repeat the initial gradient scaling experiments (cf. Figs.~\ref{fig:trainability} and~\ref{fig:escaping-BP}) comparing all models considered in this work: \texttt{mHEA}, \texttt{fHEA}, and \texttt{rPQC} (defined in Appendix~\ref{sec:ansatz}).

Fig.~\ref{fig:benchmarking-other-models-shallow} illustrates the expected behavior of the initial gradient norm at $\theta = 0$ employing various PQC models with $L=1$ layer. Gradients are computed analytically via Eq.~\eqref{eq:explicity-taylor-coeff2}. Results are averaged over 10 independent runs, each involving a randomly selected single Pauli string $H = P$ and a random direction $k \in [D]$ used in constructing the LCE transformation. As shown, the initial gradient norm for the non-LCE model decays exponentially and becomes indistinguishable from zero beyond $N = 10$ qubits. In contrast, the LCE model consistently maintains an initial gradient norm of at least 1, in agreement with the theoretical prediction of Theorem~\ref{thm:linear-clifford-encoder}.

Fig.~\ref{fig:benchmarking-other-models-H-shallow} compares initial gradient norms at $\theta=0$ with and without LCE, employing \texttt{mHEA} with $L=1$ layer. Gradients are computed analytically via Eq.~\eqref{eq:explicity-taylor-coeff2}. Each run uses a randomly sampled observable $H$, and a random LCE construction. Panel (a) of Fig.~\ref{fig:trainability} focus on uniformly random single Pauli strings $H=P$. Without LCE, constructive contributions in Eq.~\eqref{eq:explicity-taylor-coeff2} vanish with increasing $N$, leading to zero expected gradient for sufficiently large $N$. With LCE, gradient norms remain constant in $N$. Panels (b) and (c) of Fig.~\ref{fig:trainability} extend results to random observables with $|\mathcal{I}|=\lfloor\sqrt{N}\rfloor$ Pauli terms. For observables with uniformly sampled Pauli strings $P_i$ and $c_i=1$ for all $i\in\mathcal{I}$ (called random unweighted $H$), the scaling compares to the single Pauli case. For observables with uniformly sampled Pauli strings $P_i$ and i.i.d. $c_i\sim\operatorname{Unif}([-1,1])$ for all $i\in\mathcal{I}$ (called random weighted $H$), gradient norms tend to behave more randomly due to random Pauli coefficients. This is evidence that the LCE transformation remains effective for arbitrary observables.

\begin{figure*}[h]
    \centering
    \includegraphics[width=1\linewidth]{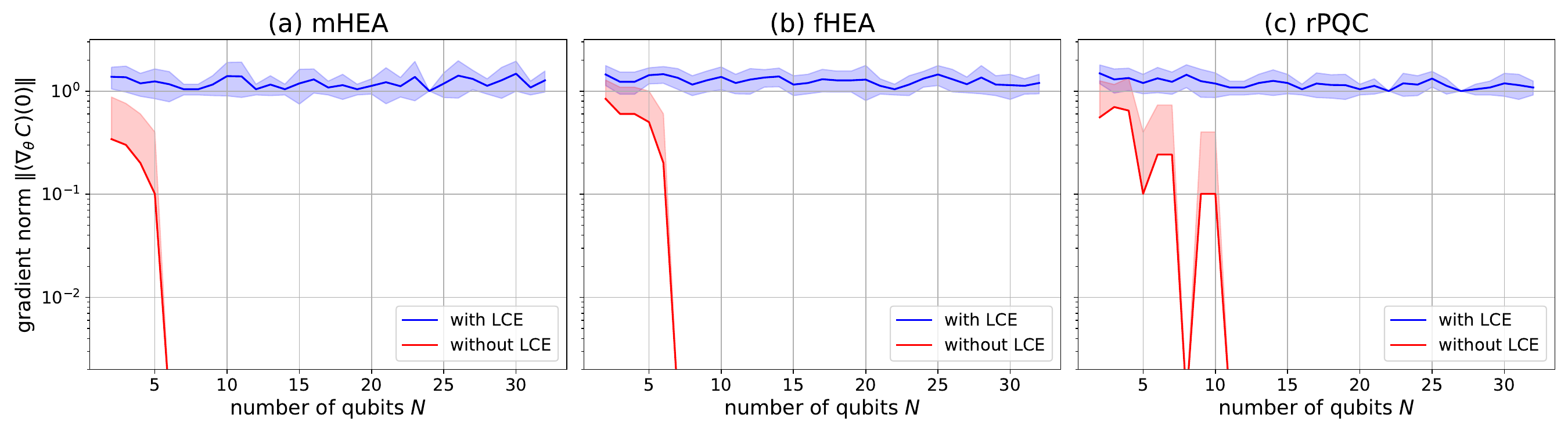}
    \caption{Scaling behavior of the initial gradient norm at $\theta=0$ for across various PQCs with $L=1$ layer. We average over $10$ samples of random single Pauli strings $H=P$, and random directions $k\in[D]$ for the LCE transformation.}
    \label{fig:benchmarking-other-models-shallow}
\end{figure*}

\begin{figure*}[h]
    \centering
    \includegraphics[width=1\linewidth]{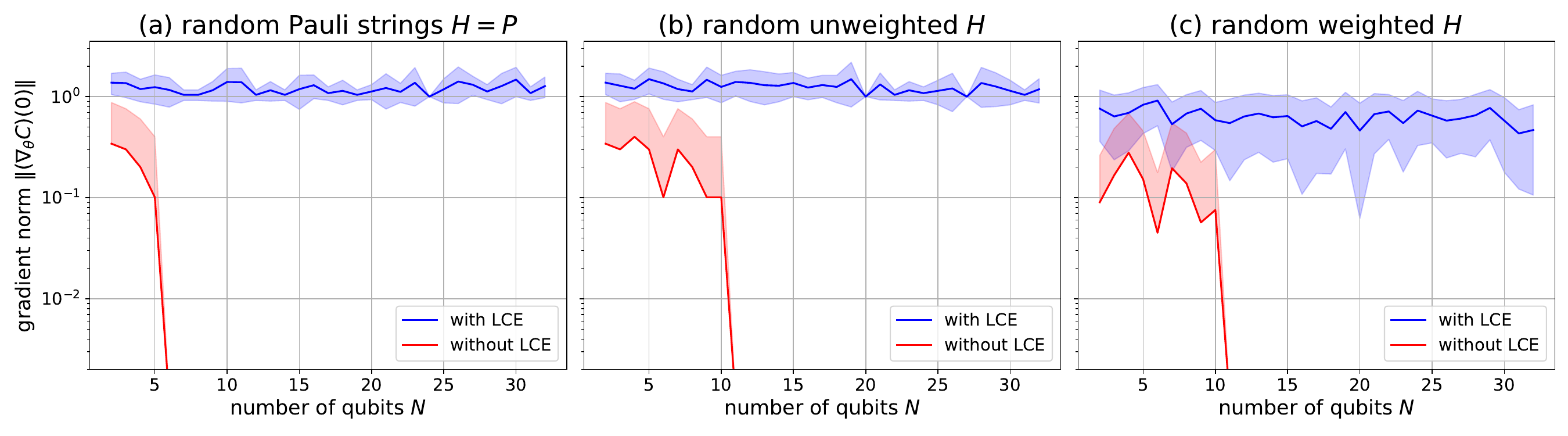}
    \caption{Scaling behavior of the initial gradient norm at $\theta=0$ for the \texttt{mHEA}-model with $L=1$ layer. The panels vary across types of random observables. The experiment highlights the effect of the LCE transformation. We average over 10 independent runs.}
    \label{fig:benchmarking-other-models-H-shallow}
\end{figure*}

Fig.~\ref{fig:benchmarking-other-models2} repeats the experiment of Fig.~\ref{fig:benchmarking-other-models-shallow} on near-Clifford patches, comparing various linear-depth PQC models, and varying patch sizes via Gaussian initialization $\sigma=D^{-r}$. Each run uses a randomly sampled parameter, a random single Pauli string observable $H=P$, and a random LCE construction. The red-shaded region corresponds to classically efficient regimes (cf. Fig.~\ref{fig:truncation-statistics}); blue-shaded to super-polynomial regimes, where gradients may still decay only polynomially. Within the constant-scaling regime (red-shaded), LCE ensures {gradient scalability}. For large $\sigma$, gradients decay exponentially. This suggests the existence of a larger patch size where gradients transition from constant decay to polynomial decay (cf. Eq.~\eqref{eq:no-barren-plateau}).

The same experiment is performed using shallow PQCs with a single layer ($L=1$) as shown in Fig.~\ref{fig:benchmarking-other-models2-shallow}. Comparing the results of linear-depth and single-layer setups reveals the impact on gradient scalability: as the number of parameters increases (i.e., in deeper circuits), the gradient norm tends to diminish more rapidly, especially for larger patch sizes.

\begin{figure*}[htbp]
    \centering
    
    \includegraphics[width=1\linewidth]{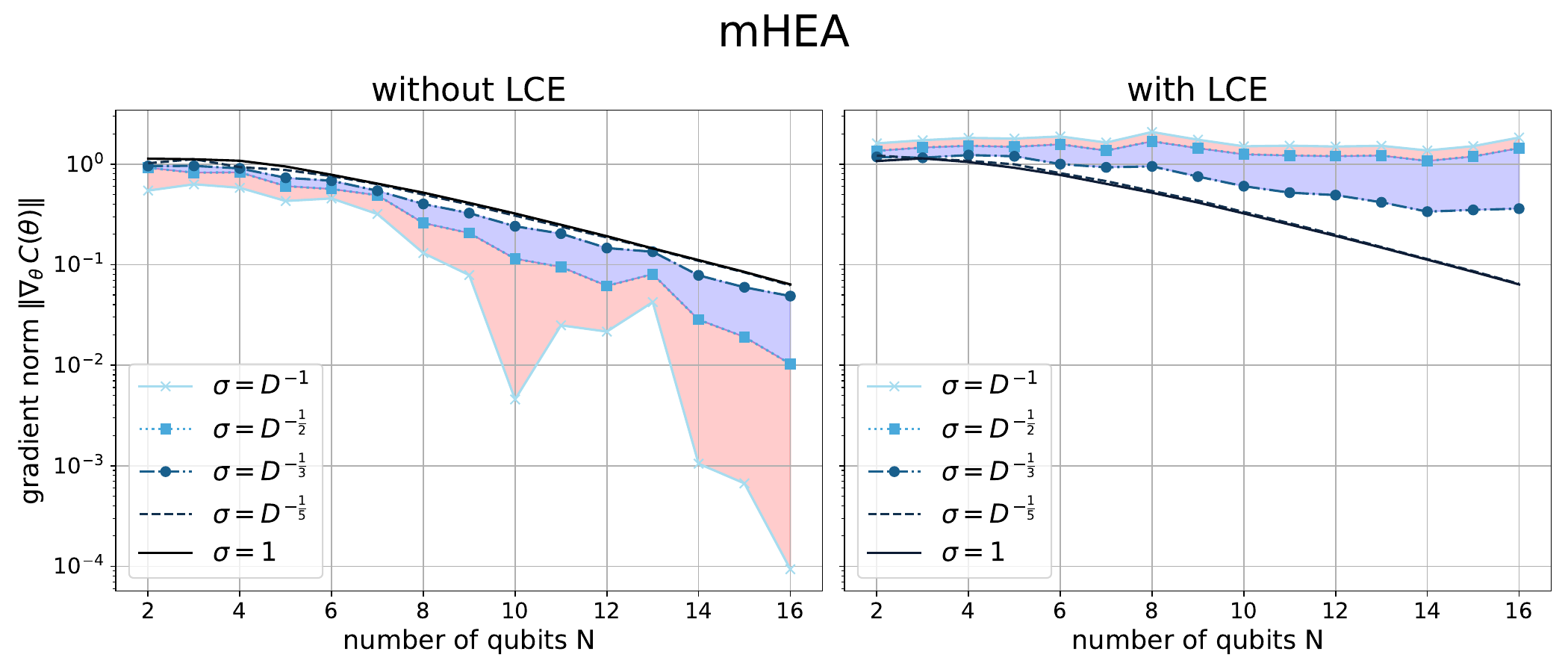}
    \includegraphics[width=1\linewidth]{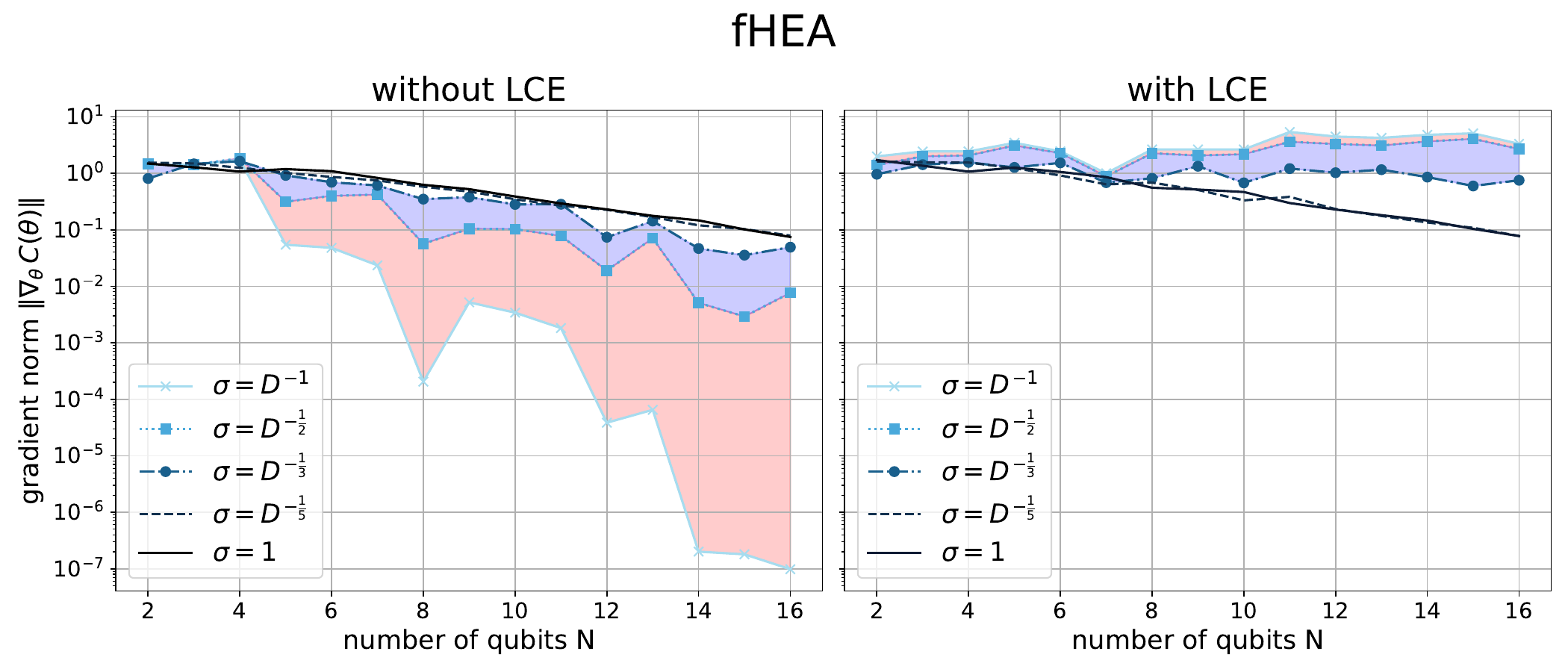}
    \includegraphics[width=1\linewidth]{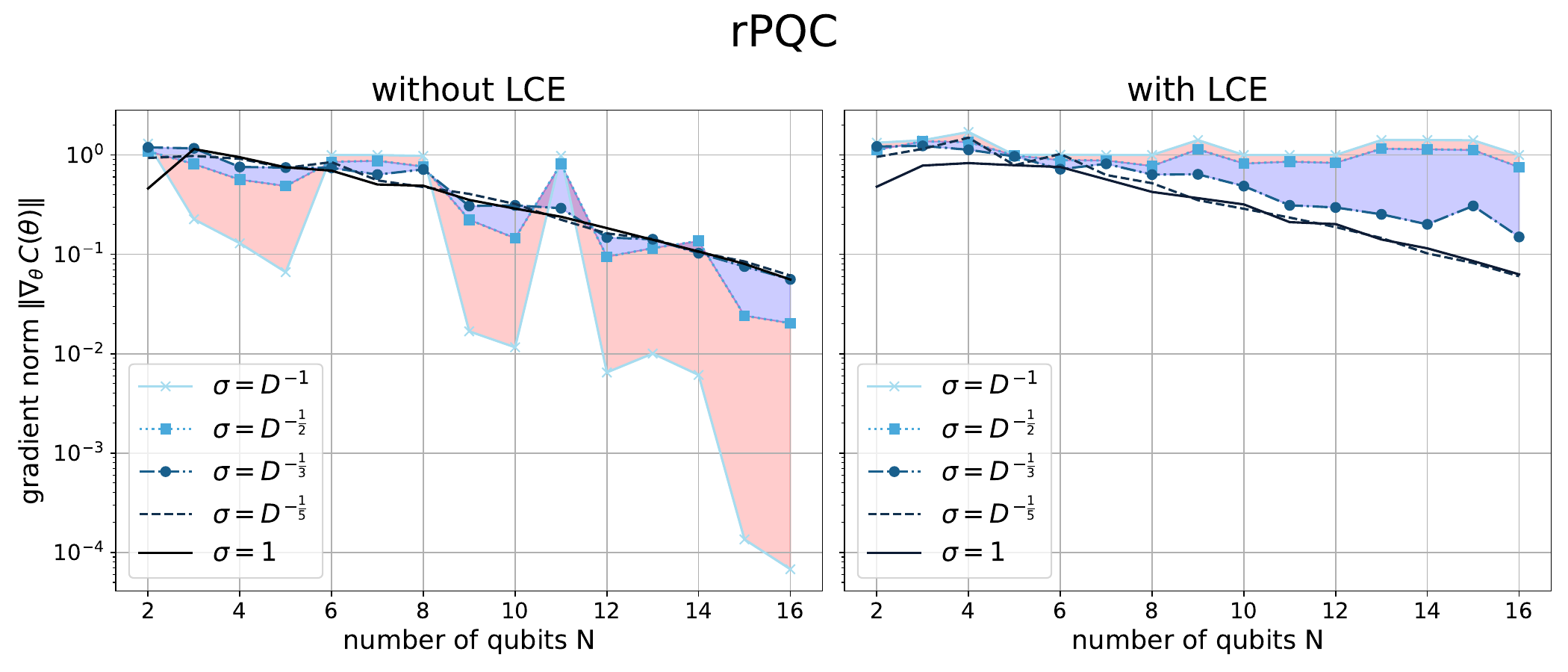}
    \caption{Scaling behavior of the gradient norm on various surrogate patches for PQC models with $L=N$ layers. We average over 50 independent runs.}
    \label{fig:benchmarking-other-models2}
\end{figure*}

\begin{figure*}[htbp]
    \centering
    
    \includegraphics[width=1\linewidth]{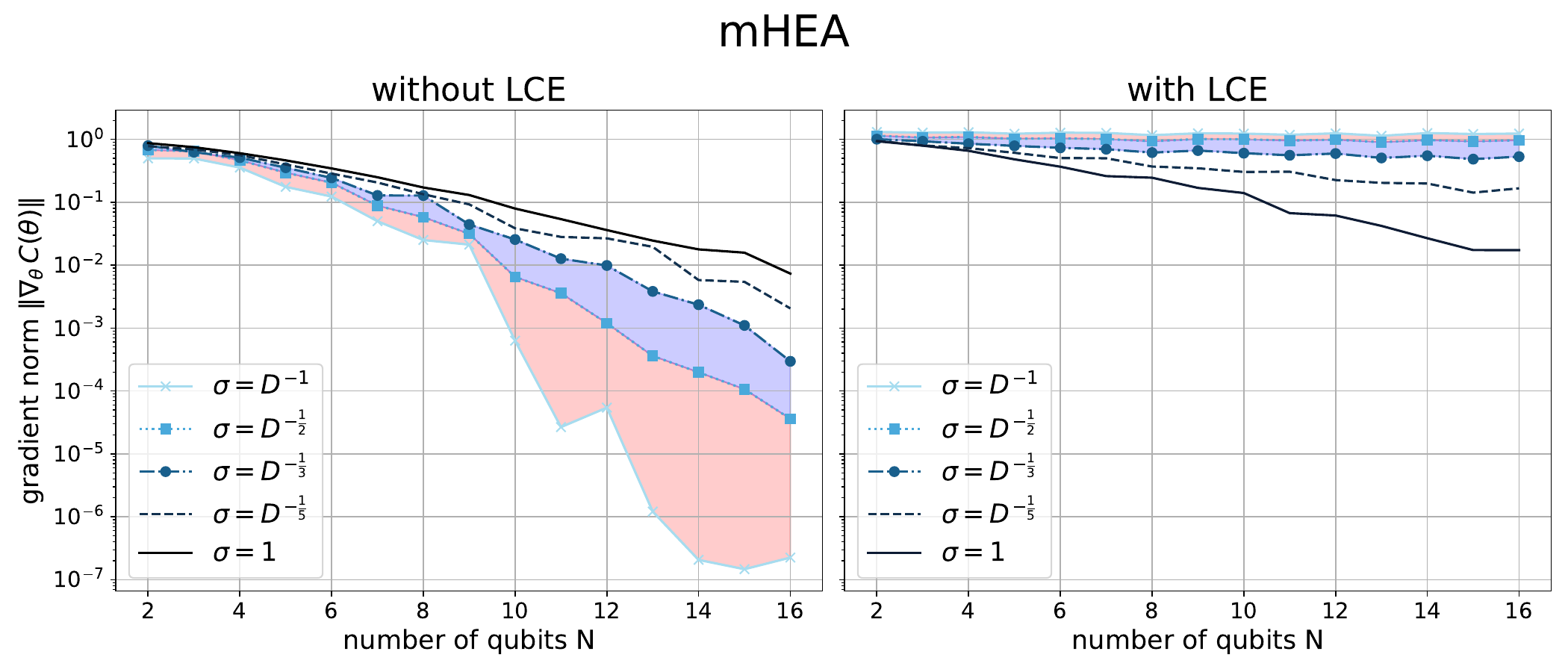}
    \includegraphics[width=1\linewidth]{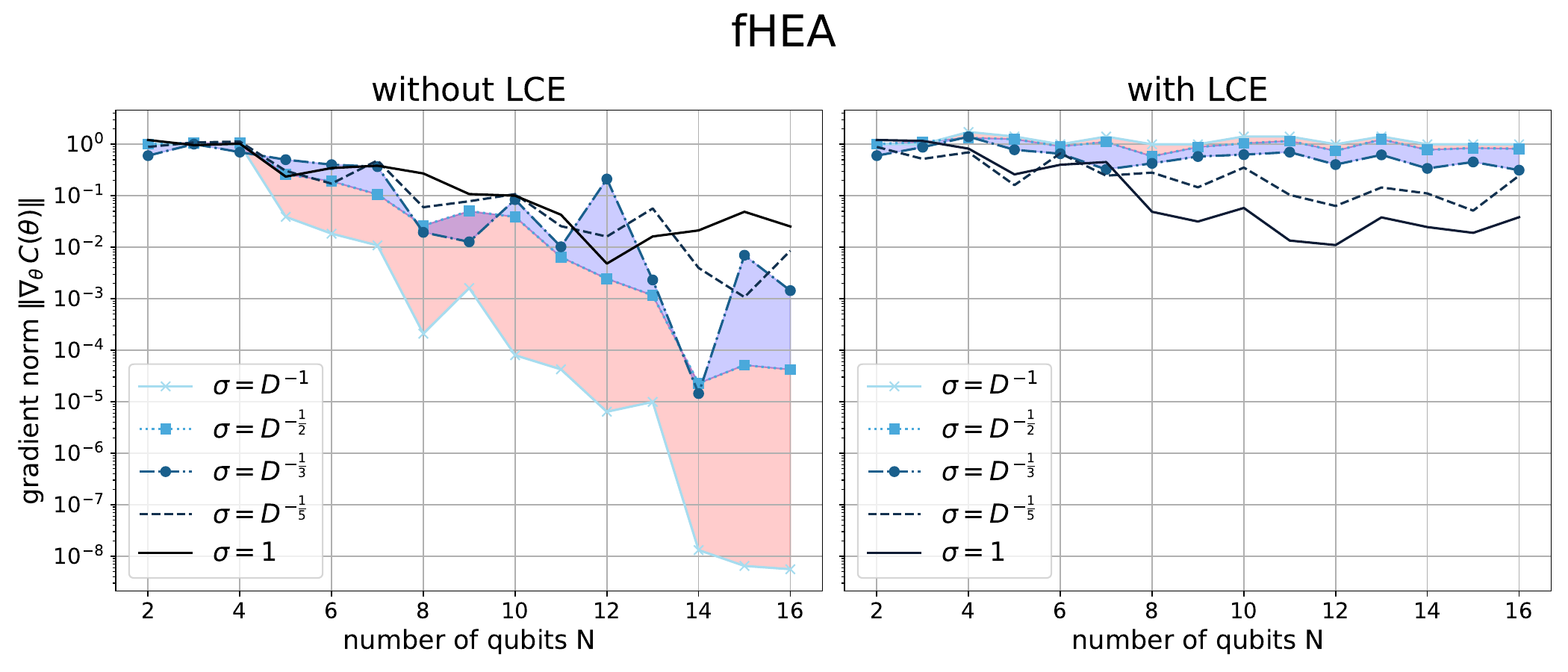}
    \includegraphics[width=1\linewidth]{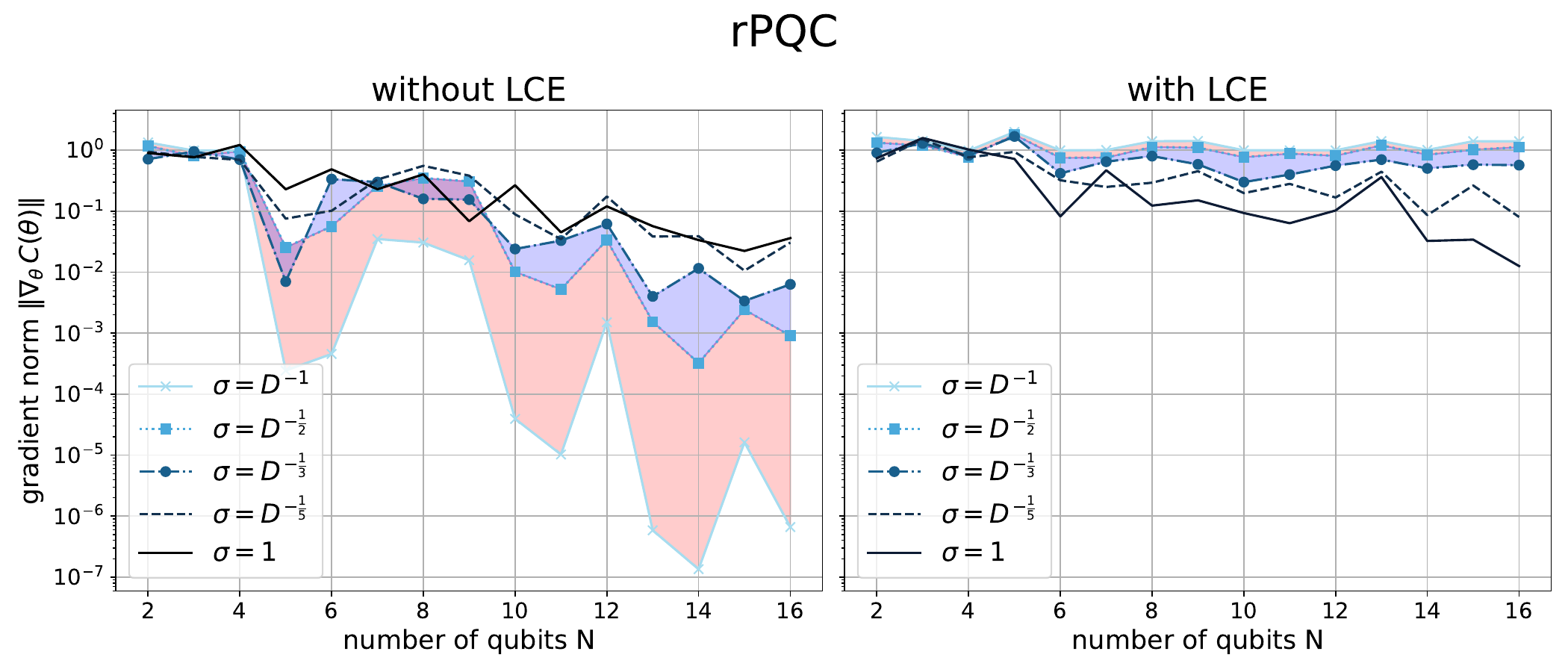}
    \caption{Scaling behavior of the gradient norm on various surrogate patches for PQC models with $L=1$ layer. We average over 50 independent runs.}
    \label{fig:benchmarking-other-models2-shallow}
\end{figure*}

\subsection{VQE Optimization with different Learning Rates and Finite Sampling Noise}

Here, we demonstrate that LCE is effective for smaller learning rates in vanilla gradient descent. Furthermore, we address the issue of finite sampling noise~\cite{kreplin2024reduction}. We use a small number of finite shots to evaluate the gradients in order to emphasize that LCE remains relevant for near-term practical applications.

Fig.~\ref{fig:VQE_learning-rates} shows VQE optimization of the global observable $Z^{\otimes N}$, employing the \texttt{mHEA} ansatz ($L=5$, $N=15$), averaging over 5 runs with Gaussian initializations $\theta\sim\mathcal{N}(0,\sigma^2I_D)$ across varying patch sizes. Optimization is performed with vanilla gradient descent, where we compare various constant learning rates. The gradients are evaluated exactly as per Eq.~\eqref{eq:explicity-taylor-coeff2}. LCE consistently yields faster convergence, independent of the learning rate. Fastest convergence is achieved with a learning rate of 0.1, and continuously slows down as the learning rate increases. While our theory guarantees trainability up to $\sigma=D^{-1/2}$, numerical results continue to demonstrate successful convergence slightly beyond, even in classically intractable regimes (cf. Theorem~\ref{thm:complexity}). LCE advantage is less pronounced as $\sigma$ increases (cf. Fig.~\ref{fig:escaping-BP}). 

Fig.~\ref{fig:VQE_shots} repeats the same VQE experiment, but this time using finite shots sampling, and a learning rate of 0.01. The improved gradient due to LCE allows fewer sampling shots. In this case, 8 shots are already sufficient for fast convergence up to the patch size $\sigma=D^{-1/3}$. Finite shots sampling explaining the increase of noise in the first three rows. Interestingly, using finite shots sampling appears to converge slightly faster compared to the exact dynamics in the last row. Hence, it may be possible that finite sampling noise may be exploited to improve the dynamics of LCE transformed VQEs. Inspecting the column containing $\sigma=D^{-1/5}$, one notices that a $15$-qubit model suffering from barren plateaus generally requires $2^{15}$ shots to accurately estimate the gradient in order to eventually achieve convergence, whereas emplyong exact gradients (cf. Eq.~\eqref{eq:explicity-taylor-coeff2}) barely enables slow convergence. These findings can be seen as evidence that LCE has the potential to enhance finite sampling efficiency in VQEs.

\begin{figure*}[h]
    \centering
    \includegraphics[width=0.94\textwidth]{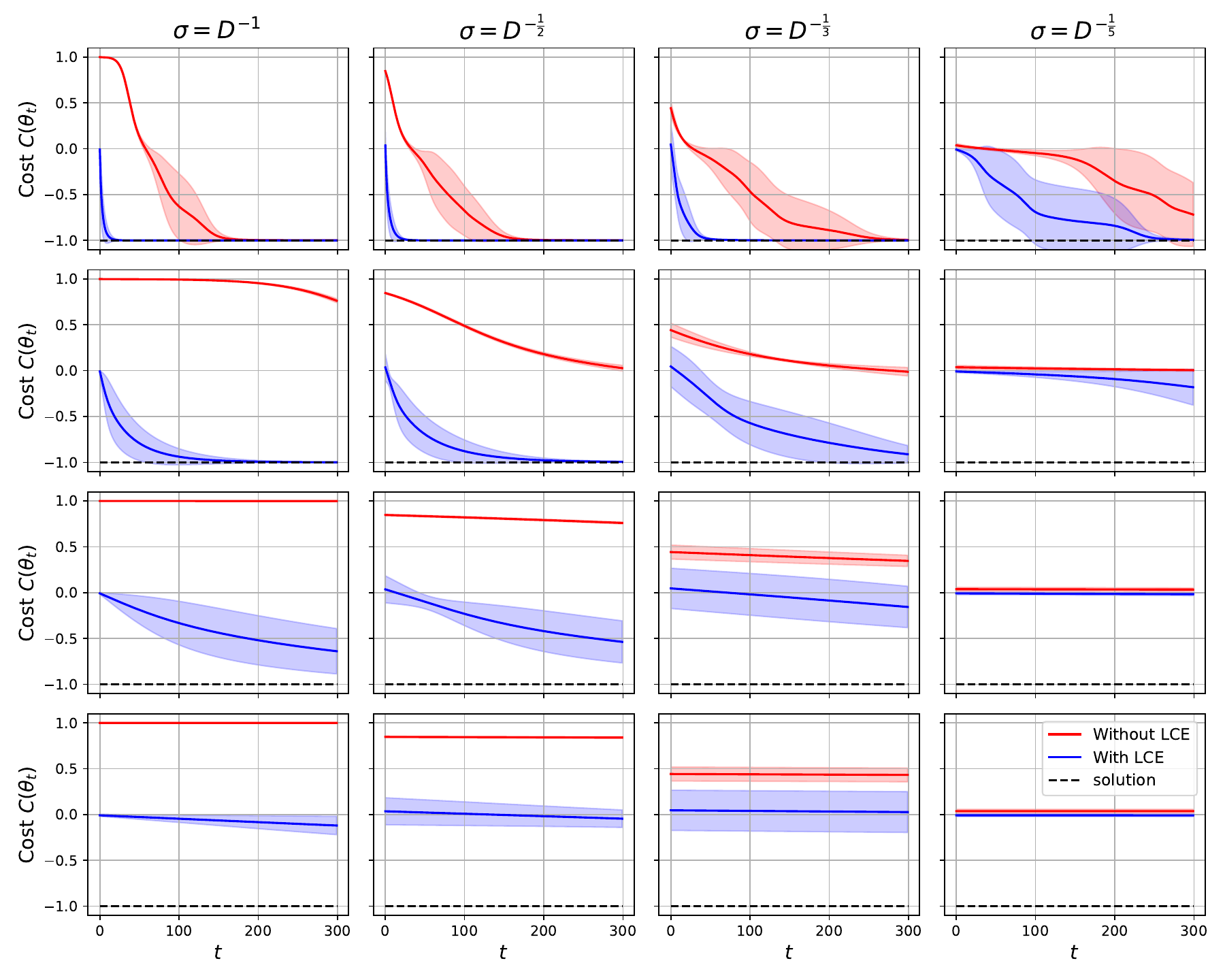}
    \caption{VQE Optimization using vanilla gradient descent with various learning rates of 0.1 (top row), 0.01 (second row), 0.001 (third row), and 0.0001 (bottom row). We evaluate exact gradients, and average over 5 runs of random Gaussian initialization across varying patch sizes. We find the ground energy of the global observable $Z^{\otimes N}$ with $N=15$, employing the \texttt{mHEA} ansatz with $L=5$ layers.}
    \label{fig:VQE_learning-rates}
\end{figure*}

\begin{figure*}[h]
    \centering
    \includegraphics[width=0.94\textwidth]{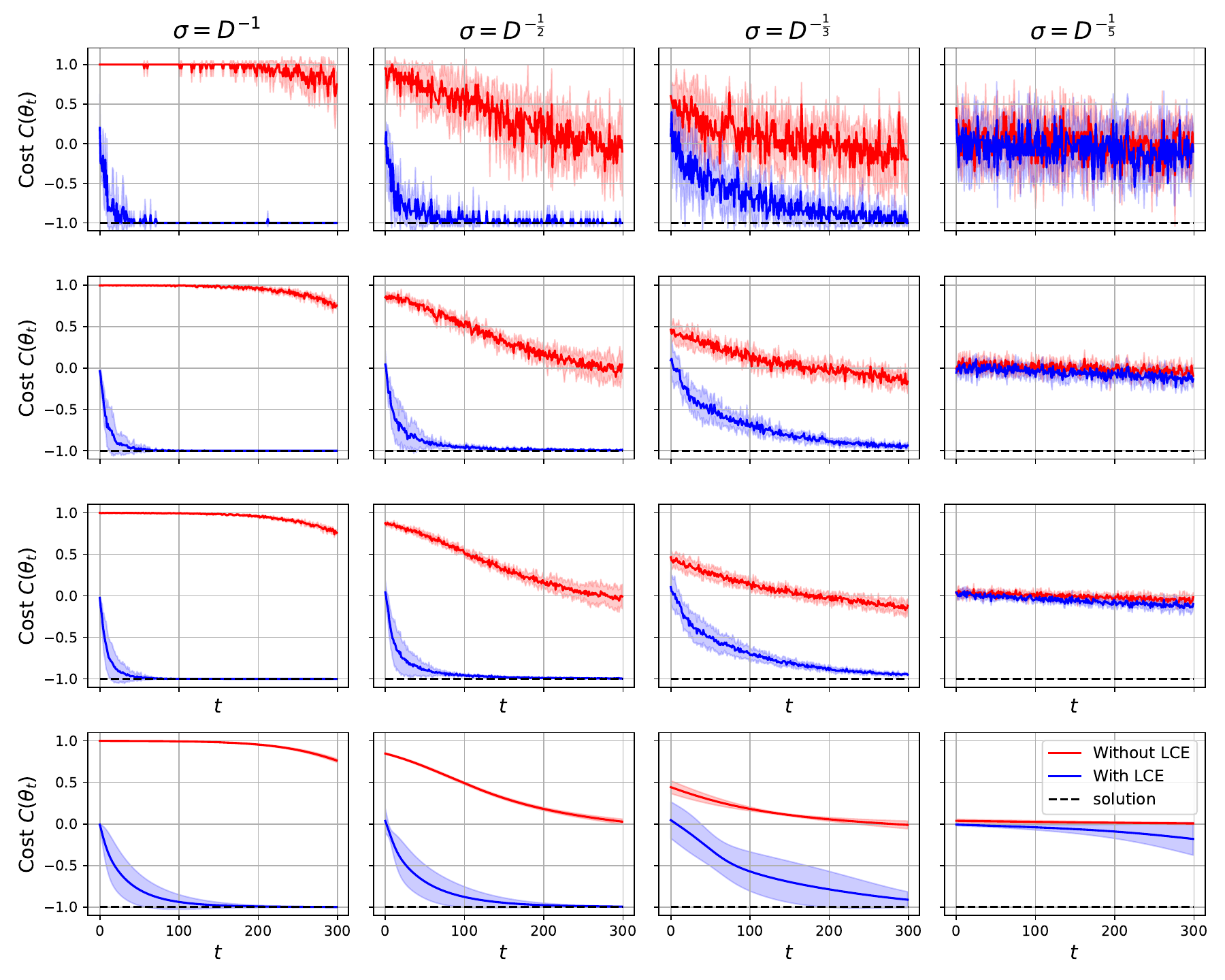}
    \caption{VQE Optimization using vanilla gradient descent with a learning rate of 0.01, estimating gradients with 8 shots (top row), 64 shots (second row), 512 shots (third row), and exact gradients (bottom row). We average over 5 runs of random Gaussian initialization across varying patch sizes. We find the ground energy of the global observable $Z^{\otimes N}$ with $N=15$, employing the \texttt{mHEA} ansatz with $L=5$ layers.}
    \label{fig:VQE_shots}
\end{figure*}

\subsection{VQE Optimization of the Heisenberg Model}
In this experiment, we leverage the benchmarked initial gradient scaling statistics to optimize the VQE of the $N$-qubit Heisenberg model~\cite{bonechi1992heisenberg}, 
\begin{equation}
    H_{\operatorname{Heisenberg}}:=\sum_{j=1}^{N-1}\left\{X_iX_{i+1}+Y_iY_{i+1}+Z_iZ_{i+1}\right\}.\label{eq:heisenberg}
\end{equation}

Eq.~\eqref{eq:heisenberg} defines a local observable and thus is less prone to the barren plateau problem~\cite{cerezo2021cost}. In order to increase the chance of inducing barren plateaus, one may consider deep circuit ansätze which result in expressivity-induced barren plateaus~\cite{holmes2022connecting}. Therefore, when employing shallow circuits to find the ground state of a local observable, the effectiveness of LCE is expected to disappear.

We compare the optimization dynamics of the 12-qubit Heiseneberg model in Eq.~\eqref{eq:heisenberg} with and without the LCE transformation. We employ \texttt{mHEA} with $L=36$ layers, and average over 5 independent runs of random Gaussian initialization of parameters across varying patch sizes. The results are presented in Fig.~\ref{fig:heisenberg-VQE}. We can see the advantage of LCE on sufficiently large patch sizes stemming from the improved initial gradient as per Fig.~\ref{fig:escaping-BP}, while the non-LCE model remains stuck on a local minimum. If the patch size is too large (e.g., $\sigma=D^{-1/5}$), then the advantage of LCE is less pronounced.

{Recalling the VQE experiment with $H_{\text{global}}=Z^{\otimes N}$ (cf. Fig.~\ref{fig:VQE-experiment} in Sec.~\ref{sec:experiments}), we note a significant discrepancy between the converge rates for $Z^{\otimes N}$ and the Heisenberg model. This is likely due to the locality of the Heisenberg observable, which is known to be barren plateaus free for shallow circuits~\cite{cerezo2021cost}. In fact, by reducing the number of layers from $L=36$ to $L=6$ as shown in Fig.~\ref{fig:heisenberg-VQE-shallow}, we can see that the non-LCE dynamics indeed tends to gets closer to the solution, while the LCE dynamics appears to get trapped in a similar energy level as in the $L=36$ experiment. These results indicate that the beneficial impact of LCE (being particularly effective for mitigating barren plateaus) is less pronounced compared to the deeper \texttt{mHEA} model with $L=36$ layers. This means that LCE should only be applied for VQEs with observables subject to barren plateaus, such as $H_{\text{global}}=Z^{\otimes N}$, ~\cite{cerezo2021cost,letcher2024tight}}. Interestingly, this experiment also suggests that the non-LCE initialization might, in some cases, begin closer to a global minimum, whereas the LCE transformation could potentially lead to entrapment in a local minimum (particularly in under-parameterized models). We hypothesize that such behavior is less prevalent in larger-scale models, and that alternative optimization strategies, such as the ADAM optimizer~\cite{kingma2014adam}, may offer a way to escape local minima.

\begin{figure*}[h]
    \centering
    \includegraphics[width=0.94\textwidth]{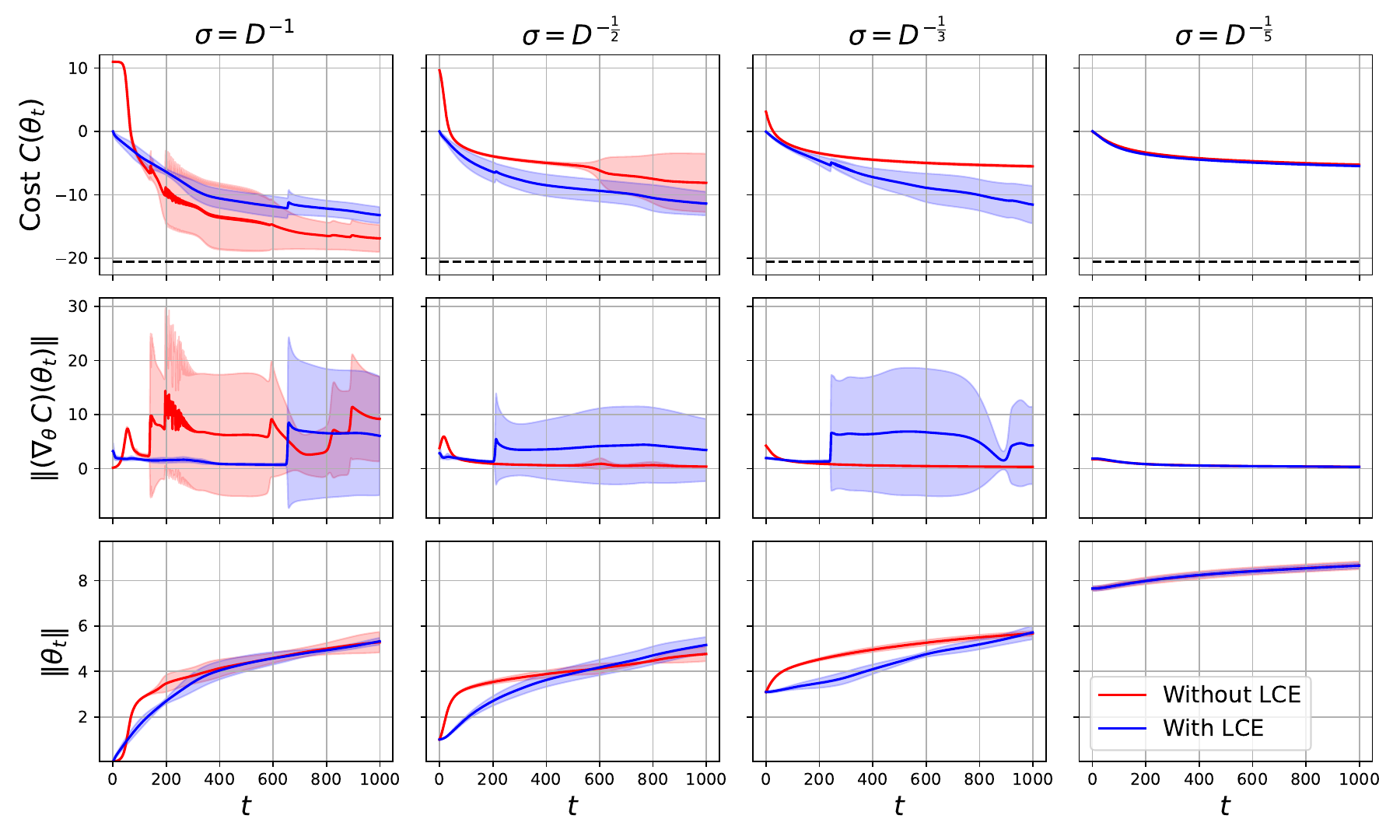}
    \caption{VQE Optimization using vanilla gradient descent with a learning rate of 0.01, averaging over 5 runs of random Gaussian initialization across varying patch sizes. We employ the \texttt{mHEA} ansatz with $L=36$ layers to find the ground energy of $12$-qubit Heisenberg model.}
    \label{fig:heisenberg-VQE}
\end{figure*}

\begin{figure*}[h]
    \centering
    \includegraphics[width=0.94\textwidth]{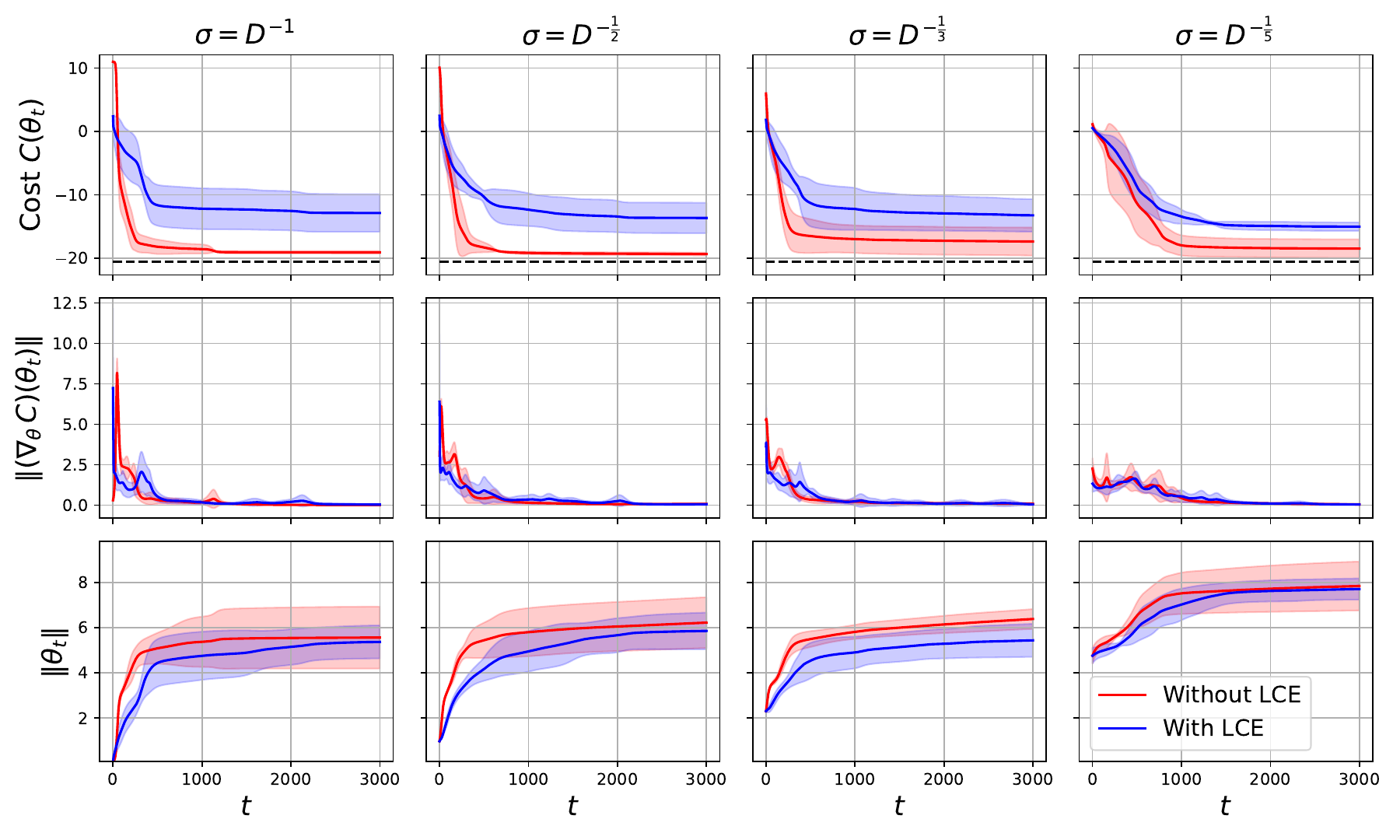}
    \caption{VQE Optimization using vanilla gradient descent with a learning rate of 0.01, averaging over 5 runs of random Gaussian initialization across varying patch sizes. We employ the \texttt{mHEA} ansatz with $L=6$ layers to find the ground energy of $12$-qubit Heisenberg model.}
    \label{fig:heisenberg-VQE-shallow}
\end{figure*}

\end{document}